\documentclass{article}
\usepackage{amstext}
\usepackage{amsmath}
\usepackage{amssymb}
\usepackage{graphicx}
\usepackage{rotating}
\usepackage{xcolor}
\usepackage{multirow}
\usepackage[pdftex,hidelinks]{hyperref}
\usepackage{cite}
\usepackage{etoolbox}
\usepackage{pdfpages}
\def\d{{\rm d}}
\def\pT{p_{\rm T}}

\makeatletter
\newcommand\AddNote[1]{\def\@addnote{#1}}
\providecommand\@addnote{}
\patchcmd{\thebibliography}
  {\list}
  {\@addnote\par\list}
  {}
  {}
\makeatother

\begin{document}
\begin{titlepage}
\pagenumbering{roman} 
\includepdf{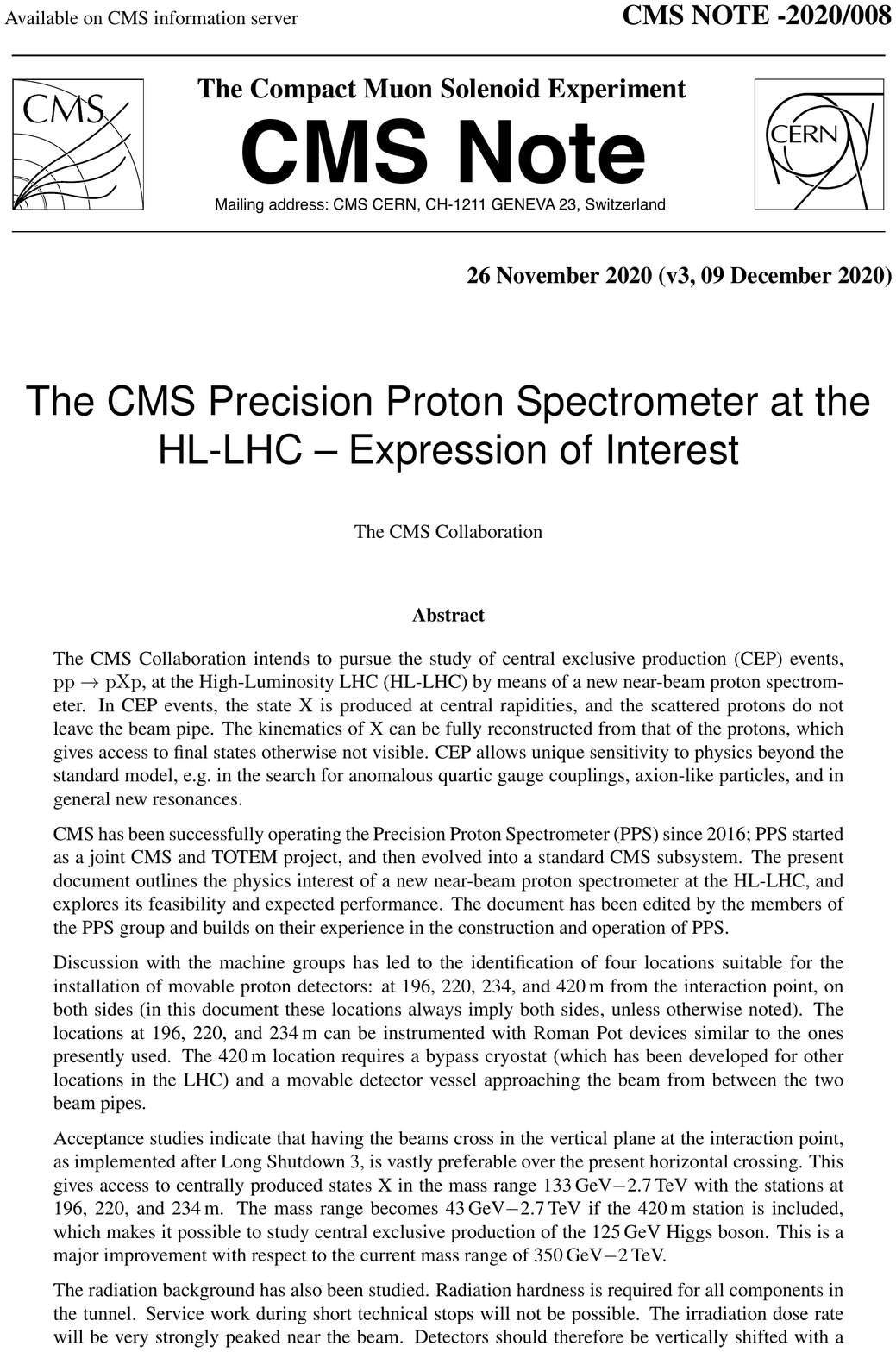}
\includepdf{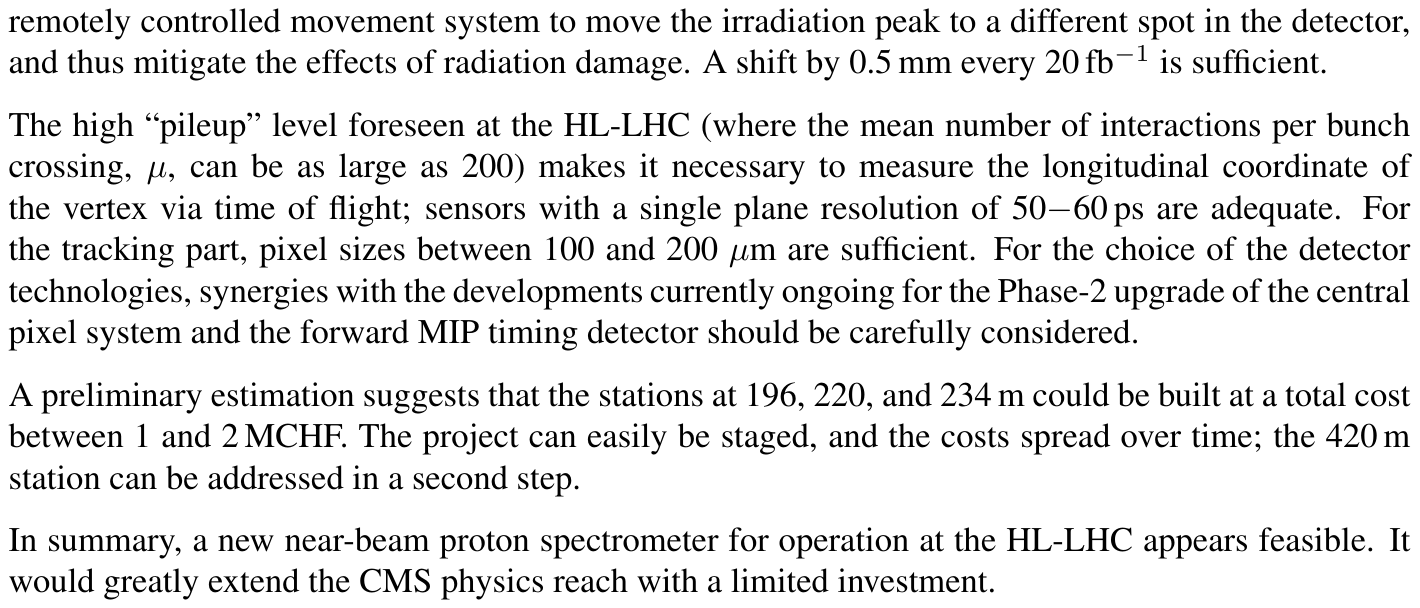}
\end{titlepage}
\pagenumbering{arabic} 
\setcounter{page}{1}
\title{{\bf The CMS Precision Proton Spectrometer at the HL-LHC -- Expression of Interest}\\
~~\\}
\date{9 December 2020}
\maketitle
\begin{abstract}
\normalsize
The CMS Collaboration intends to pursue the study of central exclusive production (CEP) events, $\rm pp \to pXp$, at the High-Luminosity LHC (HL-LHC) by means of a new near-beam proton spectrometer. In CEP events, the state X is produced at central rapidities, and the scattered protons do not leave the beam pipe. The kinematics of X can be fully reconstructed from that of the protons, which gives access to final states otherwise not visible. CEP allows unique sensitivity to physics beyond the standard model, e.g.\ in the search for anomalous quartic gauge couplings, axion-like particles, and in general new resonances.

CMS has been successfully operating the Precision Proton Spectrometer (PPS) since 2016; PPS started as a joint CMS and TOTEM project, and then evolved into a standard CMS subsystem. The present document outlines the physics interest of a new near-beam proton spectrometer at the HL-LHC, and explores its feasibility and expected performance. The document has been edited by the members of the PPS group and builds on their experience in the construction and operation of PPS.

Discussion with the machine groups has led to the identification of four locations suitable for the installation of movable proton detectors: at 196, 220, 234, and 420\,m from the interaction point, on both sides (in this document these locations always imply both sides, unless otherwise noted).
The locations at 196, 220, and 234\,m can be instrumented with Roman Pot devices similar to the ones presently used. The 420\,m location requires a bypass cryostat (which has been developed for other locations in the LHC) and a movable detector vessel approaching the beam from between the two beam pipes.

Acceptance studies indicate that having the beams cross in the vertical plane at the interaction point, as implemented after Long Shutdown 3, is vastly preferable over the present horizontal crossing. This gives access to centrally produced states X in the mass range 133\,GeV$-$2.7\,TeV with the stations at 196, 220, and 234\,m. The mass range becomes 43\,GeV$-$2.7\,TeV if the 420\,m station is included, which makes it possible to study central exclusive production of the 125\,GeV Higgs boson. This is a major improvement with respect to the current mass range of 350\,GeV$-$2\,TeV.

The radiation background has also been studied. Radiation hardness is required for all components in the tunnel. Service work during short technical stops will not be possible. The irradiation dose rate will be very strongly peaked near the beam. Detectors should therefore be vertically shifted with a remotely controlled movement system to move the irradiation peak to a different spot in the detector, and thus mitigate the effects of radiation damage. A shift by 0.5\,mm every 20\,fb$^{-1}$ is sufficient.

The high ``pileup'' level foreseen at the HL-LHC (where the mean number of interactions per bunch crossing, $\mu$, can be as large as 200) makes it necessary to measure the longitudinal coordinate of the vertex via time of flight; sensors with a single plane resolution of 50$-$60\,ps are adequate. For the tracking part, pixel sizes between 100 and 200\,$\mu$m are sufficient. For the choice of the detector technologies, synergies with the developments currently ongoing for the Phase-2 upgrade of the central pixel system and the forward MIP timing detector should be carefully considered.

A preliminary estimation suggests that the stations at 196, 220, and 234\,m could be built at a total cost between 1 and 2\,MCHF. The project can easily be staged, and the costs spread over time; the 420\,m station can be addressed in a second step.

In summary, a new near-beam proton spectrometer for operation at the HL-LHC appears feasible. It would greatly extend the CMS physics reach with a limited investment. 
\end{abstract}
\newpage
\tableofcontents
\newpage

\section{Introduction}
This Expression of Interest presents the new CMS Precision Proton Spectrometer for operation in Runs 4, 5, and 6 after the Long Shutdown 3 (LS3), i.e.\ in the High-Luminosity LHC (HL-LHC) phase. Based on the HL-LHC goal of producing a total integrated luminosity of 3000\,fb$^{-1}$~\cite{hllhc-tdr}, the expectation of the new PPS project is to collect about a factor 10 more data than the present PPS system will until LS3. \\

In the LHC Runs 1 and 2, the TOTEM experiment~\cite{totem-tdr,totem-jinst,totem-upgrade} and the CT-PPS project~\cite{ctpps}, later integrated in CMS as the PPS subdetector, have laid a solid foundation in the operation of Roman Pot (XRP) detectors very close to the LHC beams. Both beams coming from the Interaction Point 5 (IP5) were symmetrically equipped with leading-proton spectrometers. In Run 1, the phase of TOTEM standalone operation, the XRPs were only used in special runs at very low luminosities, whereas later, after LS1, uncharted territory was explored by inserting XRPs also in regular fills~\cite{ipac2016} up to the highest LHC luminosity of about $2 \times 10^{34}\,\rm cm^{-2} s^{-1}$. By the end of Run 2, PPS has collected an integrated luminosity of more than 100\,fb$^{-1}$.

The more than 15 years of design, construction and operation of movable near-beam devices and the analysis of their data have resulted in detailed experience with the challenges of performing precision tracking and time-of-flight measurements in the harsh environment a few millimetres from the beam (see Section~\ref{sec:propaganda}).

Exploitation of the present PPS system, with some upgrades and optimizations during LS2, will continue throughout Run 3 with the hope of collecting an additional 150$-$200\,fb$^{-1}$ of integrated luminosity. During LS3 the entire Long Straight Section around IP5 (LSS5) will be restructured, implying the removal of the present spectrometer. The project presented here aims at the construction of a new double-arm proton spectrometer, i.e.\ equipping again both outgoing beam lines with detector stations.\\

After a short outline of the experimental power that a proton spectrometer adds to CMS (essentially a $4 \pi$ coverage), the main physics objectives brought within reach by PPS with the data volume of HL-LHC are discussed in Section~\ref{sec:physics}. Then Section~\ref{sec:locations} introduces the new structure of LSS5 and the possibilities for adding movable near-beam detector devices in the outgoing beamlines. 
After identifying the locations best suited for detecting leading protons in view of maximising the kinematic acceptance, the first rough integration constraints, in particular the available longitudinal and transverse space, are presented. Section~\ref{sec:machine} is devoted to describing the boundary conditions imposed by the accelerator and beam parameters. Particular emphasis is put on the evolution of optics and crossing-angle during a fill and on the collimation scheme.
Based on these operational parameters and mechanical constraints, the expected physics performance is then evaluated (Section~\ref{sec:performance}). 

The orientation of the crossing-plane in IP5, horizontal or vertical, has a major impact on the performance and was thus of central importance in the development of the project. Already in December 2018, CMS therefore requested~\cite{hllhc-27coordinationgroup} to swap the crossing-planes in IP1 and IP5 relative to the present situation -- i.e.\ in IP5 the beams will cross vertically instead of horizontally. This change, also favoured by machine experts, has finally been adopted for implementation~\cite{hllhc-executivecommittee20200622}. Thus the main focus of this document naturally lies on the vertical crossing, but to underline its superiority, it was found instructive to compare some important quantities and performance numbers with the ones for the horizontal case. 

To assess the spectrometer performance in each of the favoured detector locations, the proton detection acceptance and resolution in terms of the key kinematic variables are derived as a function of the beam parameters and based on present-day assumptions on optics (Version 1.3~\cite{hllhc-optics1.3}), collimation scheme, near-beam-detector insertion rules from machine protection arguments, and detector resolution. Since the reconstruction of the proton kinematics requires an excellent knowledge of the machine optics and the detector alignment, calibration concepts are discussed.

The last part of this document deals with the technological implementation of the project. After an outline of the harsh radiation environment (Section~\ref{sec:radiation}), the tracking and timing detectors (Section~\ref{sec:detectors}), their possible trigger contribution (Section~\ref{sec:trigger}), and their movable vessels (Section~\ref{sec:vessels}) are discussed.
While specific technologies are still undecided at this stage, the basic requirements can already be defined. 
Building on the experience from PPS in Run 2 some options are discussed. \\

For the scope of this document, nominal performance numbers~\cite{hllhc-tdr} are assumed: in particular, a beam energy of 7\,TeV, a levelled peak luminosity of $5 \times 10^{34}\,\rm cm^{-2} s^{-1}$ and a mean pileup multiplicity up to 140. The ultimate performance numbers are somewhat higher (beam energy up to 7.5\,TeV, luminosity up to $7.5 \times 10^{34}\,\rm cm^{-2} s^{-1}$, mean pileup up to 200).

\newpage
\section{The PPS System before LS2 -- A Success Story}
\label{sec:propaganda}
The CMS Precision Proton Spectrometer (PPS) was born as a joint project 
of the CMS and TOTEM Collaborations, with the aim of studying Central 
Exclusive Production (CEP) in proton-proton collisions in standard LHC running conditions.\\

The idea of proton tagging at high luminosity goes back in time more 
than 15 years, with an initial CMS-TOTEM study on ``Prospects for Diffractive and Forward Physics at the LHC''~\cite{cms-totem-prospects}, 
followed in 2005$-$2008 by the FP420 project developed by a joint ATLAS 
and CMS group~\cite{fp420}. FP420 investigated 
the feasibility of discovering the Higgs Boson in CEP events with 
protons measured at about 420\,m from the interaction point. More studies on 
proton tagging in the TOTEM region were performed by CMS within the High 
Precision Spectrometer project (2009$-$2010), and finally again by CMS and 
TOTEM together; this led to the Technical Design Report for the 
CMS-TOTEM Precision Proton Spectrometer~\cite{ctpps}, which was approved by the LHCC in 2014.\\

PPS was meant to start taking data in 2017. However, the appearance of 
an enhancement in the photon-photon spectrum at around 750\,GeV in the 
2015 CMS~\cite{Khachatryan:2015qba,Khachatryan:2016hje,Khachatryan:2016yec} 
and ATLAS~\cite{Aad:2015mna,Aaboud:2016tru,Aaboud:2017yyg} data led to an acceleration 
of the program, with the first data 
taken already in 2016. If a resonance decaying to two photons had really 
been there, it would have been possible to produce it exclusively and it 
would have been visible in the invariant mass spectrum of the scattered 
protons measured with PPS.\\

Each arm of the spectrometer consists of a silicon tracker to measure 
the scattered proton trajectory, along with a timing system to determine 
its time of flight. The difference between the times of flight in the two 
arms can be translated into the vertex position along the beam line; 
comparing this with the position determined from the central part of the 
event allows the suppression of pileup background. The tracking system was 
initially based on the TOTEM silicon strip detectors~\cite{totem-jinst} and 
evolved to one consisting of 3D pixel detectors only. The timing used ultra-fast silicon and diamond detectors in 2017, and double diamond sensors in 
2018. An overview of the PPS station configuration until LS3 is given in Appendix~\ref{sec:pre-ls3-config}.\\

Timing is a key ingredient of PPS; resolutions of less than 100\,ps have 
been reached, with the potential of further improvements. The timing 
system has shown to be stable at the level of better than 10\,ps -- a 
remarkable achievement also in view of the approximately 500\,m distance between 
the two measuring stations.\\

PPS has been an eminent success. It has proven that operating a 
near-beam proton spectrometer at a high-luminosity collider is feasible 
-- something that only a few years ago was considered impossible by 
many. PPS has had no impact on the operation of the LHC in terms of 
background, heating, or impedance. Since 2016, it accumulated data in 
excess of $\rm 100\,fb^{-1}$ as a standard CMS subsystem.\\

The very first $\rm 10\,fb^{-1}$ of data collected with PPS in 2016 led to a 
paper on central (semi)-exclusive dilepton production, where the two leptons 
are detected in the central CMS apparatus, and at least one scattered 
proton is measured in PPS~\cite{jhep2018}. These 
results are important in themselves as they are the first on high-mass 
dilepton production with a proton tag. They are perhaps even more 
important because they prove that PPS works as designed. Several more 
analyses are currently in progress: on central exclusive production of 
photon pairs~\cite{DiphotonPAS}, WW pairs and $\rm t \bar{t}$ pairs; a search for new physics in the 
missing mass spectrum in $\rm p p \rightarrow p \oplus Z/\gamma + X \oplus p$ events is also ongoing.\\

A final remark is in order. The success of PPS has been made possible by 
a somewhat unusual event in high-energy physics: two independent 
collaborations, CMS and TOTEM, have joined forces to pursue a common 
physics interest. CMS and TOTEM have existed for decades as separate 
experiments, with different goals, organizational structures, and traditions. It was 
not a priori obvious that this joint venture would succeed. It however 
did, and in 2018 a Memorandum of Understanding between CERN, CMS, and 
TOTEM was signed, stipulating that all TOTEM members could become members 
of CMS if they so desired. Since then, the CMS-TOTEM Precision Proton 
Spectrometer has become the Precision Proton Spectrometer, a CMS 
subsystem on a par with all others.

\newpage
\section{Physics Perspectives} 
\label{sec:physics}
\begin{figure}[h!]
\centering
\includegraphics[width=0.4\textwidth]{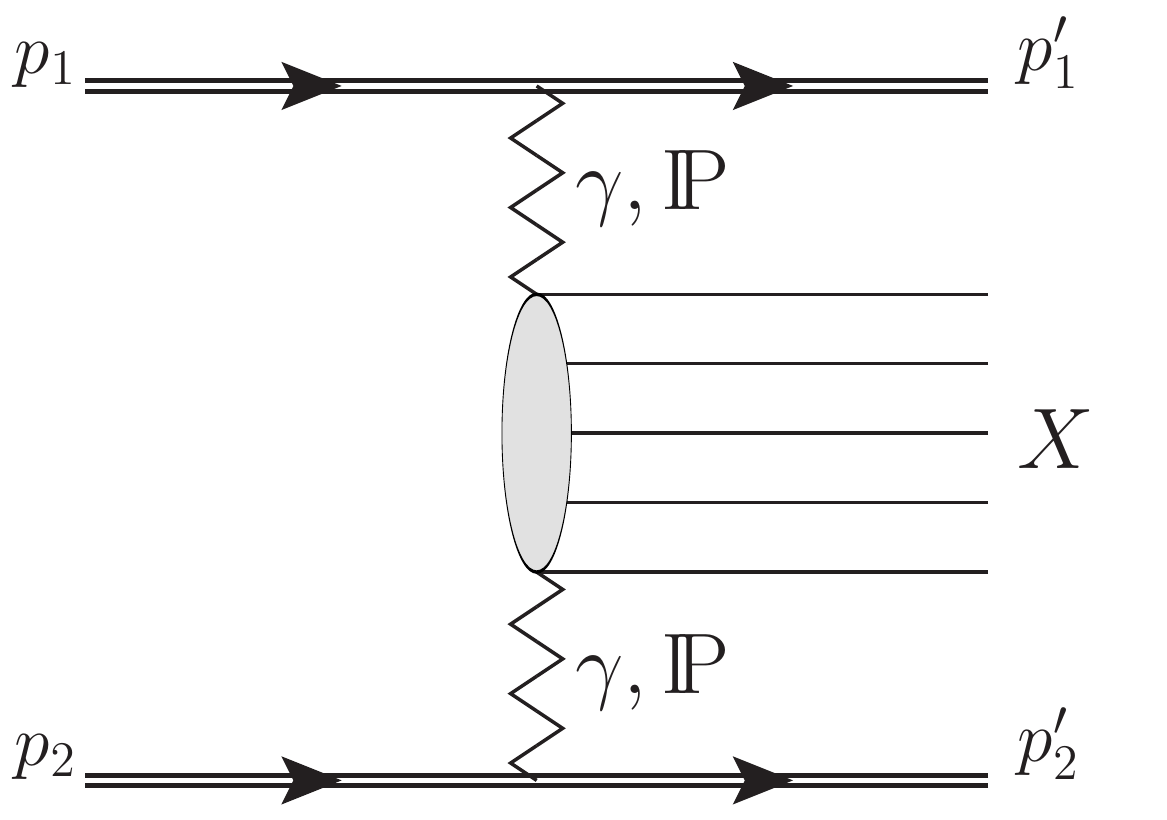}
\caption{Schematic diagram for central exclusive production, pp $\rightarrow$ pXp.}
\label{fig:DiagramsProduction}
\end{figure}

In most of the pp collisions at the LHC, the scattered protons dissociate after exchanging gluons or quarks. However, in the case of colourless exchanges, photon ($\gamma$) exchange for electromagnetic interactions or pomeron ($\rm{I\!P}$) exchange for strong interactions, interacting protons could emerge intact. The intact protons, having lost a fraction $\xi=\Delta p/p$ of their longitudinal momentum, are deflected from the proton bunch by the LHC magnets and measured in the PPS detector with unprecedented resolution. 
In hard scattering events (involving production of high $\pT$ particles), particles produced at the interaction point can be measured and reconstructed by the CMS detector. 
Combining the information from PPS with that from the central CMS detector allows the study of hard interactions in CEP processes. A schematic diagram for a CEP process is depicted in Fig.~\ref{fig:DiagramsProduction}.

For Runs 2 and 3 of the LHC, the range of invariant mass of the system ``X'' extends from approximately 350\,GeV to 2\,TeV, when both protons can be detected (``double-arm'' measurement), and above 50\,GeV when only one proton is detected by PPS (``single-arm'' measurement). In the former case, full event reconstruction is made possible by matching the proton kinematics to the kinematics of ``X'' measured by the central detector. As the exclusive standard model processes under investigation have small cross sections at high masses, $m_{\rm X}$, (more details in Section~\ref{sec:standardmodel}), most measurements will be limited by statistical uncertainties with the full Run 2+3 data set. In addition, searches for new phenomena will benefit from the higher integrated luminosities of the HL-LHC. In channels where the kinematic matching is not sufficient to adequately suppress backgrounds from uncorrelated pileup 
protons, precision timing detectors can be used to match the longitudinal vertex position of the system ``X'' with the protons, as described in Section~\ref{sec:timing}. \\

As discussed in the following chapters, the combination of all four locations under study for HL-LHC detectors (at 196, 220, 234, and 420\,m on each side of IP5, see Appendix~\ref{sec:hllhc-config} for a configuration overview) would cover an extended mass range from approximately 50\,GeV to 2.7\,TeV when both protons can be detected. In addition to the increased integrated luminosity, this will allow an expanded set of physics topics to be studied, in comparison to Runs 2 and 3. The increased upper mass range would increase the acceptance for both direct and indirect searches for beyond standard model (BSM) physics. The reduced lower mass limit, associated with a station at 420\,m from IP5, would significantly enhance the acceptance for all SM processes and, in particular, Higgs production, as well as for the production of feebly coupled BSM resonances (such as e.g. light axion-like particles). A few examples are illustrated in this section, using generator-level Monte Carlo simulations with the PPS acceptance calculated for the HL-LHC beam and optics parameters (Version 1.3~\cite{hllhc-optics1.3}) averaged over the fill (i.e.\ the middle of the ``luminosity levelling trajectory''; see Section~\ref{sec:optics} for the running scenario). All studies shown in this section have been performed for the vertical beam crossing at IP5, in which the beams cross in the $y-z$ plane rather than the $x-z$ plane, as now officially 
decided~\cite{hllhc-executivecommittee20200622}.


\subsection{Search for New Physics in the High Mass Region}

\subsubsection{Direct BSM Searches at High Masses}

Currently, several analyses are exploiting the PPS data collected during Run 2 to perform direct searches for new particles up to the upper mass acceptance limit of $\sim 2$\,TeV.
The HL-LHC acceptance of a 196\,m station would allow this search region to be extended to approximately 2.7\,TeV. A wide variety of BSM scenarios involving $\gamma\gamma$ production 
with forward protons have been explored in the theoretical literature (a first systematic study of photon-induced collisions at the LHC was discussed in~\cite{deFavereaudeJeneret:2009db}). Those in which new particles are light enough to be directly produced within the PPS acceptance typically involve either resonance production or pair production of new charged particles. Assuming no new physics is discovered during Runs 2 and 3 of the LHC, the increase in available luminosity with the PPS upgrade would improve the sensitivity to smaller couplings, 
while the configuration of the stations would expand the mass range that can be probed with forward protons.

\begin{figure}[h!]
\centering
\includegraphics[width=10cm]{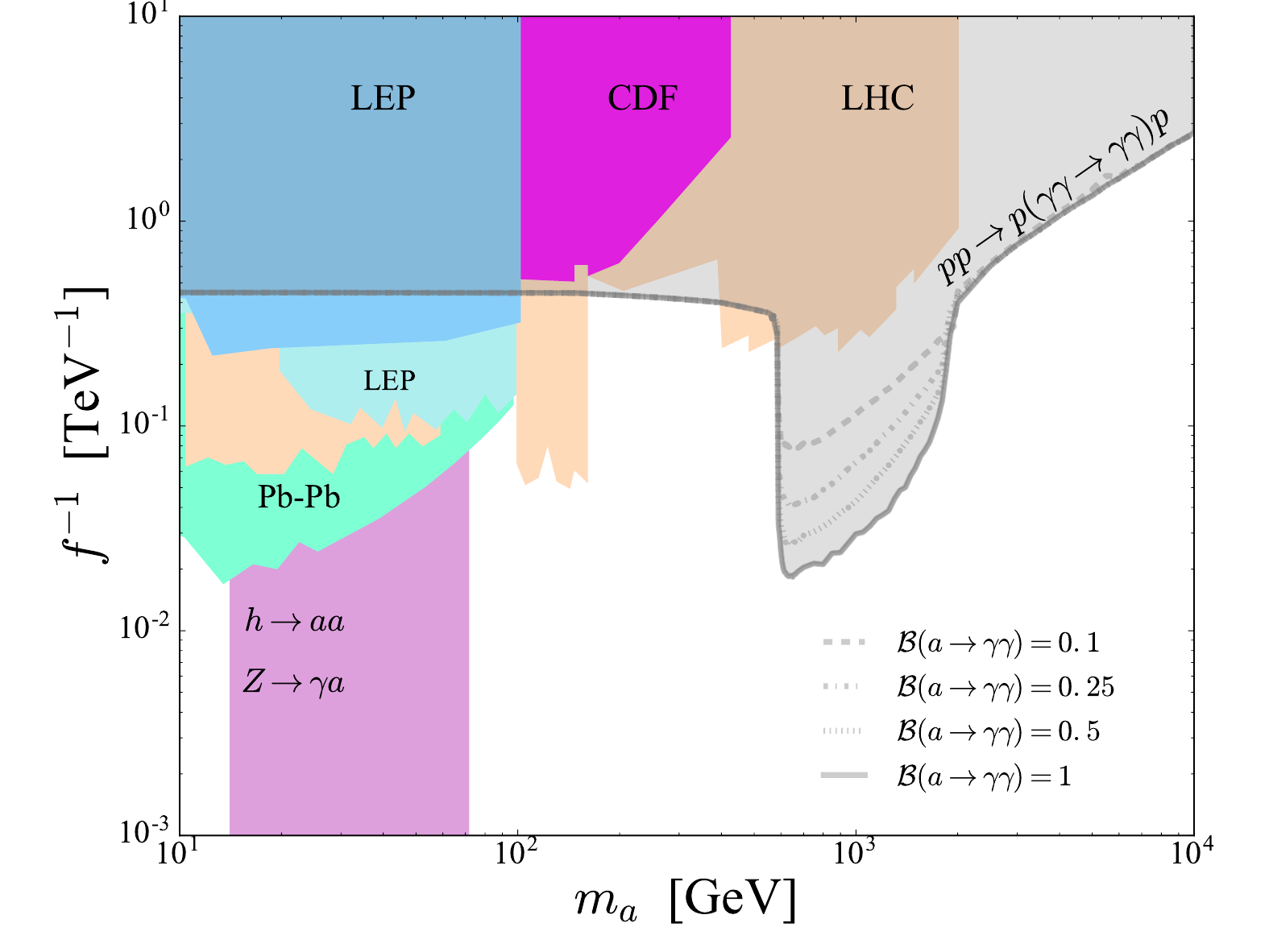}
\caption{Exclusion regions on the ALP--photon coupling ($f^{-1}$) and mass plane. The light-shaded grey regions show the expected 95\% confidence level (CL) exclusion limit for 
300\,fb$^{-1}$ in central exclusive diphoton production events for different branching fractions of the ALP into two photons~\cite{Baldenegro:2018hng}. 
The plot extends both above and below the maximum range of the PPS acceptance; the large improvement in the limits between 0.6-2.0~TeV reflects the range 
where proton detection was assumed for the selections applied in the study.}
\label{fig:highMassYY}
\end{figure}

For exclusive production with intact protons, either spin-0 or spin-2 resonances can be produced in $\gamma\gamma$ interactions~\cite{Landau:1948kw,Yang:1950rg,Olsson:1983mf}. 
This type of search is particularly 
interesting for resonances with large couplings to photons but not to gluons, which may appear in the $\rm\gamma\gamma \rightarrow X \rightarrow \gamma\gamma$ 
channel~\cite{denterria:2013,Csaki:2015vek,Harland-Lang:2016qjy,Lebiedowicz:2016lmn,Abel:2016pyc,Fichet:2015vvy,Baldenegro:2018hng}. As an example, the expected sensitivity for 
axion-like particles (ALP) is shown in Fig.~\ref{fig:highMassYY}, in the plane of the ALP mass and ALP-photon coupling $f^{-1}$, for an assumed luminosity of 300\,fb$^{-1}$.

Conversely, if a resonance is detected via decays to 
two photons, measuring the cross section with forward protons will help constrain its couplings to photons in a model-independent way~\cite{Fichet:2016pvq}.
While many of the theoretical studies of this scenario were performed in the context of the 750\,GeV diphoton enhancement, which was not confirmed, the same concepts apply more generally. 

Pair production of supersymmetric sleptons or charginos was one of the first searches considered in this production channel~\cite{Ohnemus:1993qw,Drees:1994zx,Bhattacharya:1995id}.
While the LHC has placed severe constraints on SUSY models already, the use of forward protons has recently been revisited as a possible means to improve searches
in difficult scenarios with highly compressed spectra~\cite{Beresford:2018pbt,Harland-Lang:2018hmi}, or long-lived particles~\cite{Godunov:2019jib,Schul:2008sr}.
Further studies would be needed to compare the sensitivity of these searches to other approaches. 

In other models, $\gamma\gamma$ production is expected to be particularly effective for probing pair production of particles with exotic electric or
magnetic charges. In the case of doubly charged particles~\cite{Han:2007bk,Babu:2016rcr,You:2014npa} the $\gamma\gamma$ channel is predicted to contribute a significant fraction of the total production rate of such particles, while in the case of magnetic monopoles~\cite{Baines:2018ltl,Dougall:2007tt,Kurochkin:2006jr} it is predicted to be the dominant production mechanism at the HL-LHC.

\begin{figure}[h!]
\centering
\includegraphics[width=0.4\textwidth]{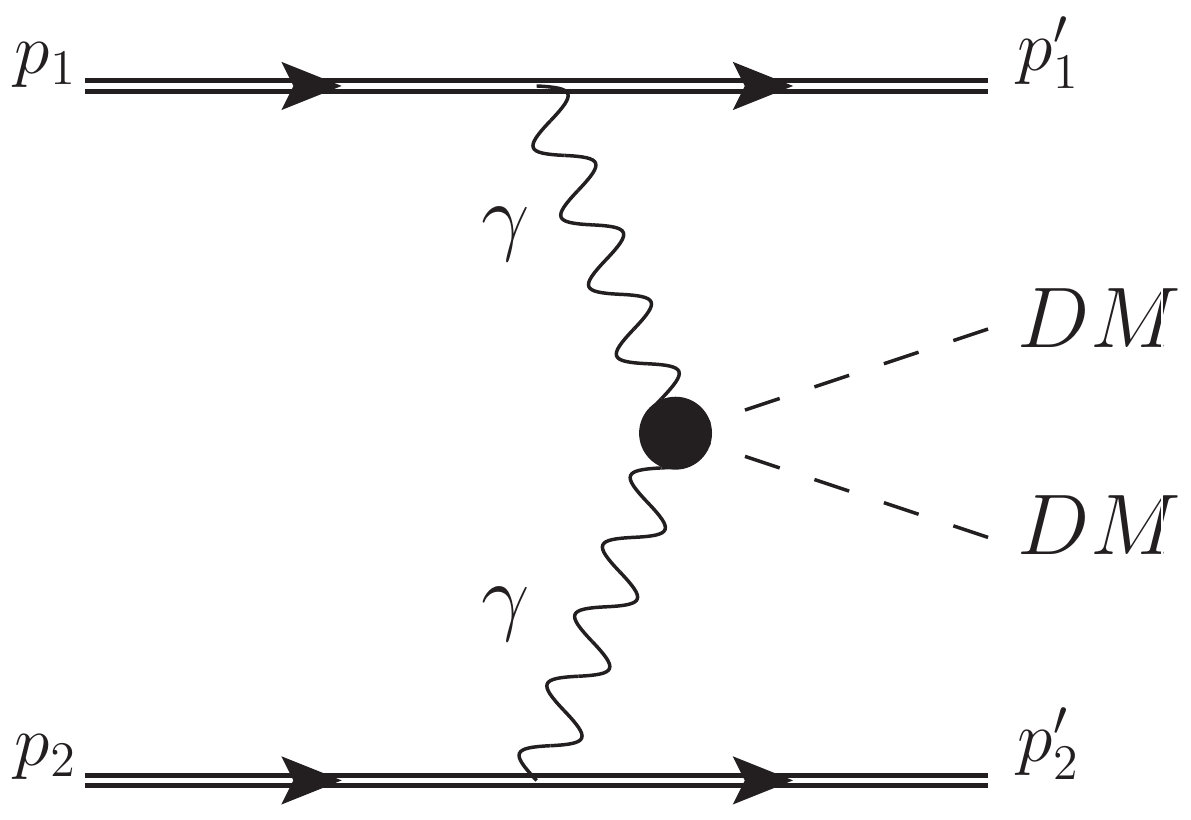}
\includegraphics[width=0.4\textwidth]{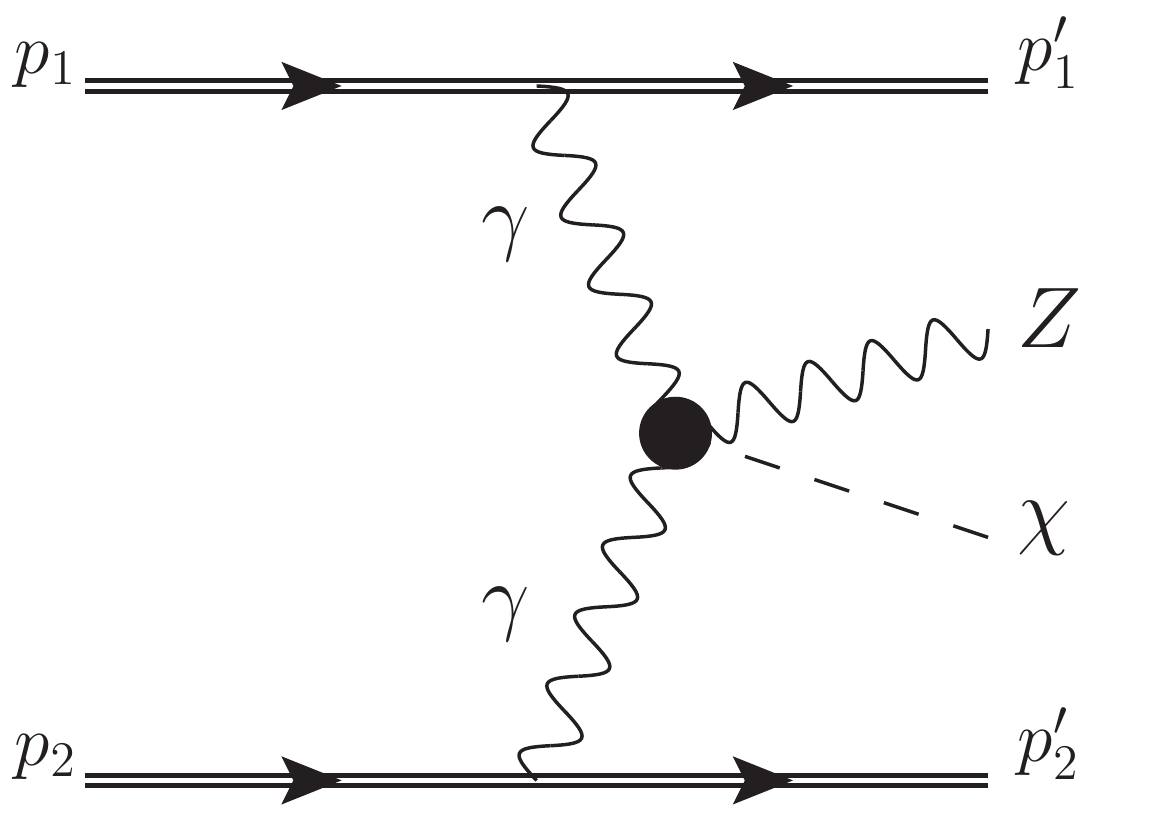}
\caption{Schematic diagrams for the central exclusive production of an invisible system without (left) and with (right) associated SM particles.}
\label{fig:DiagramsCEP_DM}
\end{figure}

Finally, the use of forward protons has been proposed to perform general searches for invisible particles via the ``missing mass'' distribution recoiling against the protons~\cite{Khoze:2017igg, Belotsky:2004ex}. Schematic diagrams for the production of new particles that do not leave any signature in the central detector are shown in Fig.~\ref{fig:DiagramsCEP_DM}.

\subsubsection{Indirect BSM Searches and Electroweak Physics at High Masses}
One of the main activities of the Run 2/3 PPS program is the study of diboson production, to search for deviations from the standard model in $\gamma\gamma$ interactions in the 
mass range of 350\,GeV$-$2\,TeV. 
Typically, deviations are quantified in terms of Anomalous Gauge Couplings or Effective Field Theories, which are intended to have minimal assumptions on the exact details of 
the new physics~\cite{Pierzchala:2008xc,Chapon:2009hh,Maniatis:2008zz,Gupta:2011be,Baldenegro:2017aen,Fichet:2014uka}. A number of specific new physics models which can provide nonresonant enhancements at high masses have also been studied in this context ~\cite{Fichet:2013ola,Fichet:2014uka,Espriu:2014jya,Delgado:2014jda}.

The regions most sensitive to new effects are generally at high mass. The HL-LHC acceptance of a 196\,m station would allow this 
to be extended to approximately 2.7\,TeV, opening a new region of phase space above the Run 2/3 acceptance limit of $\sim 2$\,TeV. 

\begin{figure}[h!]
\centering
\includegraphics[width=0.4\textwidth]{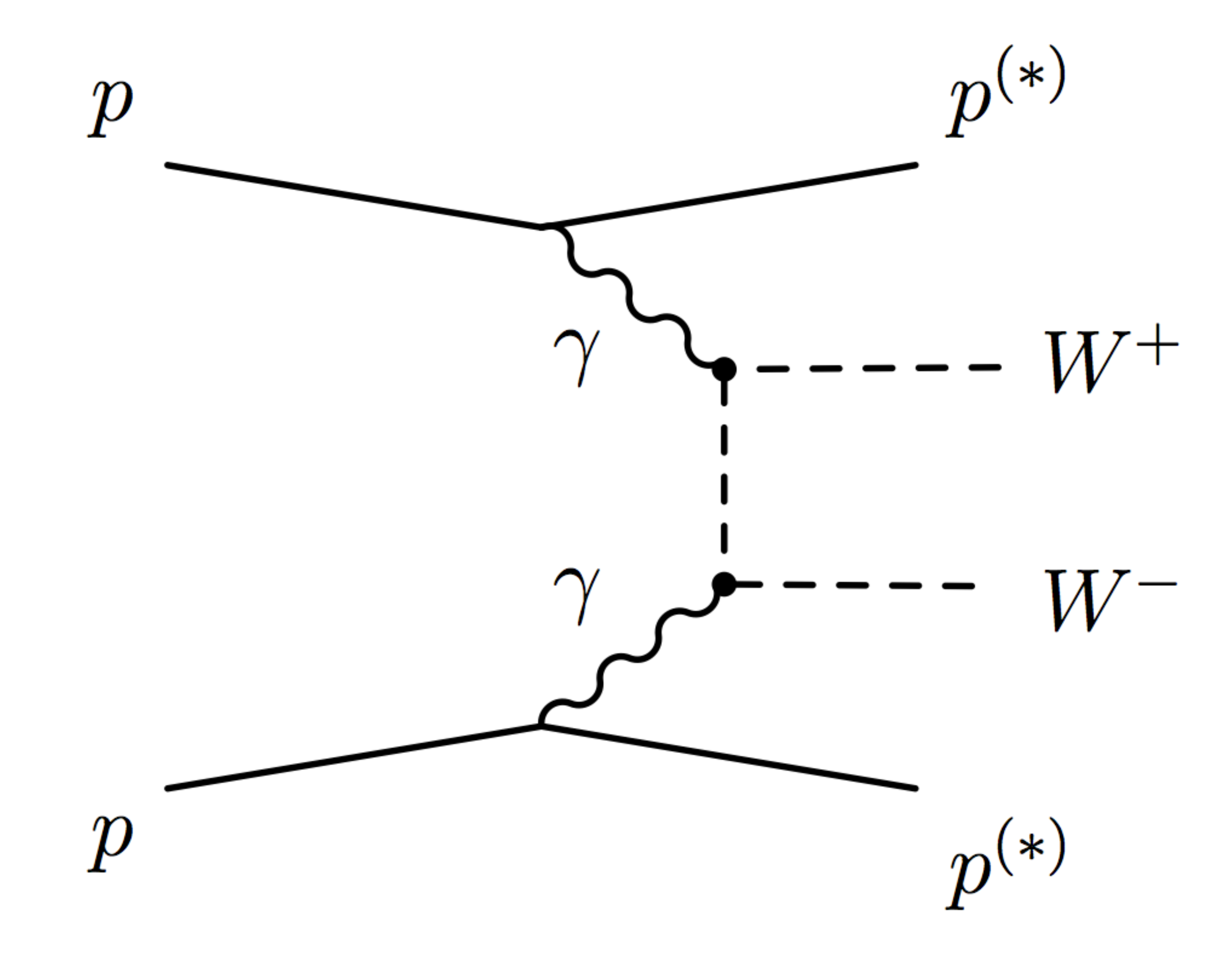}
\includegraphics[width=0.4\textwidth]{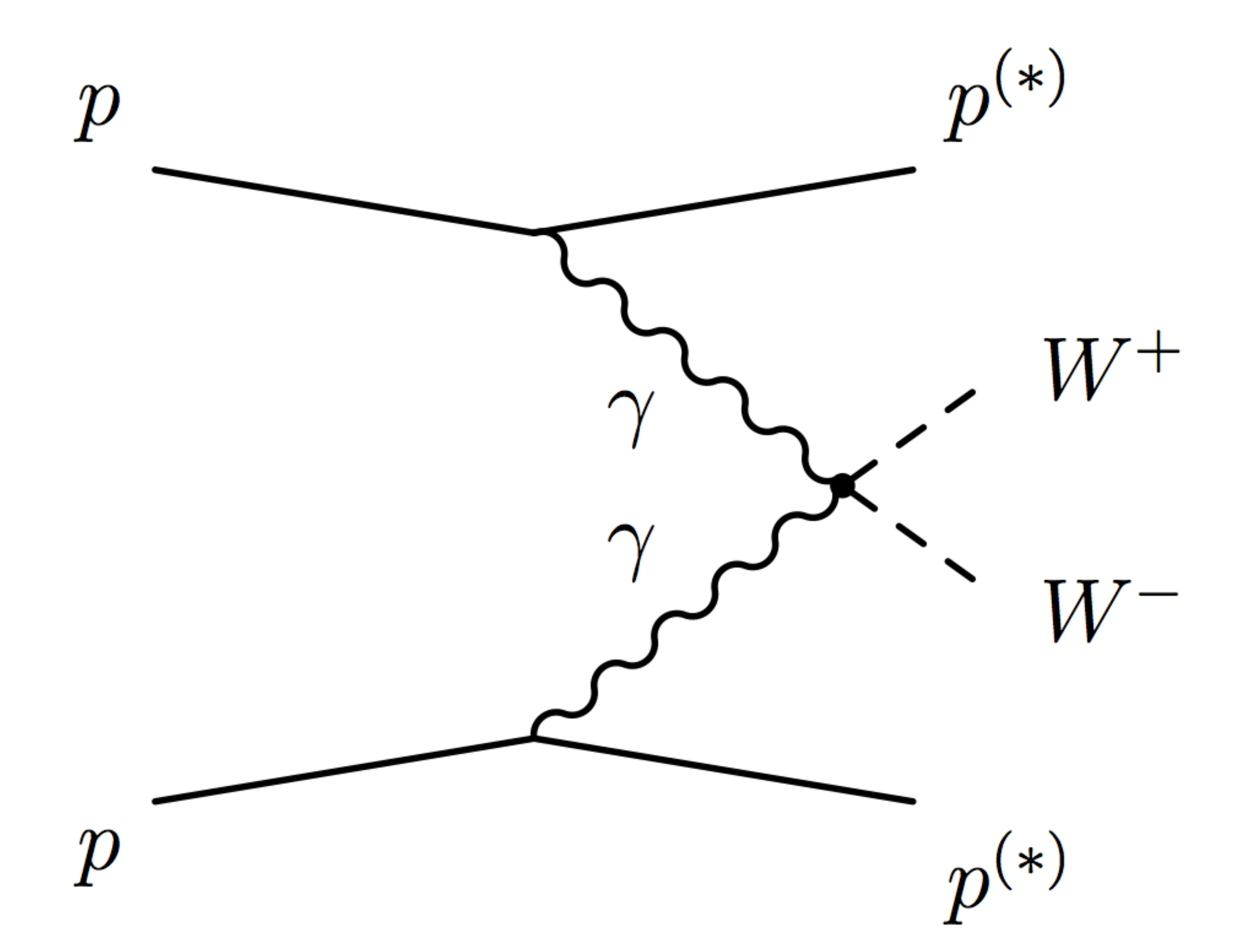}
\caption{Standard model diagrams contributing to $\rm pp \rightarrow p W^{+}W^{-} p$ production with intact protons. The t-channel diagram (left: with two triple gauge couplings) and the quartic diagram (right) are shown.}
\label{fig:smww-diagrams}
\end{figure}

\paragraph{Quartic Gauge Couplings with W Bosons\\}
~~\\
\noindent The $\rm\gamma\gamma\rightarrow W^{+}W^{-}$ process is allowed in the standard model, via diagrams with triple or quartic couplings of gauge bosons (Fig.~\ref{fig:smww-diagrams}).
This provides a good means to test the interactions of photons and W bosons at high energies, and to search for Anomalous Quartic Gauge Couplings (AQGC) or other nonresonant signals of new physics. 

During Run 1 of the LHC, this process was investigated by CMS and ATLAS without the ability to detect the scattered protons, by searching for an electron-muon vertex with 
no additional track activity~\cite{Chatrchyan:2013akv,Khachatryan:2016mud,Aaboud:2016dkv}. Due to the large backgrounds in other W decay modes, only 
the $\rm\mu^{+}e^{-}\nu_{\mu}\bar{\nu_{e}}$ final state was studied during Run 1, representing $\approx$\,2\% of the total $\rm W^{+}W^{-}$ branching fraction. In this case, large 
correction factors and systematic uncertainties were required to account for modelling of the central track multiplicity, and for the fraction of events in which one or both of the protons dissociates to an undetected system. In addition, without the constraints on the photon-photon collision energy provided by the protons, these analyses required the introduction of form factors to avoid unitarity-violating effects at very large masses, complicating the theory interpretation.


\begin{figure}[h!]
\centering
\includegraphics[width=0.75\textwidth]{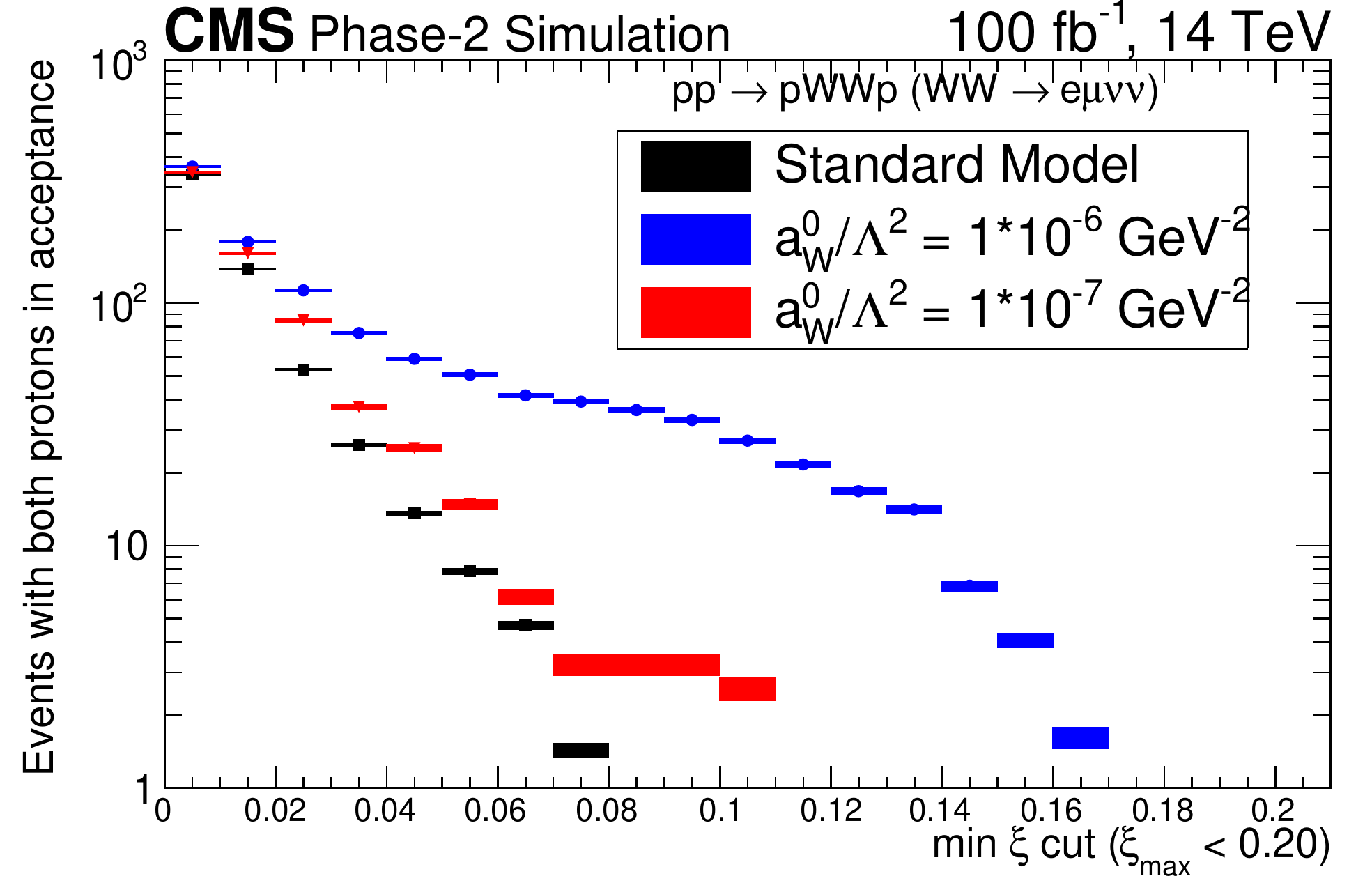}
\caption{Generator-level yields with both protons above the minimal $\xi$-acceptance value on the abscissa, and below $\xi=0.20$. 
The number of events expected per 100\,fb$^{-1}$ in the process $\rm\gamma\gamma\rightarrow WW \rightarrow \mu^{+}e^{-}\nu_{\mu}\bar{\nu_{e}}$ at $\sqrt{s} = 14\,$TeV is shown, requiring $\pT(\mu, e)>20\,$GeV, $|\eta(\mu, e)|<2.4$. 
The black histogram indicates the standard model expectation, while the blue and red histograms show the expectation for Anomalous Quartic Gauge 
Couplings with $a^{0}_{\rm W} / \Lambda^{2} = 1 \times 10^{-6}$\,GeV$^{-2}$ and $1 \times 10^{-7}$\,GeV$^{-2}$, respectively.}
\label{fig:smaqgc-crosssections}
\end{figure}

The standard model cross section is concentrated at low masses and is steeply decreasing with the diboson mass ($m_{\rm WW}$), while in the presence of anomalous couplings the cross section grows in the high-mass tails of the distribution. Because of this, the total number of $\rm \gamma\gamma\rightarrow W^{+}W^{-}$ events detected does not directly translate into the sensitivity 
to AQGC's. Instead, a measurement in the low-mass region is primarily useful to verify the standard model predictions for this process, while the AQGC sensitivity comes 
from the high-mass/high-$\xi$ (fractional momentum loss of the proton) tails, where the number of events is expected to be small. The difference in the expected yields vs.\ the minimum $\xi$ is shown in Fig.~\ref{fig:smaqgc-crosssections}, for the case of the SM only and for two examples of the SM+AQGC. As expected, 
the effect of the AQGC is apparent at large $\xi$ values, while the low-$\xi$ region is dominated by the SM contribution.

Full simulation studies of the $\rm \gamma\gamma\rightarrow W^{+}W^{-}$ process (with $\rm W^{+}W^{-} \rightarrow \mu^{+}e^{-}\nu_{\mu}\bar{\nu_{e}}$) were performed for the original 
PPS Technical Design Report~\cite{ctpps}. These indicated that, while only a few standard model events are expected within the Run 2/3 PPS acceptance in this channel, 
the Run 1 limits on unitarized Anomalous Quartic Gauge Couplings could be exceeded by almost 2 orders of magnitude. 

These early studies mainly focused on the $\mu^{+}e^{-}\nu_{\mu}\bar{\nu_{e}}$ final state, as it has favourable trigger thresholds and backgrounds, allowing easier access to the low-mass region of interest for the SM. In the high-mass search for AQGC signals, thanks to the background rejection and kinematic constraints provided by the protons,
channels with one or both W bosons decaying to jets can also be studied. This allows $>70\%$ of the $\rm W^{+}W^{-}$ decays to be used in the analysis, compared to the $\approx$\,2\% in the
$\mu^{+}e^{-}\nu_{\mu}\bar{\nu_{e}}$ channel; preliminary analyses of these channels are already in progress using Run-2 data from PPS. However, assuming no new physics, these 
analyses are expected to be limited by statistics at the end of Run 3, due to both the integrated luminosity and the more limited $\xi$ coverage of the PPS configuration in Runs 2 and 3.

\paragraph{All Neutral Anomalous Quartic Gauge Couplings\\}
~~\\
\noindent In contrast to the $\rm\gamma\gamma \rightarrow W^{+}W^{-}$ channel, triple or quartic couplings of neutral gauge bosons are forbidden in the standard model. 
Therefore processes such as $\gamma\gamma \rightarrow \gamma\gamma$, $\rm\gamma\gamma \rightarrow ZZ$, and $\rm\gamma\gamma \rightarrow Z\gamma$ can occur only 
at highly suppressed rates through higher-order diagrams in the SM, and their cross sections are difficult to measure in pp collisions. However, in scenarios 
with anomalous couplings, they can receive detectable enhancements at large masses~\cite{Gupta:2011be,Fichet:2014uka,Baldenegro:2017aen,Fichet:2013ola}. Thus any 
signal observed in these channels would be a clear indication of physics beyond the SM. 

Anomalous couplings in these processes are less explored. Only recently the first direct limits on the dimension-8 couplings of four photons, $|\zeta_{1}|<3.7 \times 10^{-13}$\,GeV$^{-4}$ and $|\zeta_{2}|<7.7 \times 10^{-13}$\,GeV$^{-4}$, were obtained~\cite{DiphotonPAS}. In Ref.~\cite{Fichet:2014uka}, simulations showed that sensitivities of the order of $10^{-14}$\,GeV$^{-4}$ could be reached in the $\gamma\gamma \rightarrow \gamma\gamma$ channel with forward proton detectors, in a scenario with 3000\,fb$^{-1}$ and pileup of 200 events per crossing. In that study a 
more limited mass acceptance was assumed for the protons, between 350 and 1700~GeV. The increased high mass limits offered by the 196\,m station in the current proposal would 
further improve the acceptance for such searches. Searches for $\gamma\gamma \rightarrow \gamma\gamma$ production using the Run 2 PPS data are in progress~\cite{DiphotonPAS}. 

\begin{figure}[h!]
\centering
\includegraphics[width=\textwidth]{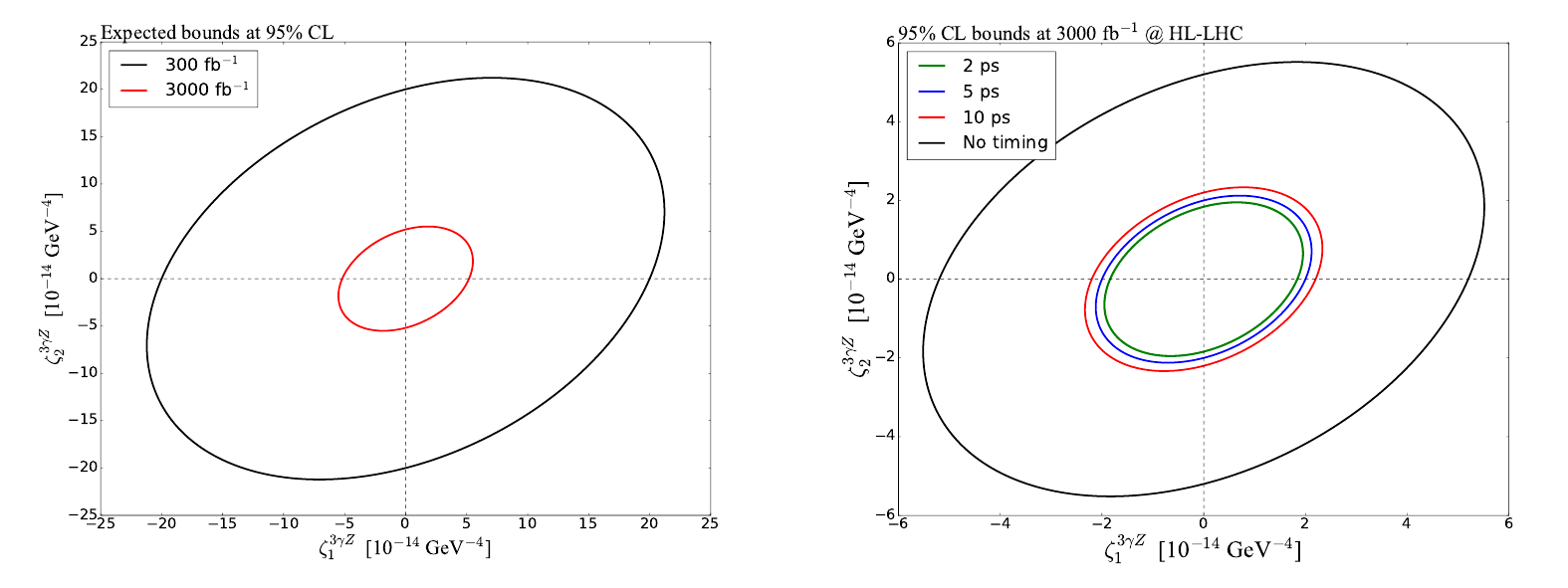}
\caption{Expected bounds on the anomalous $\rm \gamma \gamma \gamma Z$ couplings at 95\% CL with 300\,fb$^{-1}$ and 3000\,fb$^{-1}$ at the HL-LHC without time-of-flight measurement (left). Expected bounds at 95\% CL for timing resolutions of $\delta t = 2, 5, 10$\,ps at the HL-LHC (right). Figure from Ref.~\cite{hllhc-sm-yellowreport}.}
\label{fig:ExpectedZgamma}
\end{figure}

Constraints on the $\rm\gamma\gamma\gamma Z$ coupling (one-loop induced processes in the SM) can be realized via the $\rm\gamma\gamma\to\gamma Z$ scattering or directly through the SM $\rm Z\to\gamma\gamma\gamma$ decay. Since the exclusive channel is very clean ($\rm Z+\ell\bar{\ell}$), a stringent limit on the dim-8 effective operators $\zeta^{\rm 3\gamma Z}_{1}$ and $\zeta^{\rm 3\gamma Z}_{2}$ can be set. For the HL-LHC with 3000\,fb$^{-1}$, the sensitivity to the anomalous coupling at 95\% CL reaches the level of $10^{-14}$\,GeV$^{-4}$~\cite{hllhc-sm-yellowreport}. An example of expected bounds for different integrated luminosities, with and without timing detectors, is shown in Fig.~\ref{fig:ExpectedZgamma}. The contour lines in the right-hand panel correspond to time resolutions (2$-$10\,ps) that are not yet achievable with present-day detector systems. While no simulation for a more realistic performance ($\sim 20$\,ps) is available yet, a visual extrapolation can be made based on the distances between the ellipses shown: the bounds for 20\,ps will not be drastically worse than for 10\,ps.

\paragraph{Anomalous Effects in the $\tau$-Lepton Sector\\}
~\\
\noindent In addition to anomalous production of gauge bosons, indirect searches may be performed in other channels, such as $\gamma\gamma \rightarrow \tau^{+}\tau^{-}$. 
This channel is of particular interest as it is sensitive to the anomalous magnetic moment (or ``$g-2$'') of the $\tau$ lepton. Currently, measurements of 
$\gamma\gamma \rightarrow \tau^{+}\tau^{-}$ production at LEP provide some of the strongest constraints on this quantity~\cite{Abdallah:2003xd,Achard:2004jj}. Preliminary 
phenomenological studies~\cite{Atag:2010ja} found that these could be surpassed at the LHC using forward proton detectors. More detailed studies are required to 
confirm the initial estimates and to compare with the sensitivity of other proposed analyses. 

\subsection{Standard Model Processes}
\label{sec:standardmodel}

\begin{figure}[h!]
\centering
\includegraphics[width=0.6\textwidth]{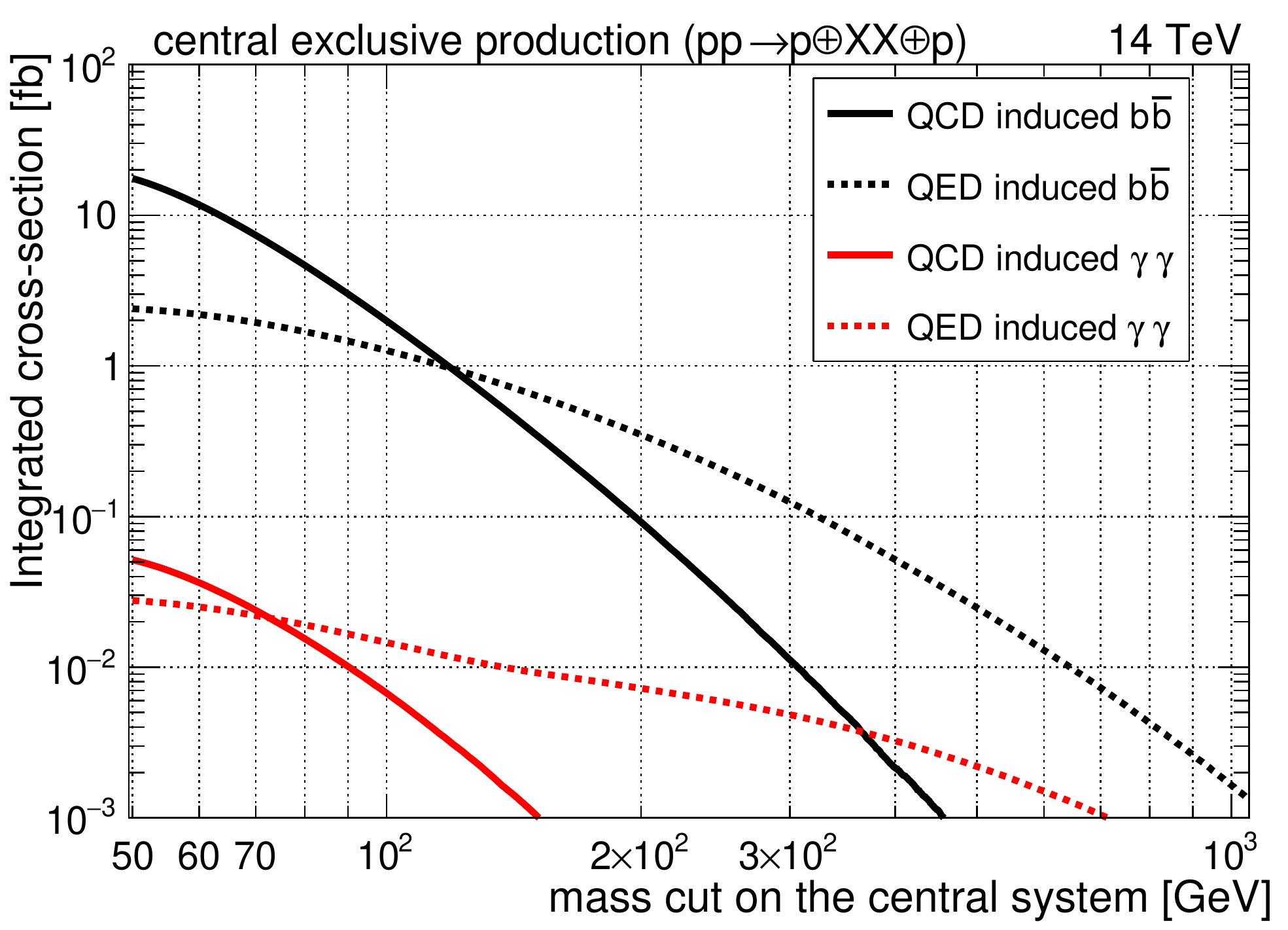}
\caption{
Integrated cross sections of different exclusive processes with intact protons at $\sqrt{s} = $14\,TeV, plotted as a function of the required minimum central system mass. Both photons or b-quarks are required to have a transverse momentum above 20\,GeV. 
}
\label{fig:Xsec_POM_PHO}
\end{figure}

The central exclusive production cross section spans several orders of magnitude, from tens of pb for di-jet production to a few fb for Higgs boson production. Three types of interaction can be considered: pomeron-pomeron ($\rm{I\!P}-\rm{I\!P}$)~\cite{Petrov:2004nx}, photon-photon ($\gamma-\gamma$) or pomeron-photon ($\rm{I\!P}-\gamma$). The last is sometimes referred to as \textit{photoproduction} and will be discussed in Section~\ref{sec:photoproduction}. Figure~\ref{fig:Xsec_POM_PHO} illustrates the production cross section of exclusive $b\bar{b}$ and $\gamma\gamma$ events as a function of the mass for different production modes.  The cross sections were obtained using SuperChic v4~\cite{Harland-Lang:2020veo} and FPMC~\cite{Boonekamp:2011ky} MC event generators for the exclusive production of pair of b-quarks and photons, respectively. In all cases, a lower cut on proton momentum loss of $\xi_{min}=0.003$ was applied. 

While QCD-induced processes are typically dominant at low masses, the photon-photon scattering is enhanced at high masses (Fig.~\ref{fig:Xsec_POM_PHO}). The visible rates of exclusive processes are suppressed because of
additional soft interactions between the spectator partons; these interactions produce particles that fill the rapidity gap and slow down the final-state protons (rescattering effects). 
This effect is quantified by the so-called rapidity gap survival probability, which ranges between about 90\% for $\gamma-\gamma$ induced processes and a few percent for $\rm{I\!P}-\rm{I\!P}$ induced processes. The survival probability has been calculated for various centre-of-mass energies and proton impact parameters, but still lacks sufficient experimental data to precisely validate the theoretical models. 

Fiducial cross sections for different standard model processes at $\sqrt{s} = 14$\,TeV for $\rm{I\!P}-\rm{I\!P}$ and $\gamma-\gamma$ production modes were evaluated with the generator-level requirements that $\pT$ be at least 20\,GeV for the final-state particles within the central detector acceptance, and that the two protons be both within the PPS acceptance. They are shown 
in Table~\ref{tab:xsec-sm-processes} for different PPS acceptance scenarios.

\begin{table}[h!]
  \begin{center}
    \caption{Fiducial cross sections of CEP of standard model processes in pp collisions at $\sqrt{s}=14$\,TeV, calculated with the FPMC generator~\cite{Boonekamp:2011ky}, using the KMR exclusive model for pomeron fluxes~\cite{Khoze:2001xm} and the Equivalent Photon Approximation (EPA) for photon fluxes~\cite{Budnev:1974de}. Gap survival probabilities of 3\% and 90\% are considered for $\rm{I\!P}-\rm{I\!P}$ and  $\gamma-\gamma$, respectively. A central detector selection cut of $\pT>20$\,GeV  on the generated objects was applied for all processes with two particles in the final state. Two scenarios for proton tagging acceptance are shown: with and without the stations at $\pm$420\,m.}
    \label{tab:xsec-sm-processes}
    \begin{tabular}{|l|c|c|c|c|}
    \hline
    \multirow{3}{*}{\textbf{Process}} & \multicolumn{4}{c|}{\textbf{fiducial cross section [fb]}} \\  
            
      & \multicolumn{2}{c}{\textbf{all stations}} & \multicolumn{2}{|c|}{\textbf{w/o 420}} \\

      & $\rm{I\!P}-\rm{I\!P}$ & $\gamma-\gamma$ & $\rm{I\!P}-\rm{I\!P}$ & $\gamma-\gamma$ \\   
        
      \hline
$\rm jj$ & $\mathcal{O}\left(10^6\right)$ &  60 & $\mathcal{O}\left(10^4\right)$  &  2 \\
$W^+W^-$ & --- &  37 & --- & 15 \\
$\mu\mu$ & ---&  46 & --- &  1.3  \\
$\rm t\bar{t}$ & --- &  0.15 & --- &  0.1 \\
H & 0.6 &  0.07 & 0 &  0 \\
$\gamma\gamma$ & --- &  0.02 & --- &  0.003 \\
    \hline
    \end{tabular}
  \end{center}
\end{table}


\subsubsection{QCD Physics}
\label{sec:qcd-physics}
Several crucial aspects of the proton structure and the strong interaction can be tested with the central diffractive production of QCD events. The study of central diffractive production is of great importance per se, as it is directly related to long range behaviour of gauge fields in the confinement region. Angular distributions of quasi elastically scattered protons provide additional information~\cite{Petrov:2004hh,Ryutin:2014eua}. Hard exclusive production of di-jet events is described in QCD as originating from $\rm gg\to gg$ (since the light quark di-jets production should be suppressed approximately with the quark mass $m^2_{\rm q}/m^2_{\rm jj}$). Therefore, it can be used as a ``gluon jet factory'' 
to study jet fragmentation with a pure sample of gluon-initiated jets. Exclusive tri-jet production is also predicted to occur at a substantial rate, and predominantly contain three gluon 
jets~\cite{Harland-Lang:2015cta}.

Additionally, although the jets should be gluon jets $\approx$99\% of the time, heavy-quark jets are less suppressed and open a door for additional tests of QCD physics. 

\subsubsection{Electroweak Physics}
\begin{figure}[h!]
\centering
\includegraphics[width=0.49\textwidth]{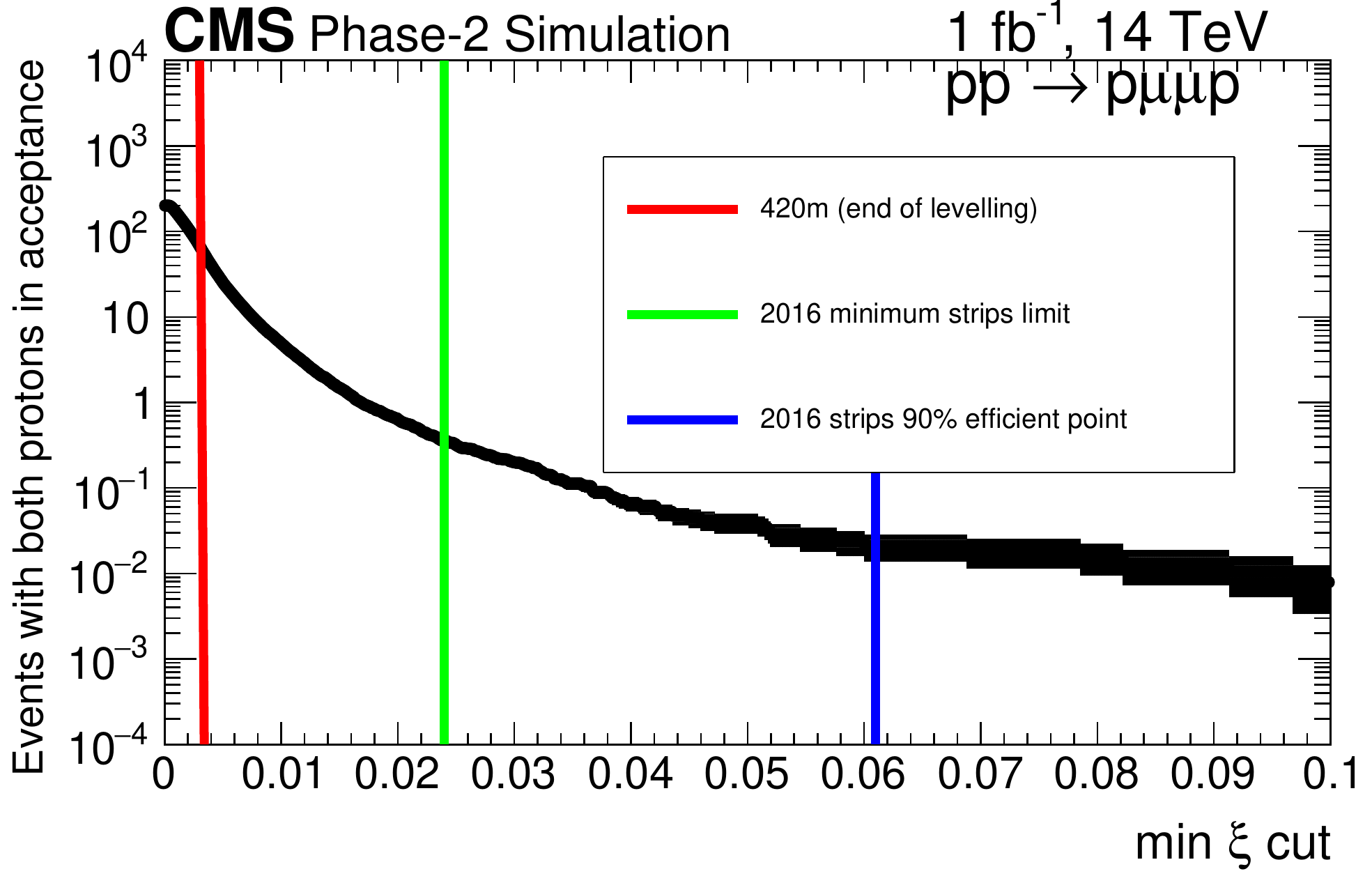}
\includegraphics[width=0.49\textwidth]{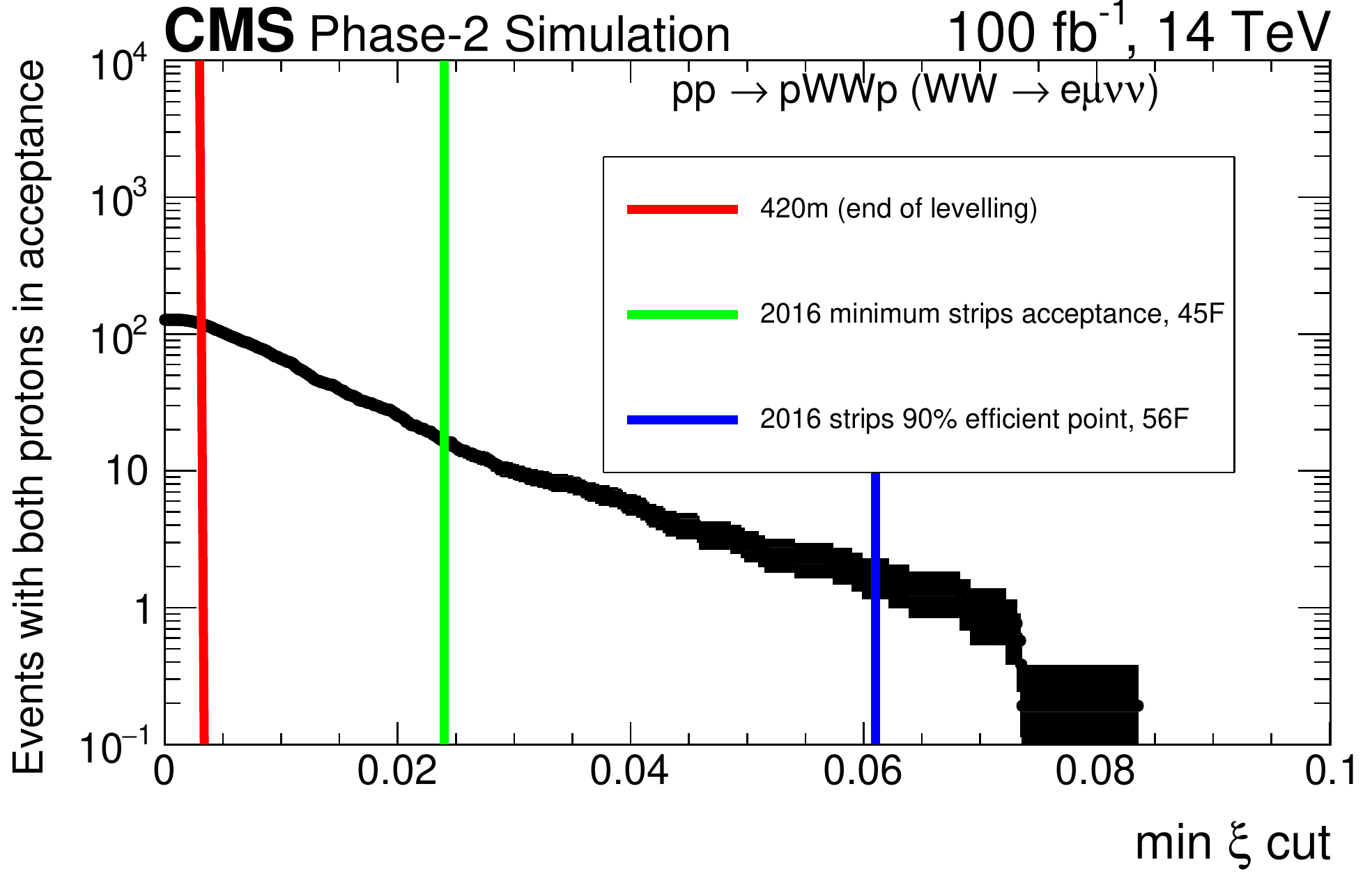}
\caption{Generator-level yields with both protons above the lower $\xi$-acceptance threshold on the abscissa, assuming 100\% efficiency. The minimum reachable $\xi$ limit for the PPS strip detector with the best acceptance (45-210-F-H, green line) and the 90\%-efficiency $\xi$ limit of the detector with the worst acceptance (56-210-F-H, blue line) during 2016 are shown for comparison with the acceptance limit of a future 420\,m detector (red line) at the HL-LHC. The accepted $\xi$ regions lie to the right of the vertical lines.
Left: $\gamma\gamma\rightarrow\mu^{+}\mu^{-}$ events per fb$^{-1}$, requiring $\pT(\mu)>20\,$GeV, $|\eta(\mu)|<2.4$, $m(\mu\mu)>40\,$GeV. Right: $\rm \gamma\gamma\rightarrow W^+W^- \rightarrow \mu^{+}e^{-}\nu_{\mu}\bar{\nu_{e}}$ events per 100\,fb$^{-1}$, requiring $\pT(\mu, e)>20\,$GeV, $|\eta(\mu, e)|<2.4$.}
\label{fig:sm-crosssections}
\end{figure}

Standard model $\rm \gamma\gamma\rightarrow \ell^{+}\ell^{-}$ production ($\ell^\pm = e^\pm, \mu^\pm$) with single proton tags was the first process measured with PPS Run-2 data and is an important calibration channel to validate the reconstruction of the protons~\cite{jhep2018,ProtonRecoDPS}. The Run 2 and 3 PPS acceptance is such that 10\,fb$^{-1}$ are enough to obtain a clear single-arm signal, while the number of double-arm events is small for even a full year of data-taking. For the HL-LHC, Fig.~\ref{fig:sm-crosssections} and Table~\ref{tab:sm-processes} illustrate the effect of adding the 420\,m station to the other three stations. It would increase the visible cross section by approximately 3 orders of magnitude and allow the collection of a significant number of double-arm events per fb$^{-1}$ in the $\mu\mu$ channel alone.
In addition, the more numerous single proton dissociation events can be used for calibration studies, though with larger theoretical uncertainties on the visible cross section. Assuming a survival probability of 0.76~\cite{Harland-Lang:2016apc}, approximately 85 single-dissociation $\mu\mu$ events per fb$^{-1}$ per arm could be expected within the acceptance, mostly in the 420\,m station. By comparison, the first PPS analysis using Run-2 data obtained approximately one event of this type per fb$^{-1}$. The 
$\rm \gamma\gamma\rightarrow e^{+}e^{-}$ process is expected to have an equal cross section within the acceptance, and would provide an another important calibration channel. 

For diboson production, $\rm \gamma\gamma\rightarrow W^{+}W^{-}$ (with $\rm W^{+}W^{-} \rightarrow \mu^{+}e^{-}\nu_{\mu}\bar{\nu_{e}}$) is a particularly clean channel that was studied in detail for the PPS TDR~\cite{ctpps}; however only a handful of standard model events are expected in 100\,fb$^{-1}$ with the Run 2 and 3 PPS acceptance. The configuration of stations considered here would substantially increase the acceptance for 2-arm events, allowing a significant measurement of the SM cross section in the $\rm \mu^{+}e^{-}$ final state, which will serve as a benchmark for diboson searches in other channels and at higher masses. 

\begin{table}[h!]
  \begin{center}
    \caption{Example standard model processes, and the rate of 2-arm events expected
at generator level within the acceptance of the four PPS stations under
consideration, in pp collisions at 14 TeV. The acceptance cuts for both the central
detector and the forward protons are applied at generator level.} 
    \label{tab:sm-processes}
    \scalebox{1.0}{
    \begin{tabular}{|l|c|l|} 
     \hline
     \textbf{Process} & \textbf{Central detector acceptance cuts} & \textbf{2-arm events} \\
      \hline
      $\gamma\gamma\rightarrow\mu\mu ~\rm(SM)$ & $\pT(\mu) > 20$\,GeV, $|\eta(\mu)|<2.4$, $m(\mu\mu)>40$\,GeV & 47 / fb$^{-1}$ \\
       $\rm \gamma\gamma\rightarrow WW \rightarrow\mu e ~\rm(SM)$ & $\pT(\mu, e) > 20$\,GeV, $|\eta(\mu, e)|<2.4$ & 72 / 100\,fb$^{-1}$\\
      \hline
    \end{tabular}
    }
  \end{center}
\end{table}

\subsubsection{Higgs Physics}
\begin{figure}[h!]
\centering
\includegraphics[width=0.4\textwidth]{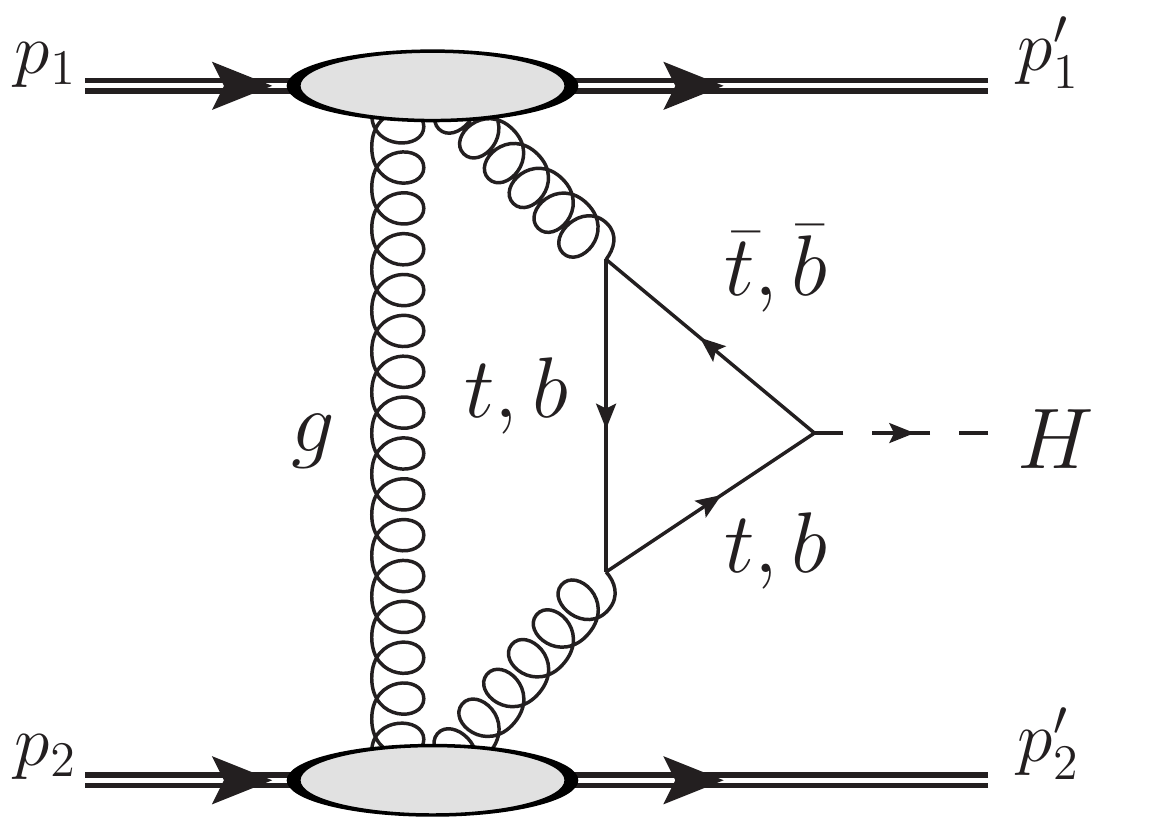}
\includegraphics[width=0.4\textwidth]{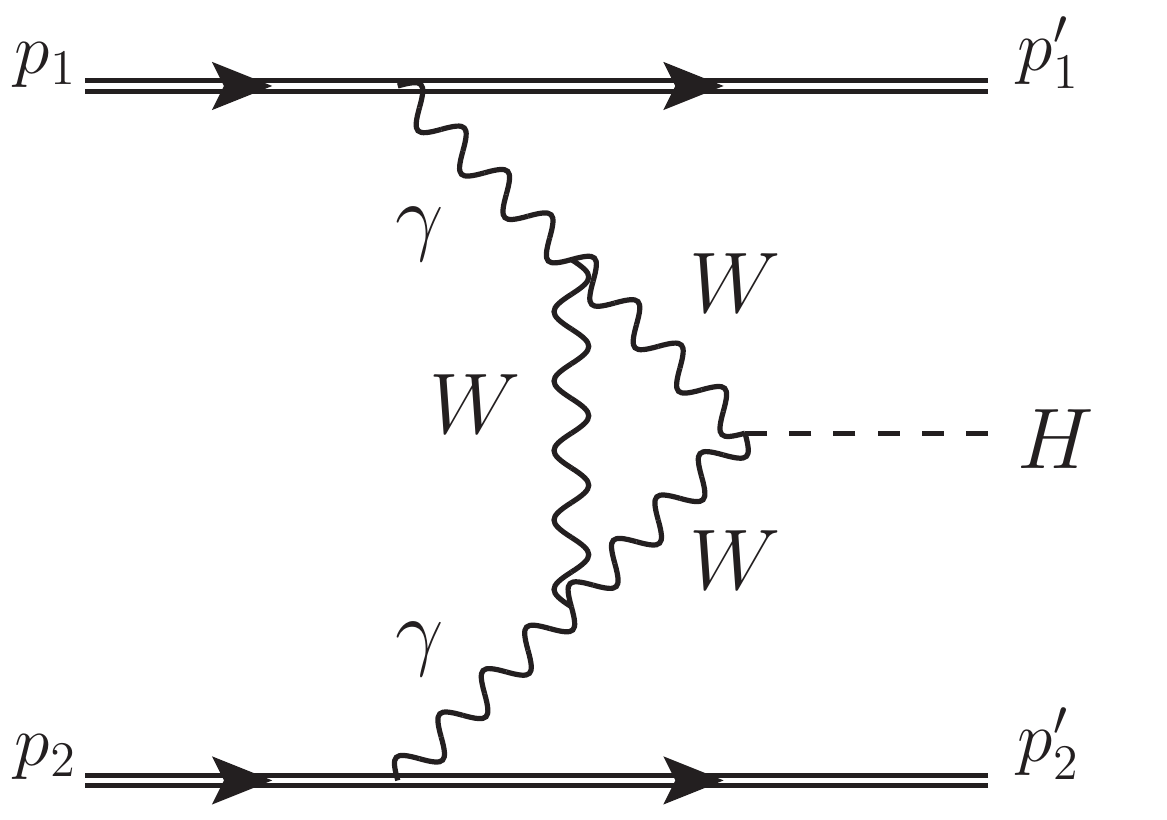}
\caption{Diagram for central exclusive $\rm pp \rightarrow pHp$ production by pomeron exchange (left) and photon-fusion (right).}
\label{fig:CEPggH}
\end{figure}

Central exclusive Higgs boson production has been extensively studied theoretically and in simulations (including the original detailed studies of the FP420 project~\cite{fp420}). In 
this case, unlike higher-mass and weakly coupled final states, gluon-gluon production 
is expected to dominate over $\gamma\gamma$ production (Fig.~\ref{fig:CEPggH}). The production is strongly constrained to $J^{PC} = 0^{++}$ final states, providing an independent determination of the quantum numbers of the Higgs boson. 

The cross section for CEP Higgs production in the SM has been evaluated by several groups. While exclusive dijet production data from the Tevatron ruled out many initial 
models, there are still several viable predictions, most based on variations of a pQCD approach. Roughly, these predict central values of the 
total cross section ranging between a few fb and a few tenths of a fb, depending on details of the survival probabilities, parton distribution functions (PDFs), Sudakov factors and other 
assumptions of the calculations~\cite{Khoze:2000cy,Petrov:2003yt,Khoze:2002py,DeRoeck:2002hk,Cudell:2010cj,Maciula:2010tv,Coughlin:2009tr,Ryutin:2012np}. 
The theoretical uncertainties of one available calculation have been estimated to be up to a factor of 25~\cite{Dechambre:2011py}. As described in many of these publications, most of the same theoretical 
uncertainties enter in the central exclusive dijet process, described earlier (Section~\ref{sec:qcd-physics}). A measurement of CEP dijets at the same energy and mass range would therefore 
remove most of the remaining theoretical uncertainties in the Higgs cross section predictions. 

For the KMR model as implemented in the FPMC Monte Carlo generator~\cite{Boonekamp:2011ky}, a 125.4\,GeV Higgs boson is predicted to have a cross section
of 1.7\,fb in central exclusive production, including the effect of an assumed 0.03 survival probability correction factor. For this mass value, the protons 
can be detected only in asymmetric events with one proton in the 420\,m acceptance and one in the 234\,m acceptance, and in symmetric events with both protons in the acceptance of a 420\,m station. Without the 420\,m stations, there is no acceptance for the SM Higgs in any combination of other stations. 
The acceptance of the 420\,m station is relatively insensitive to the crossing angle and $\beta^{*}$ levelling scheme 
employed by the LHC, while the acceptance of the 234\,m station varies more strongly throughout the levelling trajectory (i.e.\ from the beginning to the end of the fill). Taking the mean acceptance between the beginning and the end of the levelling trajectory results in a prediction of $\sim 600$ events with both protons in the acceptance per 
ab$^{-1}$, assuming a Higgs mass of 125.4\,GeV. This number includes all Higgs events produced, before any channel-dependent effects 
of triggering on and reconstructing the decay products in the central region. As noted previously, the theoretical uncertainties are potentially large. For example, using the KMR model implemented 
in the SuperChic v4 generator~\cite{Harland-Lang:2020veo}, with a modified treatment of the survival probabilities and updated PDFs, the prediction is 0.4\,fb. A more detailed discussion of the 
dependence on PDFs, survival probabilities, and higher order corrections, can be found in Ref.~\cite{HarlandLang:2013jf}.

\begin{figure}[h!]
\centering
\includegraphics[width=0.49\textwidth]{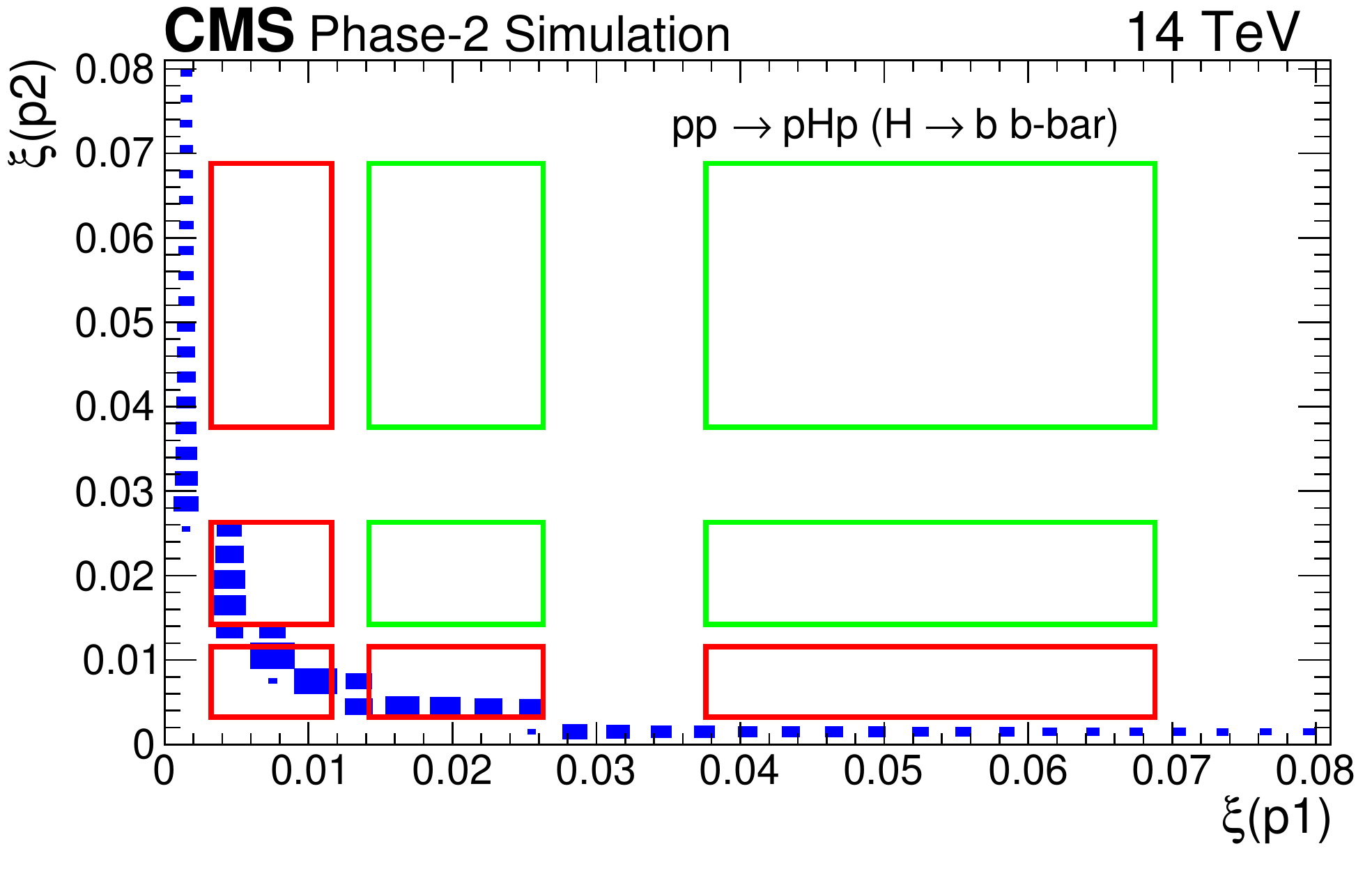}
\includegraphics[width=0.49\textwidth]{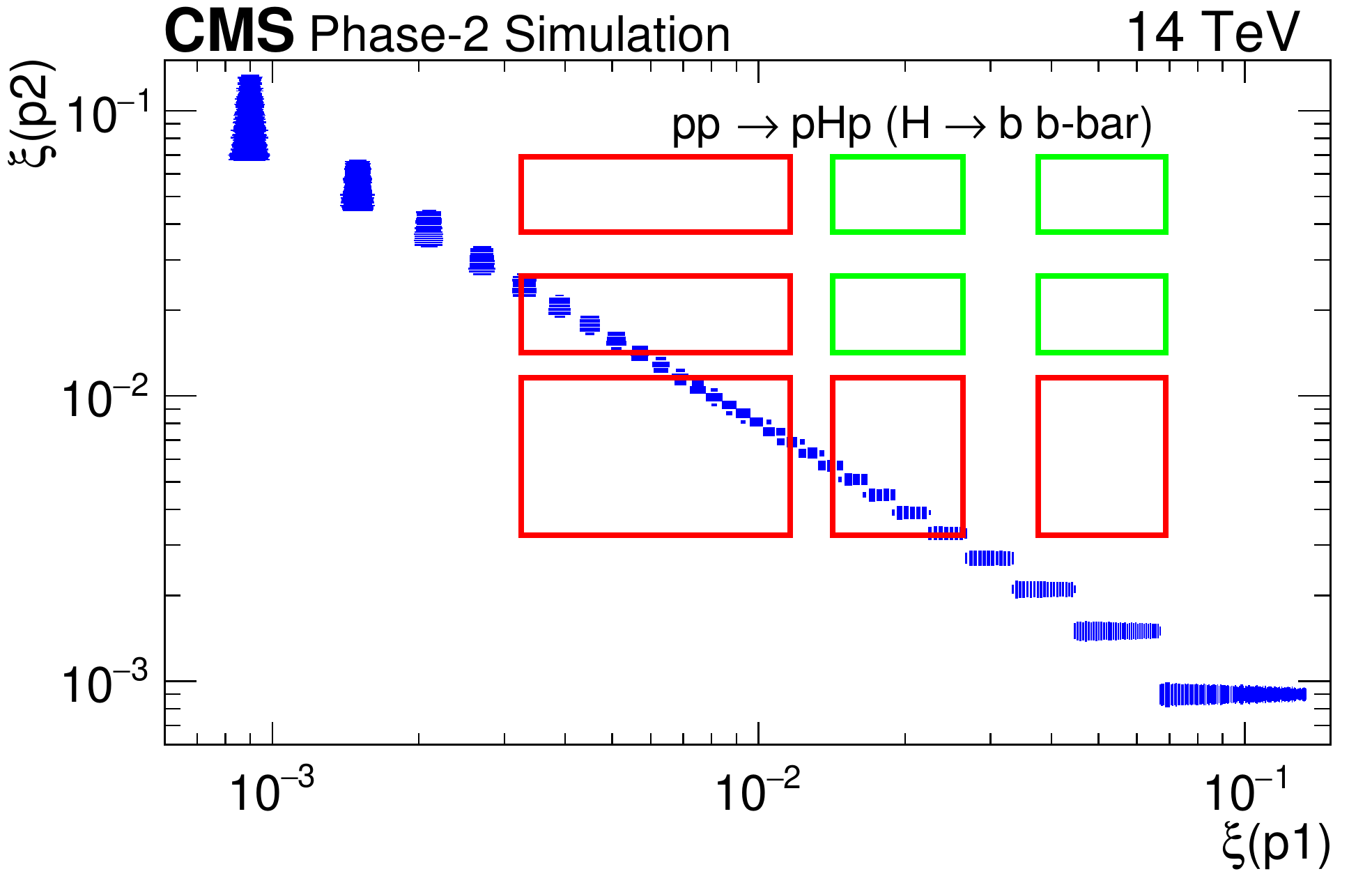}

\caption{Generator-level distributions of the $\xi$ of the outgoing protons in standard model $\rm pp \rightarrow pHp$ ($\rm H \rightarrow b \bar{b}$), after acceptance cuts. The filled blue boxes represent generated signal events, the hollow ones illustrate the projected PPS acceptance with the 220\,m and 234\,m stations alone (green), and when the 420\,m stations (red) are included in addition, in linear (left panel) and log (right panel) scale. The proton 
acceptance ranges correspond to the average between the acceptance at the beginning and end of a fill. The acceptance of the 196\,m position 
is not shown, since it does not contribute to the acceptance for the 125\,GeV Higgs boson. The acceptance requirements in the central region are applied 
to the generator-level b-quarks: $\pT(b) > 40$\,GeV and $|\eta(b)| < 2.4$.}
\label{fig:higgs-xi-distrib}
\end{figure}

Figure~\ref{fig:higgs-xi-distrib} illustrates the predicted $\xi$ distributions of the two outgoing protons in exclusive $\rm H \rightarrow b \bar{b}$ production,
compared to the acceptances of different possible combinations of PPS detector stations. 

In this example only the $\rm b\bar{b}$ channel is plotted because it provides the largest decay branching fraction. However, it should be noted that the acceptance and resolution for
the protons will be independent of the decay channel of the Higgs boson. The reconstructed proton resolution at a 420\,m station is discussed in detail 
in Section~\ref{sec:resolution}. Given the available space, the resolution of the reconstructed $\xi$ is estimated to be $\sigma(\xi) \approx 3 \times 10^{-4}$, in the approximation 
that the polar scattering angles $\theta^{*}_{x}$ and $\theta^{*}_{y}$ are equal to zero. For a Higgs boson of 125.4\,GeV produced at rapidity $\mathsf{y} = 0$, this corresponds to a 
mass resolution of $\approx 3$\,GeV. While this is worse than the central CMS detector mass resolution for the highest precision Higgs decay modes, the determination using the protons is independent of the exact final state. 

In addition, the azimuthal angular distributions of the outgoing protons are sensitive to CP-violating effects in the Higgs sector~\cite{Khoze:2004rc}.

\begin{figure}[h!]
\centering
\includegraphics[width=0.4\textwidth]{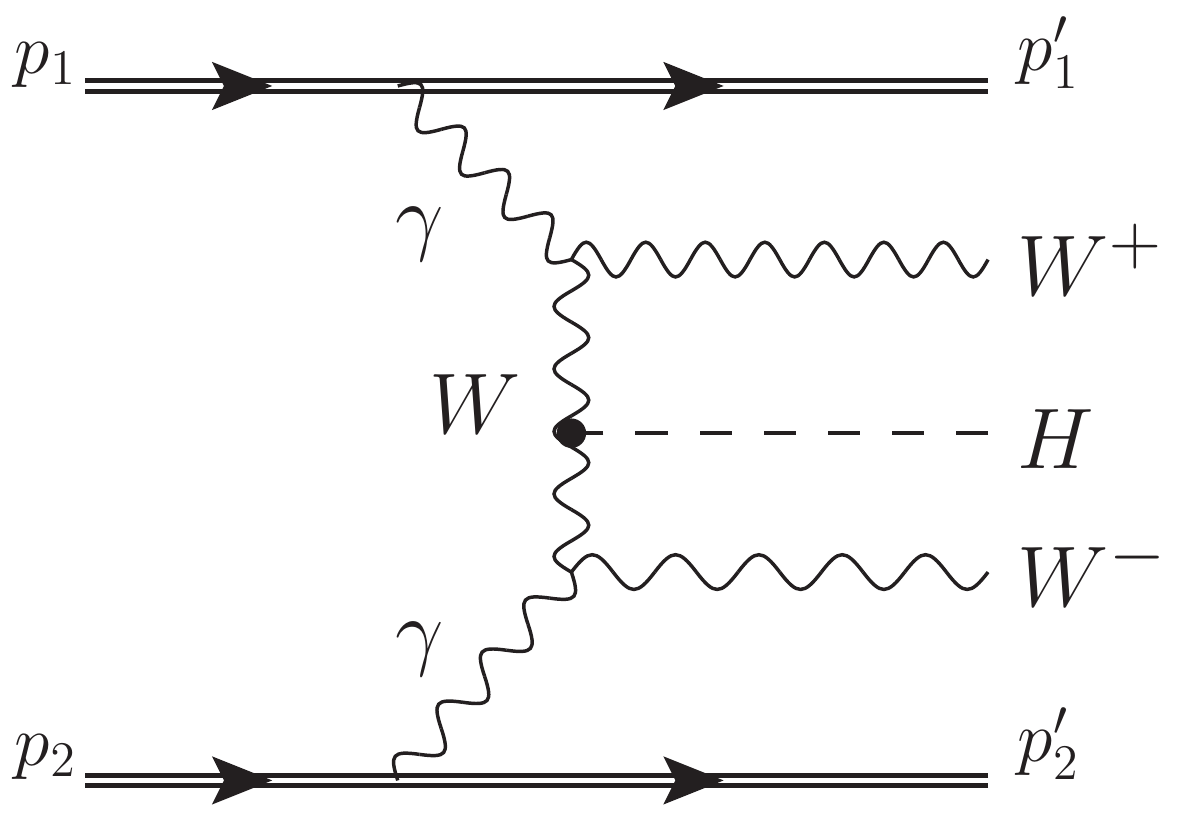}
\includegraphics[width=0.47\textwidth]{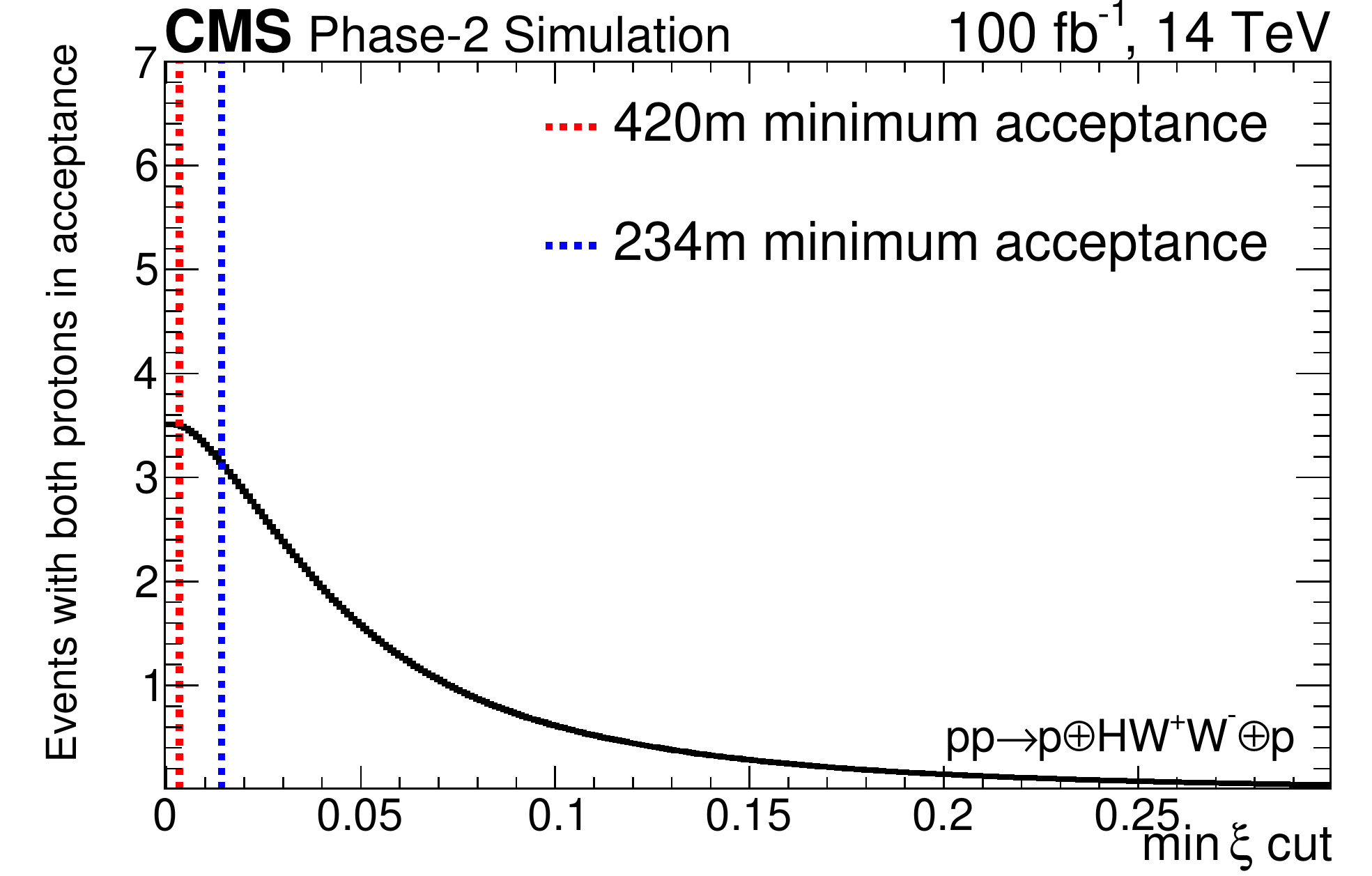}
\caption{Associated production of a Higgs boson and a $\rm W^{+}W^{-}$ pair via photon-photon fusion. Left: schematic diagram. Right: generator-level yields with both protons above the $\xi$-acceptance value on the abscissa. The coloured lines represent the minimal acceptance of the 234\,m (blue) and 420\,m (red) stations.}
\label{fig:hww}
\end{figure}

\paragraph{$\rm HW^{+}W^{-}$ production\\}
~~\\ 
\noindent
Associated production of a Higgs boson and a $\rm W^{+}W^{-}$ vector-boson pair (Fig.~\ref{fig:hww}, left) has the potential for probing the Higgs sector in CEP events in the absence of the $\pm$420\,m stations. The exclusive production cross section is estimated to be $\sigma \approx 0.04$\,fb at tree-level by MadGraph5\_aMC@NLO~\cite{Alwall:2014hca} version 2.7.2 in the Higgs Effective Field Theory (HEFT) model~\cite{Shifman:1979eb,Dawson:1993qf,Kniehl:1995tn}, using the EPA for photon fluxes~\cite{Budnev:1974de} at a fixed factorization scale of $\mu_F=$10\,GeV and assuming a gap survival probability of 0.9.
High acceptance is expected because of the large invariant mass of the central system. The expected signal yield is illustrated in Fig.~\ref{fig:hww} (right). The reconstruction of full event kinematics becomes possible using the tagged protons. In this case, the production rate of the $\rm HW^{+}W^{-}$ process can be measured regardless of the Higgs boson decay mode, by reconstructing the W bosons in semi- or fully hadronic decays, and the protons. Requiring the residual mass to be compatible with the Higgs mass would allow an 
independent measurement of the Higgs boson branching ratios.

\subsubsection{Top Physics}
The exclusive production of top-quark pairs has a relatively low cross section (of the order of tenths of a fb, see Table~\ref{tab:xsec-sm-processes} and Ref.~\cite{dEnterria:2009cwl}). The production is dominated by the $\rm\gamma\gamma\to t\bar{t}$ process since QCD contributions are rapidly vanishing at high $\rm t\bar{t}$ invariant masses. As discussed below, this reaction is potentially interesting
for SM and BSM studies.

Central exclusive production of $\rm t\bar{t}$ pairs has the potential of probing the colour-singlet exchange process in a region so far unexplored. This mechanism is suppressed in inclusive $\rm t\bar{t}$ events, where the protons mostly do not survive~\cite{Sumino:2010bv}. 
Spin correlations in exclusive $\rm t\bar{t}$ events differ from those in the inclusive channel because of the different initial-state particles. This can serve as an independent test of SM predictions.

The observation of central exclusive $\rm t\bar{t}$ events can be translated into a measurement of the quartic dimension-8 anomalous couplings ($c_{\rm\gamma\gamma tt}$), and potentially provide a stringent limit from the measured invariant mass spectrum.

\begin{figure}[h!]
\centering
\includegraphics[width=0.5\textwidth]{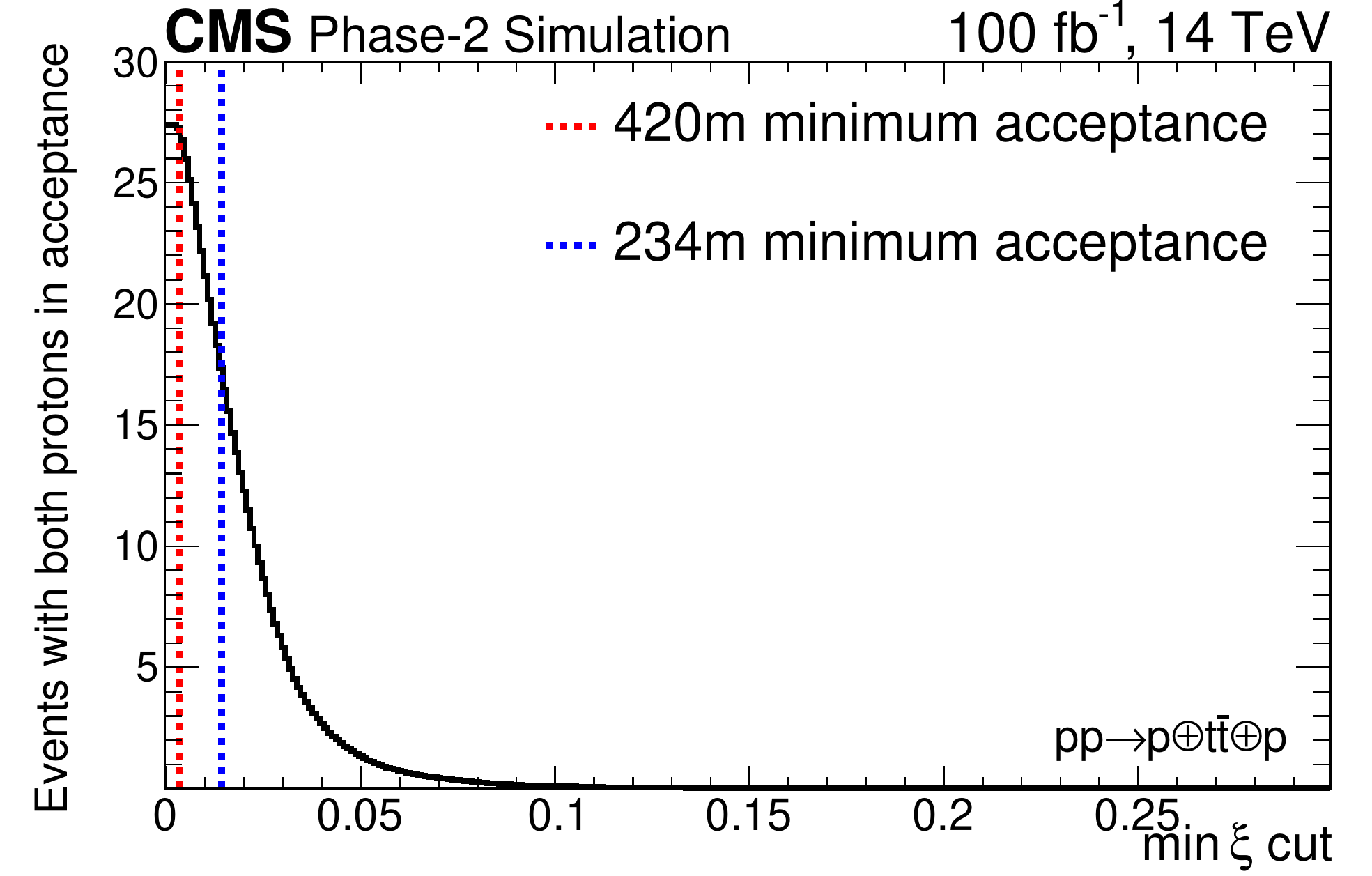}
\caption{Generator-level yields with both protons above the $\xi$-acceptance value on the abscissa for $\rm t\bar{t}$ exclusive production via photon-photon fusion. The coloured lines represent the minimal acceptance of the 234\,m (blue) and 420\,m (red) stations. 
}
\label{fig:ttbarsm-crosssections}
\end{figure}

Because of the high invariant mass (above $\sim 350$\,GeV), the acceptance of the first three stations is large. The expected signal yields are illustrated in Fig.~\ref{fig:ttbarsm-crosssections} as a function of the minimum $\xi$ value. 

The main background for exclusive $\rm t\bar{t}$ production is the inclusive $\rm t\bar{t}$ process with the two intact protons due to pileup. Although the inclusive $\rm t\bar{t}$ production cross section is $\mathcal{O}\left(10^6\right)$ times larger than the exclusive one, the exclusivity of the process and the correlation between the $\rm t\bar{t}$ system measured in the central detector and the scattered protons can be used to suppress the inclusive background.

\subsubsection{Photoproduction}
\label{sec:photoproduction}

\begin{figure}[h!]
\centering
\includegraphics[width=0.4\textwidth]{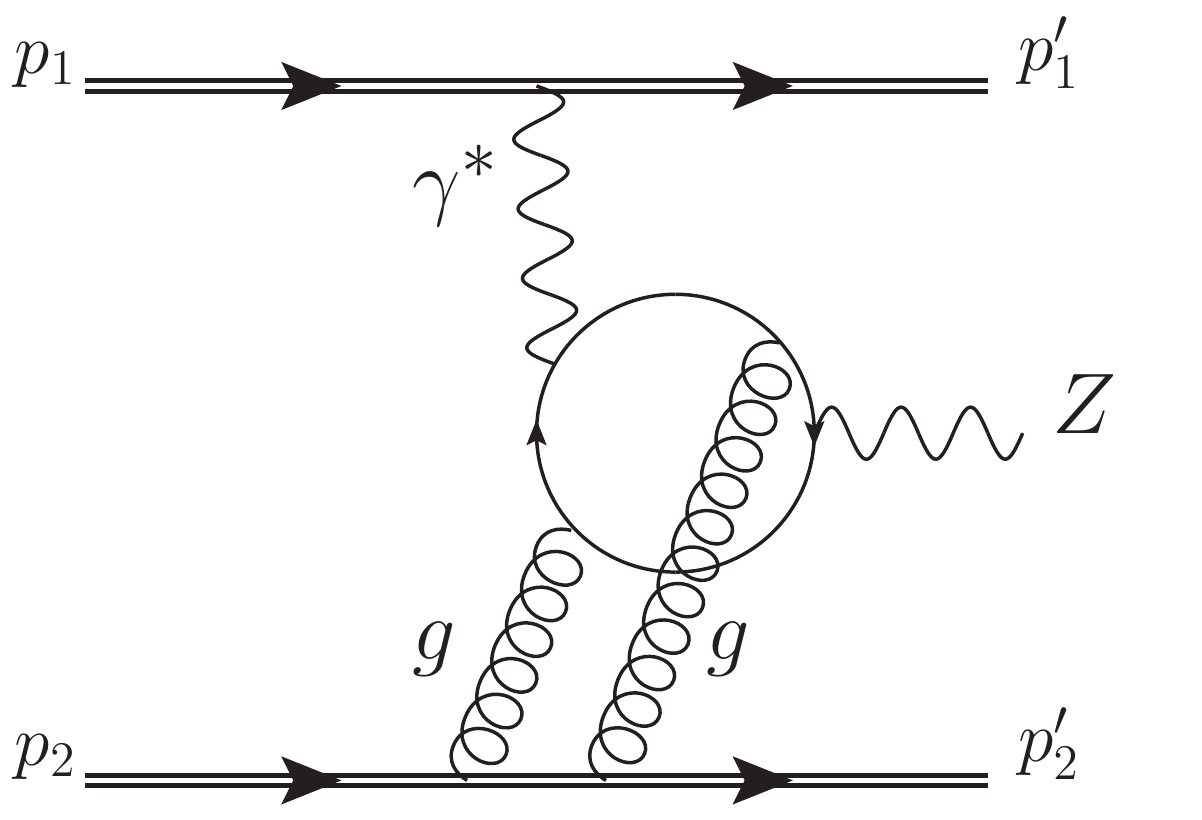}
\caption{Diagram representing exclusive Z-boson photoproduction.}
\label{fig:photoproductionZ}
\end{figure}

Single exclusive Z-production can happen only through photoproduction: $\rm\gamma^*I\!P \to Z$. A virtual photon radiated from one proton fluctuates into a $\rm q\bar{q}$ pair, which scatters 
via $\rm{I\!P}$ exchange off the other proton and then produces the vector boson (see Fig.~\ref{fig:photoproductionZ}).

The photon-emitting (Coulomb scattered) proton will have a small squared four-momentum transfer $|t|$, less than 0.1\,GeV$^2$, while the other one will have a larger $|t|$, up to 0.5\,GeV$^2$, typical for strong diffractive processes. In this case, the Z-boson will have $\pT(\rm Z)<0.7$\,GeV, the leptons from $\rm Z\to\ell\ell$ will be almost back-to-back in azimuth, and the additional charged multiplicity at the dilepton vertex will be small. 
A future PPS spectrometer with the 420\,m stations would have acceptance for exclusive Z production with both protons detected. Without the 420\,m stations, on the other hand, the lower mass acceptance limit for double proton detection is too high (between 133 and 265\,GeV, varying during the fill, see Table~\ref{tab:masslimits}), but one can tag these events using only one proton: if the Z is sufficiently boosted in the $z$-direction, the proton in that
direction has larger $\xi = \frac{m_Z}{\sqrt{s}}e^{y_{Z}}$ and falls inside the PPS acceptance. 

It should be noted that high mass central exclusive dijet production, $\rm pp\to p\oplus jj\oplus p$, can also occur via
photoproduction in a $C = -1$ state, where $C$ is the charge parity. The cross section of this process is, however, expected to be significantly lower than the corresponding one from color-singlet gg fusion.

\clearpage
\section{Locations for Near-Beam Detectors}
\label{sec:locations}

\subsection{Layout of LSS5}
\begin{sidewaysfigure}
\begin{center}
\includegraphics[width=1.0\textwidth]{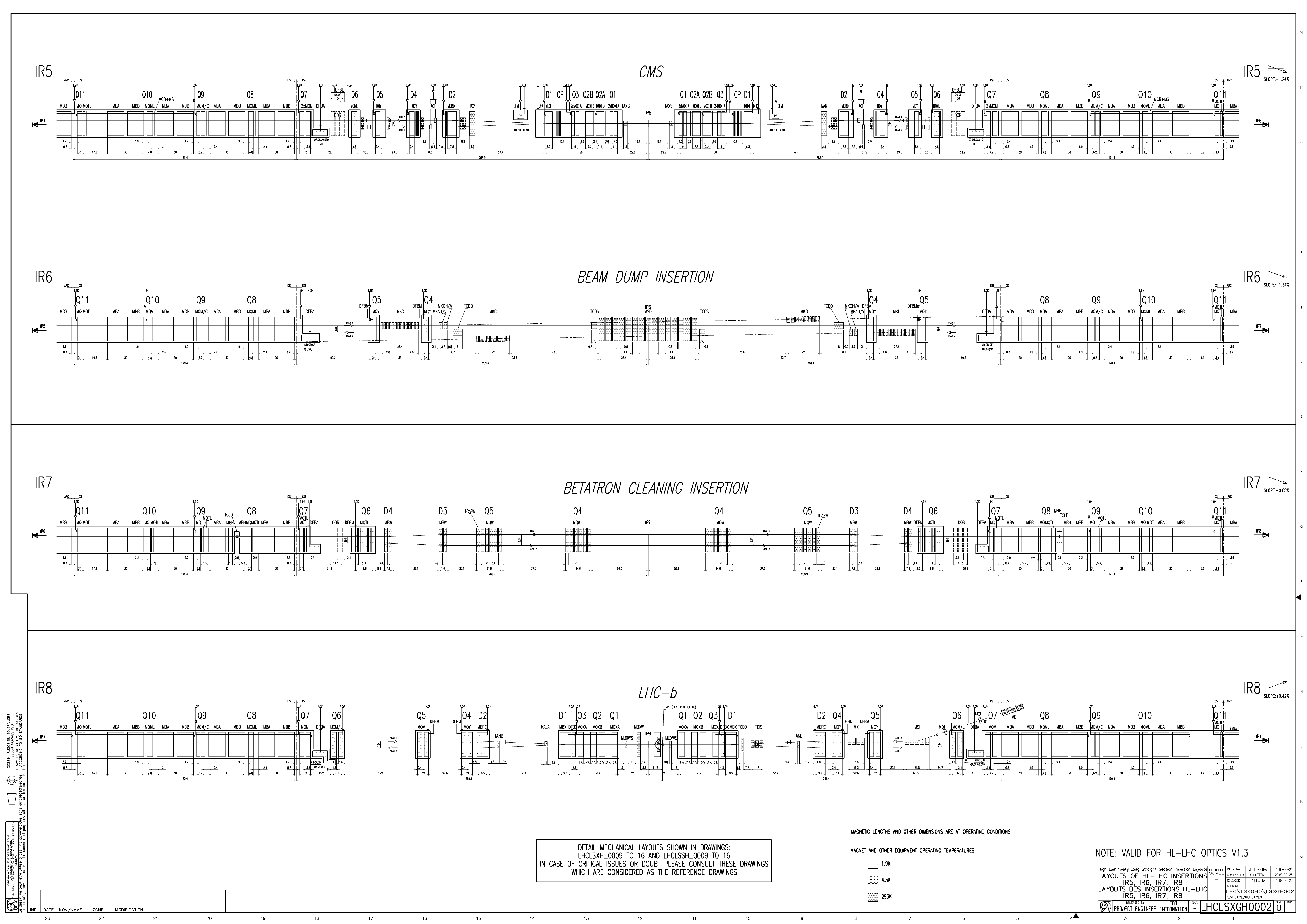}
\end{center}
\caption{Layout of Long Straight Section LSS5 (Sector 5-6) at HL-LHC~\cite{edms-layout-lss5}.}
\label{fig:layout-lss5}
\end{sidewaysfigure}

During the Long Shutdown LS3 the Long Straight Section LSS5 will be entirely redesigned. Some magnet and collimator locations will change, and crab cavities will be added. An overview drawing of the new layout is shown in Fig.~\ref{fig:layout-lss5}. The objective of the following section is to identify free space suited for the installation of movable near-beam detectors.

\subsection{Selection of Potential Detector Locations}
\label{sec:locationselection}
The search for suitable detector locations around IP5 is driven by the goal to cover the widest possible range of masses $M$ in CEP processes to be measured via the fractional momentum losses 
\begin{equation}
\label{eqn:xidefinition}
\xi_{1/2} := \frac{\Delta p_{1/2}}{p} 
\end{equation}
of the two surviving protons using the relation
\begin{equation}
\label{eqn:mass-xi1-xi2}
M^{2} = \xi_{1}\, \xi_{2}\, s \: ,
\end{equation}
where $\sqrt{s} = 14\,$TeV is the centre-of-mass energy. 
The minimum (maximum) accepted mass is obtained in the symmetric case $\xi_{1} = \xi_{2} = \xi_{\rm min(max)}$: 
\begin{equation}
\label{eqn:min-max-mass}
M_{\rm min (max)} = |\xi|_{\rm min (max)} \sqrt{s} \: .
\end{equation}
The best detector acceptance for leading protons with $\xi \ne 0$ is obtained with sensors approaching the beam horizontally, i.e. along $x$. This can be seen in the transverse hit distributions shown later in Section~\ref{sec:detectorrequirements} (Figs.~\ref{fig:xymap} and~\ref{fig:irradiationhitmap}). 
The minimum accessible $|\xi|$ at a detector location $\varsigma$~\footnote{In this document the variable $\varsigma$ is used for the longitudinal coordinate (beam direction) instead of $s$ to avoid confusion with the Mandelstam $s$.} along the beam line is given by the smallest accepted hit position $x_{\rm min}$ -- determined by the closest detector approach to the beam -- and the horizontal dispersion $D_{x}$ at that location:
\begin{equation}
\label{eqn:xi-min}
|\xi|_{\rm min}(\alpha, \beta^{*}, \varsigma) = 
\frac{x_{\rm min}}{|D_{x}(\alpha, \xi_{\rm min}, \varsigma)|}
= \frac{[n_{\rm TCT}(\beta^{*}) + \Delta n] \sigma_{x}(\beta^{*},\varsigma) + \Delta d + \delta}{|D_{x}(\alpha, \xi_{\rm min}, \varsigma)|} \: ,
\end{equation}
where $\sigma_{x}$ is the horizontal beam width depending on the optics (characterized by $\beta^{*}$, i.e.\ the value of the $\beta$ function at the IP), $n_{\rm TCT}$ is the half-gap of the tertiary collimators (TCT) as defined by the collimation scheme, $\Delta n = 3$ is the retraction of the near-beam detector vessels (e.g.\ Roman Pots) relative to the TCT position in terms of $\sigma_{x}$, $\Delta d = 0.3\,$mm is an additional safety retraction to allow for beam orbit fluctuations, and the constant $\delta$, typically 0.5\,mm, accounts for any distance between the outer housing surface closest to the beam and the sensitive detector. The dispersion $D_{x}$ depends not only on the crossing-angle $\alpha$ but also on $\xi$.
This $\xi$-dependence implies that Eq.~(\ref{eqn:xi-min}) has to be solved for $|\xi|_{\rm min}$, which is accomplished using a parameterization of $D_{x}(\alpha, \xi)$ based on simulations with MAD-X~\cite{MADX}.

Similarly to $|\xi|_{\rm min}$, the maximum accessible $\xi$ can be expressed as
\begin{equation}
\label{eqn:xi-max}
|\xi|_{\rm max}(\alpha, \beta^{*}) =  \frac{d_{A}(\beta^{*})}{|D_{A}(\alpha, \xi_{\rm max})|}\: ,
\end{equation}
where $d_{A}$ is the tightest aperture limitation -- in most cases the half gap of a TCL collimator (``Target Collimator Long'') as defined by the collimation scheme, and $D_{A}$ is the dispersion at that aperture limitation. Again, because of the $\xi$-dependence of $D_{A}$, Eq.~(\ref{eqn:xi-max}) has to be solved for $\xi_{\rm max}$ after parameterising $D_{A}(\alpha, \xi)$.

\begin{figure}[h!]
\begin{center}
\includegraphics[width=0.49\textwidth]{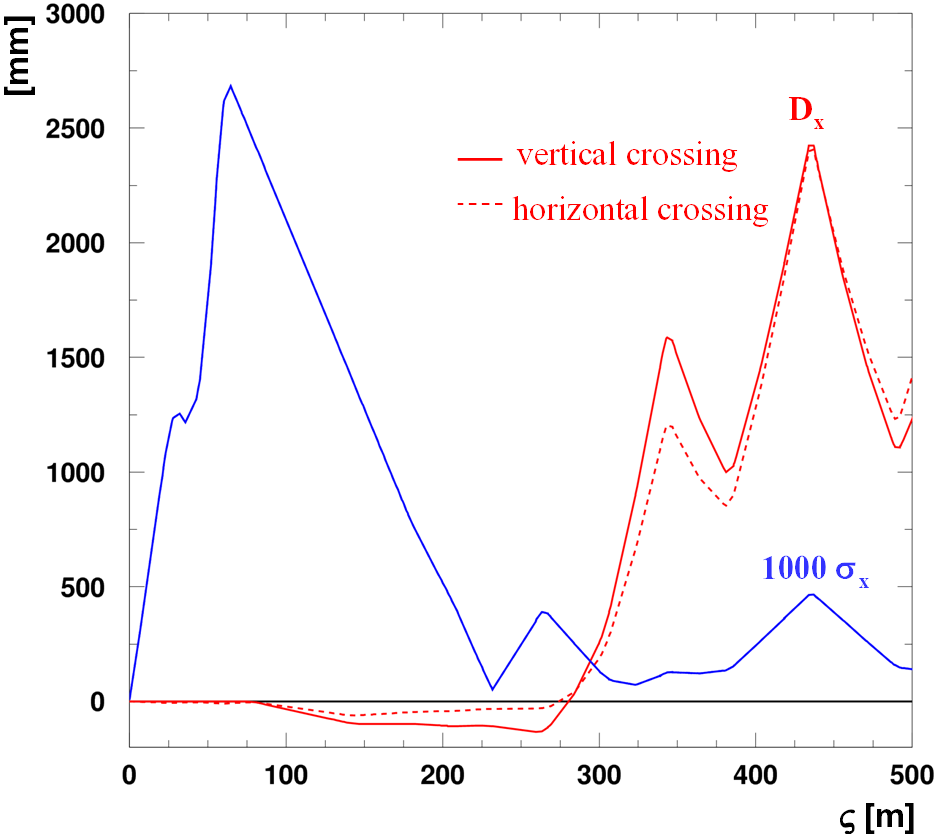}
\includegraphics[width=0.49\textwidth]{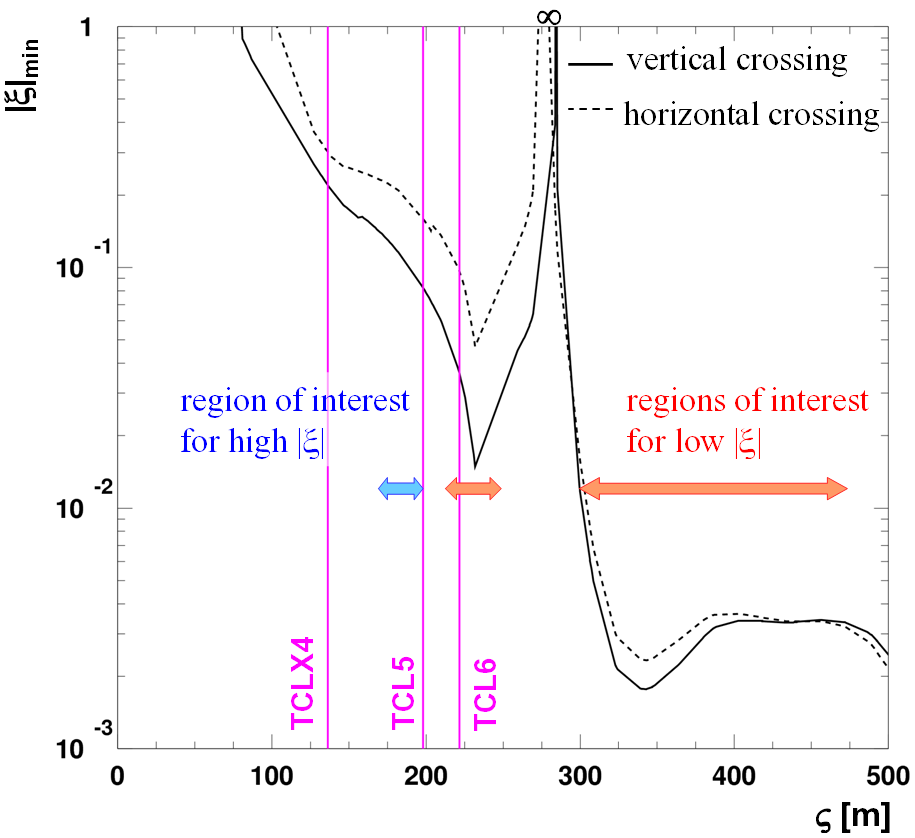}
\end{center}
\vspace{-0.5cm} 
\caption{Left: Horizontal dispersion and beam width (scaled by 1000) as a function of the distance $\varsigma$ from IP5 for Beam 1, i.e.\ in LHC Sector 5-6. Right: Minimum accepted $|\xi|$ as a function of $\varsigma$ according to Eq.~(\ref{eqn:xi-min}) for $(\alpha/2, \beta^{*}) = (250\,\rm \mu rad, 15\,cm)$ and $n_{\rm TCT} = 12.9$. 
The TCL collimator positions are indicated. In both pictures the continuous and dashed lines represent vertical and horizontal crossing in IP5, respectively. The study was done with optics version 1.3~\cite{hllhc-optics1.3}.} 
\label{fig:d-sigma-xi_vs_s}
\end{figure}

The first step of the location search is to plot the $\varsigma$-dependent quantities, $\sigma_{x}$ and $D_{x}$, along the outgoing beam line for one typical HL-LHC optics configuration (Fig.~\ref{fig:d-sigma-xi_vs_s}, left). The resulting $|\xi|_{\rm min}$ is shown in Fig.~\ref{fig:d-sigma-xi_vs_s} (right). For vertical crossing, smaller values are reached. The locations most suitable for the measurement of small $|\xi|$ values are marked in red. Closer layout inspection of the region around the minimum at 232\,m (inside the quadrupole Q6) indicated two promising locations: at 220\,m (just before ``TCL6'', the TCL collimator in LHC cell number 6 counted from IP5) and at 234\,m (after the exit of Q6). Even smaller momentum losses can be reached beyond 300\,m, but the only location with some free space lies around 
420\,m (the ``missing magnet'' region already studied previously by the FP420 project~\cite{fp420}). 

At $\varsigma \approx 270\,$m the dispersion changes sign (Fig.~\ref{fig:d-sigma-xi_vs_s}, left). This means in practice that the proton trajectories transition from $x > 0$ to $x < 0$. The implication for the potential detector location at 420\,m is that detectors need to be placed in the confined space between the incoming and the outgoing beam pipes, excluding conventional Roman Pot technology. A further complication is that in this location the beam pipes are in a cryostat, necessitating more involved engineering changes.

A region of interest for the detection of higher masses lies at 196\,m, just upstream of the collimator TCL5 that intercepts protons with large $|\xi|$ (Section~\ref{sec:maxmass}). Locations even farther upstream, before TCLX4, would give an even 
higher upper mass cut but are excluded because of the prohibitively high low-mass limit leaving no acceptance interval.\\

{\bf In summary, for the more detailed discussions in the following sections, four detector locations have been retained: 196\,m, 220\,m, 234\,m, 420\,m, each on both sides of IP5.}

\subsection{Mechanical Constraints, Available Space}
\label{sec:spaceconstraints}
In this section, preliminary information on the longitudinal and vertical space constraints is presented for each of the four detector locations. This will serve as input for the first iteration in the design of the detector stations. In particular, the available longitudinal space determines whether a station can be composed of two units for the measurement of local track angles and, if so, how long the lever arm between the units can be. The resulting resolution improvement in the kinematic reconstruction is then derived in Section~\ref{sec:resolution}.

A preliminary space reservation request for the four locations has been made by CMS~\cite{hllhc-coordinationgroup202002}.

\subsubsection{The 196\,m Station}

\begin{figure}[h!]
\begin{center}
\includegraphics[width=\textwidth]{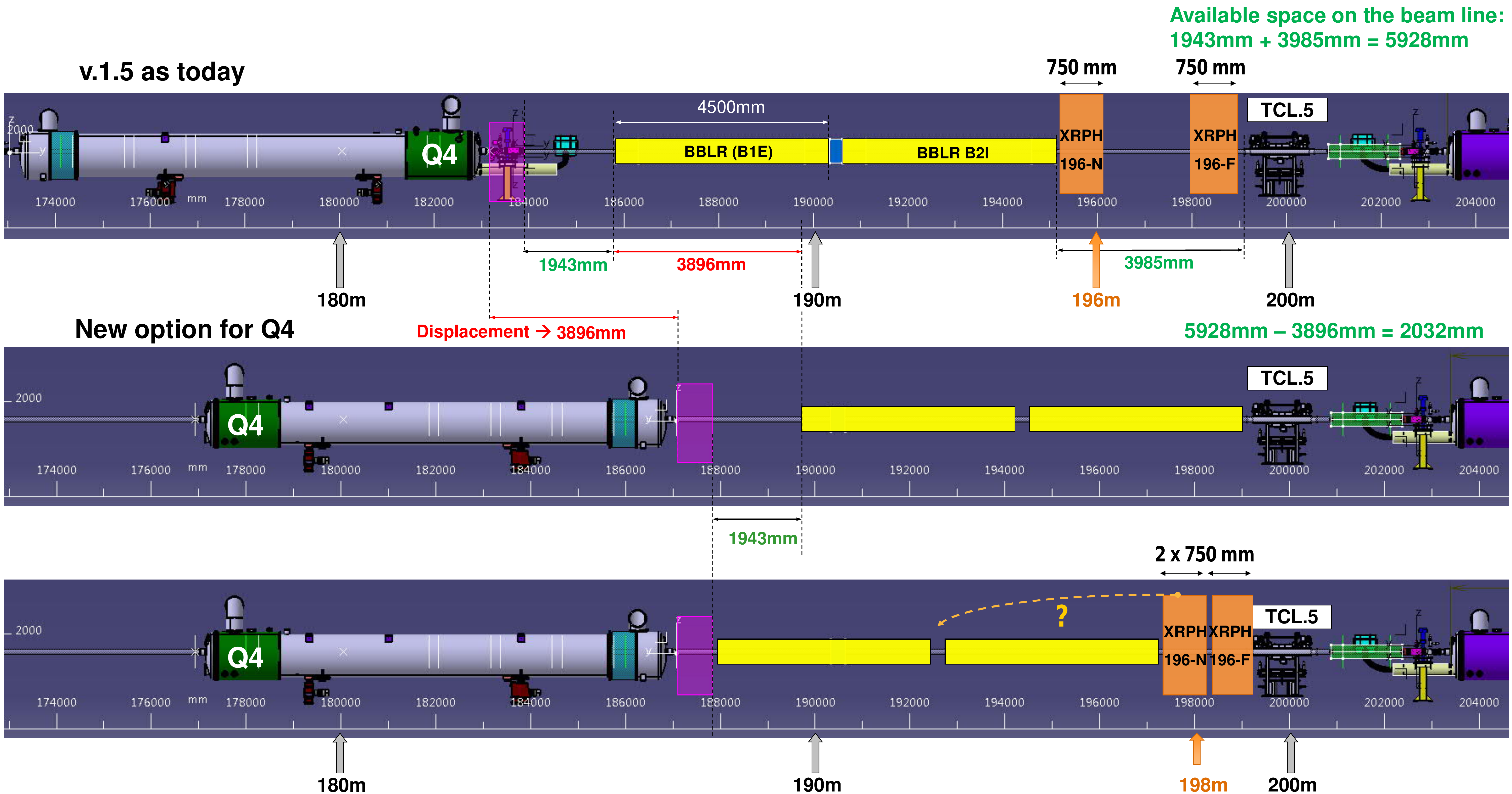}
\end{center}
\caption{Preliminary layout~\cite{slide_cabana} of the sector between Q4 and Q5 with tentative detector units, labelled as horizontal Roman Pots (XRPH). Three possible scenarios are shown: present standard layout (top), flipped-Q4 scenario (middle and bottom) with different possible locations for the BBLR~\cite{bblr}.}
\label{fig:layout196}
\end{figure}
The layout of the sector from the quadrupole Q4 to the TCL5 collimator (Fig.~\ref{fig:layout196}) is still uncertain because it might be necessary to flip and shift Q4 for irradiation reasons. Furthermore, there is a space reservation for two Long-Range Beam-Beam compensation units (BBLR)~\cite{bblr}, one per beam. 
After first discussions with the people in charge of the BBLR, it seems possible to integrate two detector units of a tentative length of about 750\,mm in all configurations. Resolution and acceptance considerations (Section~\ref{sec:resolution}) would favour a global position as close to TCL5 as possible and a lever arm of a few metres between the units. In the present scenario without Q4 flip, about 3.5\,m could be realized, whereas in the flipped scenario the reduced space of 1.9\,m would limit this lever arm. However, there are preliminary considerations about a new BBLR design where the units for the two beams would not be arranged in series but parallel. In that case the total longitudinal space needed by the BBLR would be about 5.5\,m instead of two times 4.5\,m. The two detector units could then be placed upstream and downstream of the BBLR with a lever arm of 5.5\,m. The integration discussions will continue with good chances for a satisfactory solution.

The beam pipe in this region is circular with inner and outer diameters of 80 and 84\,mm, respectively, i.e.\ the same geometry as at the present LHC. A special requirement of this sector is that all instrumentation on the beam line must be remotely movable by $\pm$(2$-$2.5)\,mm, both horizontally and vertically. Developments of movable support structures are ongoing on the machine side and should be adaptable for PPS needs.

\subsubsection{The 220\,m Station}

\begin{figure}[h!]
\begin{center}
\includegraphics[width=\textwidth]{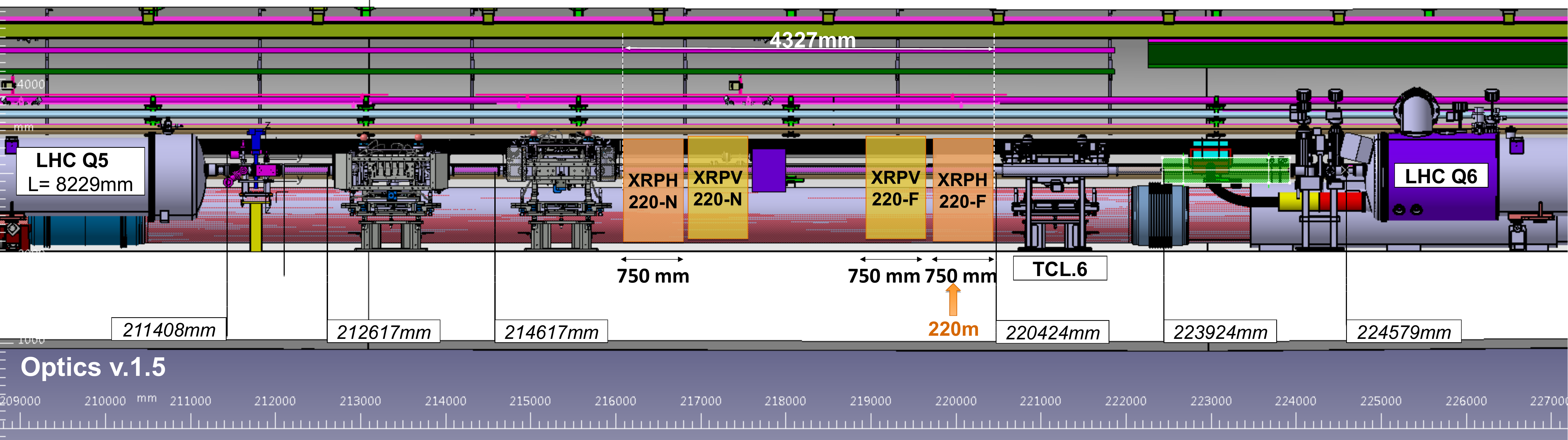}
\end{center}
\caption{Layout~\cite{slide_cabana} of the sector between Q5 and Q6 with tentative horizontal and vertical detector units (XRPH and XRPV, respectively) near 220\,m.}
\label{fig:layout220}
\end{figure}
Near 220\,m (Fig.~\ref{fig:layout220}) the space situation is the most relaxed of all the locations under consideration. A total length of 4327\,mm is potentially available. This would allow for two detector units, called 220-N (``Near'') and 220-F (``Far''), with a lever arm of 4 to 5\,m to improve the precision of track reconstruction without jeopardising the acceptance overlap of the two units (Section~\ref{sec:resolution}). There should also be space for two pairs of detector vessels approaching the beam vertically, which are needed by the present techniques for alignment and optics calibration (Section~\ref{sec:alignment}). The beam pipe in this region is circular with inner and outer diameters of 80 and 84\,mm, respectively, i.e.\ the same geometry as at the present LHC.

\subsubsection{The 234\,m Station}

\begin{figure}[h!]
\begin{center}
\includegraphics[width=\textwidth]{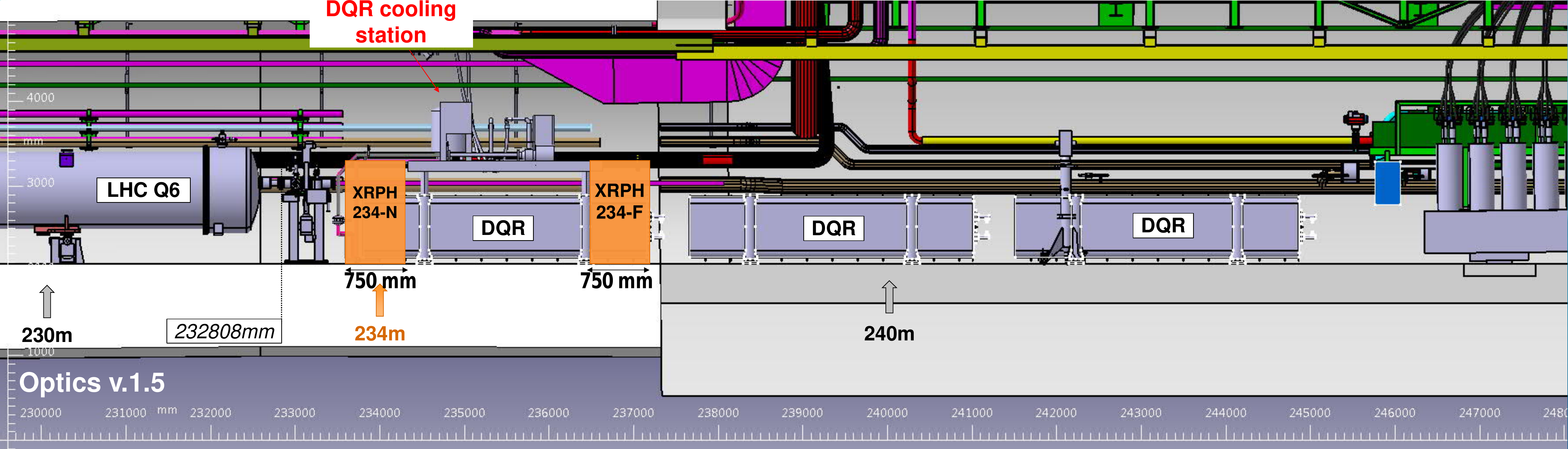}
\end{center}
\caption{Layout of the region behind the quadrupole Q6 with tentative detector units near 234\,m~\cite{slide_cabana}.}
\label{fig:layout234}
\end{figure}
In the current design of HL-LHC, the location envisaged for the 234\,m station is occupied by the first of the three DQR boxes underneath the beam pipes. However, it is probably possible to arrange these elements differently, by stacking two of them vertically, as already done in the interaction region IR7 at the present LHC. Discussions with the layout team are ongoing and look promising. This relocation would make space for two detector units (see Fig.~\ref{fig:layout234} for a preliminary idea), called 234-N and 234-F, with a lever arm of about 3\,m. The inner and outer beam pipe diameters are 80 and 84\,mm, respectively, allowing in principle the integration of conventional Roman Pot units.

\subsubsection{The 420\,m Station}

\begin{sidewaysfigure}
\begin{center}
\includegraphics[width=\textwidth]{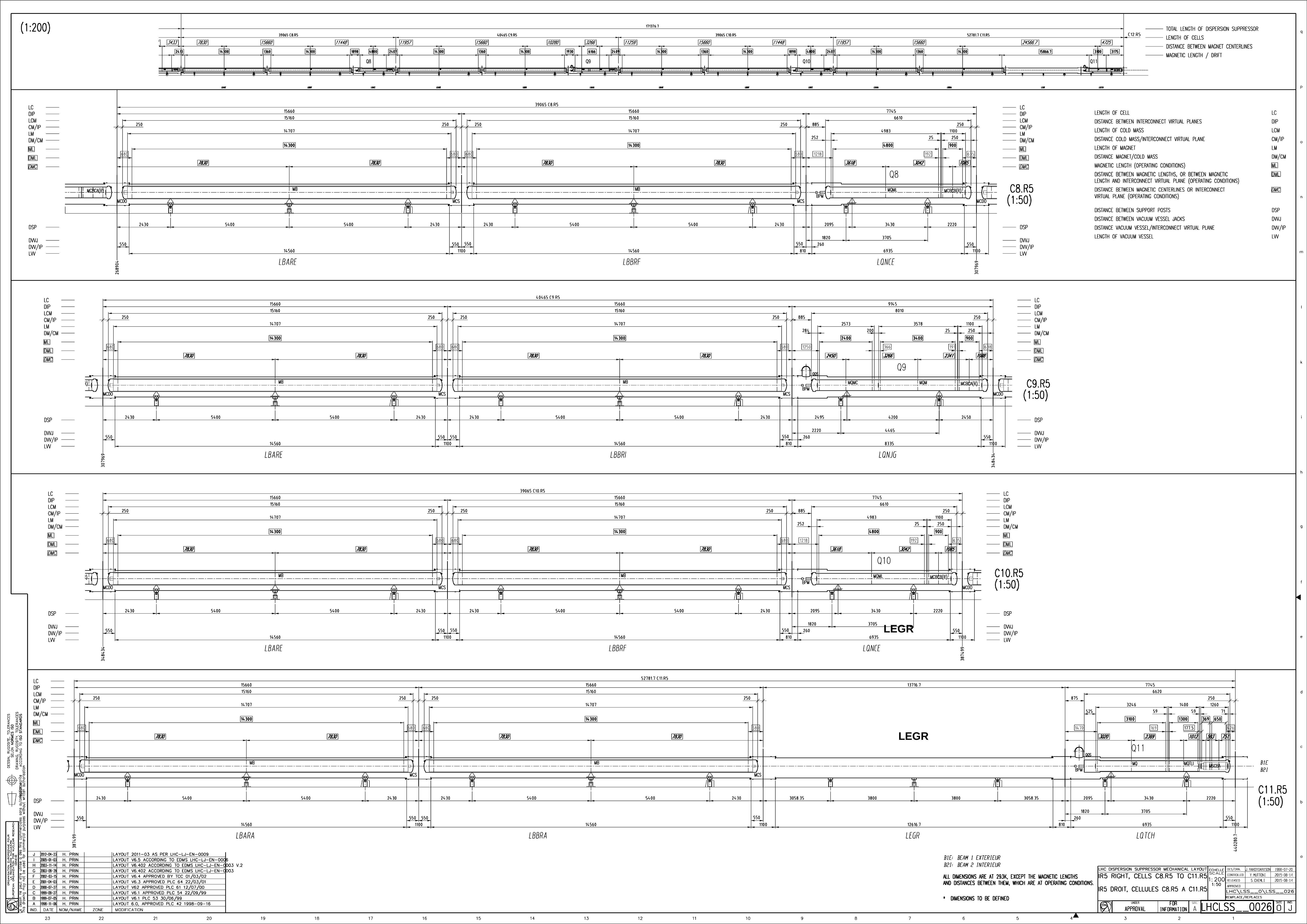}
\end{center}
\caption{Layout of cell C11R5 just behind Q10~\cite{edms-layout-420}. The region of interest for a detector station lies in the location of the empty cryostat LEGR.}
\label{fig:layout420}
\end{sidewaysfigure}
The region of interest near 420\,m (Fig.~\ref{fig:layout420}) lies in the so-called ``missing magnet'' location previously studied by the FP420 project~\cite{fp420}. It has the following characteristics:
\begin{itemize}
\item This location consists of an empty cryostat of 12.6\,m length in cell C11.R5 between the magnets Q10 and Q11. To install any equipment in the beam line requires the construction of a cryogenic bypass. A solution already exists: for the installation of the new TCLD collimators (``Target Collimator Long Dispersion suppressor'') in equivalent locations in IR2 and IR7 a connection cryostat has been developed~\cite{tcld,connectioncryostat}. Within the space created by the connection cryostat there will probably be no room for a second detector unit. Hence the kinematic reconstruction will not profit from a lever arm longer than the unit itself (see discussion in Section~\ref{sec:resolution}).
\item The positive sign of the dispersion (Fig.~\ref{fig:d-sigma-xi_vs_s}) means that scattered proton tracks run in the narrow space between the outgoing and the incoming beam pipe: the centres of the two beams have a distance of only 194\,mm. This precludes the installation of a standard Roman Pot. 
\item The beam pipes have a truncated circular shape as outlined in Fig.~\ref{fig:xymap} (page~\pageref{fig:xymap}, bottom right) with a diameter of 44.0\,mm and a vertical dimension of 34.3\,mm.
\end{itemize}
Ideas for movable detector devices compatible with these space constraints will be presented in Section~\ref{sec:vessels}.

\clearpage
\section{Machine Parameters and their Impact on Forward Physics}
\label{sec:machine}
\subsection{Beam Parameters}
Table~\ref{tab:hllhc-parameters} lists the beam parameters foreseen for HL-LHC~\cite{hllhc-tdr} in comparison with the nominal LHC parameters. 

\begin{table}[h!]
\caption{HL-LHC nominal parameters~\cite{hllhc-tdr} for 25 ns operation for two production modes of the LHC beam in the injectors.}
\label{tab:hllhc-parameters}
\includegraphics[width=1.01\textwidth]{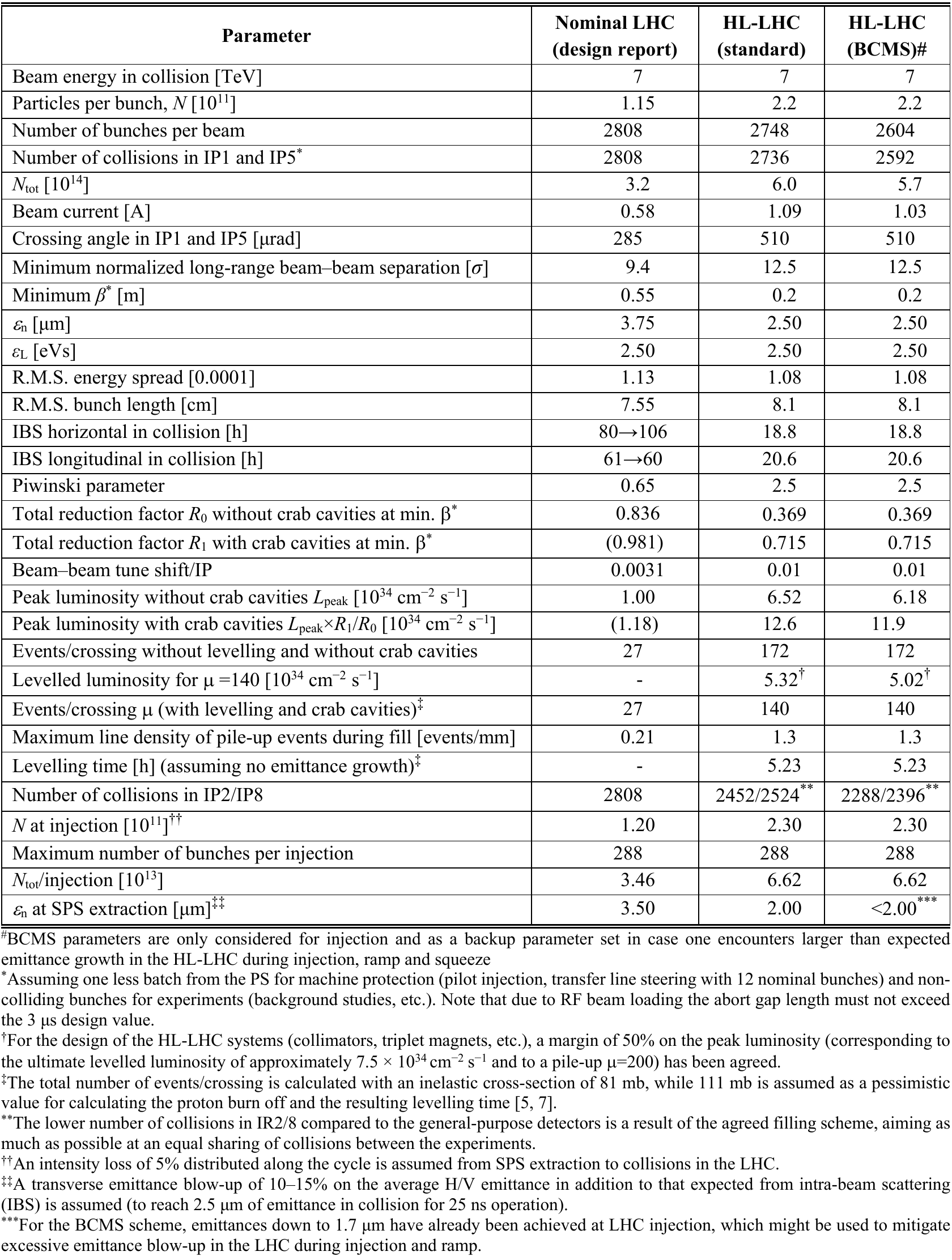}
\end{table}
For the purposes of near-beam detector instrumentation and forward physics, the following key changes relative to the present-day LHC need to be pointed out:
\begin{itemize}
\item The number of protons per bunch, $N$, and hence the beam current, increases by almost a factor 2. This implies a factor four in impedance heating. Therefore special care will have to be taken in impedance mitigation techniques and detector cooling (Section~\ref{sec:impedance}).
\item The luminosity increase leads to an increase of detector-vessel induced shower rates by the same factor. This shower development will have to be carefully studied in collaboration with the collimation and FLUKA simulation~\cite{fluka} team to avoid radiation-induced heat-load problems in machine elements downstream of the detector stations. Mitigation and monitoring are discussed in Section~\ref{sec:showers}. Another side effect of the high luminosity is the irradiation of all instrumentation in the tunnel (Section~\ref{sec:radiation}). This has to be taken into account in the choice of detector and electronics technology (Section~\ref{sec:detectorenvironment}) and the design of the vessel and movement system (Section~\ref{sec:movementsystem}).
\item The table shows minimum values for $\beta^{*}$ and maximum values of the crossing-angle, both well beyond the LHC extrema. In fact, owing to the strong luminosity levelling requirement, both values can change over a wide range during a fill, as discussed in Section~\ref{sec:optics}.
\item The beam emittance is expected to be lower, leading to thinner beams. Hence a new nominal normalized emittance value of $\varepsilon_{n} = 2.5\,\mu$m\,rad (instead of 3.5\,$\mu$m\,rad), used for all nominal beam width calculations and collimator position settings (Section~\ref{sec:collimation}), was introduced. 
\item The increased pileup ($\mu \sim 140$) imposes stringent requirements on the time-of-flight resolution for longitudinal vertex identification (Section~\ref{sec:timing}).
\item The table mentions the introduction of crab cavities. Their effect on forward protons is studied in Section~\ref{sec:crab} and found to be negligible. 
\end{itemize}

\subsection{Optics and Crossing-Angle}
\label{sec:optics}
\begin{figure}[h!]
\begin{center}
\includegraphics[width=0.9\textwidth]{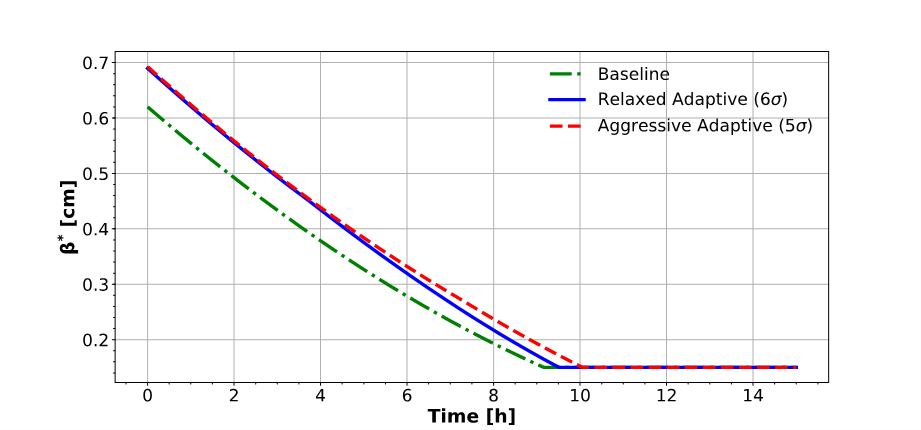}
\includegraphics[width=0.9\textwidth]{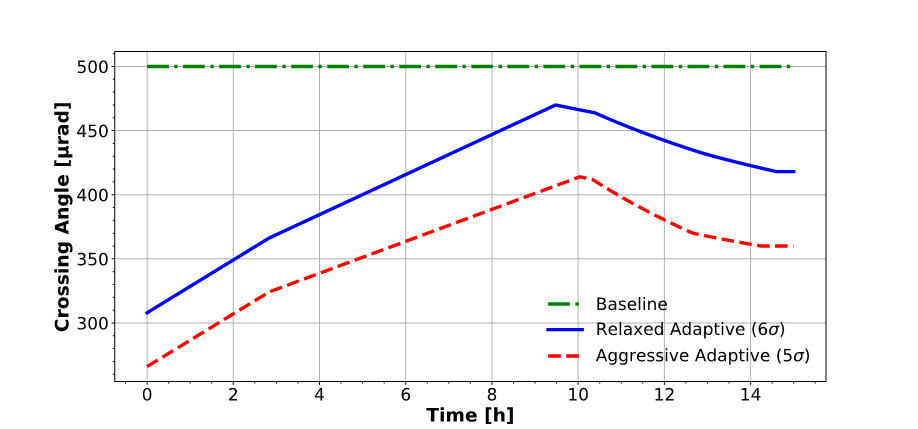}
\end{center}
\vspace{-0.5cm} 
\caption{Evolution of $\beta^{*}$ (top) and the full crossing-angle $\alpha$ (bottom) during a fill according to different luminosity levelling schemes (extracted from Ref.~\cite{levellingtalk}). A discussion of these schemes (``baseline'', ``relaxed adaptive'', ``aggressive adaptive'') would go beyond the scope of this document and is not needed for its understanding.}
\label{fig:levelling-1dplots}
\end{figure}

The optics at HL-LHC will follow the logic of the ATS optics~\cite{ats-optics} (see also Ref.~\cite{dn-19-026} for a detailed discussion of its impact on PPS) where the final part of the $\beta^{*}$ squeeze in an IP is obtained by adjusting the beta function outside the interaction region (IR) while leaving the magnet strengths in the IR unchanged. This strategy leads to a constant transport matrix from the IP to the leading-proton detectors even when $\beta^{*}$ changes during the fill. This invariance property of the optics is advantageous for forward physics because it avoids continuous optics recalibrations during fills. Experimentally, its validity has been verified in the analysis of Run-2 data. At HL-LHC this feature will be even more important because the luminosity levelling for pileup mitigation implies a variation of $\beta^{*}$ over a range of 70\,cm to 15\,cm from the beginning to the end of a fill (Fig.~\ref{fig:levelling-1dplots}, top). 

The invariance of the transport matrix elements $v_{\rm XRP}$ (magnification of the vertex position) and $L_{\rm XRP}$ (effective length translating a scattering angle into a track displacement at the detector) from the IP to a detector plane (in an XRP) can be exploited to derive the $\beta^{*}$-dependence of the beam width $\sigma_{\rm XRP}$ needed in Eq.~(\ref{eqn:xi-min}) for calculating $\xi_{\rm min}$ from machine parameters:
\begin{eqnarray}
v_{\rm XRP} &=& \sqrt{\frac{\beta_{\rm XRP}}{\beta^{*}}}\, \cos \mu_{\rm XRP} = \text{const.} \: , \label{eqn:v} \\
L_{\rm XRP} &=& \sqrt{\beta_{\rm XRP}\,\beta^{*}}\, \sin \mu_{\rm XRP} = \text{const.} \: , \label{eqn:L} 
\end{eqnarray}
where $\mu_{\rm XRP}$ is the phase advance of the beta function from the IP to the XRPs. Solving~(\ref{eqn:v}) and~(\ref{eqn:L}) for $\beta_{\rm XRP}$, one obtains
\begin{equation}
\beta_{\rm XRP} = v_{\rm XRP}^{2} \, \beta^{*} + \frac{L_{\rm XRP}^{2}}{\beta^{*}} \: ,
\end{equation}
and hence
\begin{equation}
\sigma_{\rm XRP} = \sqrt{\frac{\varepsilon_{n}\,\beta_{\rm XRP}}{\gamma}} =
\sqrt{\frac{\varepsilon_{n}}{\gamma}\left(v_{\rm XRP}^{2} \, \beta^{*} + \frac{L_{\rm XRP}^{2}}{\beta^{*}}\right)} \: ,
\end{equation}
where $\gamma$ is the Lorentz factor of the proton.
The constants $v_{\rm XRP}$ and $L_{\rm XRP}$ have to be determined with a MAD-X simulation~\cite{MADX} for any example $\beta^{*}$. If in future optics versions the strict invariance of $v_{\rm XRP}$ and $L_{\rm XRP}$ is lost, the simple analytical solution will have to be replaced with numerical techniques.

\begin{figure}[h!]
\begin{center}
\includegraphics[width=0.49\textwidth]{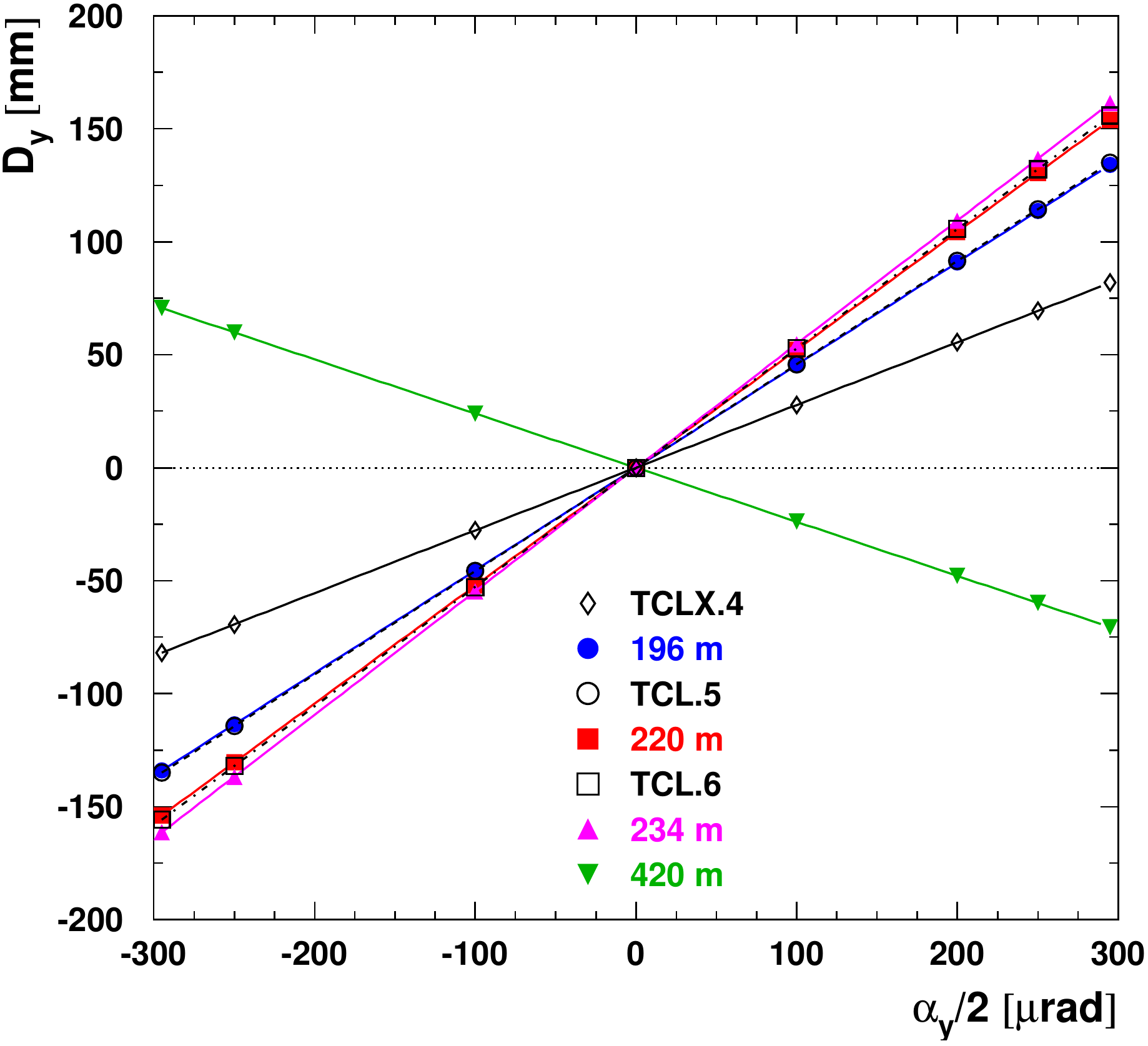}
\includegraphics[width=0.49\textwidth]{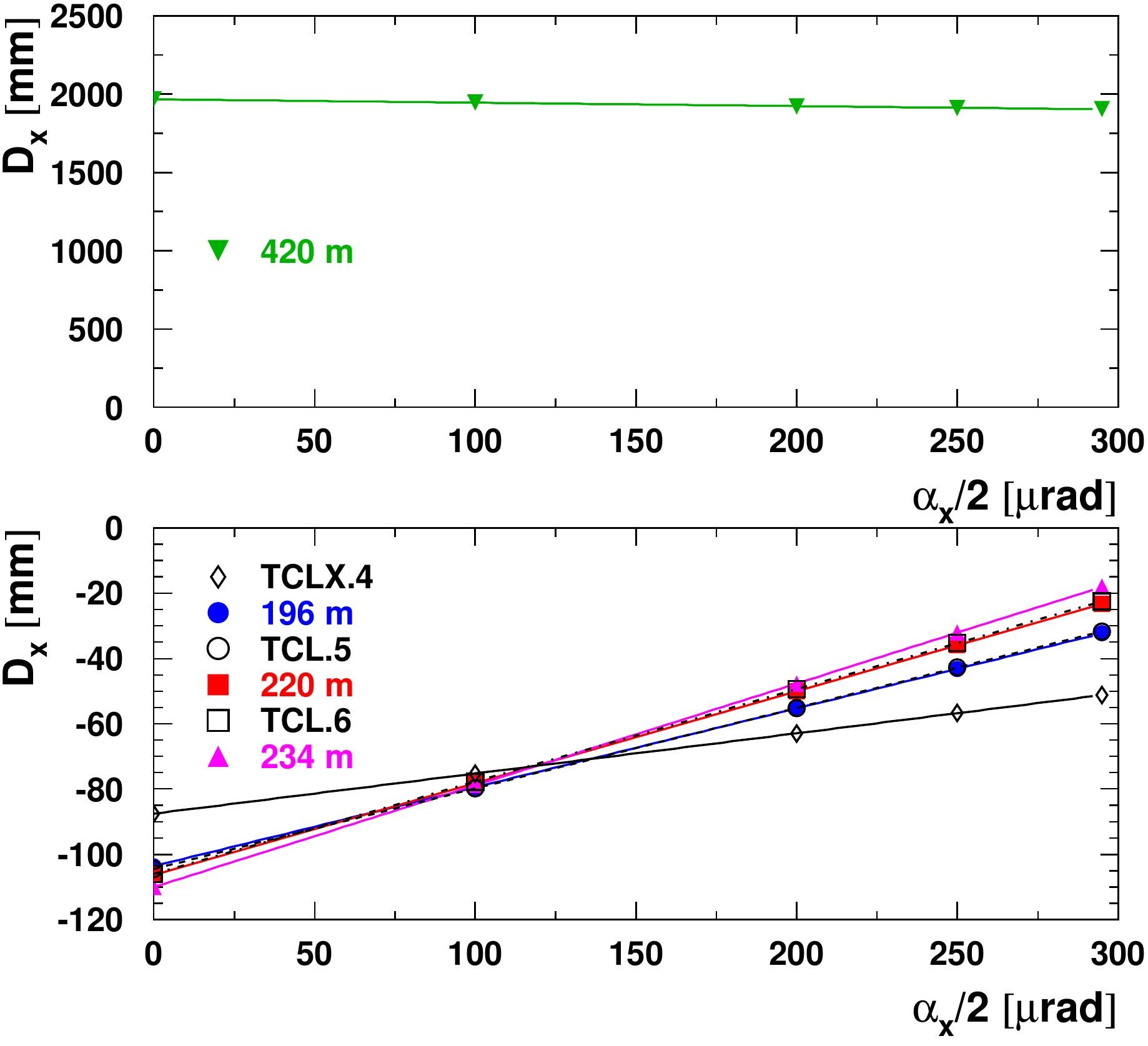}
\end{center}
\caption{Dispersion at the chosen detector locations and at the TCL collimators as a function of the crossing-angle. Left: vertical crossing; here the horizontal dispersion is independent of the crossing-angle and has the same value as shown in the right-hand plot for $\alpha_{x}=0$. Right: horizontal crossing; here the vertical dispersion (not shown) is zero. 
Because of their proximity along the beamline, the dispersion curves for TCL5 (black dashed) and the 196\,m station (blue) almost coincide; the same is true for TCL6 (black dash-dotted) and the 220\,m station (red).}
\label{fig:dispersion-vs-alpha-hori}
\end{figure}

In some luminosity levelling schemes, the crossing-angle $\alpha$ will change
simultaneously with $\beta^{*}$ (Fig.~\ref{fig:levelling-1dplots}, bottom), with a linear impact on the dispersion in the crossing-plane. The linearity is shown in Fig.~\ref{fig:dispersion-vs-alpha-hori} for HL-LHC and has been experimentally verified in Run 2. The sequence of points $(\alpha/2, \beta^{*})$ in a fill defines a levelling trajectory along which the performance of the proton spectrometer will vary.

Depending on whether the crossing-plane is horizontal or vertical, the crossing-angle variation affects the proton acceptance in near-beam detectors at different levels, as discussed in Section~\ref{sec:acceptance}. Given that at the time of the studies for this report the crossing-angle plane in IP5 (horizontal as until LS3, or vertical) had not yet been decided, both options were investigated. 

{\bf Since -- as will be concluded at the end of Section~\ref{sec:performance} -- vertical beam crossing in IP5 is overall advantageous, CMS requested it in December 2018~\cite{hllhc-27coordinationgroup}. In June 2020~\cite{hllhc-executivecommittee20200622}, this favourable scheme, vertical crossing in IP5 and hence horizontal crossing in IP1, was decided to be implemented.}

\subsection{Collimation Scheme}
\label{sec:collimation}
Like all movable devices, near-beam detectors are integrated into the collimator hierarchy of LHC's multistage collimation system where each stage has to stay in the protected shadow of its preceding stage. Experiment devices are the fourth stage in this hierarchy and thus need to be retracted by a certain distance relative to the tertiary stage, the TCTs. In Eq.~(\ref{eqn:xi-min}), giving the minimum measurable momentum loss of a proton, this retraction was expressed as a combination of a part $\Delta n$ in terms of beam widths and a fixed part $\Delta d$ in absolute distance units.

The debris-absorbing TCLs upstream of a leading proton detector are important since they usually constitute the aperture limitations defining the highest observable values of proton momentum loss $\xi$ and scattering angles $\theta_{x}$, as already seen in Eq.~(\ref{eqn:xi-max}) and discussed further in Section~\ref{sec:maxmass}.

In full generality, the distances of the collimator jaws from the beam centre may have to follow the evolution of $\beta^{*}$ during a fill. This dependence is then transferred into the proton acceptance, e.g.\ via $n_{\rm TCT}(\beta^{*})$ in Eq.~(\ref{eqn:xi-min}). The collimation strategy presently foreseen for HL-LHC~\cite{collimation} keeps the TCT and TCL gaps constant in absolute distance. This distance is defined as a multiple of the beam width at the smallest $\beta^{*}$ reachable in a fill, i.e.\ at $\beta^{*}_{0} = 15$\,cm:
\begin{eqnarray}
d_{\rm TCT} &=& 12.9\, \sigma_{\rm TCT}(\beta^{*}_{0}) \: , \\
d_{\rm TCL} &=& 14.2\, \sigma_{\rm TCL}(\beta^{*}_{0}) \: .
\end{eqnarray}
With the approximate relation, 
\begin{equation}
\sigma_{\rm TCT / TCL}(\beta^{*}) \propto \frac{1}{\sqrt{\beta^{*}}} \: ,
\end{equation}
follows
\begin{equation}
n_{\rm TCT}(\beta^{*}) = \frac{d_{\rm TCT}}{\sigma_{\rm TCT}(\beta^{*})} 
= 12.9\, \frac{\sigma_{\rm TCT}(\beta^{*}_{0})}{\sigma_{\rm TCT}(\beta^{*})} 
= 12.9\, \sqrt{\frac{\beta^{*}}{15\,\rm cm}} \: ,
\end{equation}
which is needed for calculating $|\xi|_{\rm min}$ with Eq.~(\ref{eqn:xi-min}).

\subsection{Crab Cavities}
\label{sec:crab}
\begin{figure}[h!]
\begin{center}
\includegraphics[width=0.75\textwidth]{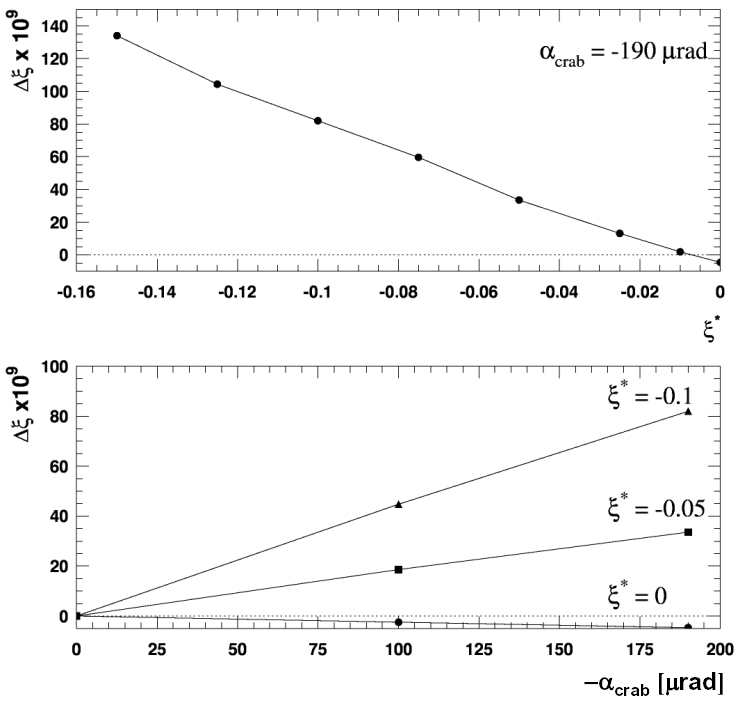}
\end{center}
\vspace{-0.5cm} 
\caption{Momentum change caused by the crab cavities to protons originating at vertex (0,0,0) and measured in the 196\,m detector station, as a function of the crabbing angle. Here a crossing-angle $\alpha/2 = 250\,\mu$rad and round optics with $\beta^{*}=15\,$cm were assumed.}
\label{fig:crab-xivsxi}
\end{figure}

\begin{figure}[h!]
\begin{center}
\includegraphics[width=0.75\textwidth]{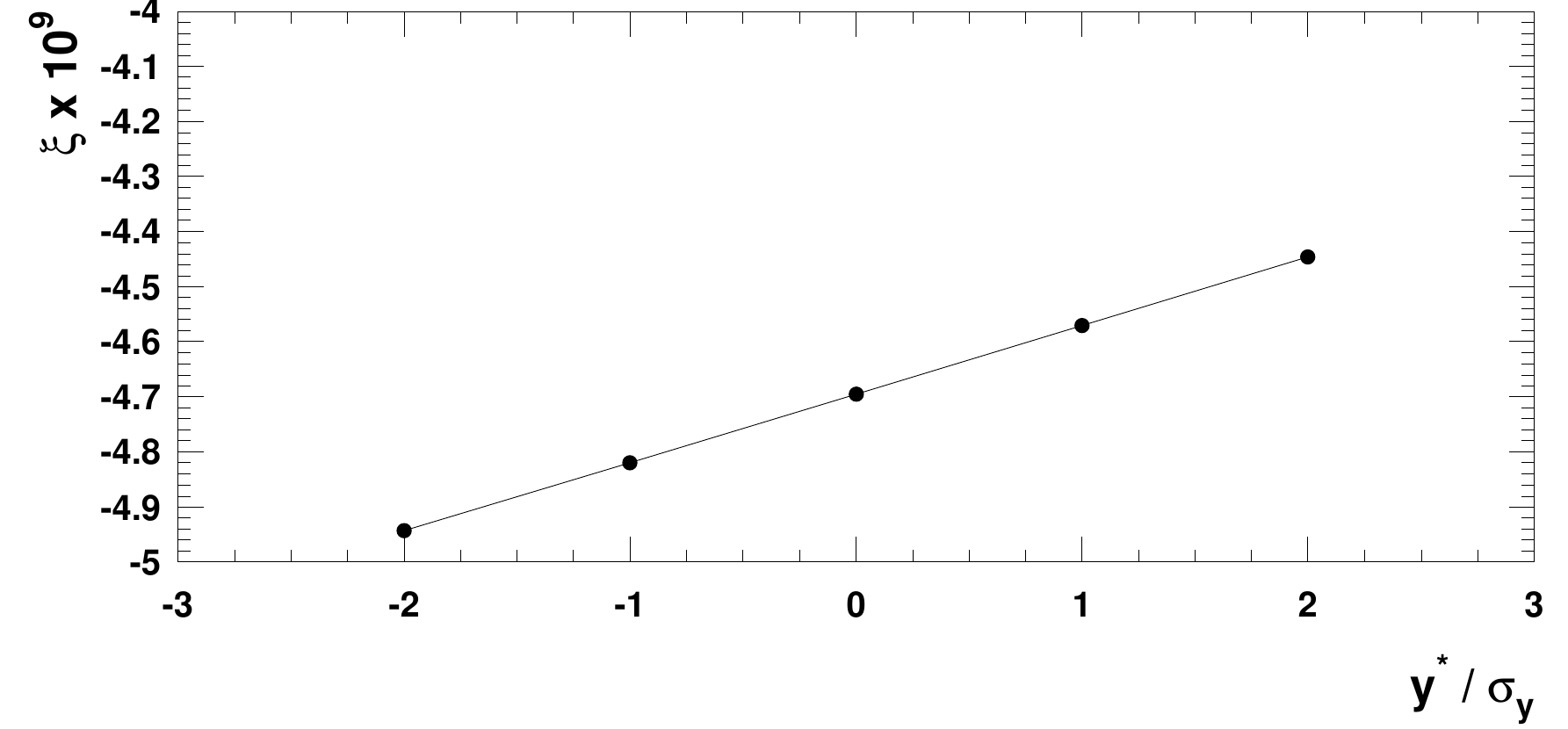}
\includegraphics[width=0.75\textwidth]{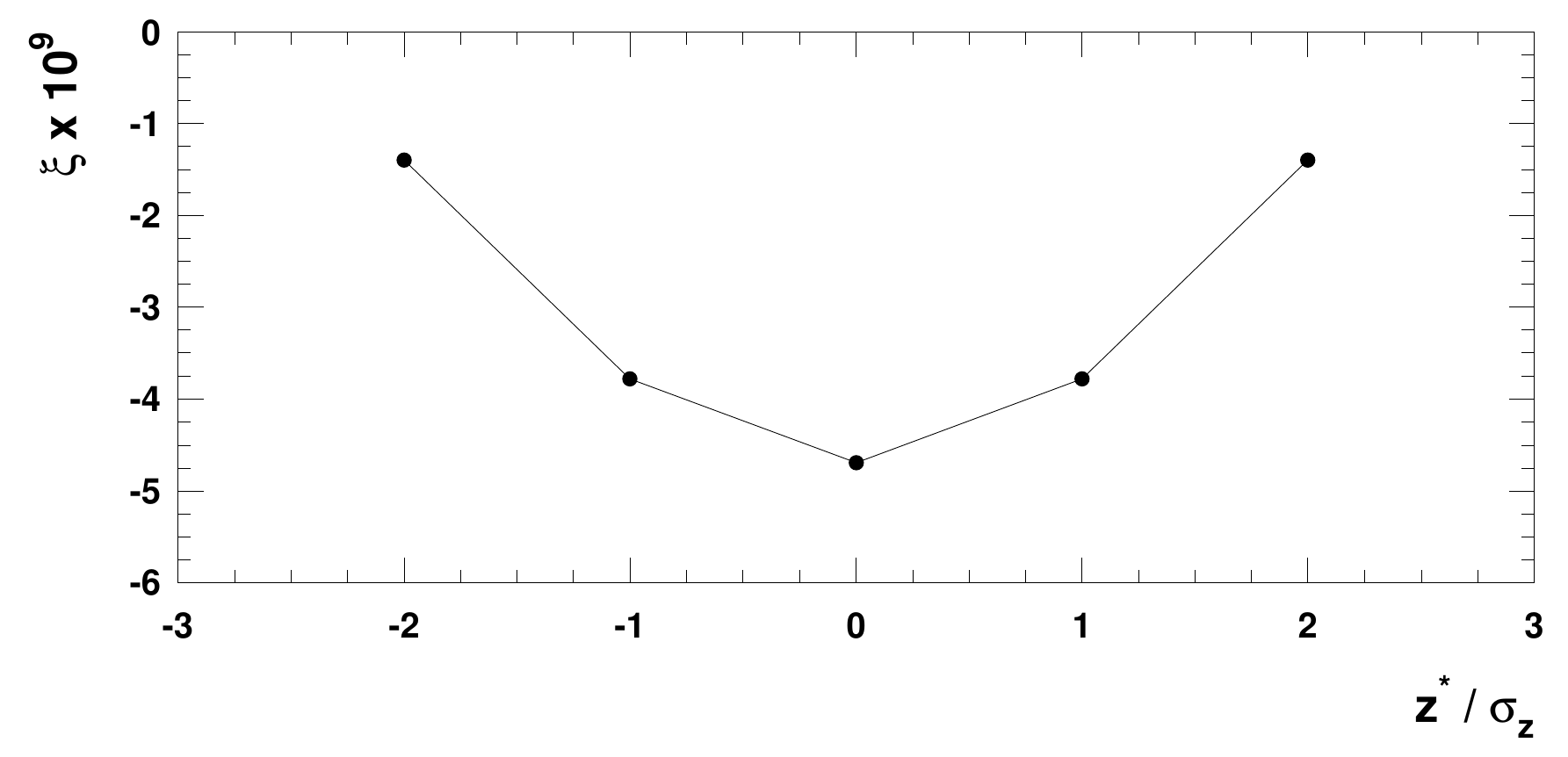}
\end{center}
\vspace{-0.5cm} 
\caption{Momentum loss caused by the crab cavities to protons originating at peripheral vertices $(0,y^{*},z^{*})$ and measured in the 196\,m detector station, for a crossing-angle $\alpha/2 = 250\,\mu$rad, a round optics with $\beta^{*}=15\,$cm, and maximum crabbing angle $\alpha_{\rm crab} = -190\,\mu$rad, as a function of the normalized vertical (top) and longitudinal (bottom) vertex positions.}
\label{fig:crab-xivsvertex}
\end{figure}

\begin{figure}[h!]
\begin{center}
\includegraphics[width=0.75\textwidth]{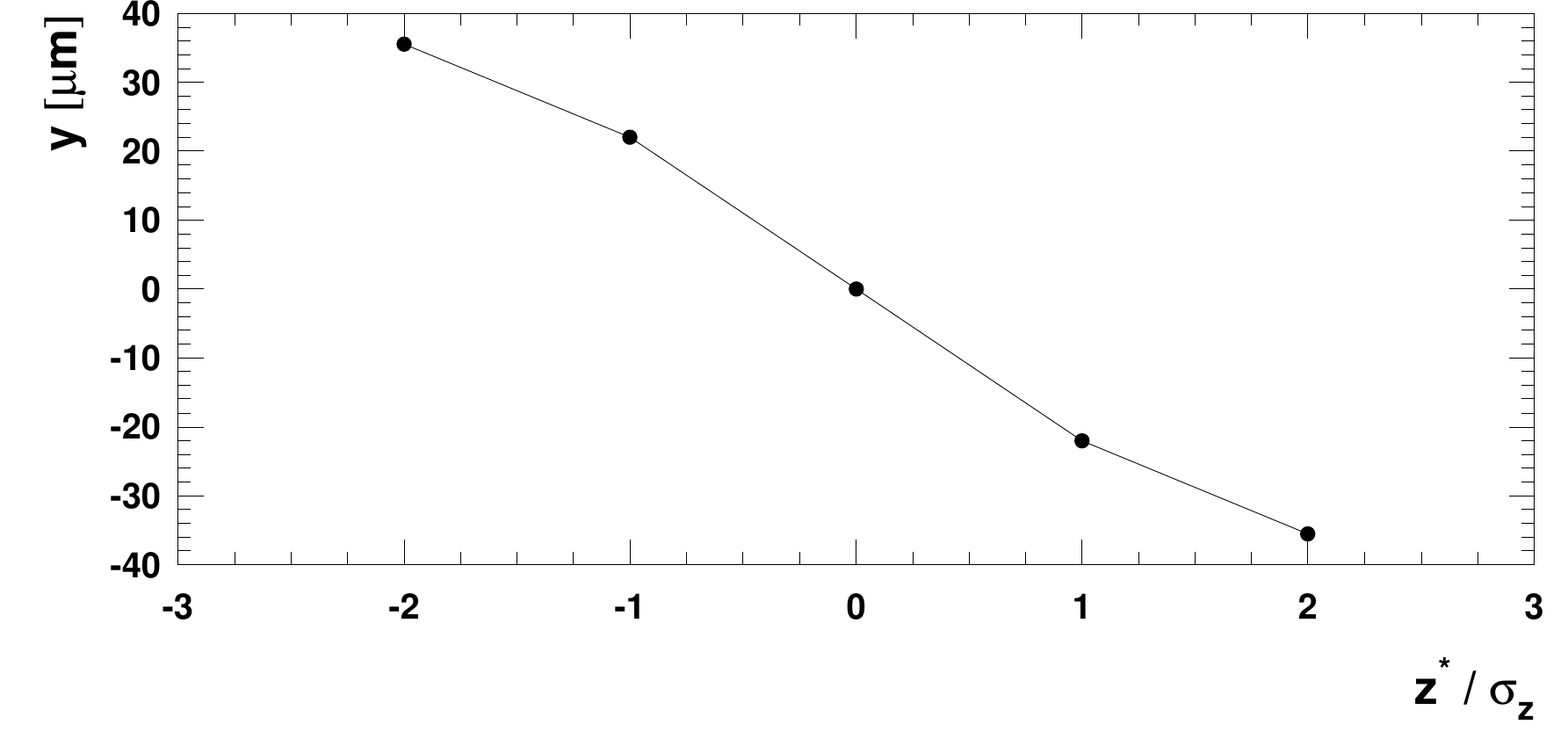}
\end{center}
\vspace{-0.5cm} 
\caption{Vertical track displacement caused by the crab cavities to protons originating at trailing vertices $(0,0,z^{*})$ and measured in the 196\,m detector station, for a crossing-angle $\alpha/2 = 250\,\mu$rad, a round optics with $\beta^{*}=15\,$cm, and maximum crabbing angle $\alpha_{\rm crab} = -190\,\mu$rad, as a function of the normalized longitudinal vertex position.}
\label{fig:crab-yvsvertex}
\end{figure}

\begin{figure}[h!]
\begin{center}
\includegraphics[width=0.75\textwidth]{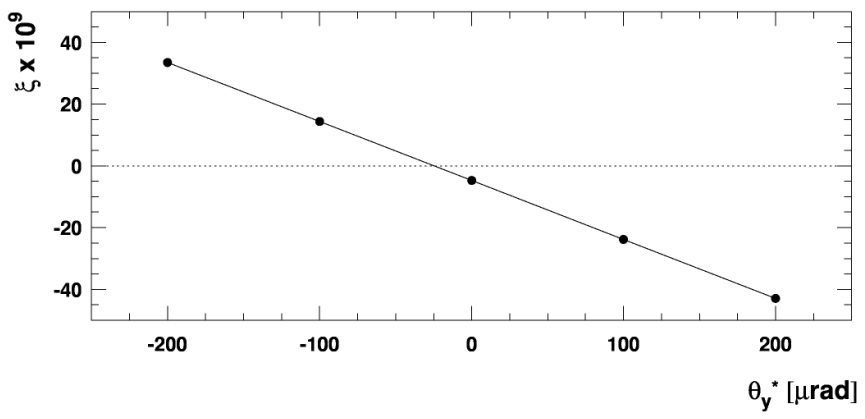}
\end{center}
\vspace{-0.5cm} 
\caption{Momentum change caused by the crab cavities to protons originating at the central vertex (0,0,0) but with a vertical scattering angle $\theta_{y}^{*}$, and measured in the 196\,m detector station, for a crossing-angle $\alpha/2 = 250\,\mu$rad, a round optics with $\beta^{*}=15\,$cm, and maximum crabbing angle $\alpha_{\rm crab} = -190\,\mu$rad.}
\label{fig:crab-xivstheta}
\end{figure}

The crab cavities (see Section~4.2 in Ref.~\cite{hllhc-tdr}), to be installed in the range 155$-$163\,m from IP5, i.e.\ between TCLX.4 and Q4, will rotate the proton bunches in the crossing plane $(z, X)$ (where $X = x,y$ for horizontal and vertical crossing, respectively) in order to compensate for the luminosity loss due to the crossing-angle $\alpha$ via the geometrical factor $\frac{\sigma_{X}}{\sqrt{\sigma_{X}^{2}\cos^{2}\frac{\alpha}{2}+\sigma_{z}^{2}\sin^{2}\frac{\alpha}{2}}}$. They therefore have an impact on the proton transport matrix that had to be studied with MAD-X~\cite{MADX}. The studies were performed for the case of a vertical crossing angle $\alpha/2 = 250\,\mu$rad and a maximal vertical crabbing angle $\alpha_{\rm crab} = -190\,\mu$rad. The following effects were observed:

\begin{itemize}
\item For protons originating in the centre of the vertex distribution, $(x^{*},y^{*},z^{*}) = (0,0,0)$, the only visible effect is a tiny modification of the momentum loss, proportional to the crabbing angle and linearly dependent on the original $\xi^{*}$ (Fig.~\ref{fig:crab-xivsxi}). The resulting $\Delta \xi$ is at most of the order $10^{-7}$, i.e.\ negligible.
\item Protons originating from the periphery of the vertex distribution in the vertical plane ($y^{*} \ne 0$ or $z^{*} \ne 0$) are affected by the bunch rotation functionality of the crab cavities. They suffer a tiny momentum loss of the order $10^{-9}$ (Fig.~\ref{fig:crab-xivsvertex}). The trailing protons ($z^{*} \ne 0$) also receive a vertical displacement by a few tens of microns (Fig.~\ref{fig:crab-yvsvertex}). Again, both effects were found to be proportional to the crabbing angle. 
\item Protons with a vertical scattering angle suffer a tiny momentum shift of the order $10^{-8}$ (Fig.~\ref{fig:crab-xivstheta}). 
\item The optical functions, $v_{y}$ and $L_{y}$, are unaffected.
\item All quantities in the non-crossing direction ($x$ in the case studied here) are unaffected.
\end{itemize}

In summary, all proton transport effects induced by the crab cavities are very small and can be ignored.

\clearpage

\section{Kinematic Acceptance and Resolution in the Chosen Locations}
\label{sec:performance}
Based on the accelerator and beam parameters reported in the previous section, the expected spectrometer performance is derived, assuming that all four preferred detector locations are instrumented. 

The kinematic properties of the leading protons in CEP events can be expressed in terms of the variables $\xi$, $\theta_{x}^{*}$, $\theta_{y}^{*}$, $x^{*}$, $y^{*}$, $z^{*}$. 

The momentum loss $\xi$ and its acceptance range have already been introduced and discussed in Section~\ref{sec:locations} as a criterion in the search for appropriate PPS detector locations. The following treatment goes one step further by looking at the acceptance for the two leading protons not individually but combined in terms of the centrally produced mass according to Eq.~(\ref{eqn:mass-xi1-xi2}), and later also in terms of the production rapidity of that mass.

In addition, the scattering angle components $\theta_{x}^{*}$ and $\theta_{y}^{*}$ (or equivalently the transverse proton momenta $p_{x}^{*}$ and $p_{y}^{*}$) are considered.

The transverse vertex, $(x^{*}, y^{*})$, will be neglected here since its distribution will only be a few microns wide, far too narrow to be resolved with forward physics detectors.

Finally, the longitudinal vertex position, $z^{*}$, will be discussed only later in Section~\ref{sec:detectors}, in the context of vertex identification via time-of-flight measurements.

\subsection{Acceptance}
\label{sec:acceptance}

\subsubsection{Minimum Mass}
\begin{figure}[h!]
\begin{center}
\includegraphics[width=0.495\textwidth]{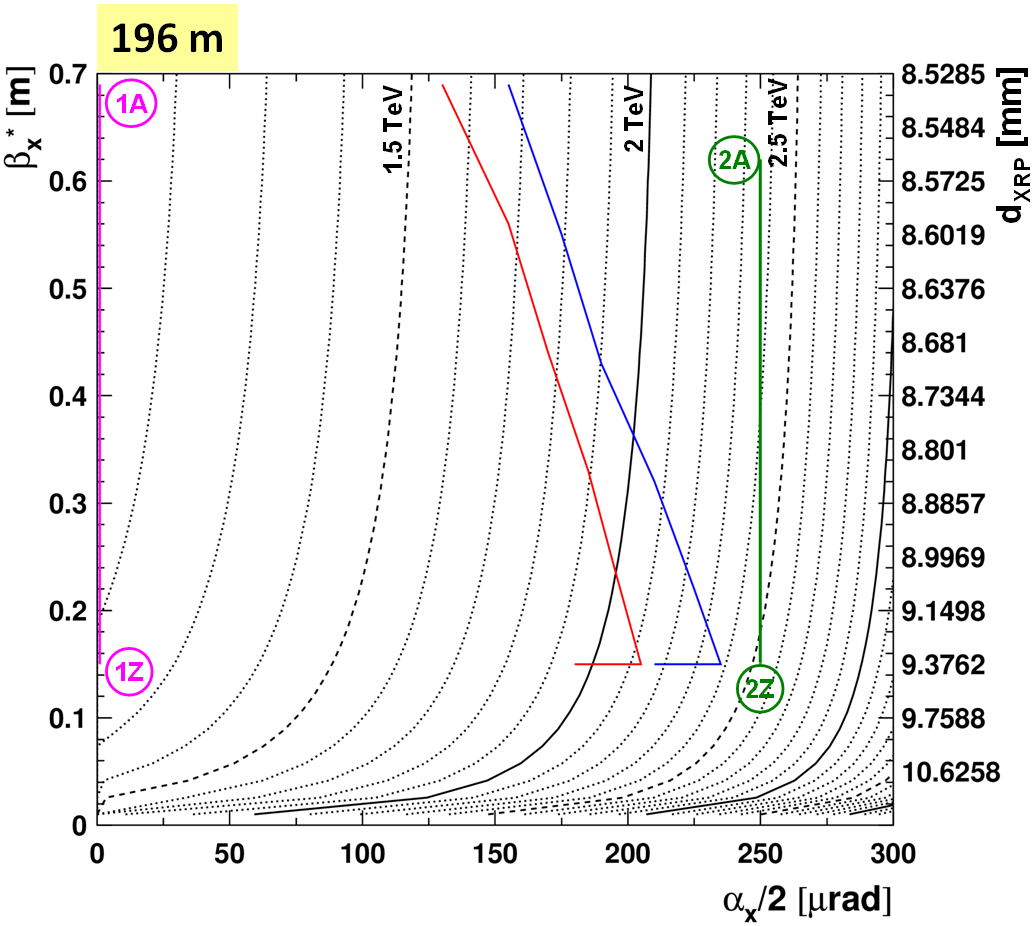}\hfill
\includegraphics[width=0.495\textwidth]{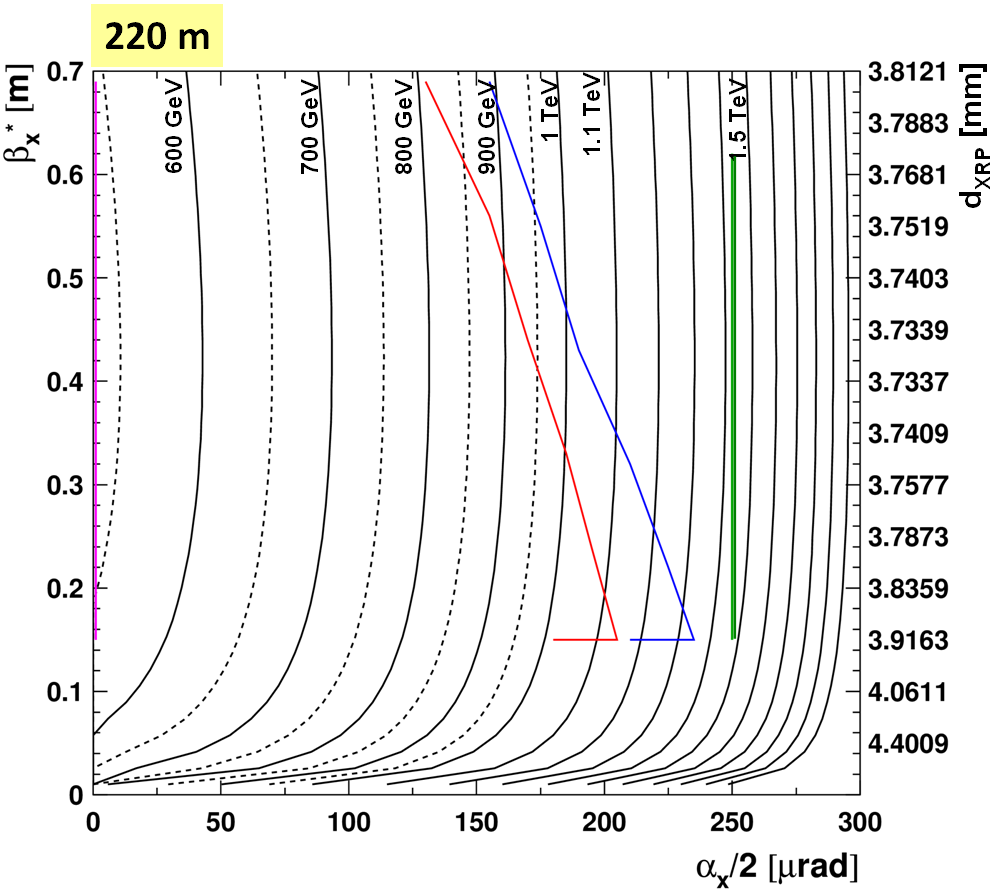}
\includegraphics[width=0.495\textwidth]{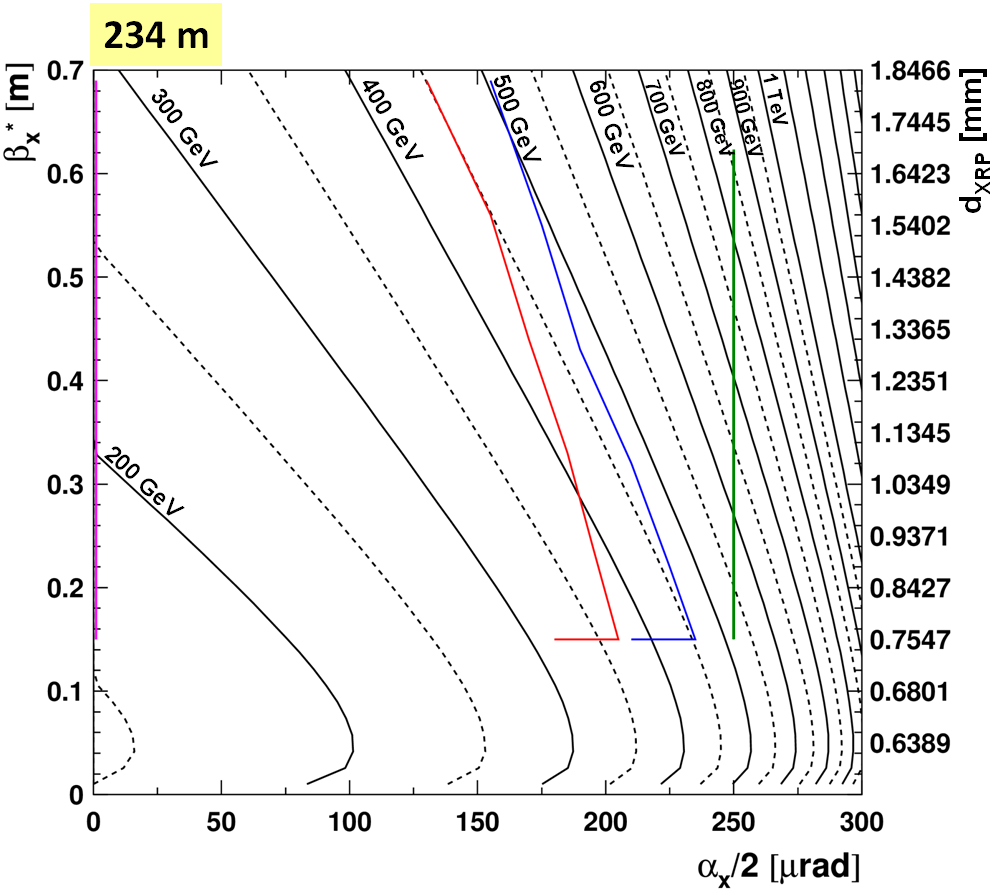}\hfill
\includegraphics[width=0.495\textwidth]{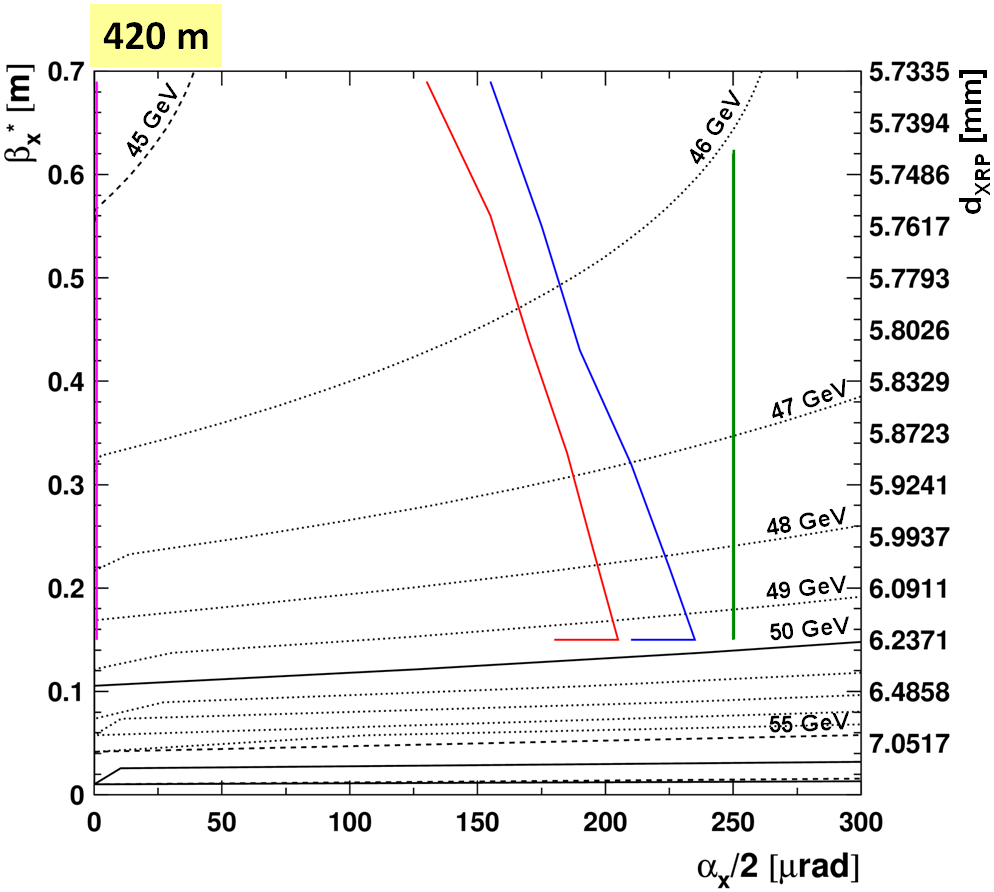}
\end{center}
\vspace{-0.5cm} 
\caption{Contour lines for the minimum accepted mass $M_{\rm min} = |\xi|_{\rm min} \sqrt{s}$ in the crossing-angle/optics parameter space $(\alpha_{x}/2, \beta^{*})$. On the right-hand ordinate the XRP approach distance is calculated from $\beta^{*}$. The coloured lines represent luminosity-levelling trajectories from Fig.~\ref{fig:levelling-1dplots} with the same colour code in case of horizontal crossing in IP5. For vertical crossing, the violet line represents any of the trajectories.
The labels (1A) -- (2Z) in the first panel define the trajectory start and end points used in Figs.~\ref{fig:m-y} and~\ref{fig:m-acceptance}.
}
\label{fig:m-min}
\end{figure}

The minimum central diffractive mass accepted at a location $\varsigma$ for given $\alpha$ and $\beta^{*}$ can be calculated using Eqs.~(\ref{eqn:mass-xi1-xi2}) and~(\ref{eqn:xi-min}). For simplicity, symmetric optics in the two beams, i.e.\ equal $|\xi|_{\rm min}$, are assumed:
\begin{equation}
M_{\rm min} = |\xi|_{\rm min}\,\sqrt{s} \: .
\end{equation}
The result of this calculation, contour lines of $M_{\rm min}$ in the beam parameter space $(\alpha/2, \beta^{*})$, is shown in Fig.~\ref{fig:m-min} for the four chosen detector locations. The luminosity-levelling trajectories from Section~\ref{sec:optics} (Fig.~\ref{fig:levelling-1dplots}) are drawn, too. The start point at the beginning of the fill is always at the maximum $\beta^{*}$ value.

From these graphs the following conclusions are drawn:
\begin{itemize}
\item The main driving factor for the minimum mass is the dispersion, which in turn is fully determined by the crossing-angle. The optics (via $\beta^{*}$) play a minor role.
\item If the 420\,m location can be instrumented, the minimum mass is about 50\,GeV with only a very weak dependence on the optics, the crossing-angle, and its plane (horizontal or vertical).
\item Without the 420\,m location, vertical crossing gives a much better low-mass acceptance (starting at 160\,GeV) than horizontal crossing (starting at 515\,GeV). 
\end{itemize}

\subsubsection{Maximum Mass}
\label{sec:maxmass}
In order to calculate the maximum accepted mass with Eqs.~(\ref{eqn:min-max-mass}) and~(\ref{eqn:xi-max}), the aperture limitations have to be identified for each crossing-angle.
In the case of vertical beam crossing in IP5, both horizontal and vertical apertures may impose limitations, whereas in the case of horizontal crossing there is no substantial vertical dispersion and hence no acceptance loss from the vertical aperture. The detailed discussion here is only given for vertical crossing as it has been decided for IP5 at HL-LHC; for the simpler horizontal case only the final result is reported.

\begin{figure}[h!]
\begin{center}
\includegraphics[width=0.495\textwidth]{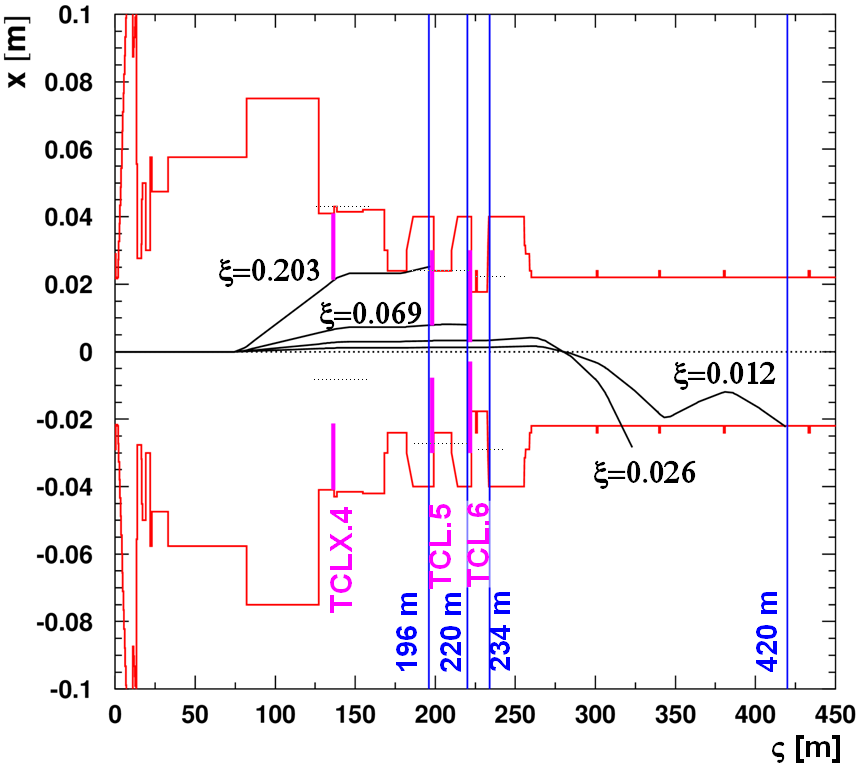}
\includegraphics[width=0.495\textwidth]{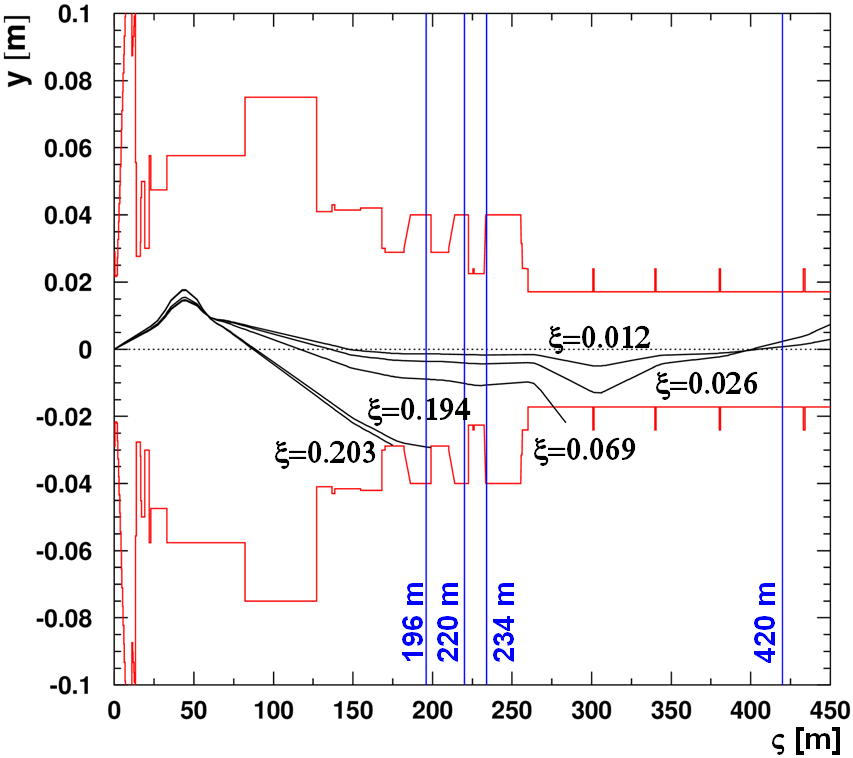}
\end{center}
\caption{Horizontal (left) and vertical (right) aperture models with example trajectories hitting an element because of their $\xi$-values (see text). This figure corresponds to a vertical crossing-angle  $\alpha_{y}/2 = 250\,\mu$rad.}
\label{fig:aperturemodel}
\end{figure}
\begin{figure}[h!]
\begin{center}
\includegraphics[width=0.495\textwidth]{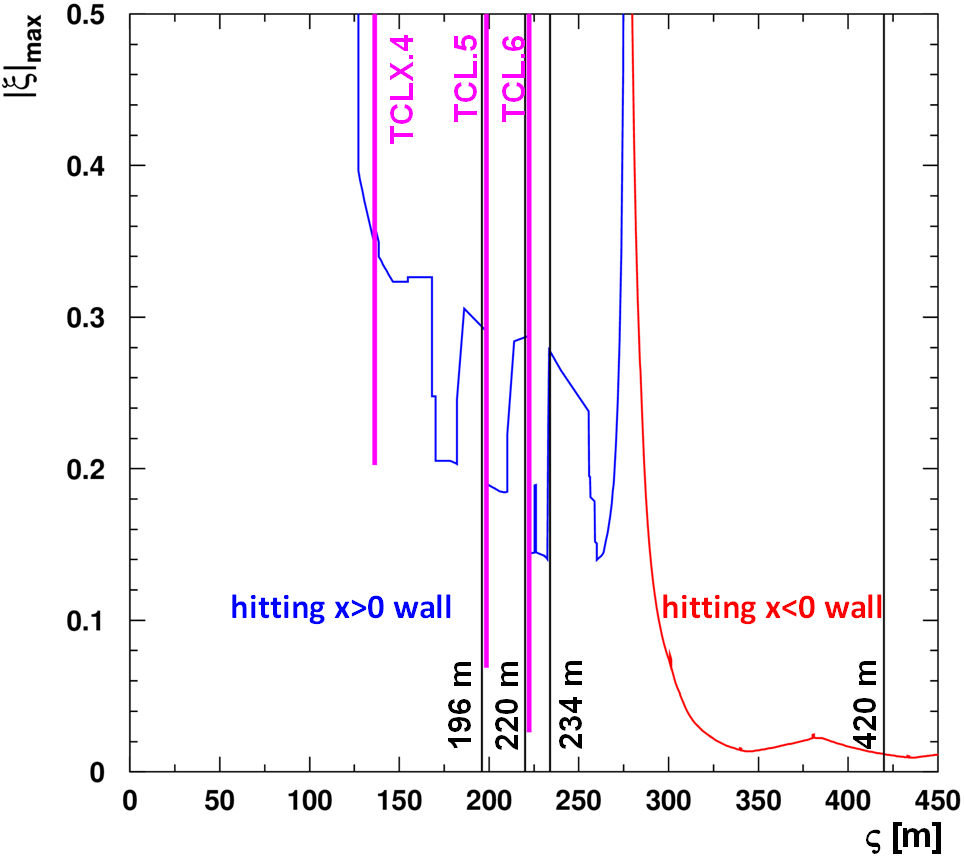}\hfill
\includegraphics[width=0.495\textwidth]{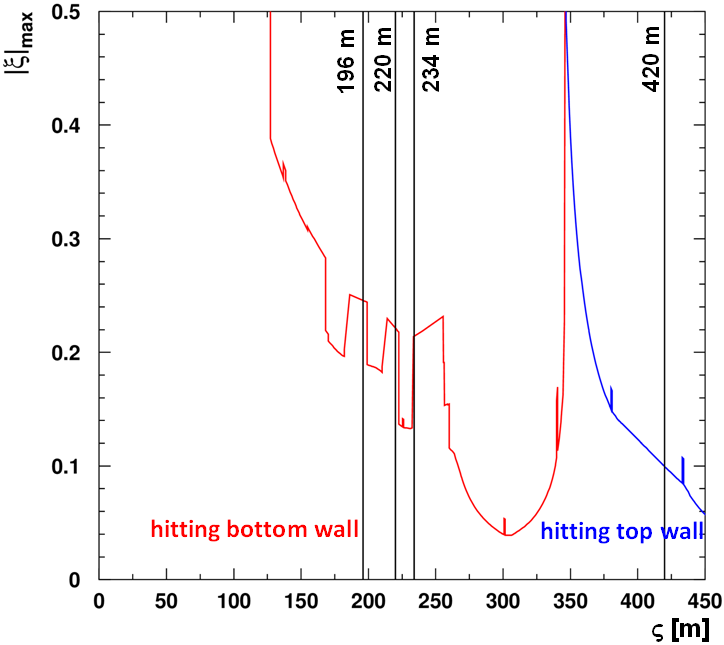}
\end{center}
\caption{Maximum value of $|\xi|$ for which a proton trajectory can just pass through the horizontal (left) and vertical (right) aperture at the longitudinal position $\varsigma$ for a vertical crossing-angle $\alpha_{y}/2 = 250\,\mu$rad. The blue and red colours denote which of the aperture walls (left/right, top/bottom) are hit.}
\label{fig:ximax-vs-s}
\end{figure}

Figure~\ref{fig:aperturemodel} shows the horizontal and vertical aperture model of the IR5 region along with projections of a few example trajectories for $\alpha_{y}/2 = 250\,\mu$rad. In particular, those $\xi$-values were chosen that just hit the edges of the TCL collimators, the predominant aperture limitations. The figure also shows two exceptions where other elements constitute the limitations:
\begin{itemize}
\item For potential locations beyond 315\,m (i.e.\ the 420\,m location relevant in the scope of this document), the last TCL collimator, TCL6, is not the tightest aperture limitation. TCL6 imposes a limit at $\xi = 0.026$ whereas the beam pipe just before 420\,m cuts already at $\xi = 0.012$. The upper mass limit of the 420\,m station is therefore independent of the collimation scheme.
\item The vertical aperture of Q4 and its corrector magnets cuts at $\xi = 0.194$ whereas TCLX.4 has its bottleneck only at $\xi = 0.203$. This vertical limitation is only dominant for crossing-angles greater than 240\,$\mu$rad.
\end{itemize}
An instructive visualization of the aperture limitations is shown in Fig.~\ref{fig:ximax-vs-s}. At each point along the beamline, the horizontal and vertical apertures are directly converted into the $\xi$-limit at that point. For a given detector location the dominant limit can be read off as the deepest minimum upstream.

\begin{figure}[h!]
\begin{center}
\includegraphics[width=0.495\textwidth]{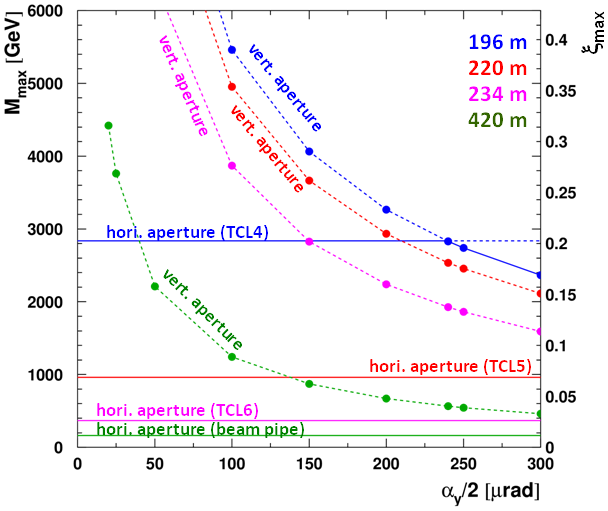}
\includegraphics[width=0.495\textwidth]{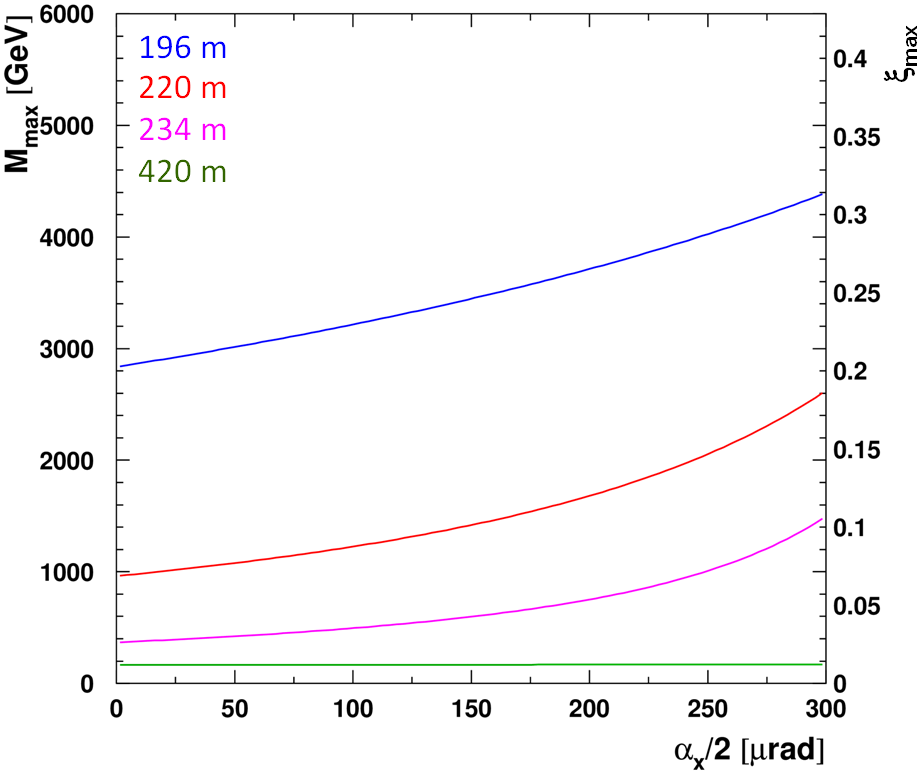}
\end{center}
\vspace{-0.5cm} 
\caption{Maximum accepted central mass for each detector location as a 
function of the crossing-angle. Left (vertical crossing): both horizontal and
vertical apertures contribute to the mass limits. The continuous lines denote the most restrictive, i.e.\ dominant, limitations.
Right (horizontal crossing): only the horizontal apertures contribute.} 
\label{fig:m-max}
\end{figure}
Figure~\ref{fig:m-max} shows the upper mass cutoff as a function of the crossing angle in both the vertical and the horizontal crossing cases. 
As discussed before, for vertical crossing the most dominant limitations come from the horizontal aperture and for all locations, except 420\,m, this horizontal aperture is limited by the TCL collimators. 
At 420\,m, on the other hand, the beam-pipe absorbs protons with $|\xi| > 0.012$.
The highest masses are accepted by the unit at 196\,m: up to 2.7\,TeV for vertical crossing, and up to 4 TeV for horizontal crossing.

\subsubsection{Mass-Rapidity Acceptance}
\label{sec:mass-rapidity}
\begin{figure}[h!]
\begin{center}
\includegraphics[width=0.49\textwidth]{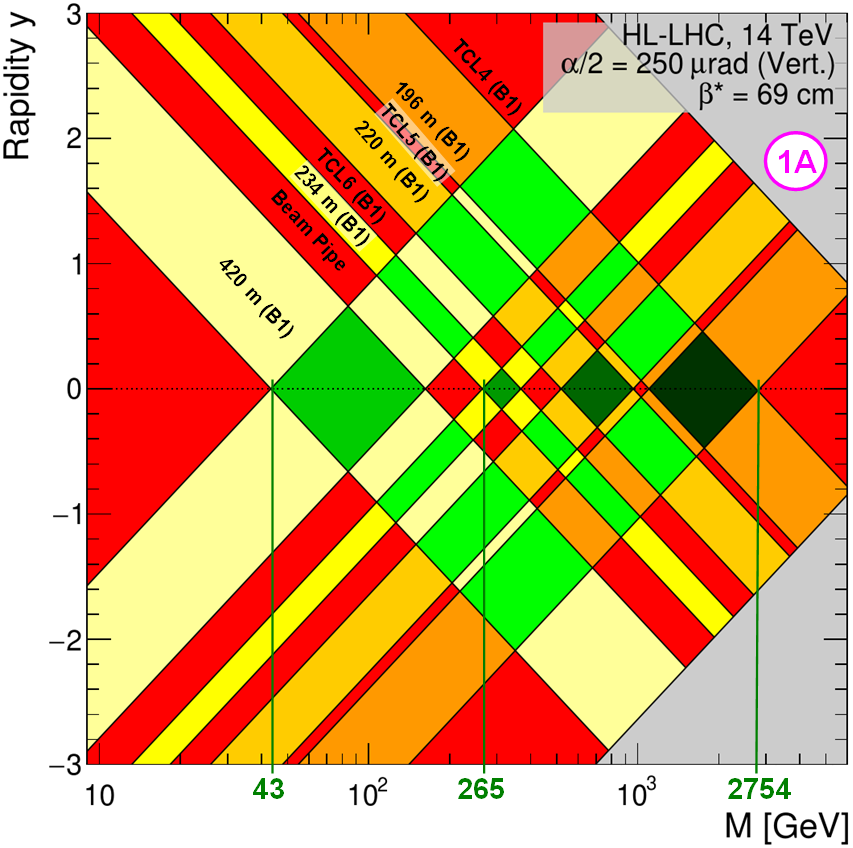}\hfill
\includegraphics[width=0.49\textwidth]{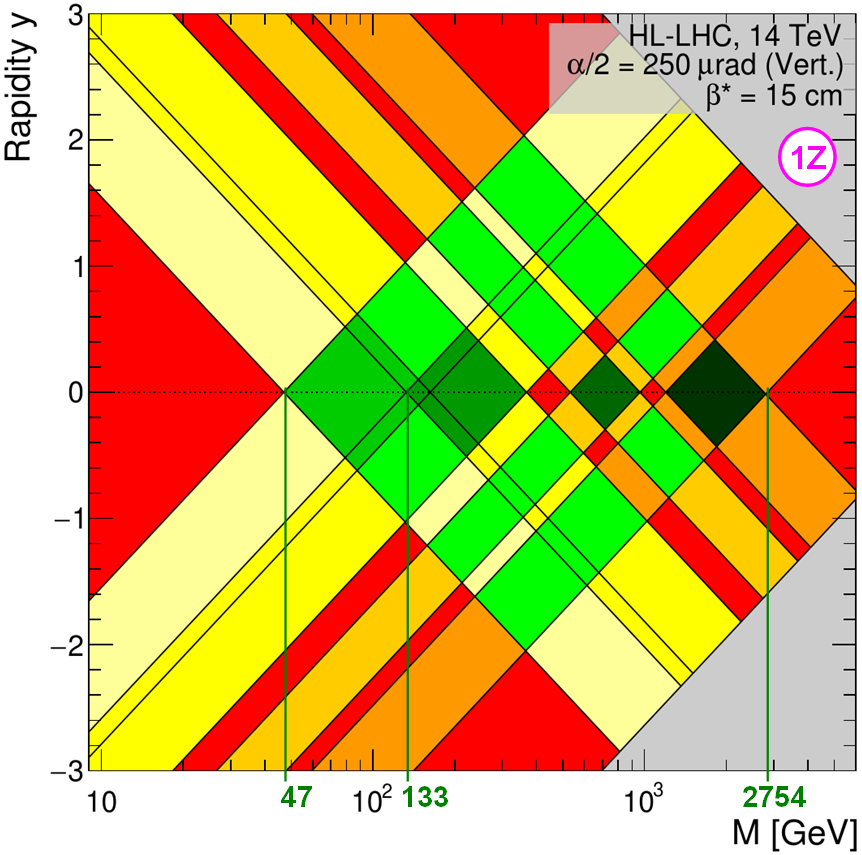}
\includegraphics[width=0.49\textwidth]{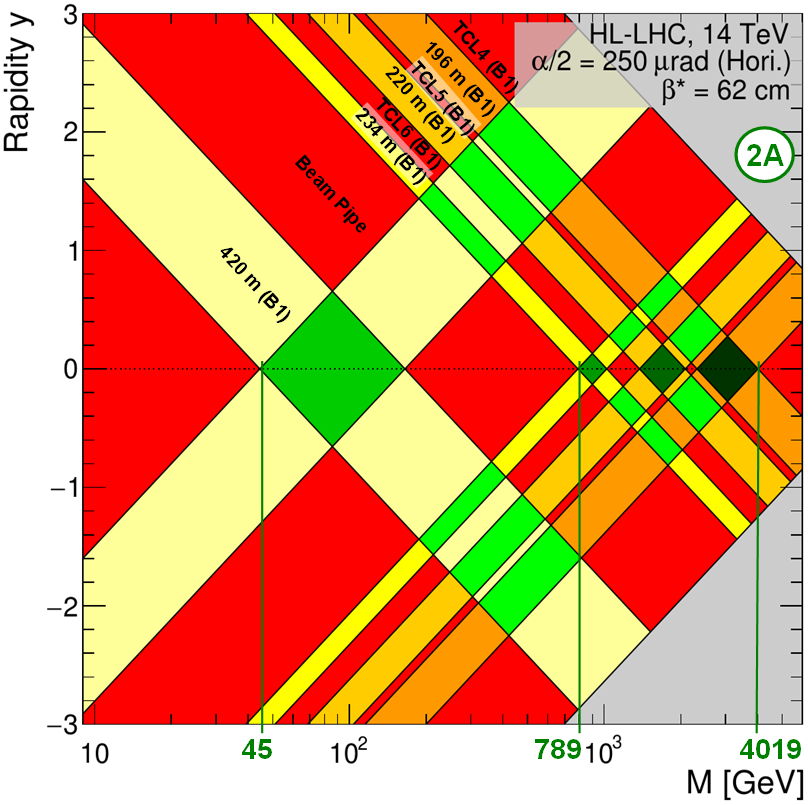}\hfill
\includegraphics[width=0.49\textwidth]{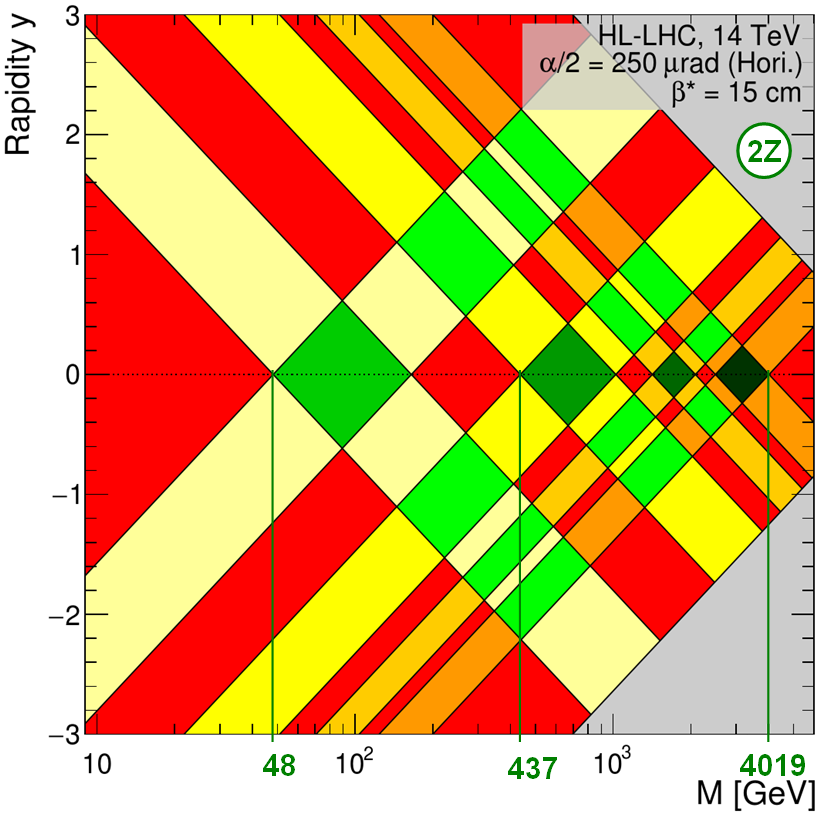}
\includegraphics[width=0.3\textwidth]{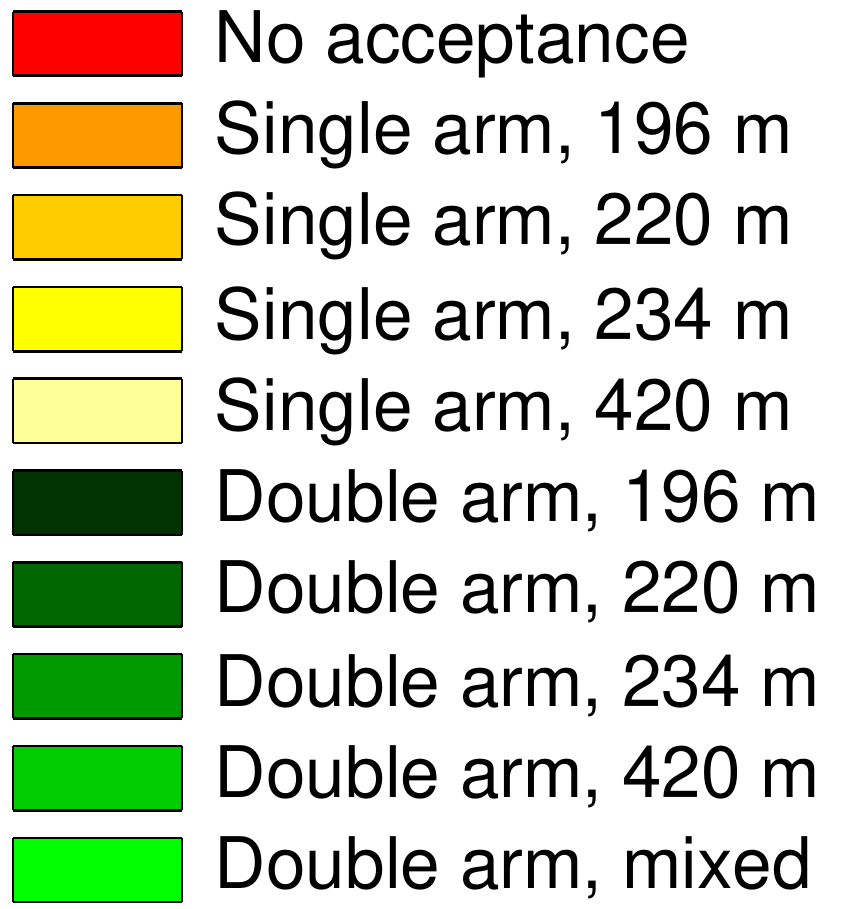}
\end{center}
\caption{Acceptance for the protons from CEP events in the mass-rapidity plane of the centrally produced system. The yellow/orange colour tones mark single-arm proton acceptance, the green tones mark double-arm acceptance. The labels (1A), ..., (2Z) denote the same points on the levelling trajectories in the $(\alpha_{x}/2, \beta_{x}^{*})$ plane as defined in Fig.~\ref{fig:m-min}. Top: start and end point of any levelling trajectory for a vertical crossing angle in IP5, bottom: start and end point of the baseline levelling trajectory for horizontal crossing in IP5.} 
\label{fig:m-y}
\end{figure}

\begin{figure}[p]
\begin{center}
\includegraphics[width=0.49\textwidth]{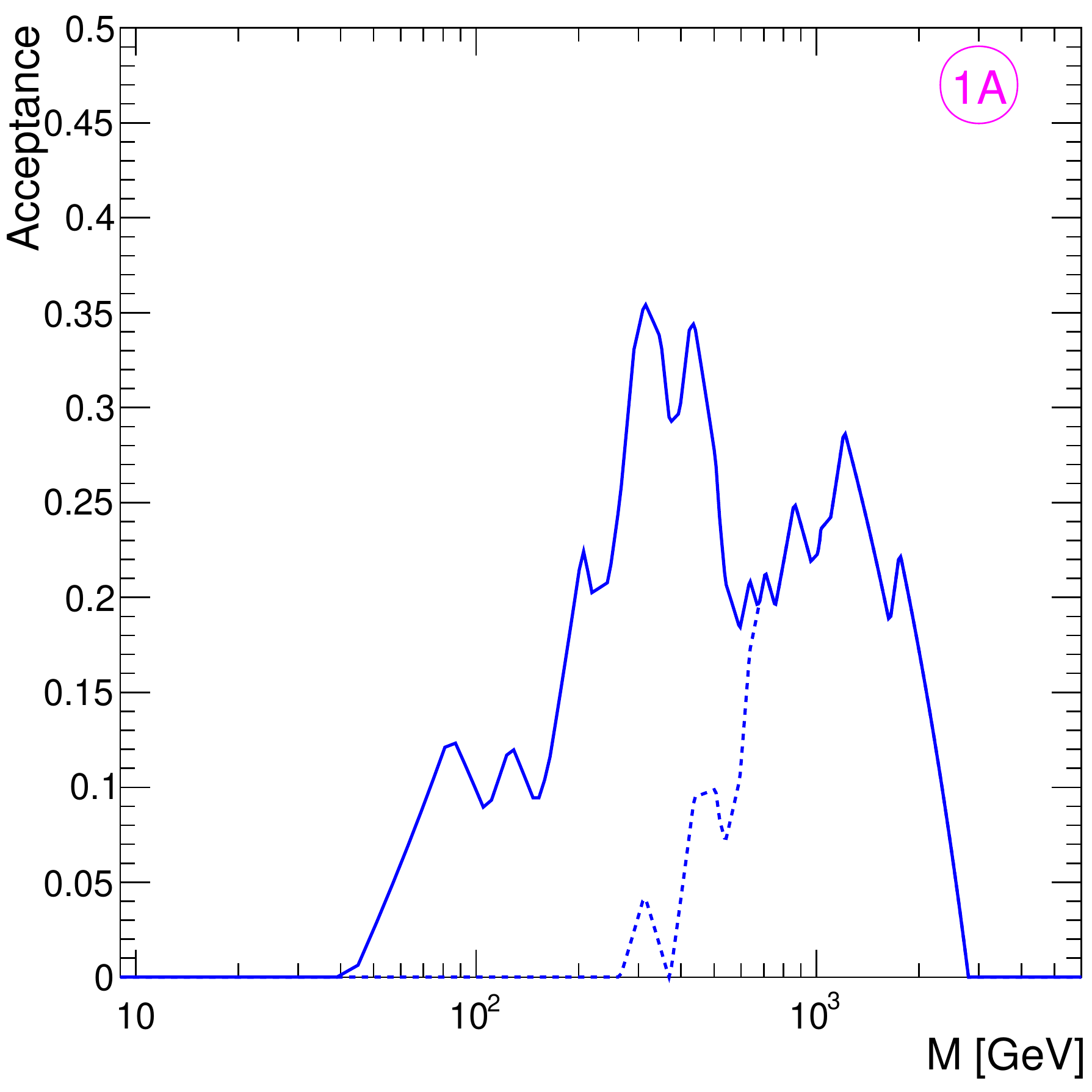}\hfill
\includegraphics[width=0.49\textwidth]{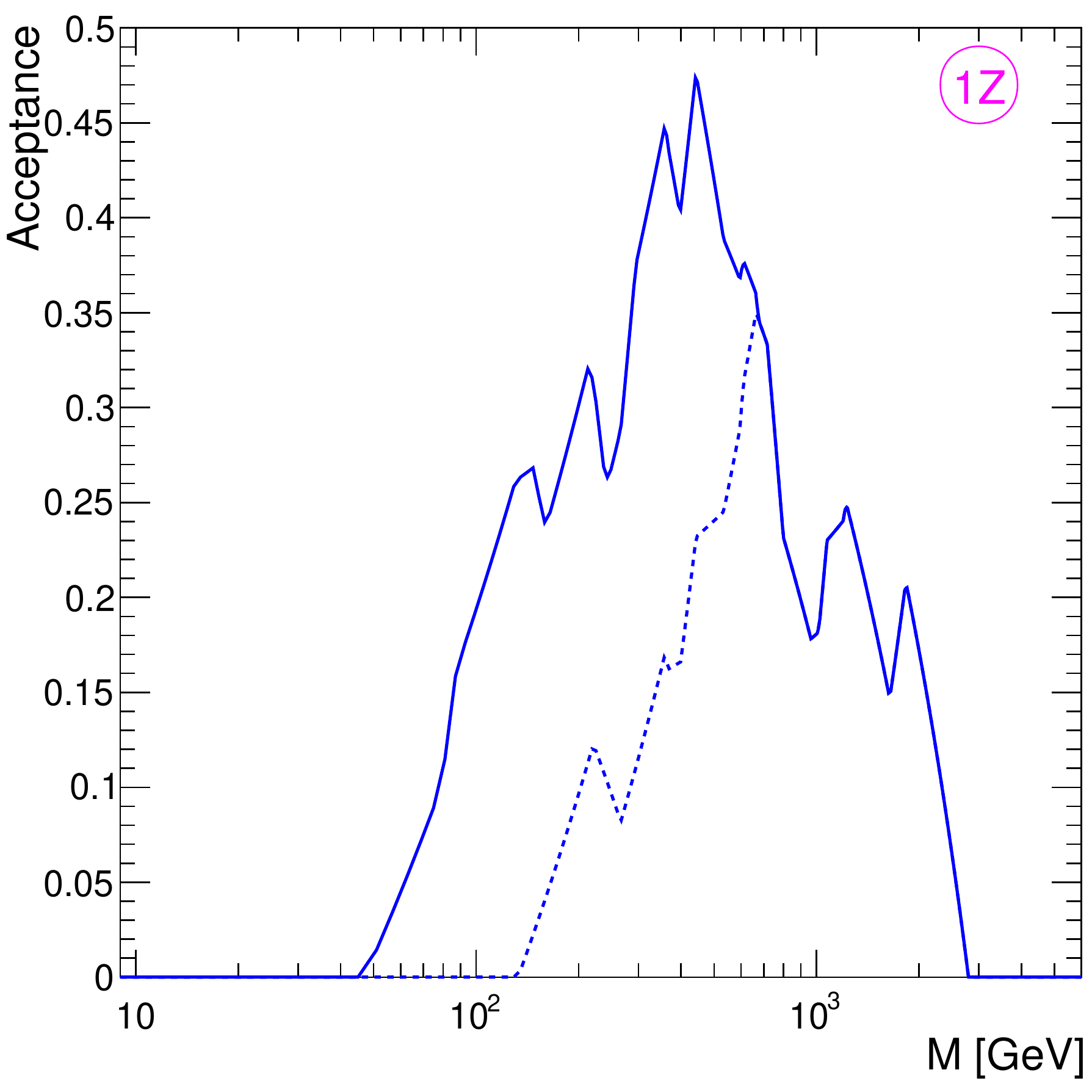}
\includegraphics[width=0.49\textwidth]{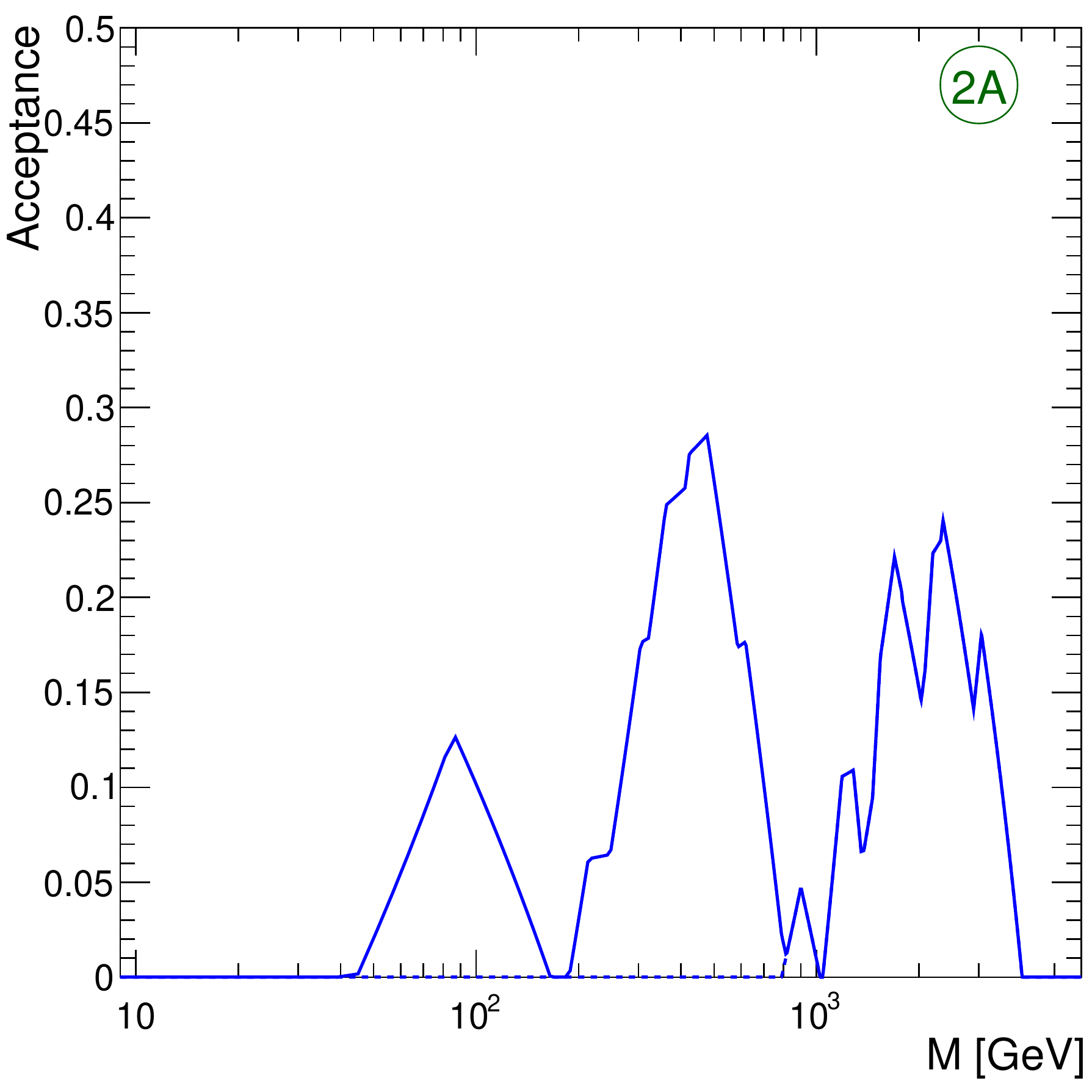}\hfill
\includegraphics[width=0.49\textwidth]{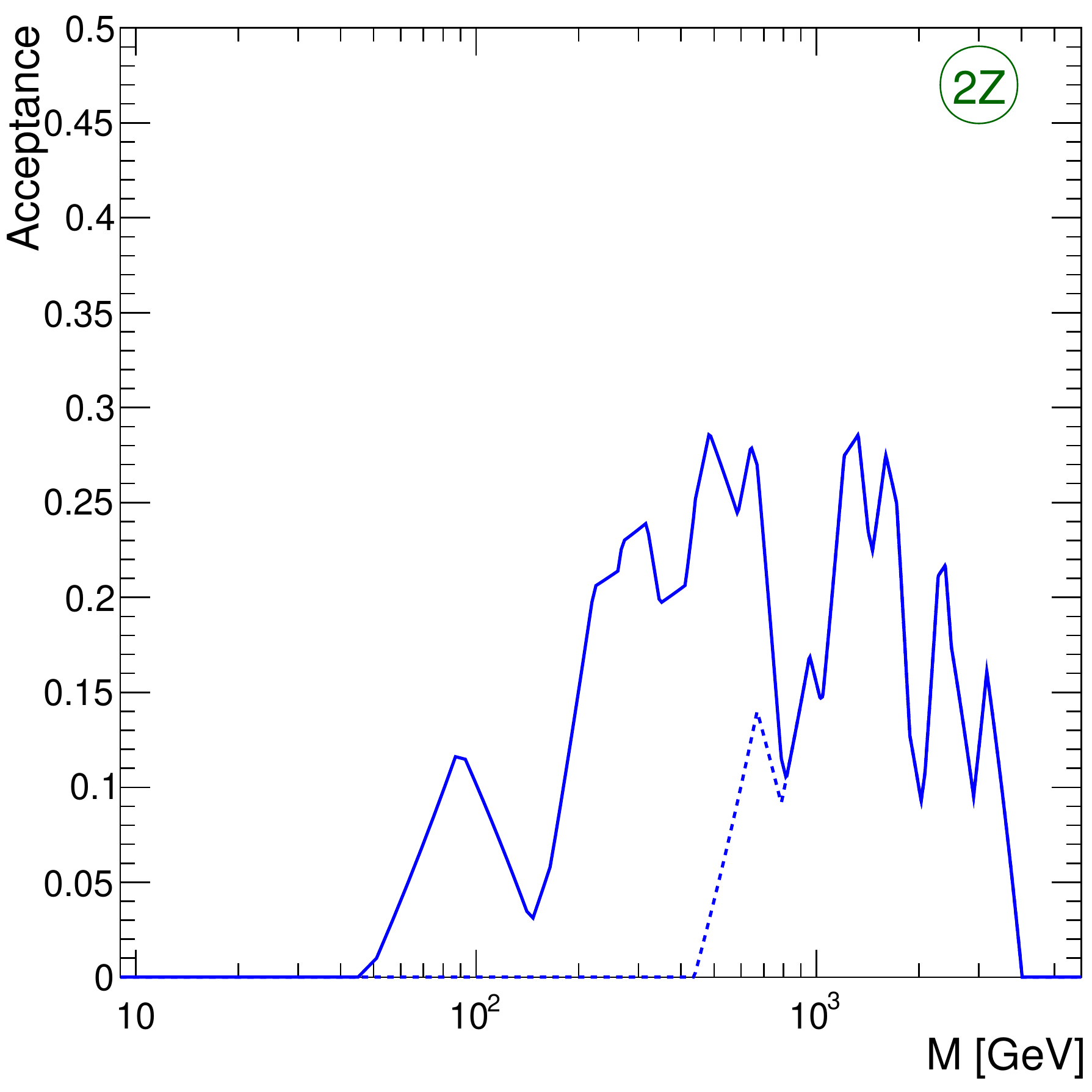}
\end{center}
\caption{Projection of the $(M, \mathsf{y})$ acceptance on the mass axis under the approximation of flat rapidity distributions, adding up all the double-arm areas of Fig.~\ref{fig:m-y} for the same points in the $(\alpha, \beta^{*})$ beam parameter space. Top: vertical crossing, bottom: horizontal crossing. The dashed lines represent a configuration without the 420\,m stations.}
\label{fig:m-acceptance}
\end{figure}

The CEP acceptance for a given point in the beam parameter space $(\alpha, \beta^{*})$ can be visualized by drawing for every instrumented detector location the $|\xi|$-acceptance bands -- whose limits are calculated as discussed in the previous section -- in the mass-rapidity plane of the centrally produced system:
\begin{equation}
\left(\ln \frac{M}{\sqrt{s}}, \mathsf{y}\right) = \left(\frac{1}{2}(\ln |\xi_{1}| + \ln |\xi_{2}|), \frac{1}{2}(\ln |\xi_{1}| - \ln |\xi_{2}|)\right) \: .
\end{equation}
Figure~\ref{fig:m-y} shows these $(M,\mathsf{y})$ contour plots for the start and end points of the two extreme levelling trajectories defined in Fig.~\ref{fig:m-min}: 
points (1A) and (1Z) for any trajectory with vertical crossing in IP5, points (2A) and (2Z) for the ``Baseline'' trajectory with horizontal crossing. The projections on the mass axis, under the approximation of uniform rapidity distributions, are given in Fig.~\ref{fig:m-acceptance}.

Table~\ref{tab:masslimits} lists the $\xi$ and mass limits for the four stations and for both crossing-angle orientations.

The following observations are made:
\begin{itemize}
\item In most cases, the acceptance zones of the four detector locations are non-overlapping and separated by gaps. Only at the end of the fill, in the case of vertical crossing, there is some overlap between the 234\,m and 420\,m stations.
For horizontal crossing, the gaps would be much wider than for the vertical crossing that has been chosen.
\item Although the double-arm acceptance has mass gaps at central rapidities, the mixed acceptance zones combining different detector units in the two arms of the experiment (e.g.\ 420\,m left + 234\,m right) fill some of these mass gaps by providing acceptance at forward rapidities. 
\item Combined with other stations on the opposite side, the 420\,m station contributes very significant acceptance also at intermediate masses (up to 670\,GeV for vertical crossing).
\item The gaps between the acceptances of 196\,m, 220\,m, and 234\,m can potentially be closed by opening TCL5 and TCL6 a little further if allowable from machine protection arguments. On the other hand, the gap between 234\,m and 420\,m is caused by the beam pipe at $\varsigma > 300\,$m limiting the aperture (Section~\ref{sec:maxmass}). It could only be closed by adding a detector unit near 300\,m where, however, no free space is available.  
\item Without the 420\,m stations, the minimum mass for vertical crossing is 133\,GeV, whereas for horizontal crossing it would be 437\,GeV, i.e.\ significantly worse.
\item Large-mass acceptance, above 1\,TeV, is contributed by the 196\,m stations, also in combination with the 220\,m stations. For vertical crossing, the upper mass limit will be shifted from about 2\,TeV for the pre-LS3 PPS system to 2.7\,TeV at HL-LHC. For horizontal crossing, the absolute maximum would be 4\,TeV, but in a very small rapidity range, i.e.\ with small acceptance.  
\end{itemize}
{\bf Based on the large acceptance gaps for horizontal crossing and, in particular, the high minimum mass without the 420\,m stations, this project expressed, already early in its development, a strong preference for vertical crossing in IP5. CMS adopted this preference and requested vertical crossing after LS3~\cite{hllhc-27coordinationgroup}. In June 2020, the decision was taken to implement the vertical crossing~\cite{hllhc-executivecommittee20200622}.}

\begin{table}
  \begin{center}
    \caption{Lower and upper $\xi$ limits, and minimum and maximum central mass accepted by each station at rapidity $\mathsf{y} = 0$. The top and bottom blocks represent vertical (officially chosen for implementation) and horizontal crossing (for comparison), respectively. The ranges in the minimum values indicate the beginning and the end of the levelling trajectories, (1A) to (1Z) and (2A) to (2Z). 
}
    \label{tab:masslimits}
    \begin{tabular}{|l|c|c|c|c|}
    \hline
    \multicolumn{5}{|c|}{Vertical Crossing-Angle}\\
    Station & $|\xi_{\rm min}|$ & $|\xi_{\rm max}|$ & $M_{\rm min}$ [GeV] @ $\mathsf{y} = 0$ & $M_{\rm max}$ [GeV] @ $\mathsf{y} = 0$\\
    \hline
    196\,m & 0.0786$-$0.0856 & 0.1967 & 1100.87$-$1197.80 & 2754.27 \\
    220\,m & 0.0371$-$0.0381 & 0.0688 & 519.89$-$533.18 & 962.70 \\
    234\,m & 0.0189$-$0.0095 & 0.0263 & 264.96$-$132.80 & 368.11 \\
    420\,m & 0.0031$-$0.0034 & 0.0116 & 43.38$-$47.04  & 162.66 \\
    \hline
    \end{tabular}

    \begin{tabular}{|l|c|c|c|c|}
    \hline  
    \multicolumn{5}{|c|}{Horizontal Crossing-Angle}\\
    Station & $|\xi_{\rm min}|$ & $|\xi_{\rm max}|$ & $M_{\rm min}$ [GeV] @ $\mathsf{y} = 0$ & $M_{\rm max}$ [GeV] @ $\mathsf{y} = 0$\\
    \hline
    196\,m & 0.1654$-$0.1779 & 0.2871 & 2316.15$-$2490.07 & 4018.94 \\
    220\,m & 0.0984$-$0.1014 & 0.1488 & 1377.48$-$1419.13 & 2083.04 \\
    234\,m & 0.0564$-$0.0312 & 0.0732 & 789.48$-$437.07 & 1024.60 \\
    420\,m & 0.0032$-$0.0034 & 0.0118 & 44.55$-$48.20 & 165.28 \\
    \hline
    \end{tabular}
  \end{center}
\end{table}

\newpage
\subsubsection{Scattering Angle}
\begin{figure}[h!]
\begin{center}
\includegraphics[width=0.8\textwidth]{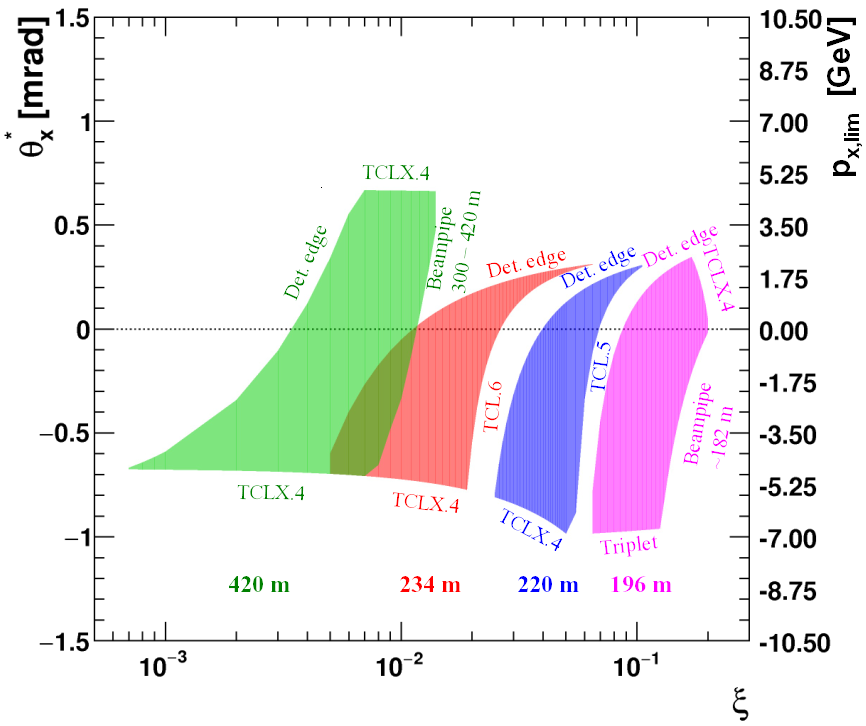}
\includegraphics[width=0.8\textwidth]{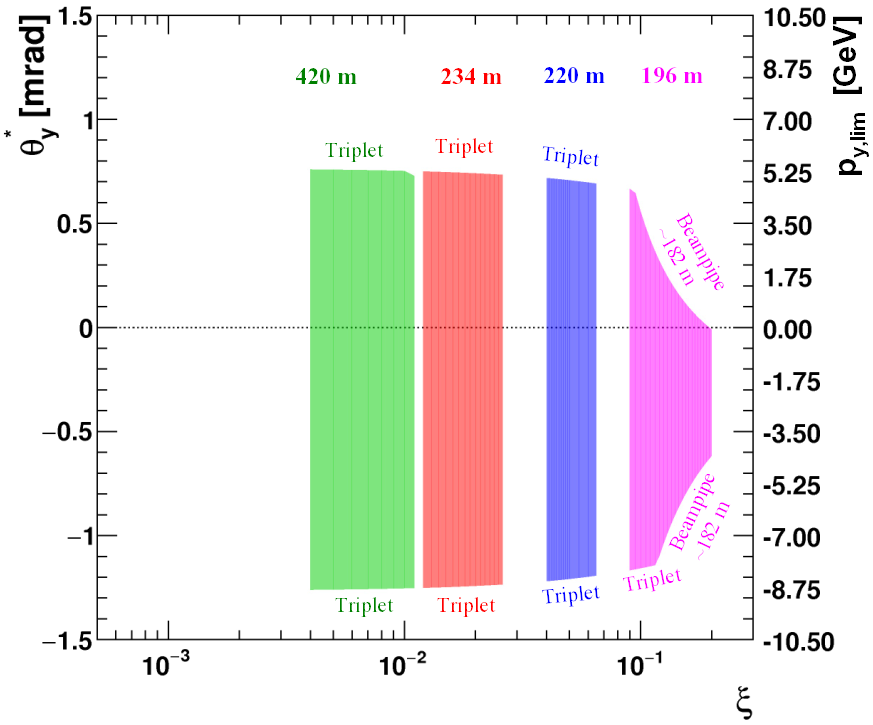}
\end{center}
\caption{Two-dimensional single-proton acceptance in the $(\xi, \theta_{x}^{*})$ plane for $\theta_{y}^{*} = 0$ (top), and in the $(\xi, \theta_{y}^{*})$ plane for $\theta_{x}^{*} = 0$ (bottom), for $\alpha_{y}/2 = 250\,\mu$rad (vertical crossing) and $\beta^{*} = 15\,$cm. The right-hand axes translate the scattering angles into the corresponding transverse momenta.}
\label{fig:theta-acceptance}
\end{figure}

Up to this point, all kinematic acceptance studies have neglected the scattering angle for simplicity. This section presents the results of an extended aperture study based on MAD-X simulations: for a series of values of $\xi$, the minimum and maximum values of $\theta_{x}^{*}$ and $\theta_{y}^{*}$ are calculated for which a proton can still pass through all beamline elements upstream of the detector station. Figure~\ref{fig:theta-acceptance} shows these results for the case of vertical crossing in IP5. The up-down asymmetry in the $\theta_{y}^{*}$ acceptance are a consequence of the vertical dispersion. Note also that for nonzero $\theta_{x}^{*}$ the gaps between the $\xi$-regions accepted by the four detector stations are reduced. This study still does not offer full generality because situations where both components $\theta_{x}^{*}$ and $\theta_{y}^{*}$ are nonzero are not treated. If the beam pipe had a rectangular cross section, the limits on $\theta_{x}^{*}$ and $\theta_{y}^{*}$ would be independent and could be applied separately to each component. However, the typical rectellipsoid shape\footnote{A rectellipse is a hybrid shape formed by superimposing a rectangle and an ellipse, resulting in a rectangle with rounded corners.} of most beam-pipe elements, leads to slight violations of the independence of the two components in the rather rare cases when both are close to their extrema (i.e.\ in diagonal track directions when the rounded corners of the beam pipe are hit). 

In the physics performance studies performed so far (Section~\ref{sec:physics}), the effect of the scattering angle on the acceptance have not yet been taken into account. 
The distribution of signal protons will be centered at a scattering angle of zero. In the case of $\gamma\gamma$ processes, the RMS of the distribution is expected to be $\sim 0.08$\,mrad at the upper end of the $\xi$ range, decreasing for smaller $\xi$ values. Therefore, neglecting the scattering angle $\theta^{*}_{x}$ will have only a small impact on the physics acceptance studies, and neglecting $\theta^{*}_{y}$ will have essentially no impact, apart from the beam pipe limitation on the upper edge of the 196\,m acceptance.

\clearpage
\subsection{Reconstruction and Resolution}
\label{sec:resolution}
\begin{figure}[h!]
\begin{center}
\includegraphics[width=0.495\textwidth]{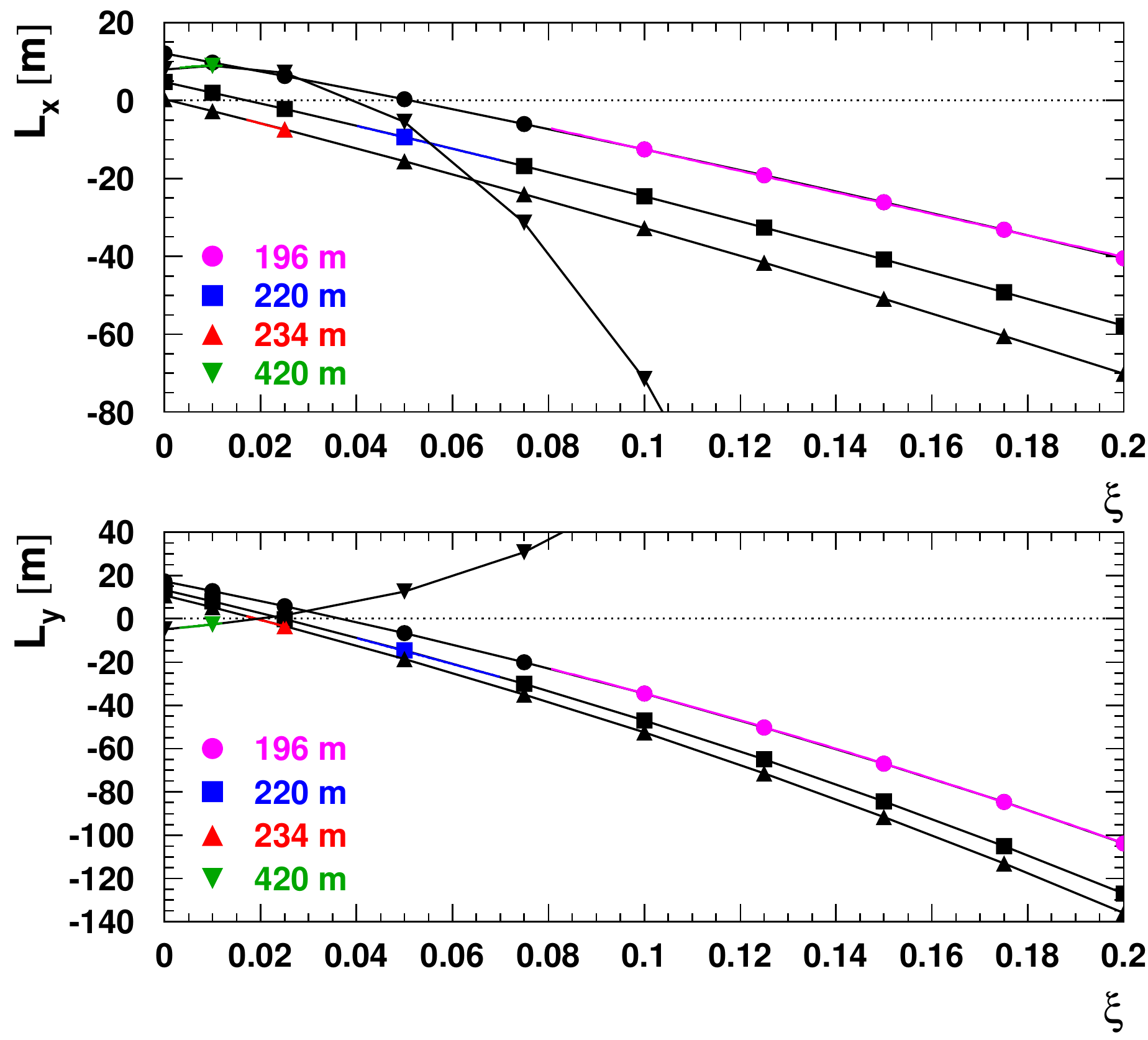}
\includegraphics[width=0.495\textwidth]{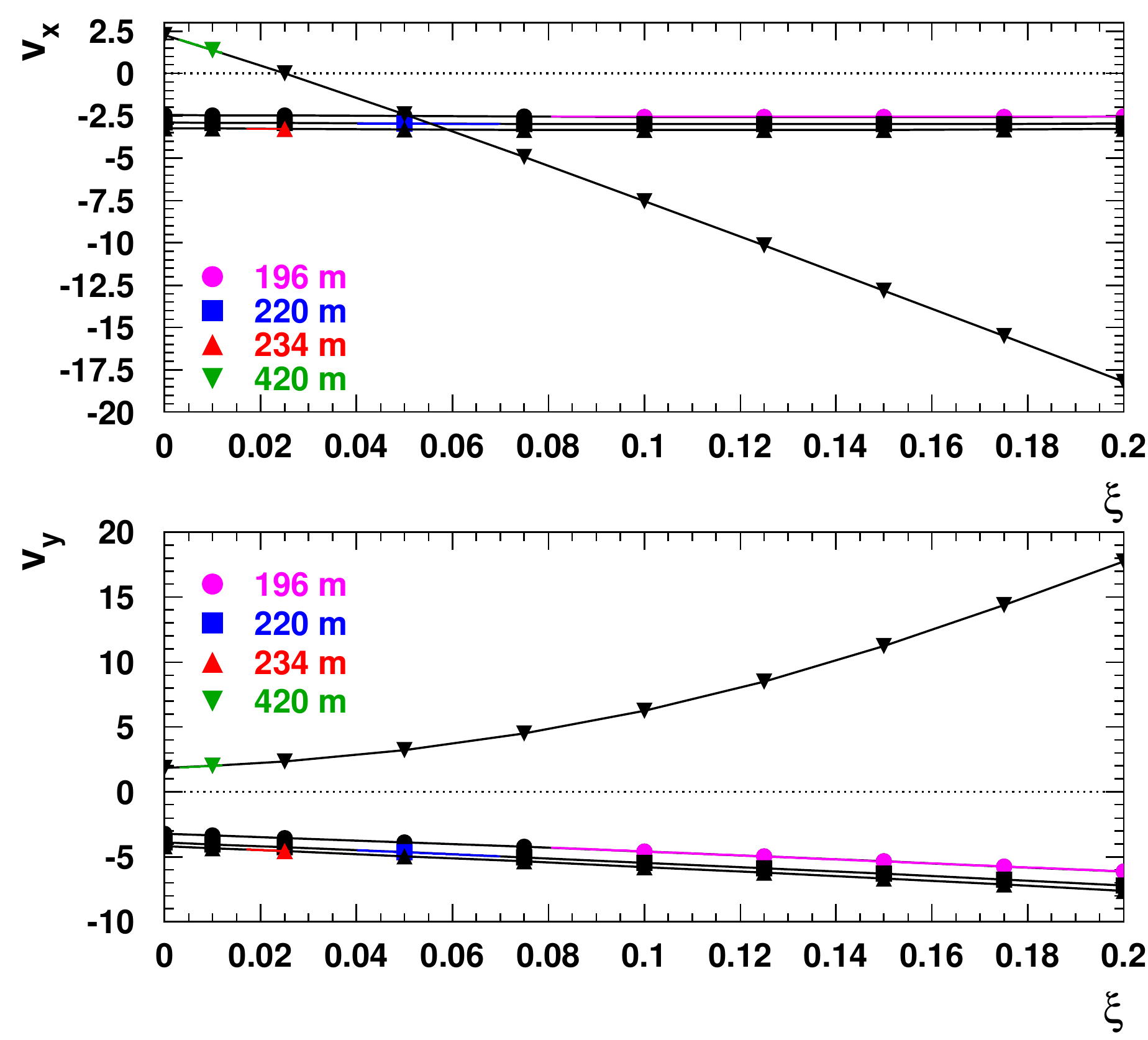}
\end{center}
\caption{$\xi$-Dependence of the effective length (left) and the magnification (right) in $x$ (top) and in $y$ (bottom) at the four detector locations. The coloured sections lie within the acceptance (for vertical beam crossing).}
\label{fig:optfuncvsxi}
\end{figure}

\begin{figure}[h!]
\begin{center}
\includegraphics[width=0.495\textwidth]{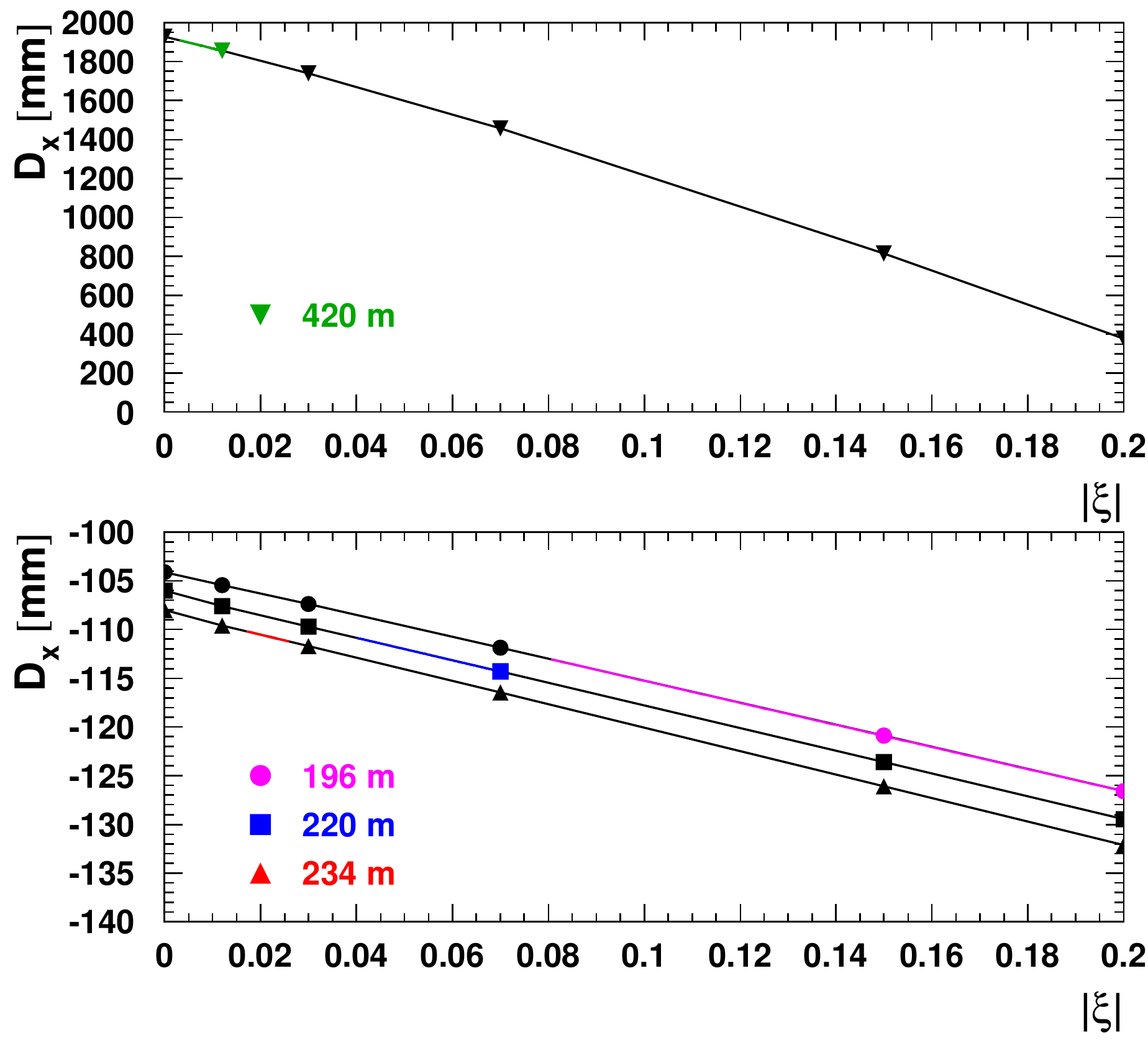}
\includegraphics[width=0.495\textwidth]{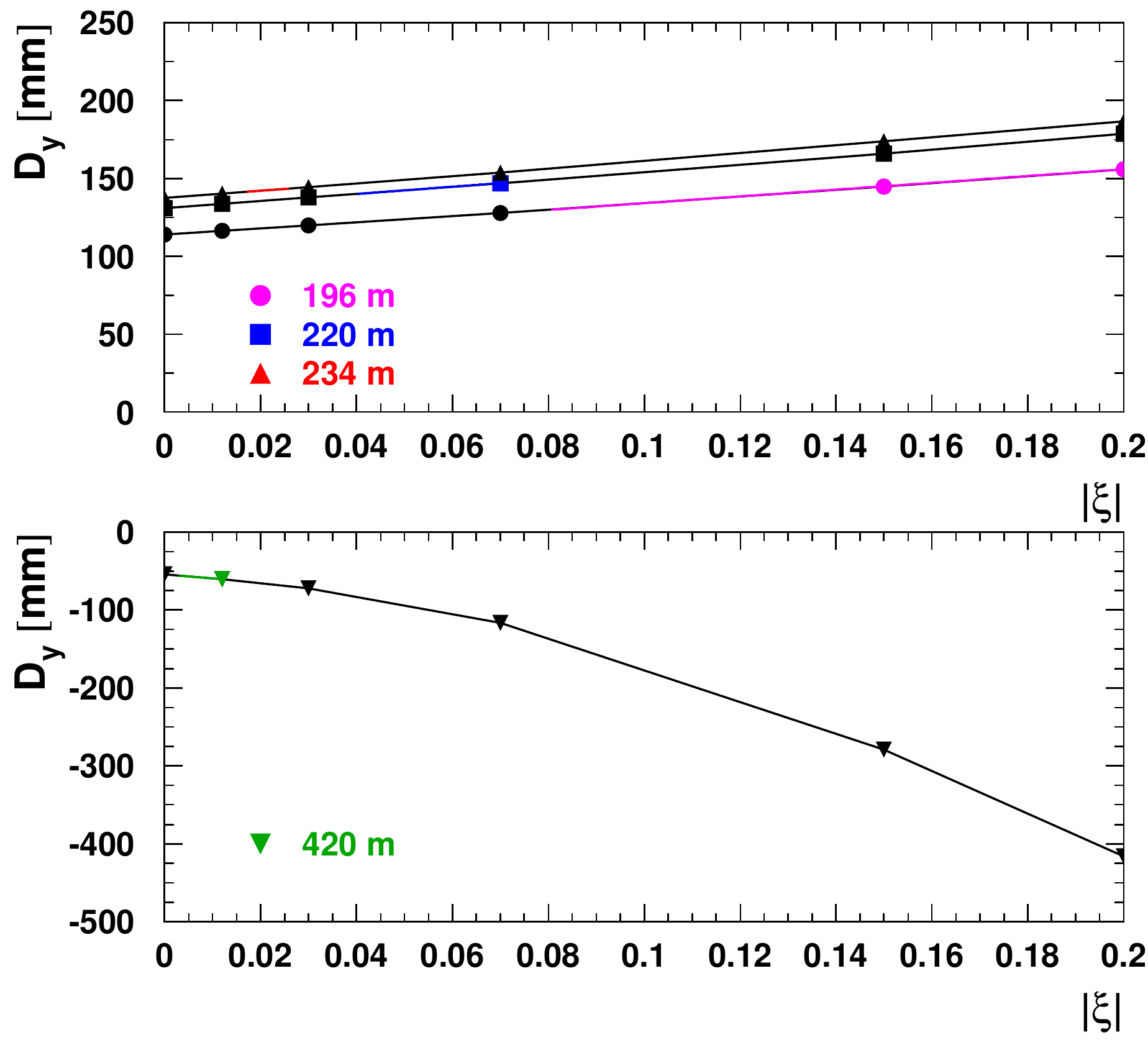}
\end{center}
\caption{$\xi$-Dependence of the horizontal (left) and vertical (right) dispersion for vertical crossing with $\alpha_{y}/2 = 250\,\mu$rad.}
\label{fig:dispersionvsxi}
\end{figure}

The goal of forward detector event analysis is the reconstruction of the scattering kinematics using the leading proton track segments measured hundreds of metres away from the interaction point. The objective of this section is not to describe the most advanced reconstruction techniques applied in real data analysis~\cite{ctpps-opticstalk20180821,dn-19-026} (see also Ref.~\cite{thesis-niewiadomski} for the first development of the concepts), but
\begin{itemize}
\item to investigate the possibility to resolve the kinematic variables of the forward protons,
\item to establish the ideal lever arm between the two units of each detector station, which then has to be confronted with the available space discussed earlier in Section~\ref{sec:spaceconstraints}.
\end{itemize}
The trajectory of a proton with fractional momentum loss $\xi$, Eq.~(\ref{eqn:xidefinition}), emerging at a scattering angle $(\theta_x^*, \theta_y^*)$ from a vertex at IP5 with transverse position $(x^*,y^*)$, is described in linear approximation by the transport matrix equation~\cite{opticspaper,totemnote2017002} 
\begin{equation}
                \vec{d}(\varsigma)=T(\varsigma)\cdot\vec{d}^{*}\,, 
                \label{eqn:proton_trajectories}
\end{equation}
where $\vec{d}(\varsigma)=\left(x,\theta_x,y,\theta_y,\xi\right)^{T}(\varsigma)$ is the coordinate vector in the position $\varsigma$ along the beam line, and $\vec{d}^{*}$ is the corresponding vector at the vertex.
The transport matrix
\begin{equation}
                T=\left(
                                \begin{array}{ccccc}
                                                v_x         & L_x     & m_{13}   & m_{14}  & D_x  \\
                                                \frac{dv_x}{ds}        & \frac{dL_x}{ds}    & m_{23}   & m_{24}  & \frac{dD_x}{ds} \\
                                                m_{31}      & m_{32}  & v_y      & L_y     & D_y  \\
                                                m_{41}      & m_{42}  & \frac{dv_y}{ds}     & \frac{dL_y}{ds}    & \frac{dD_y}{ds} \\
                                                0           & 0       & 0        & 0       & 1
                \end{array} 
                \right)
                \label{eqn:transport_matrix}
\end{equation}
is defined by the optical functions: the vertex magnification $v_{x/y}$ (see Eq.~(\ref{eqn:v})), the effective length $L_{x/y}$ (see Eq.~(\ref{eqn:L})), and the dispersion $D_{x/y}$ (Section~\ref{sec:locationselection}). For the nominal LHC optics the coefficients $m_{13}$, ... $m_{42}$, which couple the $x$ and $y$ coordinates, are zero.
In the following treatment only the case of vertical beam crossing is considered, where both dispersion components $D_{x}$ and $D_{y}$ are nonzero. 
It is important to note that all optical functions depend on $\xi$ (Figs.~\ref{fig:optfuncvsxi},~\ref{fig:dispersionvsxi}), which introduces nonlinearities into the equation system. In practice, these nonlinearities are dealt with by either an analytical, iterative approach, or by performing a global $\chi^{2}$ fit to the entire transport matrix equation with the kinematic variables as free parameters~\cite{ctpps-opticstalk20180821,dn-19-026}.

The reconstruction objective -- the determination of $\vec{d}^{*}$ -- can be achieved if
\begin{enumerate}
\item the optical functions and their $\xi$-dependences are known, and
\item a sufficient number of independent measurements is performed on each track.
\end{enumerate}
Condition 1 requires a magnetic model of the LHC and data-constrained calibration techniques (Section~\ref{sec:opticscalibration}).

How is Condition 2 quantified, and how can it be fulfilled? Each individual measurement point at a detector contributes one equation per coordinate~\cite{totem-tdr}:
\begin{eqnarray}
x(\varsigma) = v_{x}(\varsigma) x^{*} + L_{x}(\varsigma) \theta_{x}^{*} + D_{x}(\varsigma) \xi \label{eqn:transporteqnarray1} \\
y(\varsigma) = v_{y}(\varsigma) y^{*} + L_{y}(\varsigma) \theta_{y}^{*} + D_{y}(\varsigma) \xi \: . \label{eqn:transporteqnarray2}
\end{eqnarray}
The explicit measurement of a local angle $\theta_{x/y}$ can be replaced with the equivalent measurement of a second position.

The reconstruction of all five variables constituting $\vec{d}^{*}$ would require to measure each track projection in at least three points, providing six equations: three for $x$, three for $y$. Because of the missing acceptance overlap between the four station locations defined in the previous chapters, each station has to perform the full kinematic measurement in its respective $\xi$-range. This would ideally mean to accommodate three detector units in each station region, which is impracticable for reasons of space, integration, and cost. However, given the very small vertex distribution widths $\sigma(x^{*})$ and $\sigma(y^{*})$ for the strongly squeezed low-$\beta^{*}$ optics at hand\footnote{ To give examples, at $\beta^{*}=0.7\,$m and nominal emittance, $\sigma(x^{*}) = \sigma(y^{*}) = 15\,\mu$m, whereas at $\beta^{*}=0.15\,$m it measures only 7\,$\mu$m.}, and given the magnification functions $v_{x/y}$ of only $\mathcal{O}(1)$, the transverse vertex reconstruction would not be precise enough for matching the leading proton with the central event. Therefore an appropriate approach is to reconstruct only $\xi$, $\theta_{x}^{*}$, and $\theta_{y}^{*}$, for which two detector units with adequate lever arm are sufficient, and to assign a systematic uncertainty for the neglected vertex terms. 
Vertex identification, on the other hand, is performed via time-of-flight measurements providing the longitudinal vertex position.

After setting $x^{*} = y^{*} = 0$, Eqs.~(\ref{eqn:transporteqnarray1}) and~(\ref{eqn:transporteqnarray2}), evaluated at two detector units in locations $\varsigma_{1}$ and $\varsigma_{2}$, can be solved~\cite{ctpps-opticstalk20180821,dn-19-026}:
\begin{eqnarray}
\label{eqn:transporteqnsolution1}
\xi_{\rm from~x} = \frac{x_{1}L_{x,2}-x_{2}L_{x,1}}{D_{x,1}L_{x,2}-D_{x,2}L_{x,1}} \:, \quad 
\theta_{x}^{*} = \frac{x_{1}D_{x,2}-x_{2}D_{x,1}}{D_{x,2}L_{x,1}-D_{x,1}L_{x,2}} \:, \\
\label{eqn:transporteqnsolution2}
\xi_{\rm from~y} = \frac{y_{1}L_{y,2}-y_{2}L_{y,1}}{D_{y,1}L_{y,2}-D_{y,2}L_{y,1}} \:, \quad 
\theta_{y}^{*} = \frac{y_{1}D_{y,2}-y_{2}D_{y,1}}{D_{y,2}L_{y,1}-D_{y,1}L_{y,2}} \:.
\end{eqnarray}
In the presently considered case of vertical beam crossing, the variable $\xi$ can be reconstructed in two independent ways: from the horizontal and from the vertical track coordinates.

By error propagation it follows that:
\begin{eqnarray}
\label{eqn:xi-thetax-resol}
\sigma(\xi_{\rm from~x}) = \sigma(x) \frac{\sqrt{L_{x,1}^{2}+L_{x,2}^{2}}}{|D_{x,1}L_{x,2}-D_{x,2}L_{x,1}|} \:, \quad 
\sigma({\theta_{x}^{*}}) = \sigma(x) \frac{\sqrt{D_{x,1}^{2}+D_{x,2}^{2}}}{|D_{x,1}L_{x,2}-D_{x,2}L_{x,1}|} \:, \\
\label{eqn:xi-thetay-resol}
\sigma(\xi_{\rm from~y}) = \sigma(y) \frac{\sqrt{L_{y,1}^{2}+L_{y,2}^{2}}}{|D_{y,1}L_{y,2}-D_{y,2}L_{y,1}|} \:, \quad 
\sigma({\theta_{y}^{*}}) = \sigma(y) \frac{\sqrt{D_{y,1}^{2}+D_{y,2}^{2}}}{|D_{y,1}L_{y,2}-D_{y,2}L_{y,1}|} \:,
\end{eqnarray}
where $\sigma(x)$ and $\sigma(y)$ are the spatial resolutions of a detector unit (e.g.\ a package of several planes of semiconductor detectors).
A better determination of $\xi$ can be obtained by taking a weighted average 
    of the reconstructed values obtained with the horizontal and vertical coordinates only.
The simultaneous reconstruction of $\theta^{*}_{x/y}$ and $\xi$ improves with increasing denominator $D_{x/y,1}L_{x/y,2}-D_{x/y,2}L_{x/y,1} \ne 0$, which shows the importance of the lever arm between the two detector units of a station. The decisive criterion, however, is not the absolute length of the lever arm but the change of $L_{x/y}$ and $D_{x/y}$ over this distance. In fact, since this denominator is a difference of two numbers of roughly the same magnitude, the result is very sensitive to slight changes in the optical functions and their evolution along the beamline.

\begin{figure}[h!]
\begin{center}
\includegraphics[width=0.495\textwidth]{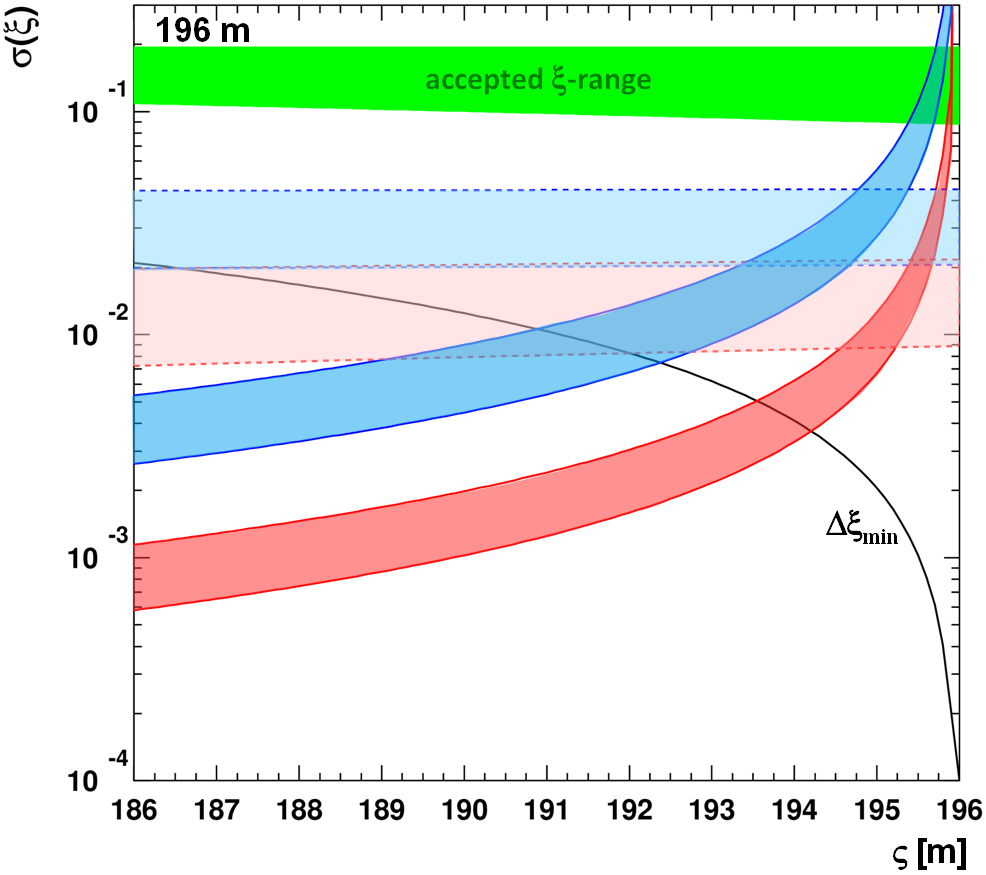}
\includegraphics[width=0.495\textwidth]{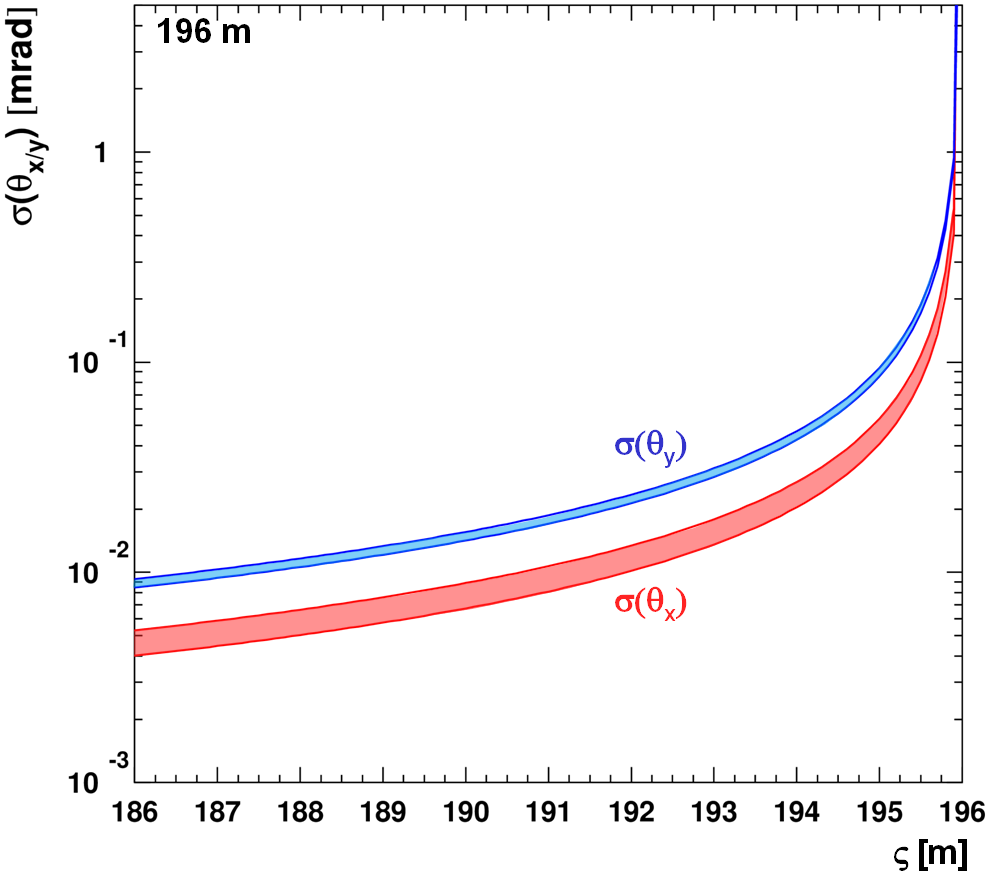}
\end{center}
\caption{Lever arm study for the 196\,m station for vertical 
crossing in IP5 with $\alpha/2 = 250\,\mu$rad and $\beta^{*} = 0.15\,$m, assuming a spatial resolution of 20\,$\mu$m per unit. One unit is placed at the fixed position 196\,m, whereas the position of the second unit varies along the abscissa of the plot. Dark red and blue bands delimited by continuous lines: resolution in $\xi$ (left panel) and $\theta^{*}_{x/y}$ (right panel) for reconstruction via $x$ (Eqs.~(\ref{eqn:transporteqnsolution1}), (\ref{eqn:xi-thetax-resol})) and $y$ (Eqs.~(\ref{eqn:transporteqnsolution2}), (\ref{eqn:xi-thetay-resol})), respectively. 
Light red and blue bands delimited by dashed lines: $\xi$ reconstruction via $x$ or $y$ but assuming $\theta^{*}_{x/y} = 0$ (Eqs.~(\ref{eqn:xi-without-thetax}) to~(\ref{eqn:xiresol-without-thetay})). Black line: deterioration of the acceptance limit $|\xi|_{\rm min}$ with increasing lever arm.}
\label{fig:resolstudy196}
\end{figure}

\begin{figure}[h!]
\begin{center}
\includegraphics[width=0.495\textwidth]{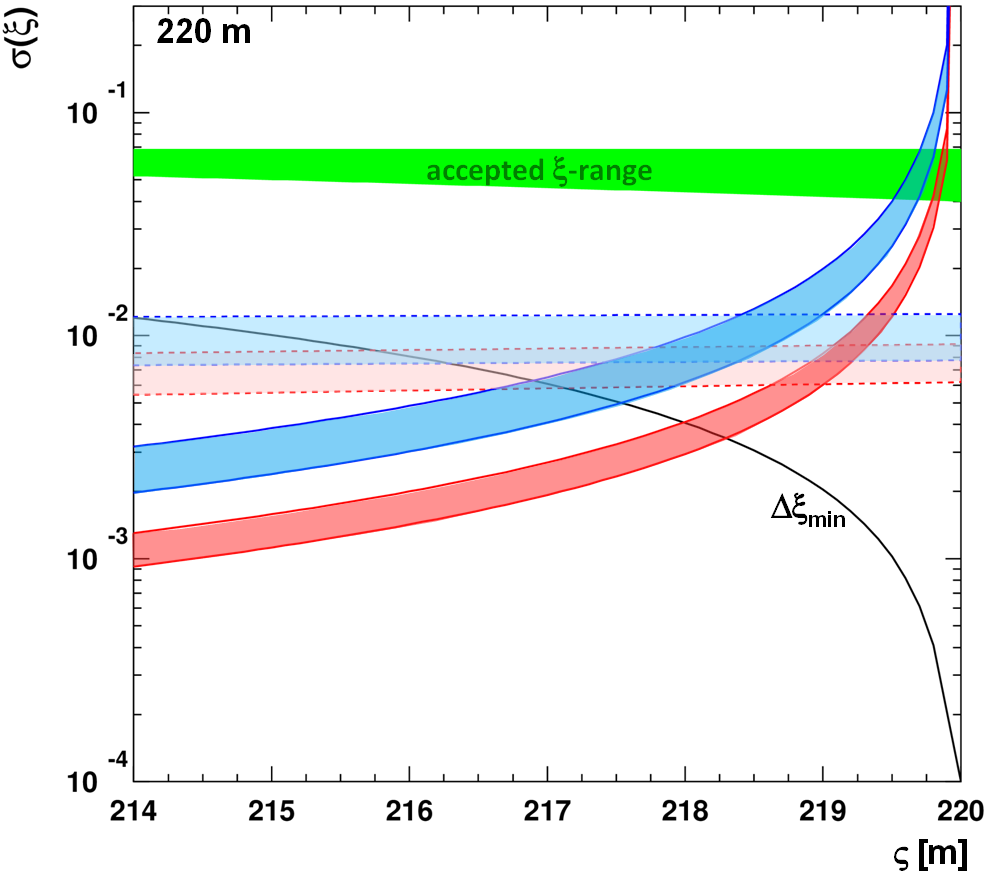}
\includegraphics[width=0.495\textwidth]{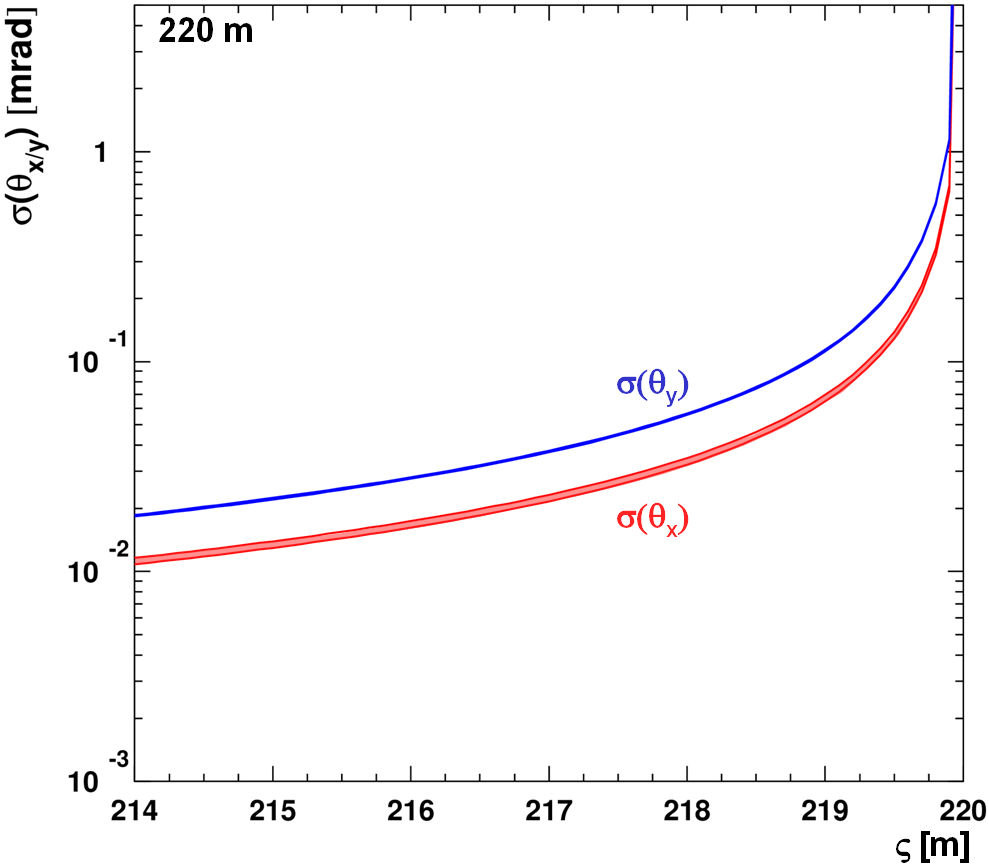}
\end{center}
\caption{Lever arm study for the 220\,m station. See caption of Fig.~\ref{fig:resolstudy196}.}
\label{fig:resolstudy220}
\end{figure}

\begin{figure}[h!]
\begin{center}
\includegraphics[width=0.495\textwidth]{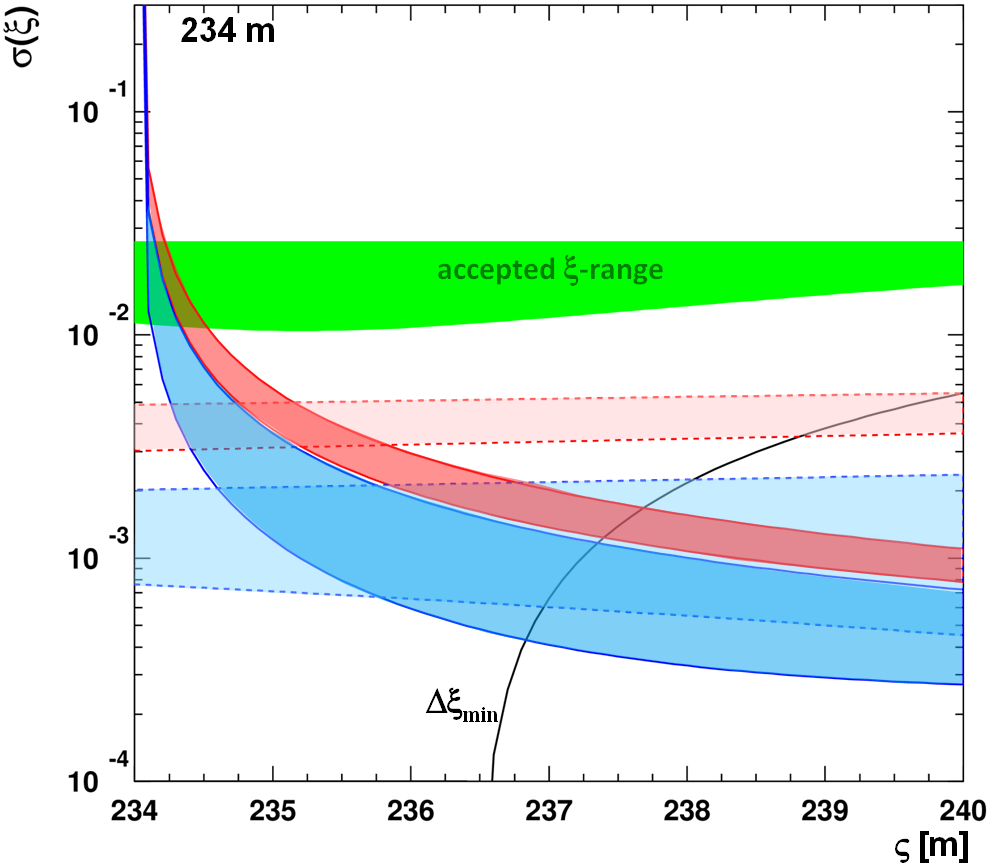}
\includegraphics[width=0.495\textwidth]{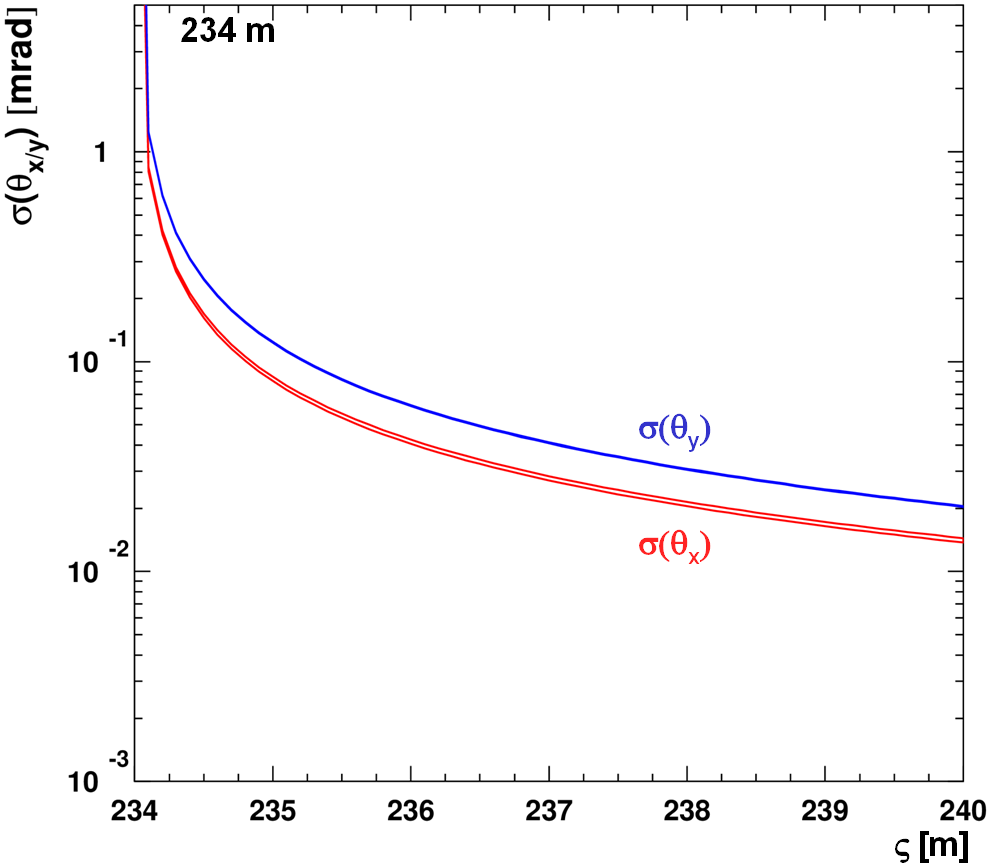}
\end{center}
\caption{Lever arm study for the 234\,m station. See caption of Fig.~\ref{fig:resolstudy196}.}
\label{fig:resolstudy234}
\end{figure}

\begin{figure}[h!]
\begin{center}
\includegraphics[width=0.495\textwidth]{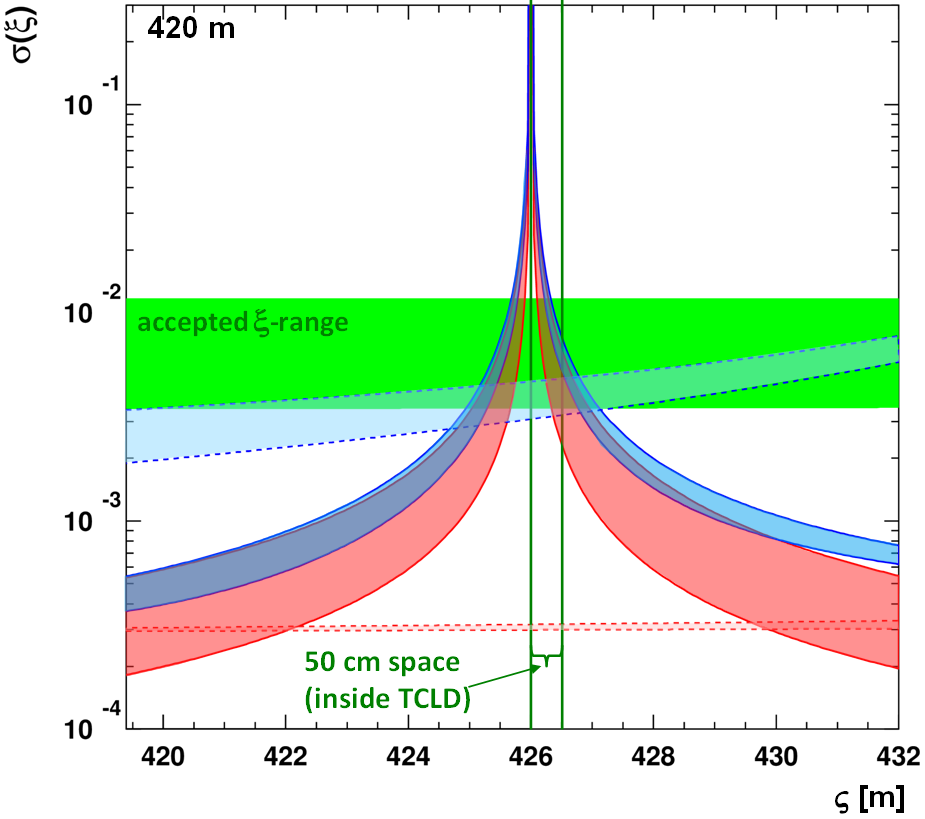}
\includegraphics[width=0.495\textwidth]{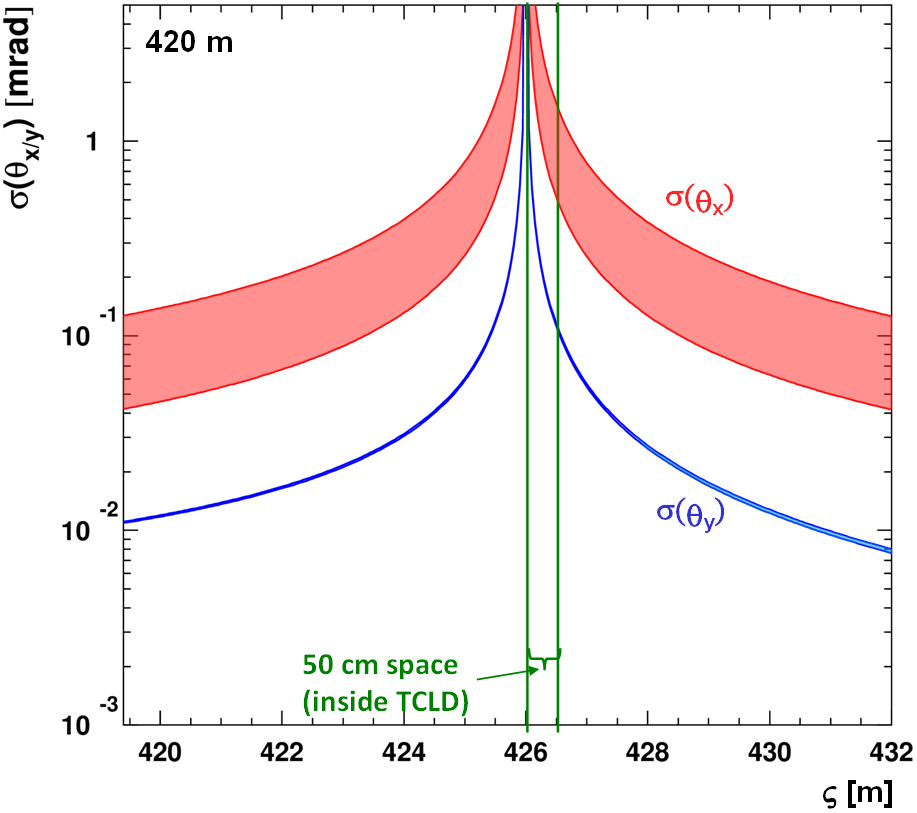}
\end{center}
\caption{Lever arm study for the 420\,m station. See caption of Fig.~\ref{fig:resolstudy196}. The first detector unit has been placed at 426\,m, the centre of the present empty cryostat. The change of $|\xi|_{\rm min}$ is negligible in this station.}
\label{fig:resolstudy420}
\end{figure}

Figures~\ref{fig:resolstudy196} to~\ref{fig:resolstudy420} show the $\xi$ and $\theta^{*}_{x/y}$ resolutions from Eqs.~(\ref{eqn:xi-thetax-resol}) and~(\ref{eqn:xi-thetay-resol}) 
as a function of the lever arm. Since the resolutions depend on $\xi$, they are represented by bands corresponding to the range $(|\xi|_{\rm min}, |\xi|_{\rm max})$. The calculations shown in the graphs also include additional terms for the smearing from the non-reconstructed vertex. Having very little impact and being lengthy, they are not explicitly discussed in this text.

The reconstruction is only usable if the resolution $\sigma(\xi)$ is smaller than the value of $|\xi|$ itself, i.e.\ for lever arms having the resolution curve below the green band of accepted $\xi$ values. For increasing lever arm the resolution improves, but at the same time the acceptance overlap between the two units deteriorates because the dispersion changes along $\varsigma$, as shown by the black $\Delta |\xi|_{\rm min}$ curve. As long as $\Delta |\xi|_{\rm min} < \sigma(\xi)$, this deterioration is below significance, so the ideal compromise between acceptance and resolution optimization lies in the region where $\Delta |\xi|_{\rm min} \approx \sigma(\xi)$. To compare the $\theta^{*}_{x/y}$ resolutions with the acceptance, see Fig.~\ref{fig:theta-acceptance}: the typical acceptance limit is roughly $|\theta^{*}_{x/y}| < 0.7\,$mrad; hence the scattering angle can be resolved, except at 420\,m.

In situations where the lever arm cannot -- for space reasons -- be made long enough to fulfill the condition $\sigma(\xi) < |\xi|$, as for example at 420\,m, it may be necessary to give up on reconstructing the angles $\theta^{*}_{x/y}$ by assuming them to be zero and assigning a systematic uncertainty on $\xi$ for this approximation. In this case, the reconstructed $\xi$ is simply given by
\begin{eqnarray}
\label{eqn:xi-without-thetax}
\xi_{\rm from~x} = \frac{1}{2} \left(\frac{x_{1}}{D_{x,1}} + \frac{x_{2}}{D_{x,2}} \right) \\
\label{eqn:xi-without-thetay}
\xi_{\rm from~y} = \frac{1}{2} \left(\frac{y_{1}}{D_{y,1}} + \frac{y_{2}}{D_{y,2}} \right)
\end{eqnarray}
with uncertainties
\begin{eqnarray}
\label{eqn:xiresol-without-thetax}
\sigma^{2}(\xi_{\rm from~x}) = \frac{1}{4} \left(\frac{1}{D_{x,1}^{2}} + \frac{1}{D_{x,2}^{2}}\right) \sigma^{2}(x) + \frac{1}{4} \left(\frac{L_{x,1}}{D_{x,1}}+\frac{L_{x,2}}{D_{x,2}} \right)^{2} \sigma^{2}(\theta_{x}^{*})\\
\label{eqn:xiresol-without-thetay}
\sigma^{2}(\xi_{\rm from~y}) = \frac{1}{4} \left(\frac{1}{D_{y,1}^{2}} + \frac{1}{D_{y,2}^{2}}\right) \sigma^{2}(y) + \frac{1}{4} \left(\frac{L_{y,1}}{D_{y,1}}+\frac{L_{y,2}}{D_{y,2}} \right)^{2} \sigma^{2}(\theta_{y}^{*}) \: .
\end{eqnarray}
The angular uncertainty at the vertex, $\sigma(\theta_{x/y}^{*})$, has two contributions:
\begin{itemize}
\item the angular beam spread: $\sigma_{\rm spread}(\theta_{x/y}^{*}) = \sqrt{\frac{\varepsilon_{n}}{\beta^{*}\,\gamma}}$ amounting to 50\,$\mu$rad for $\beta^{*} = 15\,$cm; 
\item the rms width of the scattering angle distribution. Since this is process-dependent, we use the inclusive central diffractive distribution $\frac{\rm d\sigma_{rm diff}}{{\rm d}\theta_{x/y}} \propto {\rm e}^{-B p^{2}\theta_{x/y}^{2}}$ for simplicity:\\ $\sigma_{\rm cross-sec.}(\theta_{x/y}^{*}) = \frac{1}{\sqrt{2B}}$, amounting to $\sim 45\,\mu$rad for a diffractive slope $B$ of the order of 5\,GeV$^{-2}$. 
\end{itemize}
The performance of this alternative reconstruction is also shown in Figs.~\ref{fig:resolstudy196} to~\ref{fig:resolstudy420} as bands delimited by dashed lines. 



Comparison with the available space (Section~\ref{sec:spaceconstraints}) leads to the conclusion that a sufficiently long lever arm can be accommodated in all stations except at 420\,m: even though the present empty cryostat LEGR extends from 419.4\,m to 432\,m, the available space in the new connection cryostat designed for the TCLD collimators in points 2 and 7 -- and a possible option for accommodating the 420\,m station (see Section~\ref{sec:420m-integration}) -- is only about 50\,cm long, not sufficient to simultaneously resolve  $\xi$ and $\theta^{*}_{x/y}$ with satisfactory resolution. The $\xi$ reconstruction might have to proceed via Eq.~(\ref{eqn:xi-without-thetax}). Since $|\xi|_{\rm min}$ is almost perfectly flat between 419.4\,m and 432\,m, the detector station can be placed in any position preferred from integration arguments.

%

\subsection{Calibrations}
\label{sec:calibrations}
A prerequisite for the reconstruction of kinematic variables as described in the previous section is the knowledge of
\begin{itemize}
\item the optical functions,
\item the positions and angular orientations of the detectors (alignment).
\end{itemize}
While the details of the calibration strategy for HL-LHC are still to be defined, experience from the Runs 1 and 2 provides a collection of tools and methods that can be built on. The present section will list and describe these basic concepts.

\subsubsection{Special Calibration Fills}
\label{sec:calibrationfills}
A lesson learnt in Runs 1 and 2 is the importance of special calibration fills with very low beam current (typical three nominal bunches per beam) enabling the insertion of the Roman Pots as close as about 5\,$\sigma$ from the beam centre. Furthermore, at low beam currents the TCL collimators that cut away events at the upper end of the $\xi$ scale can be left open.
Such configurations give access to a much wider $\xi$ range than the typical 15\,$\sigma$ in standard high-luminosity fills. In particular, structures originating from the $\xi$ dependence of the optical functions become visible. 

\begin{figure}[h!]
\begin{center}
\includegraphics[width=0.9\textwidth]{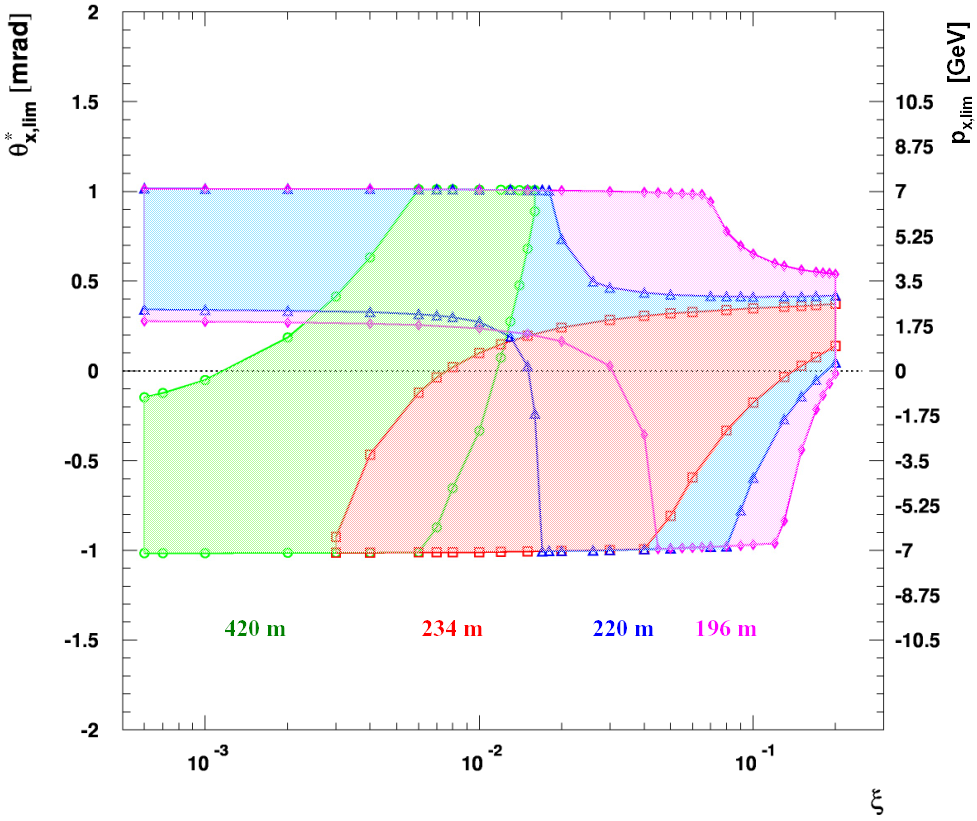}
\end{center}
\caption{Acceptance in $\xi$ and $\theta^{*}_{x}$ in special calibration fills with movable detector vessels inserted to $5\,\sigma$ from the beam centre and TCL collimators retracted. Vertical crossing in IP5 with $\alpha/2 = 250\,\mu$rad is assumed.}
\label{fig:xi-thetax-calib}
\end{figure}
At HL-LHC, special fills with close detector insertions and open TCLs would also address a specific difficulty of the forward detector configuration presented in this document: the largely missing kinematic overlap between the four detector stations (see in particular Fig.~\ref{fig:theta-acceptance}). Figure~\ref{fig:xi-thetax-calib} shows the improved $(\xi, \theta^{*}_{x})$ acceptance for an insertion distance of $5\,\sigma$ and open TCLs. At small scattering angles $|\theta^{*}_{x}|$, where -- because of the steep exponential ${\rm d\sigma/d}t$ -- most events are collected, there are now $\xi$-acceptance overlaps
\begin{itemize}
\item between the 420\,m and the 234\,m stations,
\item between the 234\,m, the 220\,m, and the 196\,m stations.
\end{itemize}
These overlaps will allow cross-calibrations between the different stations.

In practice, the calibration data collection will -- like in Runs 1 and 2 -- be performed in the same fills as the beam-based alignment, a standard procedure of the collimation system establishing the correct hierarchy within the different collimation stages~\cite{collimator-bba}. Typically, it is sufficient to take calibration data at about three different crossing-angles, $\alpha$, and one value of $\beta^{*}$, provided that the $\beta^{*}$-independence of the transport matrix (Section~\ref{sec:optics}) is maintained as an optics property. For each configuration $(\alpha, \beta^{*})$ about three hours of data collection are needed. Because of the special collimator and XRP position settings these operations are performed in the beam mode ``ADJUST'', not in ``STABLE BEAMS'', preventing some detector systems (e.g.\ the tracker) from running.

\subsubsection{Alignment}
\label{sec:alignment}
The operational and analysis experience with XRP data from the LHC Runs 1 and 2 has established a sequence of various alignment techniques that will lay a solid foundation also for HL-LHC, with some modifications due to differences in instrumentation and acceptances. The present-day strategy comprises the following steps (details in Section~3.4 of Ref.~\cite{performancepaper}, Section~2.7 of Ref.~\cite{ctpps}, Ref.~\cite{totemnote2017001}, Section~3 of Ref.~\cite{jhep2018}). Only the calibration fills require a special machine configuration, as discussed in Section~\ref{sec:calibrationfills}, whereas the alignment steps in physics fills are software procedures carried out with regular offline data and hence transparent to CMS operations.
\begin{enumerate}
\item {Alignment in calibration fills:}

\smallskip
A special low-intensity fill allowing the XRPs to approach the beam centre to a distance of only about $5\,\sigma$ is used for the first alignment stages:
\begin{enumerate}
\item Beam-based alignment of the XRP detector housing relative to the beam centre:\\
The beam is scraped with the primary collimators (TCP) in IR7 to develop a sharp edge. Then each XRP is moved in small steps (approximately 10\,$\mu$m) until it touches the edge of the beam, which generates a spike in the rate observed in the beam-loss monitors (BLM) downstream of the respective XRP. At this point, the XRP is at the same distance (in units of beam $\sigma$) as the TCP. With knowledge of the beam width $\sigma$ at the XRP the absolute distance of the XRP housing from the beam centre can be deduced. In the present PPS system this step is performed for the horizontal and the vertical XRPs.

Since in the present-day XRPs the distance between the detector edge and the beam-facing window of the XRP housing is not sufficiently well known, this step does not yet provide the absolute alignment between beam and detector, but it provides a first approximation.
\item Track-based alignment of horizontal and vertical detectors relative to each other:\\
This step takes advantage of the overlap between horizontal and vertical detectors at the close approach distance in calibration fills. It determines the relative positions of the detectors and their rotations about the beam axis.
\item Horizontal detector alignment relative to the beam using elastic events:\\
The close approach of the vertical detectors to the beam gives access to a sufficient number of elastic events. The symmetries in their hit distributions in the detector planes can be exploited to determine the horizontal beam position.
\item Vertical detector alignment relative to the beam using diffractive events:\\
The vertical beam position with respect to the sensors is determined by fitting a straight line to the $y$ coordinate of the maximum of the hit distribution as a function of $x$ and extrapolating it to $x = 0$.
\end{enumerate}
The uncertainties of this procedure amount to 5\,mrad for rotations, 50\,$\mu$m for horizontal shifts, and 75\,$\mu$m for vertical shifts. These numbers apply to the alignment of a single pot with respect to the beam. The relative alignment of two XRP units is further constrained by the zero angle of the beam axis and has an uncertainty of the order of 10\,$\mu$m.
\item Alignment transfer to physics fills:

\smallskip
Since the beam position can change from fill to fill, a detector alignment has to be repeated for each fill. In physics fills, characterized by very high intensities and hence a strict collimator hierarchy, the horizontal XRPs are much farther away from the beam (typically $15\,\sigma$), and the vertical XRPs -- irrelevant for inelastic events at low $\beta^{*}$ -- are not inserted at all. This excludes the steps (1b) and (1c) above. Instead, properties of the hit distributions -- uniquely determined by the kinematic distributions of the physics processes and by the optics -- are exploited.
\begin{enumerate}
\item Horizontal alignment:\\
The normalized $x$-coordinate distribution of the hits in the detector plane relative to the beam is directly given by the $\xi$-distribution and hence invariant. Any shifts of the beam along $x$ can therefore be obtained by comparing the $x$ distribution of a physics fill with the one of a calibration fill.
\item Vertical alignment:\\
Like in step (1d).
\end{enumerate}
The precision of this alignment transfer is about 100\,$\mu$m, but again, like in the calibration fills, the relative alignment between XRP units has a much smaller uncertainty (about 10\,$\mu$m).
\end{enumerate}
The special constraints of HL-LHC will require modifications in some steps of this procedure. While details of the alignment strategy are still to be defined, some ideas to explore can already be listed.
\begin{description}
\item[$\text{\textmd{1(HL-LHC).}}$] Calibration fills:
\renewcommand\theenumi {\alph{enumi}}
\renewcommand\labelenumi {(\theenumi)}
\begin{enumerate}
\item The beam-based alignment could be improved and sped up by installing button Beam Position Monitors (BPMs) in the jaws of the movable detector vessels as it is done for the collimators. Furthermore, a precise measurement of the sensor position relative to the housing would enable already the beam-based alignment to determine the beam position relative to the detector, reducing the importance of steps (1b) and (1c).
\item Under standard low-$\beta^{*}$ optics conditions, vertical detectors have no significant acceptance for protons with $\xi \ne 0$ and are therefore not needed for physics runs. They only serve for the short calibration runs collecting elastically scattered protons. Given this limited use and the space constraints in the machine, 
not all locations will be equipped with vertical units. The present idea is to equip only the 220\,m station with one or two vertical units. The steps (1b) (track-based alignment relying on horizontal-vertical overlap) and (1c) (alignment using elastic events) would then be performed only for the 220\,m station. Taking advantage of the kinematic overlap between the stations in the calibration fills (Section~\ref{sec:calibrationfills}), one could then transfer the alignment from the 220\,m station to the others.
\item The alignment step (1d) based on hit-map properties could be extended, taking advantage of the strong vertical dispersion. During a calibration fill the vertical crossing-angle could be inverted, so that hit maps for both signs could be recorded. For each crossing-angle sign a straight line would be fitted to the $y$ coordinate of the maximum of the hit distribution as a function of $x$: $y_{+\alpha}(x)$ and $y_{-\alpha}(x)$. The extrapolated point where $y_{+\alpha}(x) = y_{-\alpha}(x)$ is the beam position. The additional data taking at the opposite crossing-angle would increase the length of the calibration run by about two hours.
\end{enumerate}
\item[$\text{\textmd{2(HL-LHC).}}$] Physics fills:
\begin{enumerate}
\item The alignment step (2a) would again be performed in each physics fill.
\item The alignment step (2b) would again be performed in each physics fill.
\item An absolute horizontal alignment can also be done by fixing the $\xi$ scale with a standard-model process, e.g.\ dilepton production where the mass is measured by the central detector (Section~\ref{sec:standardmodel}). However, because of the steeply falling cross section as a function of $|\xi|_{\rm min}$ (Fig.~\ref{fig:sm-crosssections} left) this method will only be available for the low-mass stations (420\,m and 234\,m).
\end{enumerate}
\renewcommand\theenumi {\arabic{enumi}}
\renewcommand\labelenumi {\theenumi.}
\end{description}

\subsubsection{Optics Calibration}
\label{sec:opticscalibration}
Like for the alignment, there is a well-established procedure for calibrating the optical functions using proton data. The technique is described in Refs.~\cite{totemnote2015001,opticspaper,totemnote2017002,dn-19-026,ctpps-opticstalk20180821,frici-ppsgeneral201909},~\cite{performancepaper} (Section~3.3),~\cite{jhep2018} (Section~4.1). The optics calibration is based on a collection of data-driven tools to constrain the optical functions, $L_{x}, L_{y}, D_{x}, D_{y}$, and their $\xi$-dependence. The HL-LHC strategy will be largely based on the same principles, even though a few new complications still need to be solved.
\begin{enumerate}
\item The effective lengths $L_{x}(\xi = 0)$ and $L_{y}(\xi = 0)$ are constrained based on the measured distributions of track positions and local angles for elastic events, exploiting the collinearity of the two protons. These constraints can then be used in a multiparameter fit to correct the nominal machine optics for imperfections and deviations in the magnetic lattice that are not directly measured. The nominal tolerances of the magnetic lattice properties are taken into account in this ``optics rematching'' technique.

For the design of an HL-LHC PPS spectrometer it has to be pointed out that the measurement of local angles for elastic events requires two pairs of vertical detector units. As discussed earlier, vertical detectors cannot be implemented in each of the four stations but probably only at 220\,m. The constraints determined there will have to be used to also correct the effective lengths at the 196\,m station upstream. The downstream stations, at 234 and 420\,m, cannot fully profit from this correction since the tracks have to pass further quadrupoles before reaching them. This problem still needs to be addressed.
\item The $\xi$-dependence of $L_{x}$ is determined from the width of the $x$-distribution in a detector plane: $\sigma_{x}(\xi) = L_{x}(\xi) \sigma_{\theta_{x}^{*}} \oplus v_{x}(\xi) \sigma_{x^{*}}$, where the width $\sigma_{\theta_{x}^{*}}$ of the scattering-angle distribution does not depend on $\xi$, and the vertex term is small with only a weak $\xi$-dependence. This is a very new technique~\cite{frici-collabo201909,frici-ppsgeneral201909} and has not yet been tested for $L_{y}$ but should be applicable. At present, the $\xi$-dependence of $L_{y}$ is taken from the MAD-X simulation. A data-driven cross-check can be made by verifying that the reconstructed scattering-angle distribution $(\theta_{x}^{*}, \theta_{y}^{*})$ is round as required by the kinematics. A special case at HL-LHC will be the 420\,m station where the $\xi$-dependence of $L_{x}$ and $L_{y}$ within the acceptance interval is weak (Fig.~\ref{fig:optfuncvsxi}).

This method has been developed for PPS and does not require vertical detector units.
\item The calibration of the dispersion $D_{x}(\xi), D_{y}(\xi)$ takes advantage of the zero-crossing \mbox{$L_{y}(\xi_{0}) = 0$}, which is visible as a narrow waist in the hit distribution in the detector plane because all protons with $\xi = \xi_{0}$ pass through (apart from the vertex smearing) the same point $\left(x(\xi_{0}), y(\xi_{0})\right)$. At this point, the horizontal and vertical dispersions are fixed to $D_{x} = x(\xi_{0}) / \xi_{0}$ and $D_{y} = y(\xi_{0}) / \xi_{0}$, respectively.
Note, however, that in normal HL-LHC physics runs the point $\left(x(\xi_{0}), y(\xi_{0})\right)$ will not be within the detector acceptance. This underlines the importance of the calibration runs, where the detectors are close enough to the beam that the waist is visible in all stations, except at 420\,m, where this waist lies beyond the upper $\xi$ limit (Fig.~\ref{fig:xymap-calib}). These special runs will also profit from an acceptance overlap between the stations, allowing cross-calibrations.

Alternatively or as a cross-check, the $\xi$-distributions in the left and the right spectrometer arm can be compared and brought to congruence by adjusting the dispersion.
\item The $\xi$-dependence of the dispersion is presently taken from the MAD-X simulations.
\item A final verification of the dispersion can be performed using dilepton events where the mass is measured by the central detector (Section~\ref{sec:standardmodel}). 
\end{enumerate}

\clearpage
\section{Radiation Environment}
\label{sec:radiation}
\begin{figure}[h!]
\begin{center}
\includegraphics[width=\textwidth]{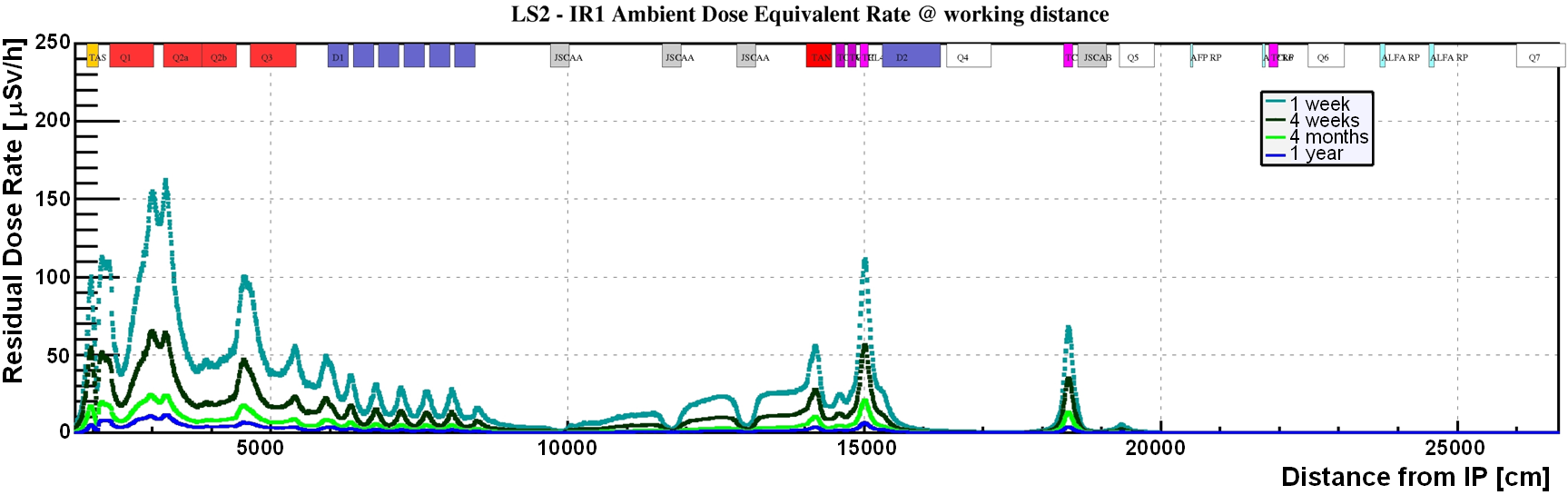}
\end{center}
\caption{Simulated radiation environment in Sector 1-2 at the end of Run 2~\cite{radiationmaps-adorisio}. The different curves represent the cooling times specified in the legend. The situation in Sector 5-6 is expected to be very similar. The local maximum at $\sim 180$\,m corresponds to the collimator TCL5 (the layout differs from HL-LHC shown in Fig.~\ref{fig:radiation_run4}). TCL6 does not show any peak because it was open in the simulation.}
\label{fig:radiation_ls2}
\end{figure}

\begin{figure}[h!]
\begin{center}
\includegraphics[width=\textwidth]{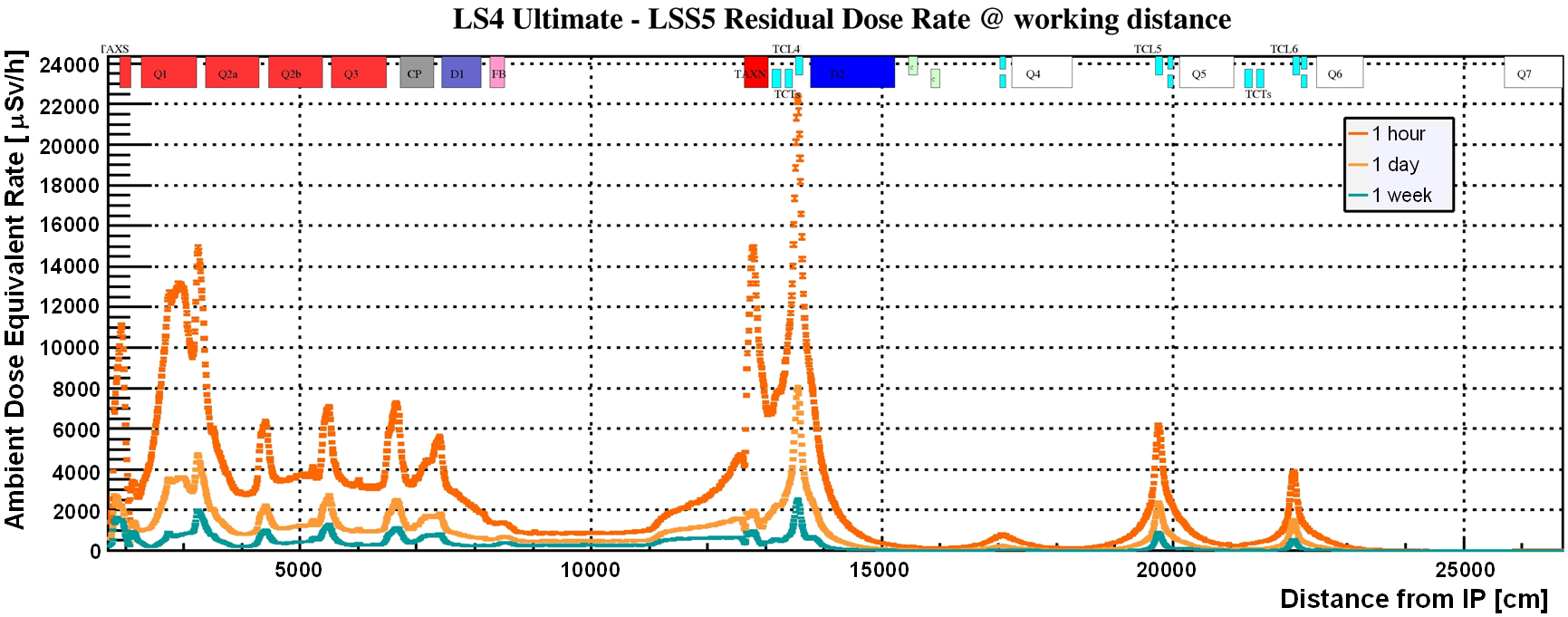}
\includegraphics[width=\textwidth]{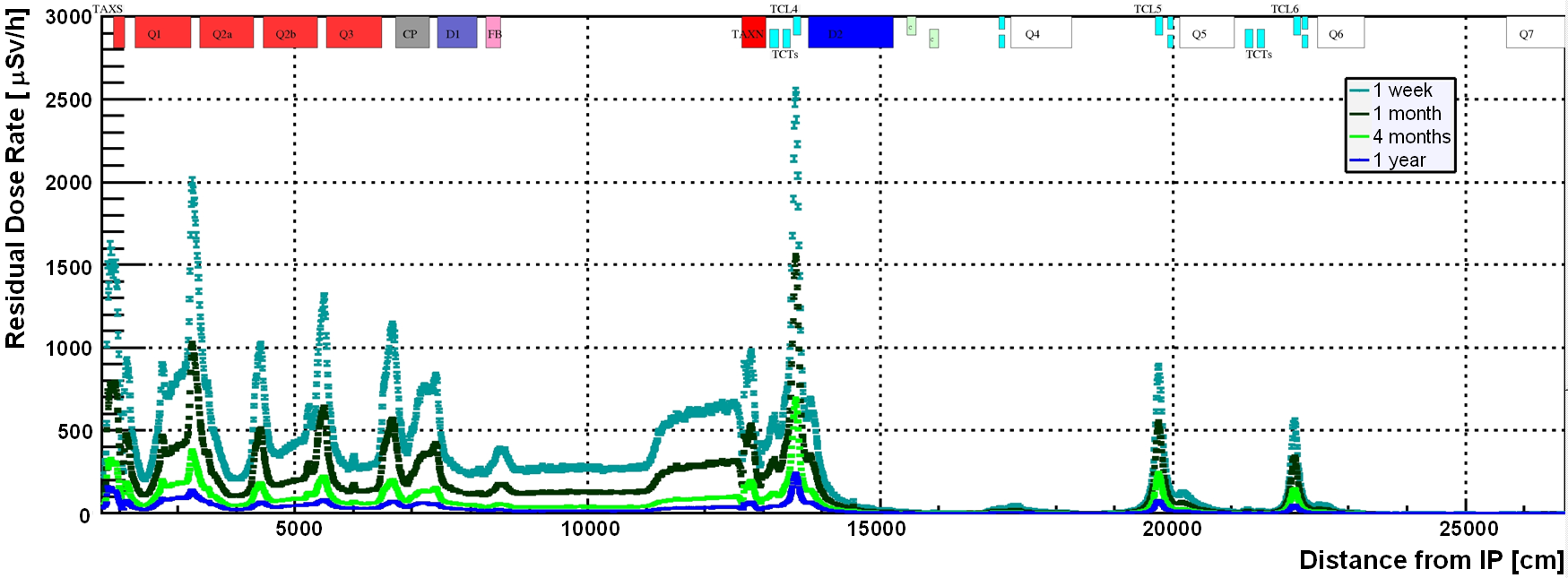}
\end{center}
\caption{Simulated radiation environment in Sector 5-6 at the end of Run 4~\cite{radiationmaps-adorisio}. The different curves in the two panels show the dose rates after different cooling times. The local maxima at $\sim 197$\,m and $\sim 222$\,m correspond to the collimators TCL5 and TCL6.}
\label{fig:radiation_run4}
\end{figure}

\begin{figure}[h!]
\begin{center}
\includegraphics[width=\textwidth]{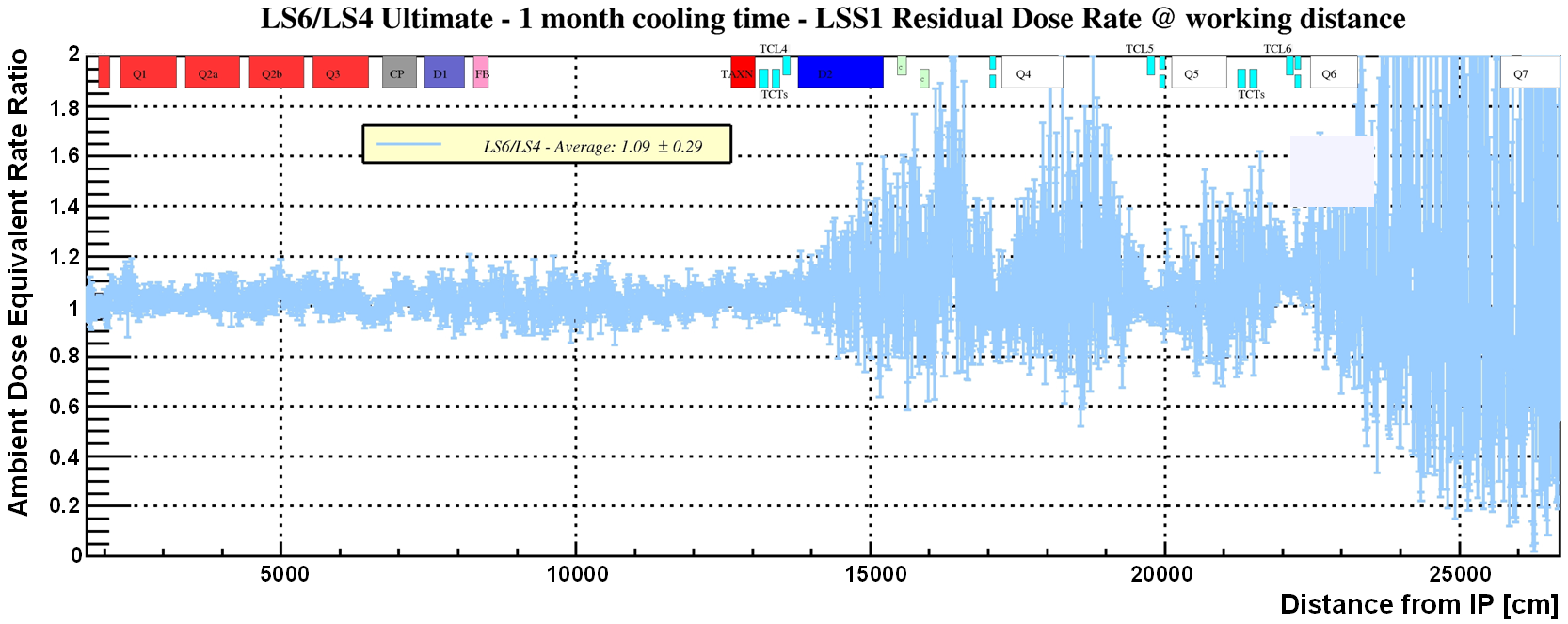}
\end{center}
\caption{Ratio of the simulated dose rates in Sector 1-2 after LS6 and after LS4 for a cooling time of one month~\cite{radiationmaps-adorisio}. The situation in Sector 5-6 is expected to be similar.}
\label{fig:radiationratio_ls6-ls4}
\end{figure}

The increased luminosity at HL-LHC leads to a harsh radiation environment with consequences on equipment irradiation and maintenance planning. FLUKA simulations, benchmarked with measurements in Run 2, were carried out for the following configurations~\cite{radiationmaps-adorisio}:
\begin{itemize}
\item LHC until LS3, assuming integrated luminosities of 144 and 180\,fb$^{-1}$ for Run 2 and 3, respectively;
\item HL-LHC, assuming the ultimate performance case with a total integrated luminosity of 900\,fb$^{-1}$ for each of the 3-year periods of Runs 4, 5, and 6. 
\end{itemize}
The results were expressed in terms of residual dose rates in the Long Straight Sections of the high-luminosity interaction points 1 and 5 at the end of a running period after certain cool-down times; see Fig.~\ref{fig:radiation_ls2} for the end of Run 2 and Fig.~\ref{fig:radiation_run4} for the end of Run 4. The ratio between the residual rates after Run 6 and after Run 4 with a cooling time of one month (Fig.~\ref{fig:radiationratio_ls6-ls4}) is close to~1, indicating a saturation.
 
A common observation is that the hottest regions of interest for forward physics are: 
\begin{itemize}
\item around the neutral beam absorber (TAN at pre-LS3 LHC, TAXN at HL-LHC) and TCL4: only relevant because service personnel need to pass through this area to access the detector locations;
\item the TCL5 location: not used for detectors at present LHC but very close to the 196\,m location at HL-LHC;
\item the TCL6 location: very close to the 220\,m station in both LHC and HL-LHC layouts.
\end{itemize}
Comparison of the dose rates near TCL5 between LS2 and LS4 shows an increase by a factor 8.6, which is of the same order as the ratio in integrated luminosities, 6.3.
Since in the pre-LS3 simulation TCL6 was assumed fully open, a direct comparison with the spike at HL-LHC is not possible, but a similar increase as at TCL5 can be assumed.

The TCL6 area was already very hot in 2016 and 2017, requiring the construction of a movable shield for maintenance work. To arrive at the same dose rate as at the pre-LS3 LHC after one week, a cooling time of about 17 months would be needed. While shielding can help to some extent, it is clear that interventions during the short Technical Stops are excluded and work can only be performed in the Year-End Technical Stops after long waiting times still to be quantified. In particular, detector package replacements during a running season, as performed in the pre-LS3 PPS during the short Technical Stops, will be impossible. 
Hence all instrumentation from detectors and electronics to the movement system will have to be designed maintenance-free with particular focus on radiation hardness. A more detailed analysis of the expected detector irradiation has to take into account the strongly peaked hit distribution (Section~\ref{sec:detectorenvironment}). The dose is concentrated on a very small spot in the detector area. This fact was exploited in the past by shifting detectors vertically in technical stops during the year, in order to displace the most exposed spot and thus extend the detector lifetime. At HL-LHC such mechanical position adjustments will have to be automated with remote control. A piezo-electric movement system is being introduced during LS2 for the tracking detector packages to be operational in Run~3.
\pagebreak
\clearpage

\section{Detectors}
\label{sec:detectors}

\subsection{Requirements}
\label{sec:detectorrequirements}

\subsubsection{Sensitive Area}
\label{sec:sensitivearea}
\begin{figure}[hp!]
\vspace*{-5mm}
\begin{center}
\includegraphics[width=0.42\textwidth]{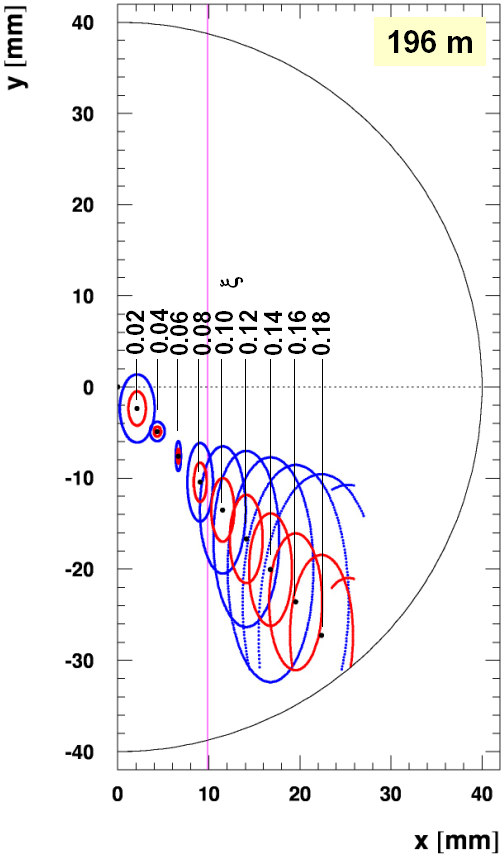}
\includegraphics[width=0.42\textwidth]{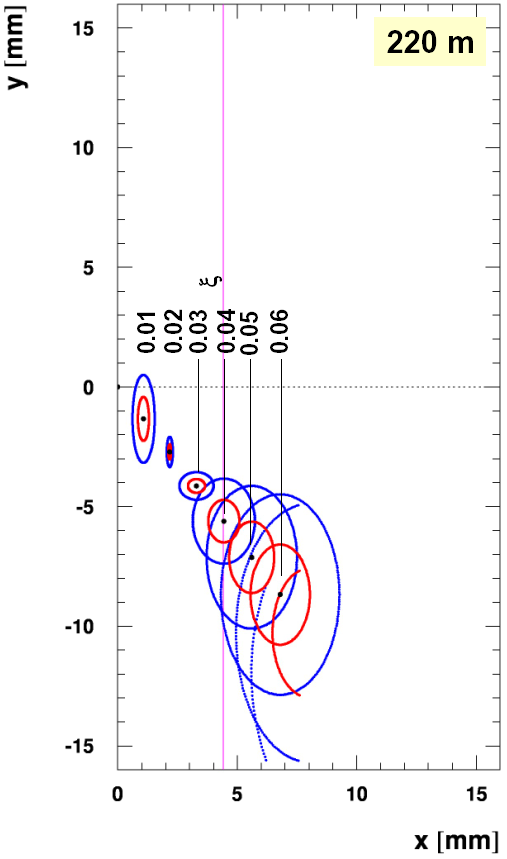}

\includegraphics[width=0.42\textwidth]{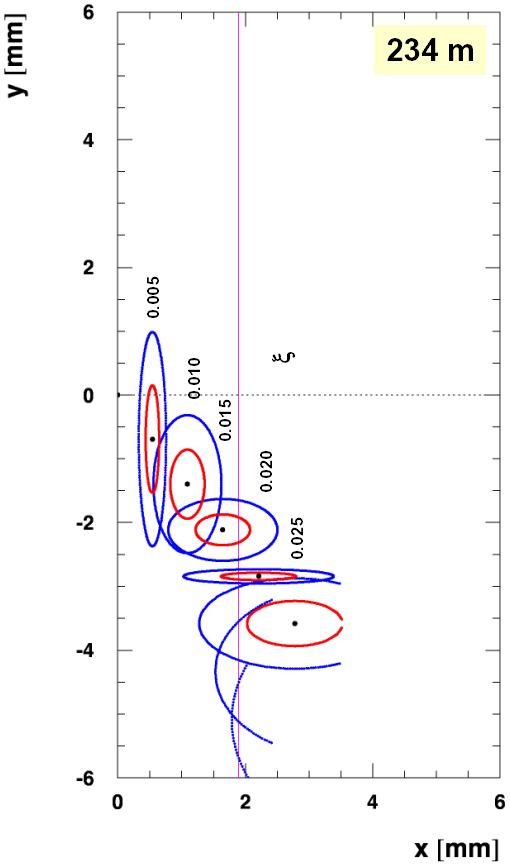}
\includegraphics[width=0.42\textwidth]{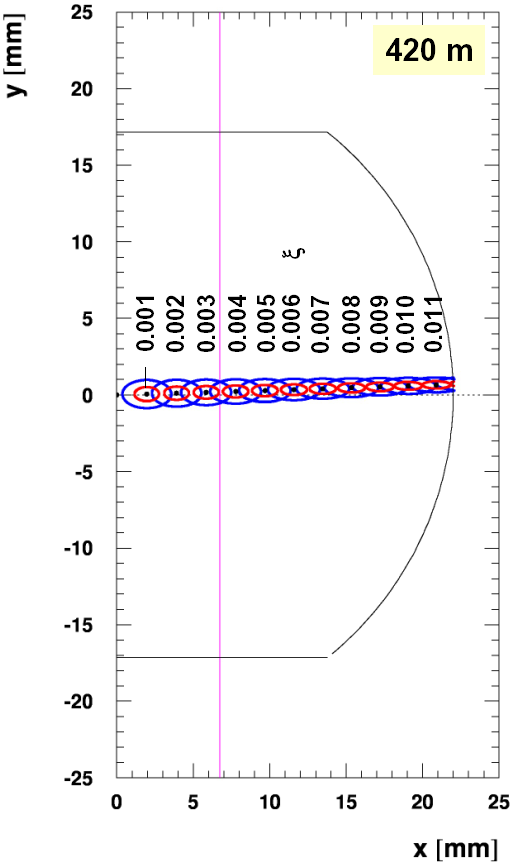}
\end{center}
\vspace*{-5mm}
\caption{Mapping of the kinematic variables $(\xi, \theta^{*})$ on the detector planes in the four locations for vertical crossing and $\alpha/2 = 250\,\mu$rad. Several discrete values of $\xi$ are shown. The black points are for $\theta = 0$. The red and blue ellipses are for $\theta^{*} = 100$ and $200\,\mu$rad, respectively, with varying azimuth. Only points within aperture acceptance are shown. TCL and detector position settings correspond to regular physics runs. The violet vertical lines show the detector edges near the beam; the sensitive area lies to their right. The black outer lines represent the shapes of the beamlines at the detector stations.
}
\label{fig:xymap}
\end{figure}

\begin{figure}[hp!]
\vspace*{-5mm}
\begin{center}
\includegraphics[width=0.42\textwidth]{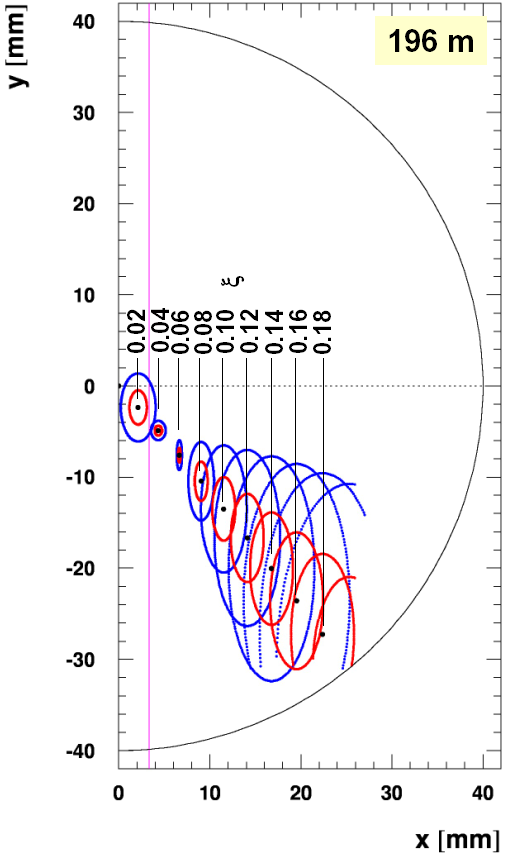}
\includegraphics[width=0.42\textwidth]{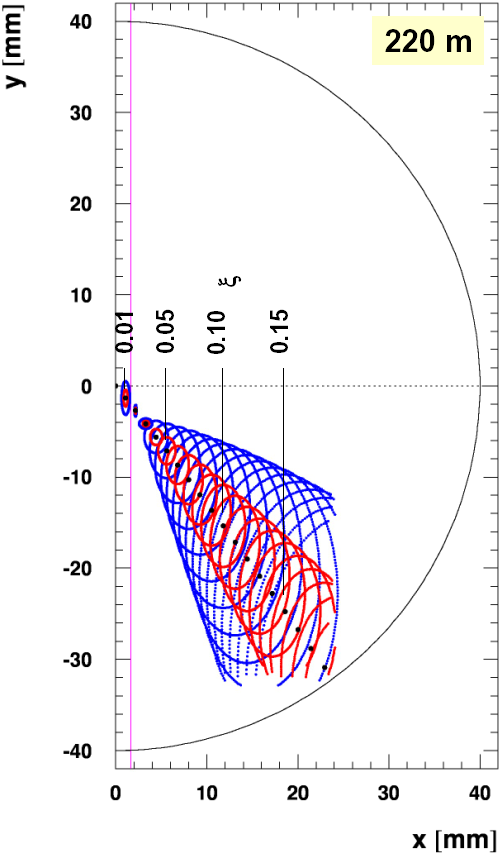}

\includegraphics[width=0.42\textwidth]{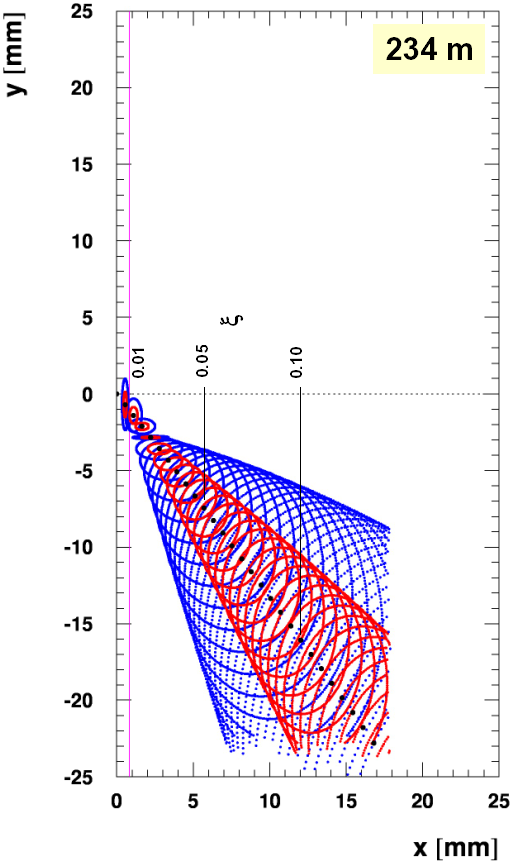}
\includegraphics[width=0.42\textwidth]{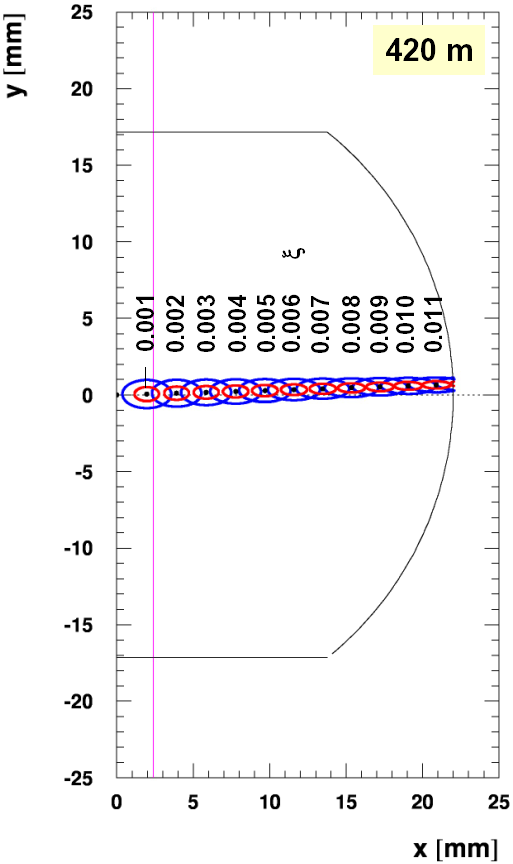}
\end{center}
\vspace*{-5mm}
\caption{Like Fig.~\ref{fig:xymap}, but for the detector positions and open TCL settings of calibration runs.
}
\label{fig:xymap-calib}
\end{figure}

For an optimum acceptance in the kinematic variables $\xi$ and $\theta_{x/y}^{*}$ the detectors should not impose any limitations beyond those defined by the beamline aperture between IP5 and the detector location. Hence the detector size has to be big enough to intercept all tracks passing the aperture. Figures~\ref{fig:xymap} and~\ref{fig:xymap-calib} show the mapping of $(\xi, \theta^{*}, \phi^{*})$ on the scoring planes at the four station locations for regular physics runs and for calibration runs, respectively. Each pair $(\xi, \theta^{*})$ corresponds to an ellipse in the detector plane. Only the accepted portions of the ellipses are drawn. The underlying cutoffs are displayed in Figs.~\ref{fig:theta-acceptance} and ~\ref{fig:xi-thetax-calib}.
The inversion of the vertical crossing-angle -- foreseen to happen every year -- flips the sign of the $y$ coordinate of each track point. The detectors therefore need to cover twice the vertical extent of the hit population drawn.

To have full coverage of the kinematics not only in regular physics runs but also in the calibration runs providing overlaps between the stations, the detectors should cover at least the following areas ($x \times y$), where the factor two in the $y$-dimension accounts for the periodic flip of the vertical crossing-angle:
\begin{itemize}
\item 196\,m: 24\,mm $\rm \times (2 \times 32\,mm$),
\item 220\,m: 23\,mm $\rm\times (2 \times 33\,mm$),
\item 234\,m: 17\,mm $\rm\times (2 \times 24\,mm$),
\item 420\,m: 20\,mm $\rm\times (2 \times 2\,mm$).
\end{itemize}
Note, however, that these sizes do not take into account that for diluting the spots of most intense irradiation, vertical shifts of the detector packages on a regular basis should be envisaged, which would require larger vertical dimensions; see the discussion in the next section.

\subsubsection{Environment}
\label{sec:detectorenvironment}
The most important environmental challenge for the detectors is the intense, highly non-uniform radiation background, dominated by Single Diffractive protons, followed by inelastic collision debris (not yet studied for HL-LHC). The number of single-diffractive protons per unit area impacting on the detector plane in the point $(x, y)$ has been calculated from the simplified differential cross section 
\begin{equation}
\frac{{\rm d^{3}}\sigma_{\rm SD}}{{\rm d}\xi \, {\rm d}t \, {\rm d}\phi} = \frac{A}{2\pi} \frac{1}{|\xi|} {\rm e}^{-B|t|} \quad {\rm or} \quad 
\frac{{\rm d^{3}}\sigma_{\rm SD}}{{\rm d}\xi \, {\rm d}\theta_{x}^{*} \, {\rm d}\theta_{y}^{*}} = \frac{A\,p^{2}}{\pi} \frac{1}{|\xi|} {\rm e}^{-B\,p^{2}(\theta_{x}^{*\,2}+\theta_{y}^{*\,2})} \: ,
\label{eqn:SD-diff-cross-section}
\end{equation}
where $A = B \, \sigma_{\rm SD} / \ln \frac{\xi_{\rm max}}{\xi_{\rm min}}$ with $\sigma_{\rm SD} = 8\,$mb, $B = 10\,\rm GeV^{-2}$ and the usual cutoffs $|\xi|_{\rm min} = \frac{m_{p}^{2}}{s}$,  $|\xi|_{\rm max} = 0.3$.
Then, using the optical transport functions for $x$ and $y$ coordinates as delta functions, one obtains the following equation:

\begin{equation}
\label{eqn:transformationintegral}
\frac{{\rm d^{2}}\sigma_{\rm SD}}{{\rm d}x \, {\rm d}y} = 
\int_{|\xi|_{\rm min}}^{|\xi|_{\rm max}} {\rm d}\xi \int_{-\infty}^{\infty} {\rm d}\theta_{x}^{*} \int_{-\infty}^{\infty} {\rm d}\theta_{y}^{*} \, \frac{{\rm d^{3}}\sigma_{\rm SD}}{{\rm d}\xi \, {\rm d}\theta_{x}^{*} \, {\rm d}\theta_{y}^{*}} \, 
\delta[x - (D_{x}\xi + L_{x} \theta_{x}^{*})]
\delta[y - (D_{y}\xi + L_{y} \theta_{y}^{*})] \: .
\end{equation}

\begin{figure}[hp!]
\vspace*{-5mm}
\begin{center}
\includegraphics[width=0.45\textwidth]{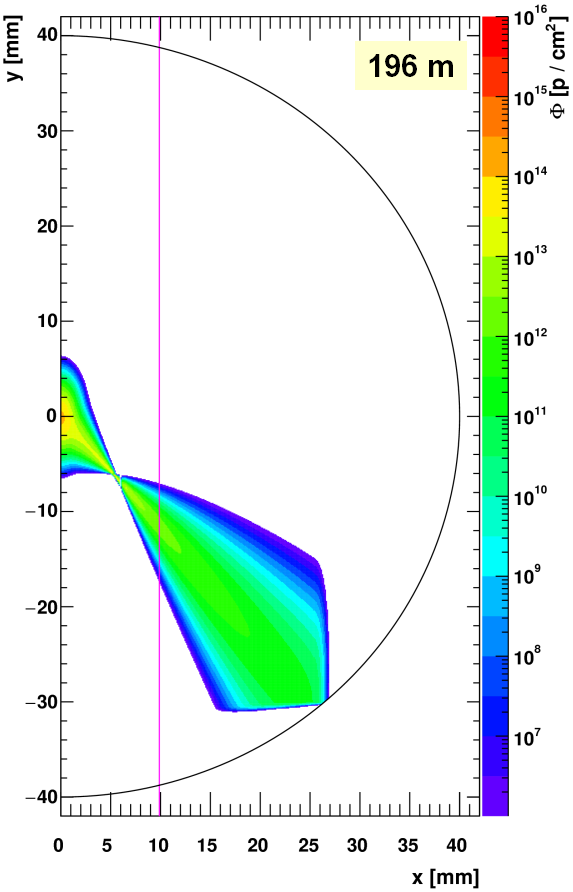}
\includegraphics[width=0.45\textwidth]{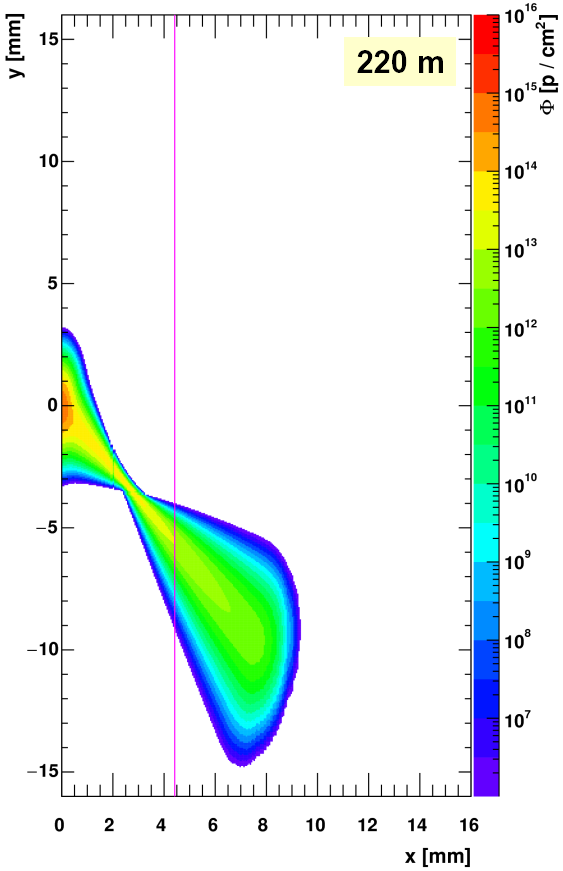}

\includegraphics[width=0.45\textwidth]{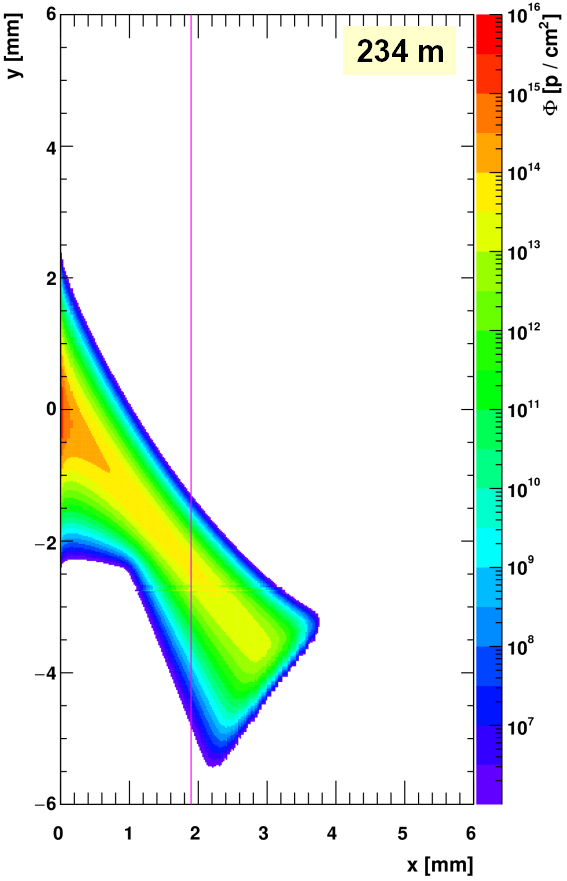}
\includegraphics[width=0.45\textwidth]{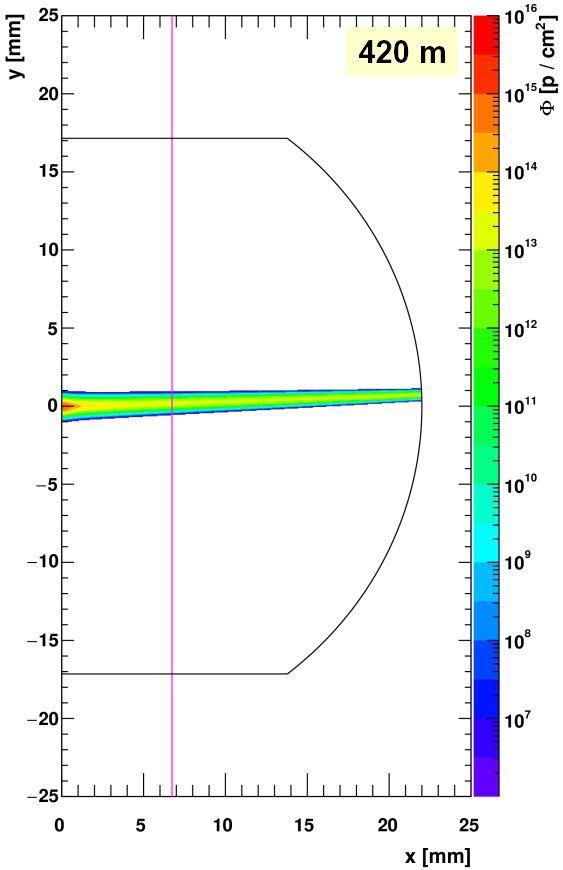}
\end{center}
\vspace*{-5mm}
\caption{Hit maps for single diffractive events accumulated for an integrated luminosity of 1\,fb$^{-1}$. The colour scale represents the fluence $\Phi [\rm p / cm^{2}]$ calculated using the transformation integral~(\ref{eqn:transformationintegral}). The violet vertical lines mark the positions of the detector edges near the beam in regular physics runs; the sensitive areas lie to their right. The black outer lines represent the shapes of the beamlines at the detector stations.}
\label{fig:irradiationhitmap}
\end{figure}
Figure~\ref{fig:irradiationhitmap} shows that the diffractive proton fluence is concentrated in a narrow band peaking on a line starting at the beam and continuing through the detector plane with a local slope ${\rm d}y / {\rm d}x (\xi) = D_{y}(\xi) / D_{x}(\xi)$. On this line, the intensity is highest near the beam and in the focal point where $L_{y}(\xi) = 0$. For the 234\,m station, this point lies within the acceptance even in normal physics runs, as can be seen in the kinematic map in Fig.~\ref{fig:xymap} (the very flat ellipse in the bottom left panel). At the 196\,m and 220\,m stations, it is only visible in the short calibration runs where no significant dose is received. At 420\,m, this point lies outside the aperture and is never visible. Hence the 234\,m station has the highest fluence peak. Table~\ref{tab:maximumflux} lists the points of maximum fluence in all detector stations. As an independent cross-check, the fluence maps have also been calculated by generating the kinematic variables $(\xi, \theta_{x}^{*}, \theta_{y}^{*})$ weighted with the differential cross section~(\ref{eqn:SD-diff-cross-section}) and then transforming them into hit positions $(x, y)$ using the transport equations. Since the two methods suffer from different computational imprecisions, both results are given in Table~\ref{tab:maximumflux} to indicate the systematic uncertainty from the calculation technique. The error due to the simplicity of the single-diffraction model will be much larger.

To assess the effect of one year of HL-LHC on the detectors, the fluence was also scaled to the foreseen yearly integrated luminosity of 300\,fb$^{-1}$ (last column in Table~\ref{tab:maximumflux}). For the station at 234\,m, fluences of the order $10^{16}\,\rm p / cm^{2}$ are expected per year. This is a challenge for detector development, but upcoming detector generations are likely to cope with it. As a partial mitigation of the problem the most exposed spot can be moved over the detector area to dilute the load. For this purpose a remote-controlled vertical movement mechanism is foreseen. The vertical detector dimension needs to be adapted accordingly. At 196\,m and -- to a lesser extent -- at 220\,m the hit distribution fills a substantial part of the beam pipe area, which makes vertical movements difficult, but the peak fluences in these stations are an order of magnitude lower than at 234\,m. 

\begin{table}[h!]
  \begin{center}
    \caption{Points of maximum proton fluence $\Phi$ (after 1 and 300\,fb$^{-1}$) extracted from the data of Fig.~\ref{fig:irradiationhitmap}. The values in brackets are the results of the alternative method via $(\xi, \theta_{x}^{*}, \theta_{y}^{*})$ generation and tracking (see text).  
}
    \label{tab:maximumflux}
    \begin{tabular}{|l|c|c|c|c|}
    \hline  
    Station & $x_{\rm peak}$ [mm] & $y_{\rm peak}$ [mm] & $\Phi [\rm p / cm^{2}]$ (1\,fb$^{-1}$) & $\Phi [\rm p / cm^{2}]$ (300\,fb$^{-1}$) \\
    \hline
    196\,m & 9.9 & $-11.6$ & $0.18 ~(0.19) \times 10^{13}$ & $ 5.4 ~(5.7) \times 10^{14}$\\
    220\,m & 4.5 & $-5.7$ & $0.98 ~(0.99) \times 10^{13}$ & $ 2.9 ~(3.0) \times 10^{15}$ \\
    234\,m & 2.3 & $-2.7$ & $4.7 ~(4.4) \times 10^{13}$ & $ 1.4 ~(1.3) \times 10^{16}$ \\
    420\,m & 6.8 & 0.2 & $2.0 ~(2.0) \times 10^{13}$ & $ 6.0 ~(6.0) \times 10^{15}$ \\ 
    \hline
    \end{tabular}
  \end{center}
\end{table}

\begin{figure}[h!]
\begin{center}
\includegraphics[width=0.6\textwidth]{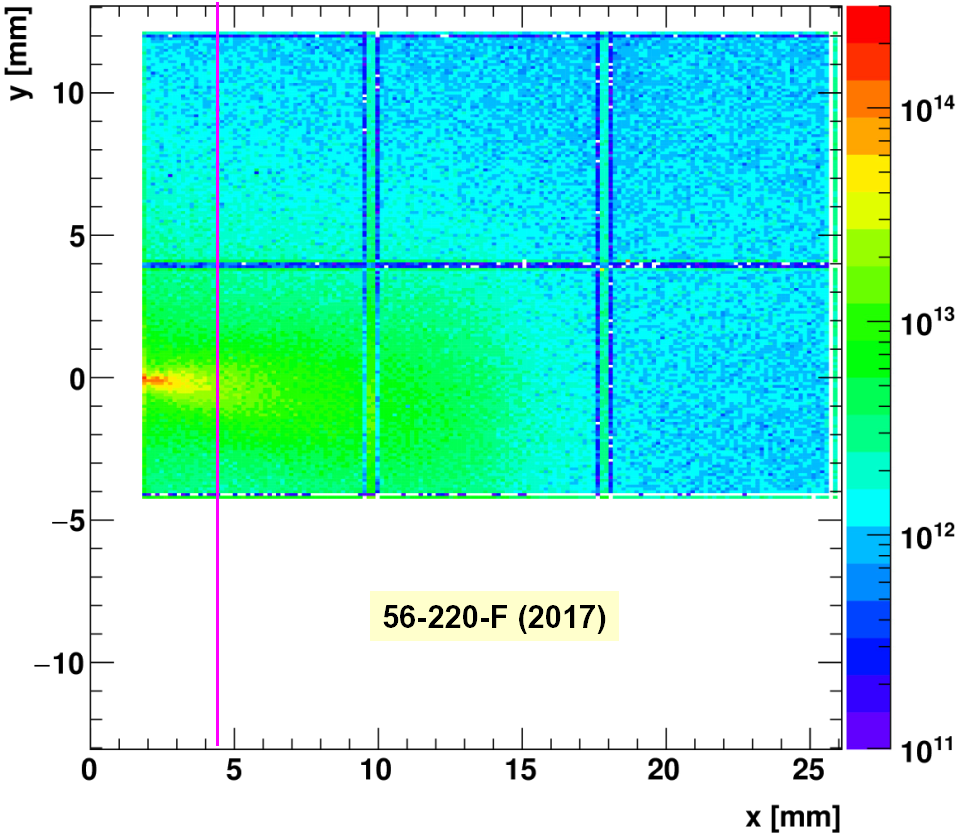}
\end{center}
\caption{Measured hit map from Run 2 (2017) in the horizontal XRP unit ``56-220-F'', pixel plane 2. The map taken with zero-bias trigger for a sample with recorded integrated luminosity of 0.14\,nb$^{-1}$ was scaled to 1\,fb$^{-1}$ for direct comparison with the calculation for HL-LHC (Fig.~\ref{fig:irradiationhitmap} top right). The colour scale represents the fluence $\Phi [\rm p / cm^{2}]$. The violet line shows where the detector edge would be located at HL-LHC; the sensitive area lies to its right.}
\label{fig:hitmap2017scaled}
\end{figure}

The above calculation was benchmarked with pixel hitmaps recorded in 2017 ($\alpha/2 = 150\,\mu$rad, $\beta^{*} = 0.4\,$m) with the XRP unit ``56-220-F-H'' (nomenclature: Sector 5-6, unit 220\,m Far Horizontal). The result for one detector plane after scaling to an integrated luminosity of 1\,fb$^{-1}$ is shown in Fig.~\ref{fig:hitmap2017scaled}. The two most obvious differences to the corresponding map for HL-LHC (Fig.~\ref{fig:irradiationhitmap} top right) are 
\begin{itemize}
\item the slope of the dominant band due to the vertical dispersion at HL-LHC, almost negligible in Run 2,
\item the diffuse cloud of points present in the LHC map due to machine backgrounds, not included in the HL-LHC calculation.
\end{itemize}
Nevertheless, the fluences at the point ($x = 4.42$\,mm, $y = 0$), i.e.\ the detector edge position at HL-LHC, agree within a factor 2: $2.10 \times 10^{13}\, \rm p / cm^{2}$ (from data taken in 2017, averaged over the six detector planes in the XRP ``56-220-F-H'') versus $0.98 \times 10^{13}\, \rm p / cm^{2}$ (HL-LHC calculation). This shows that despite differences between the optical functions in Run 2 and at HL-LHC, the order of magnitude of particle rates in a given detector position is similar.

The predicted hit maps can also be used to calculate the occupancy per unit area and bunch crossing, which in turn can contribute one of the arguments for determining the necessary detector segmentation. Table~\ref{tab:occupancy} shows the peak fluence per bunch crossing and the resulting occupancy for three example segmentations. In the Run-2 measurement (2017) the peak fluence per bunch crossing in the 220\,m detectors was 14 times higher than expected for HL-LHC in the same location. This is caused by the much smaller detector distance from the beam centre due to the different optics: 1.9\,mm in 2017 and 4.4\,mm at HL-LHC. 

\begin{table}[h!]
  \begin{center}
    \caption{Second column: peak fluence scaled to the integrated luminosity of one bunch crossing for a luminosity of $5 \times 10^{34}\,\rm cm^{-2}s^{-1}$ (with levelling) and $k = 2736$ colliding bunches (the beam parameters given in Table~\ref{tab:hllhc-parameters}). Third to fifth columns: occupancy per detector pad for three example segmentations. The last line represents the measurement in 2017 with the XRP ``56-220-F-H''.
}
    \label{tab:occupancy}
    \begin{tabular}{|l|c|c|c|c|}
    \hline  
    Station & $\Phi [\rm p / cm^{2}]$ (BX) & \multicolumn{3}{c|}{Occupancy}\\
            &             & $\rm 50 \times 50\,\mu m^{2}$ & $\rm 100 \times 100\,\mu m^{2}$ & $\rm 200 \times 200\,\mu m^{2}$\\
    \hline
    196\,m & 2.9 & $7.3 \times 10^{-5}$ & $2.9 \times 10^{-4}$ & 0.0012 \\
    220\,m & 16 & $4.0 \times 10^{-4}$ & $1.6 \times 10^{-3}$ & 0.0064 \\
    234\,m & 75 & $1.9 \times 10^{-3}$ & $7.5 \times 10^{-3}$ & 0.030 \\
    420\,m & 32 & $8.0 \times 10^{-4}$ & $3.2 \times 10^{-3}$ & 0.013 \\ 
    \hline
    220\,m (2017) & 221 & \multicolumn{3}{l|}{$\rm 150 \times 100\,\mu m^{2}$: Occupancy = 0.03}\\
    \hline 
   \end{tabular}
  \end{center}
\end{table}
This study indicates that at all four locations a segmentation with pads as large as $200\,\mu$m can be afforded while keeping the occupancy at a similar level as observed in Run 2 (2017). At the 196\,m and 220\,m stations, where large areas need to be covered by detectors, this solution would keep the number of channels manageable. At 234\,m, where the occupancy is highest, it might be prudent to reduce the pad size to $100\,\mu$m, given that no backgrounds beyond single diffraction have been considered so far. 

\subsubsection{Tracking Requirements}
The occupancy study in the previous section yielded pad sizes between 100 and $200\,\mu$m. Considering the upper end of that range, the resolution of a $200\,\mu$m pad without charge sharing and without plane inclination would be $\rm 200\,\mu m / \sqrt{12} = 58\,\mu m$. However, the present pixel detectors with a horizontal (vertical) cell pitch of 150\,$\mu$m (100\,$\mu$m) and a plane inclination of 18.4$^{\circ}$ have a spatial resolution of $\sim 25\,\mu$m per plane, which is clearly better than the conservative estimate of pitch/$\sqrt{12}$. 

With four to six planes per unit the total resolution per detector stack will be better, depending on the projective alignment of the pixels in different planes. This is fully adequate for the reconstruction of the proton kinematics. 

\subsubsection{Timing Requirements}
\label{sec:timing}
\begin{figure}[h!]
\begin{center}
\includegraphics[width=0.49\textwidth]{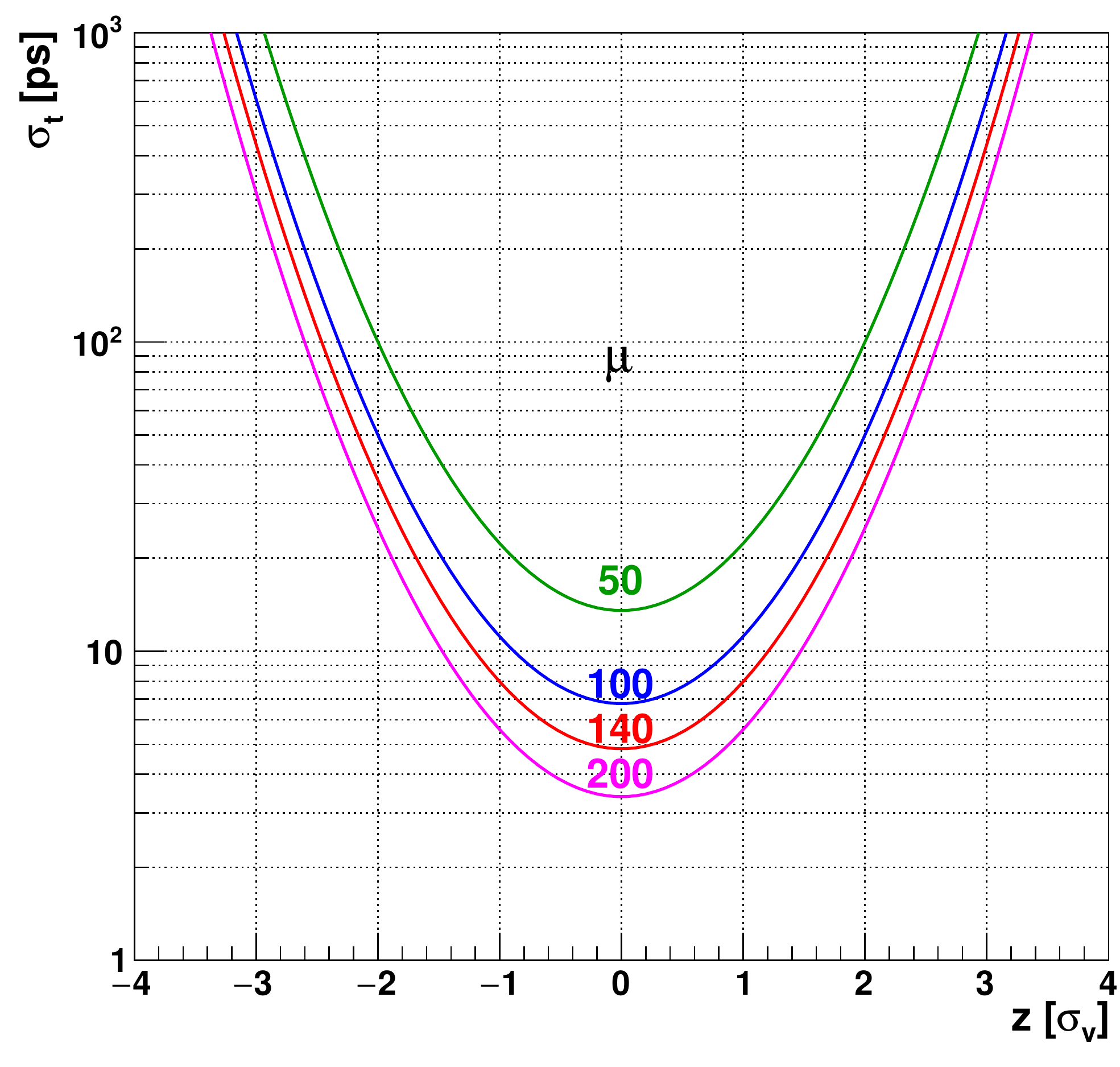}
\includegraphics[width=0.49\textwidth]{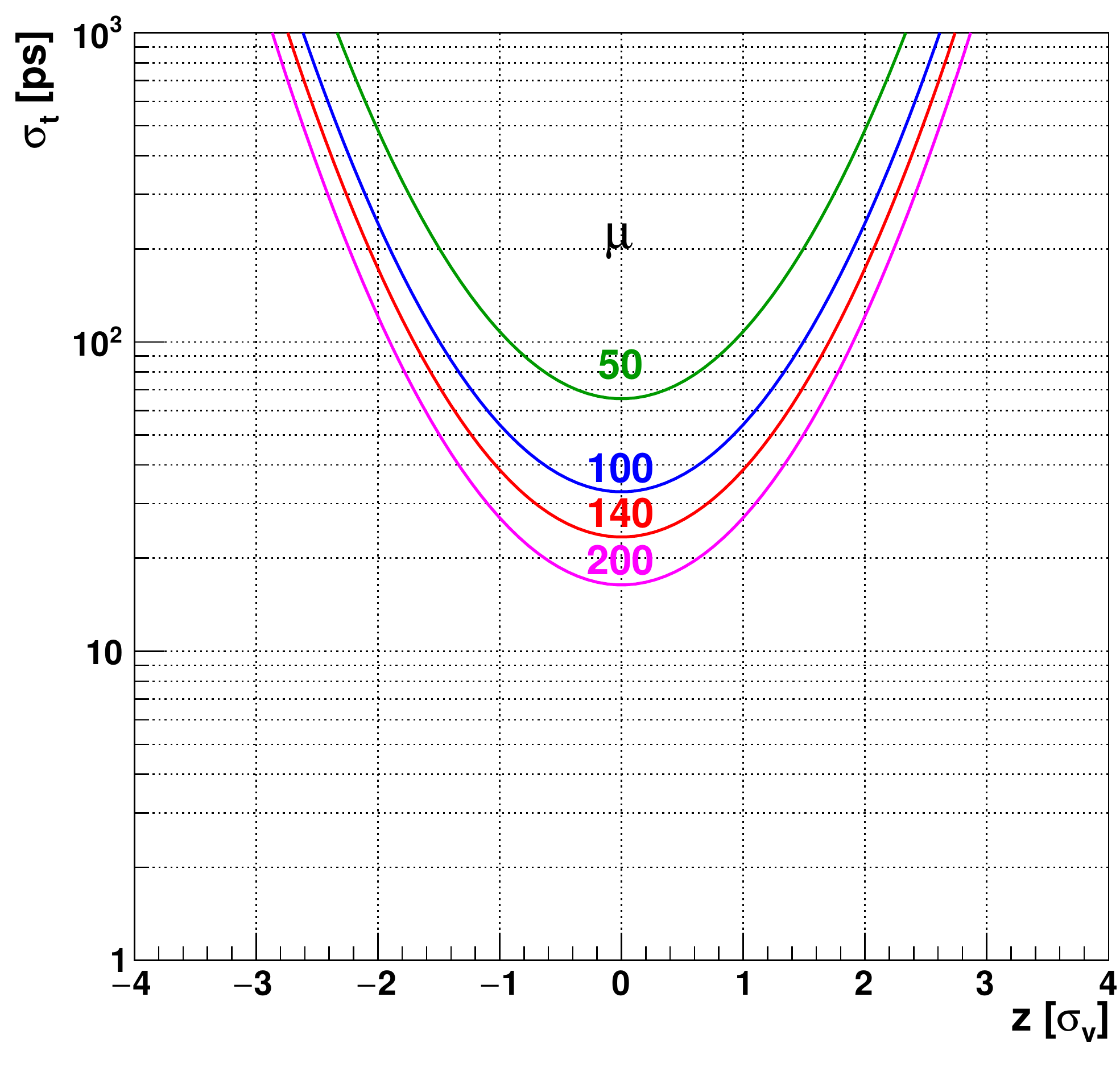}
\vspace*{-5mm}
\end{center}
\caption{Time resolution required per spectrometer arm to resolve the mean vertex distance at a position $z$ (in units of the longitudinal vertex width $\sigma_{\rm v}$) from the IP centre. Four different pileup multiplicities are shown: $\mu = 50$ (LHC Run 2), 100, 140 (nominal HL-LHC performance), and 200 (ultimate HL-LHC performance). Left: for standalone PPS timing. Right: combining the PPS timing with the MTD system, selecting a time-slice of $\pm 50\,$ps around the central bunch crossing time.}
\label{fig:timeresol}
\end{figure}

The required time resolution is given by the linear density of vertices at the IP which in turn is determined by the longitudinal bunch profile and the pileup multiplicity $\mu$. To first order, the bunch profile is Gaussian within $\pm 2 \sigma$ from the centre with a width of $\sigma_{\rm b} = 0.270\,$ns, corresponding to 8.1\,cm~\footnote{All calculations in this section are carried out in natural units with $c = 1$.} (Table~\ref{tab:hllhc-parameters} and Ref.~\cite{bunchprofile}) under the approximation of a constant bunch length throughout the fill. The mean pileup multiplicity is expected to be as high as 140 at standard HL-LHC performance and up to 200 at ultimate performance. With this input the linear vertex density at the longitudinal position $z$ (relative to the ideal IP) and the time $t$ (relative to the crossing time of the bunch centres) can be derived as
\begin{equation}
\rho_{\rm v} (z, t) = \frac{\d^{2}\mu}{\d z \,\d t} = \frac{\mu}{\left(\sqrt{2 \pi} \frac{\sigma_{\rm b}}{\sqrt{2}}\right)^{2}} 
\, {\rm e}^{-\frac{z^{2}}{\sigma_{\rm b}^{2}}} \, {\rm e}^{-\frac{t^{2}}{\sigma_{\rm b}^{2}}} \: .
\label{eqn:vertexdensity}
\end{equation}
This vertex density distribution is a two-dimensional Gaussian with spatial and temporal widths 
\begin{equation}
\sigma_{\rm v} = \sigma_{\rm b} / \sqrt{2} \: .
\end{equation}
After time integration (from $-\infty$ to $\infty$ for simplicity, without using vertex time information in central CMS) the mean vertex distance as a function of $z$ is given by
\begin{equation}
\langle \Delta z \rangle(z) = \frac{1}{\rho_{\rm v} (z)} 
= \frac{\sqrt{2 \pi}\,\sigma_{\rm v}}{\mu}\, {\rm e}^{\frac{z^{2}}{2\,\sigma_{\rm v}^{2}}} \: .
\label{eqn:vertexdistance}
\end{equation}
The reconstructed vertex position is given by the difference of the times of flight in the two arms, $t_{1}$ and $t_{2}$:
\begin{equation}
\label{eqn:timedifference}
z_{\rm v} = \frac{1}{2} (t_{2}-t_{1})
\end{equation}
To resolve the mean vertex distance $\Delta z$, each arm of the proton spectrometer needs a time resolution 
\begin{equation}
\sigma_{t} = \sigma(t_{1,2}) < \sqrt{2} \, \langle \Delta z \rangle \:.
\end{equation}
Figure~\ref{fig:timeresol} shows $\sigma_{t}$ as a function of $z$ for four different values of $\mu$. 

As the graphs demonstrate, for $\mu = 50$ a time resolution of 10\,ps is sufficient to resolve vertices even in the densest part of the bunch crossing region, whereas at HL-LHC pileup multiplicities of 140 and 200, resolutions of 4.8\,ps and 3.4\,ps would be needed, respectively, which has not yet been achieved. Present detector technologies can reach a single-plane resolution of $\sim 50 - 60$\,ps (Section~\ref{sec:diamonds}). With two movable vessel units containing together about 10 planes, the single-arm resolution would be about $\sigma_{t} = $15$-$20\,ps. In such a scenario, vertices would only be resolved in the periphery of the distribution. To give an example, for $\mu = 140$ (200) a single-arm time resolution $\sigma_{t} = 20\,$ps would resolve vertices outside $\pm 1.68\,(1.88)\,\sigma_{\rm v}$, i.e.\ 9\% (6\%) of all vertices. However, for most physics analyses it is not necessary to fully resolve the entire ensemble of vertices of a given bunch crossing:
\begin{itemize}
\item The selection of a specific central event topology reduces the number of vertices relevant for combination with leading protons. 
\item The CMS Phase-2 upgrade for HL-LHC foresees a thin MIP timing detector layer, MTD~\cite{cms-mtd-tdr}, between the tracker and the calorimeters, divided into a barrel and two endcap sections. With this new detector an additional vertex separation in the time dimension, i.e.\ by the vertex time, $t_{\rm v}$, rather than its position, $z_{\rm v}$, will become possible. 
The time integration of the density in Eq.~(\ref{eqn:vertexdensity}) can then be carried out in slices compatible with the MTD time resolution between 30 and 60\,ps (without and with radiation damage, respectively), reducing the effective linear vertex density in each time slice. A time slice of $\pm 50\,$ps around the peak density time point would relax the single-arm time resolution requirement to 23\,ps and 16\,ps for $\mu = 140$ and 200, respectively. This lies within the achievable range. To assign the two protons to the correct time slice, the vertex time, $t_{\rm v}$, has to be calculated from PPS as well. This is done by using the sum of the times of flight:
\begin{equation}
\label{eqn:timesum}
t_{2} + t_{1} = 2 t_{\rm v} + 2 z_{\rm d} \:,
\end{equation}
where $\pm z_{\rm d}$ are the detector positions in the two arms.
\end{itemize}
Note that Eq.~(\ref{eqn:timesum}) can also be used to suppress background from strongly out-of-time particles in the two beams, e.g.\ debunched beam halo, by eliminating events where the calculated $t_{\rm v}$ is several $\sigma_{\rm v}$ away from the peak of the vertex time distribution.

\subsubsection{Combination of Tracking and Timing}
Given that 
\begin{itemize}
\item in all locations, except at 220\,m, the space available for integrating movable detectors is very limited,
\item in regular operation there is no acceptance overlap between the four locations,
\item the number of movable devices should be kept at a minimum, both for impedance minimization and for limiting the complexity of the system and its operation,
\end{itemize}
it is not possible to have separate movable units for tracking and timing detectors. The most elegant solution is to have detectors performing both measurements at the same time. Developments at present are exploring pixel detectors that will allow a simultaneous measurement of the position and the timing; they will be briefly described in Section~\ref{sec:3D}.
The concurrent use of diamond detectors for timing and coarse tracking is possible. An alternative initial solution would be to have in each movable vessel two different types of detectors dedicated to only tracking and only timing, respectively. The disadvantage would be the doubling of the total number of detector planes which would increase the longitudinal space needed and thus limit the reduction of the overall dimensions of the movable devices.

\subsection{Detector Technologies}
As already stated in Section~\ref{sec:sensitivearea}, the dimensions of the detectors of the new PPS spectrometer are very small, and each layer considered will have to cover an area between 12\,cm$^2$ at 196\,m and a few cm$^2$ at 420\,m.

While the search for suitable technologies will proceed in all directions, there are already some candidates for which present operational experience at the LHC exists and whose performance can be expected to improve until the Long Shutdown 3. 
Needless to say, all possible synergies with the developments currently
ongoing for the Phase-2 upgrade of the central pixel system and for the
forward MIP timing detector will be carefully evaluated.

\subsubsection{Diamond Detectors}
\label{sec:diamonds}
Diamonds detectors are presently used by PPS for timing measurements~\cite{diamond-detectors, upgradeWS-timeresol, diamond-dpgnote}. 
The Double Diamond (DD) configuration, i.e.\ a double-layer of detectors 
connected to the same amplifier input, is the best performing option for 
diamond detectors aimed at timing and limited tracking. A first measurement in a test beam~\cite{jinst-berretti} showed that a time precision of 50\,ps per DD plane could be obtained for a readout pixel size of $0.55 \times 4.2$\,mm$^{2}$ when read out with a waveform digitizer (SAMPIC~\cite{sampic}) or an oscilloscope. 

The present timing precision for the DD used by PPS during Run 2 at the 
highest luminosity shows that the improvement due to the double layer 
configuration is not as significant as expected~\cite{diamond-dpgnote,diamond-dpgnote2}. A detailed check was done at a test beam for sensors coming from the Run 2. The reasons of the poorer precision have been analysed and understood; here they are briefly explained.

These measurements show:
\begin{itemize}
\item The loss due to radiation damage of the detector is minimal (loss of a few \% and limited to a small area near the most irradiated point) and can be ruled out as the main cause of the large difference in precision observed during Run 2.

In fact, after the end of Run 2, the diamond detectors that had integrated the 100\,fb$^{-1}$ were mounted on a new hybrid and examined at a test beam.
The detector pixel that had been closest to the LHC beam was mapped with a silicon tracker. It exhibited a small region of the order of a mm$^2$ with a reduced efficiency but still $>\,95\%$~\cite{bossini-Siena, diamond-desytestbeam}. The mapping in Fig.~\ref{fig:diam-rad-dam-map} shows a decrease in the rise time of the pulse and a minor decrease in pulse amplitude.
\begin{figure}[h!]
\begin{center}
\includegraphics[width=0.99\textwidth]{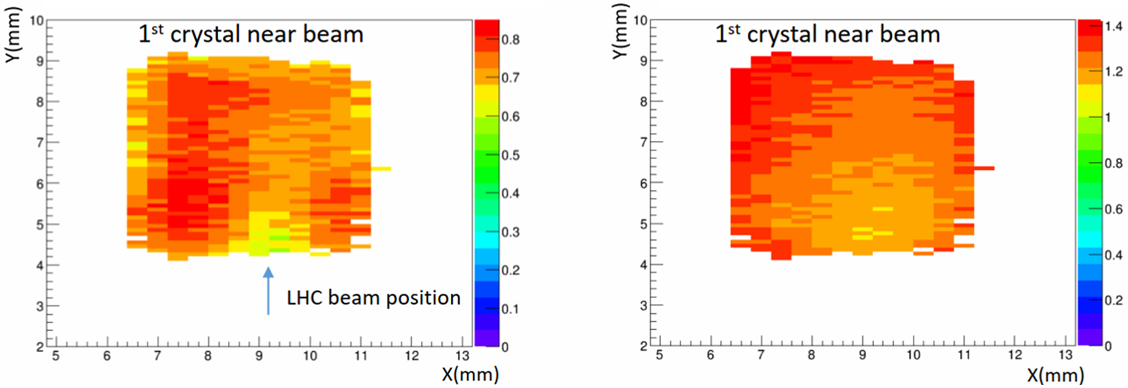}
\vspace*{-5mm}
\end{center}
\caption{Average signal amplitude (left) and rise time (right) as a function of the position in the irradiated crystal that was closest to the LHC beam during Run2, as mapped on a test beam in 2019. The sensor was mounted on a new hybrid with non-irradiated electronics. In this test setup the $x$ and $y$ coordinates are swapped relative to the LHC configuration.}
\label{fig:diam-rad-dam-map}
\end{figure}
This agrees with measurements at earlier radiation damage tests in the CERN irradiation facility~\cite{upgradeWS-irradiation}. 
These tests had indicated that single crystal diamond detectors exposed to fluences up to $5 \times 10^{15}$ protons/cm$^{2}$ (corresponding to $3 \times 10^{15}\,\rm n_{\rm eq} / cm^{2}$), which is reached after 100\,fb$^{-1}$, exhibit a slightly lower signal amplitude and a shorter rise time but without a significant deterioration of the time resolution. 
A possible increase of the operation high voltage can be applied, if necessary, to recover signal amplitude and extend the lifetime.

This result, obtained with never irradiated hybrids, indicates that the decrease in timing precision with increased irradiation is essentially due to the damage of the amplifier of the pads closer to the circulating beam.

These hybrids were initially developed for the conditions foreseen for the TOTEM experiment: a new hybrid is being designed with a better optimization and will be ready for Run 3.
\item Another cause of loss of precision is the fact that the arrival time of the signal was measured with the existing PPS High Performance TDC (HPTDC~\cite{hptdc}): the signal from the hybrids was routed via a NINO chip~\cite{nino} where the signal coupling could not be perfectly optimized, which deteriorated the resolution.

After a further re-optimization of the hybrid and the NINO connections, a resolution of 50$-$60\,ps per plane is expected for Run 3.
\end{itemize}
Furthermore, some technical aspects of the present XRP detector vessel and its movement and cooling systems have to be improved.
In fact, some observed shortcomings limiting the performance were induced by non-detector related problems.
The detectors and the on-board electronics are operated under a secondary vacuum of 8$-$10\,mbar and cooled at a temperature of $-12\,^{\circ}$C in the very small volume of the Roman Pot.

It was observed that for the nominal signal amplifier gain, in a real LHC run, crossboard pickup led to larger noise oscillations, forcing the reduction of the preamplifier gain to obtain stable operation, which brought a deterioration of the precision.

In addition, for some detectors, HV induced discharges forced the reduction of the operating high voltage, again resulting in a worse resolution, bringing the overall timing precision to 100$-$150\,ps.

Work is in progress to remove these limitations for Run 3. However, the secondary vacuum and the operating temperature of $-12\,^{\circ}$C will continue to be required.\\

In conclusion, if diamonds are chosen as detectors for the new PPS spectrometer and each station is equipped with eight measuring planes designed with the present technology and optimized for these new running conditions, the single-arm resolution will then be around 20\,ps.

With improved detectors and electronics the goal of $\lesssim$10\,ps per arm with this technology does not seem out of reach.

\subsubsection{3D Pixel Detectors}
\label{sec:3D}
3D silicon pixel detectors are presently used by PPS for tracking~\cite{pixel-detectors,pixel-dpgnote}. They operate at a temperature of about $-20\,^{\circ}$C in the secondary vacuum of the XRP. The spatial resolution is already adequate for the future PPS spectrometer: $\sim 25\,\mu$m for 150\,$\mu$m wide pixels with an inclination of 20$^{\circ}$ between the perpendicular to the detector plane and the beam. The present technology, with its so-called 200\,$\mu$m wide ``slim edges'', has an insensitive margin of only $\sim 50\,\mu$m because the depletion region extends significantly into the edge area. This leads only to minor acceptance losses at the lower end of the $|\xi|$ or mass range. Typically, an increase of the insensitive margin width by 0.1\,mm increases the minimum accepted mass by about 10\,GeV. An option for further improving the ``edgelessness'' would be 3D detectors with active edges~\cite{3d-active_edges}.

All acceptance numbers given in the earlier sections of this document take into account a distance $\delta = 0.5$\,mm from the outer surface of the movable vessel to the point in the detector where full sensitivity is reached, which comfortably accommodates the $\sim 50\,\mu$m wide insensitive margin.

The radiation hardness of these detectors was measured in irradiation tests and operationally observed in Run 2~\cite{pixel-dpgnote}.

The present sensors can cope with a uniform irradiation of $5 \times 10^{15}$ protons/cm$^{2}$ (corresponding to $3 \times 10^{15}\,\rm n_{\rm eq} / cm^{2}$) reached after 100\,fb$^{-1}$ for the optics and XRP configuration of Run 2.  
Future sensors, the same as used in the Phase 2 CMS Inner Tracker upgrade, are required to withstand $2 \times 10^{16}\,\rm n_{\rm eq} / cm^{2}$.

The radiation field in the PPS detector is highly non-uniform, showing a sharp peak around the very localized regions crossed by the huge flux of diffractive protons.
This leads to a strong reduction of the detection capability in an area of the order of less than a millimetre while a large fraction of the detector area is still working properly.

For the PPS pixels in Run 2 the main problem was traced to the strongly non-uniform irradiation exposure of the electronics chip bonded to the detector, the PSI46 chip developed for the CMS Pixel detector.

It is possible to cope with this localized damage during the run by periodically moving the detector planes vertically inside the XRP vessels, i.e.\ tangentially to the beam. In Run 2 this strategy of diluting the irradiation peak has been implemented by shifting the detector packages manually during short technical stops of the LHC. For Run 3 a remotely controlled vertical movement system based on piezoelectric motors has been developed. 
PPS experience from operation in Run 2 indicates that the present detector electronics need to be displaced by about 0.5\,mm after having integrated approximately 20\,fb$^{-1}$ of luminosity in a specific running position near the beam. For a total integrated luminosity of 300\,fb$^{-1}$ per year at HL-LHC, this strategy would imply 15 movement steps of 0.5\,mm or a total displacement amplitude of 7.5\,mm per year. The peak fluences in Table~\ref{tab:maximumflux} (last column) would thus be divided by a factor 15 and reach a manageable level.

Furthermore, new pixel read-out chips, more radiation-hard than the ones currently used, are being developed for CMS and ATLAS on the HL-LHC horizon. PPS will profit from these developments.\\

Thanks to the unique structure of 3D detectors, it is possible to have a 
uniform field in the depletion layer; by choosing small electrode 
spacing, very short drift times can then be achieved without reducing 
the substrate thickness, thus preserving the signal amplitude. This makes the detector architecture potentially interesting also for timing. 

Very first measurements~\cite{3d-timing-sherwood} on prototype detectors with 50\,$\mu$m electrode spacing, a non-optimized electronics, and using a $^{90}$Sr source already indicated time resolutions between 30 and 180\,ps per plane depending on the signal amplitude. A later experiment~\cite{kramberger_2019} with $^{90}$Sr electrons confirmed the 30\,ps resolution. 

Recent measurements~\cite{3d-timing-vertex19,3d-timing-IPRD19} on specially shaped 3D cells with laser beams and 270\,MeV pion beams yielded time resolutions of 20 and 30\,ps, respectively.

If these results are confirmed, this detector technology, capable of 
providing simultaneous tracking and timing, can become available rapidly 
for the very small dimensions and the small number of the detectors 
needed for the new PPS.

\subsubsection{Planar Silicon Detectors}
Prototypes of the Low Gain Avalanche Detectors, LGAD or more simply 
Ultra-Fast Silicon Detectors, UFSD~\cite{ufsd-detectors}, have been used in Run 2. They served mainly for timing measurements in special TOTEM+CMS runs at $\beta^{*} = 90\,$m and low pileup ($\mu \sim 0.5$), but a few planes were also used in PPS timing XRPs, complementing the diamond detectors. 

This technology could be explored further when a new generation of more radiation hard detectors, now under development for the CMS MTD, will be available.

\subsection{Reference Timing System}
The clock distribution system presently used by PPS can be reimplemented for HL-LHC. It is described in Refs.~\cite{totem-upgrade} (Section 6.3.2) and~\cite{ctpps} (Section 5.7.2). This system also offers the possibility of correlating the proton timing information with the one of the upgraded central detectors.

The system has been used for PPS timing detectors since September 2017.

The linearity of the HPTDC used for the PPS readout has been checked by leveraging the tunable phase shift.

The optical clock distribution jitter is of the order of 1\,ps (measurement limited by the scope bandwith).

Since the present system is synchronized to LHC via the Beam1 reference, the beam cogging (phase delay between the two beams used to centre the interaction point in IP5) can introduce a bias or shift in time.

The phase drift is a long term one and oscillates in a range of about 10\,ps.

\clearpage
\section{Trigger Strategy}
\label{sec:trigger}
The trigger strategy is expected to largely follow the lines of the present PPS system, which does not provide a Level 1 (L1) trigger but can contribute to the High Level Trigger (HLT). However, as already discussed in Ref.~\cite{ctpps} (Chapter 6), there are options for L1 proton triggers capable of reducing event rates, based on proton multiplicity or the longitudinal vertex position obtained from the timing detectors.
Furthermore, the TOTEM XRP tracking detector trigger transmitted via a fast electrical cable~\cite{electricaltrigger} could be adapted to PPS use. The increased CMS L1 latency time in Run 4~\cite{cmstriggerupgrade} would open up this possibility even for the most distant station at 420\,m.

\clearpage
\section{Movable Detector Vessels and Machine Integration}
\label{sec:vessels}
The development of movable detector vessels and their integration into the machine can build on a wealth of experience gathered in the LHC Runs 1 and 2. For three out the four preferred locations, advanced versions of the Roman Pot technology, developed by the TOTEM experiment and later upgraded~\cite{totem-upgrade} for high-luminosity operation by PPS, are conceivable. The fourth location, near 420\,m from IP5, suffers from special space and integration constraints and will need a specific development.

The present section discusses the most important guidelines for further optimizations and new developments. In the end, the presently known candidate technologies are listed.

\subsection{Impedance Effects and their Mitigation}
\label{sec:impedance}
Like every beamline element with resistive material around the beam, near-beam detectors contribute to the impedance acting on the beam current and causing two kinds of effects:
\begin{itemize}
\item Effects on the beam: wake fields induced by the passing proton bunches perturb subsequent bunches and thus lead to longitudinal and transverse beam instabilities. To limit these effects, an impedance budget has been defined for the whole LHC ring. Each beamline element is only allowed to contribute a certain fraction of this budget.
\item Local impedance heating, i.e.\ the transfer of power from the beam to the  wall of a cavity. 
\end{itemize}
In the original design of the TOTEM XRPs impedance reduction was less important because operation had been foreseen only in special runs with very low beam currents. Thus the original box-shaped XRPs exhibited both an abrupt aperture restriction (when inserted) and cavities around the detector boxes (even when retracted). The only impedance mitigation implemented was the installation of ferrite plates around the detector boxes~\cite{impedance-run1} in order to dampen cavity-induced resonances. In the first high-luminosity test insertions of the XRPs at the end of Run 1 these measures were found to be insufficient, as significant heating was observed at the XRPs, despite the active cooling of the detector packages, resulting in ferrite outgassing and vacuum degradation.

Consequently, to make the XRP system compatible with continuous high-luminosity operation in Run 2, more detailed studies and tests were performed during LS1, resulting in the following system upgrades~\cite{impedance-run2}:
\begin{itemize}
\item All existing XRPs were equipped with Radio Frequency (RF) shields filling most of the cavities between the cylindrical flange walls and the box-shaped detector housings. This removed the most dangerous high-amplitude resonances below 1.4\,GHz. 
\item The new XRPs for timing detectors were designed with cylindrical rather than box-shaped geometry, avoiding cavities almost entirely. 
\item Like in other machine elements, the ferrites were replaced with a new type made of a better material with a higher Curie temperature. They were baked out under vacuum to minimize outgassing.
\item The mounting location of the ferrites was moved to the region most affected by impedance heating; the ferrite geometry was adapted accordingly. 
\end{itemize}
In Run 2, the XRPs were inserted regularly in almost all physics fills. During operation the temperature in several points on the XRPs as well as the machine vacuum in the respective sectors were continuously monitored. The temperature stayed within comfortable limits, no vacuum degradation and no beam instabilities were observed. This proved the effectiveness of the impedance mitigation measures and the adequacy of the cooling system.

For the HL-LHC, where the current will increase by a factor of two and hence the heating by a factor of four, the LHC impedance team sees ``no indication that the current PPS Roman Pot system would be an issue with HL-LHC beams, provided cooling is sufficient''~\cite{impedance-benoit-upgradeWS}. For the design of a new system the following points need to be kept in mind:
\begin{itemize}
\item The number of near-beam devices should be kept at the minimum necessary for the physics.
\item For beam stability, the impedance is most critical in the initial fill phase before collisions because of the absence of Landau damping. Therefore it is important to be able to keep the devices as far from the beam as possible during this phase, ideally totally retracted. 
\item Devices that cannot be fully retracted should not have abrupt aperture steps but tapers with an angle smaller than 11$^{\circ}$.
\item Bellows have to be shielded.
\item The design should avoid resonant cavities below 2\,GHz.
\item The device surfaces facing the beam should be coated with a good conductor.
\item The cooling has to be strong enough to cope with the increased beam current and hence heating power. 
\end{itemize}
During all preparations a close cooperation with the Impedance Working Group will be maintained. For all new devices impedance simulations have to be carried out, profiting from tools that have significantly improved since LS1. For this purpose, the modelling of impedance heating was recently benchmarked with temperature probe measurements taken during an LHC fill in 2015~\cite{teofili-talk}. 

\subsection{Showers}
\label{sec:showers}
The insertion of movable devices close to the beam generates showers since beam halo and collision debris scatter off the vessel and detector material. While these processes will have to be studied in detail with FLUKA, as previously done for the Run-2 PPS configuration, some mitigation and monitoring guidelines can already be given.
\begin{itemize}
\item The material exposed to the beam halo has to be minimized. In particular, 
the windows separating the secondary vacuum of the detector vessels from the primary beam vacuum have to be as thin as possible, but thick and strong enough to limit the deformation in case of an accidental secondary vacuum loss which could lead to the dangerous scenario of a window bulging into the beam. 
\item It is advantageous to have TCL collimators downstream of movable devices to absorb debris generated by the latter. For the 196\,m and 220\,m stations the collimators TCL5 and TCL6 will play that role. For the 234\,m and 420\,m stations the present layout does not foresee any downstream TCLs. FLUKA studies will address possible problems from showers. Note, however, that in 2018 the PPS system operated at the highest Run-2 luminosities with TCL6 fully open and did not induce any radiation problem for the downstream elements.  
\item After each detector station (and ideally even after each unit) a BLM has to be installed to monitor losses generated by the movable device. This is also needed for the beam-based alignment.
\end{itemize}

\subsection{Technologies for Movable Devices}
\subsubsection{The ``Warm'' Locations (196\,m, 220\,m, and 234\,m)}
The three non-cryogenic locations are easiest to equip with detectors:
\begin{itemize}
\item No cryogenic bypass is needed.
\item The dispersion is negative, implying that scattered protons with $\xi<0$ arrive on the side radially away from the LHC ring centre, not inwards where the second beam pipe lies at only 194\,mm distance (beam centre to beam centre). Hence the transverse space constraints are quite relaxed.
\end{itemize}
Since the beam pipes in these regions have the same diameters as in the present-day LHC, it would be possible to install Roman Pots very similar to the existing ones. However, one will nevertheless take advantage of the opportunity to improve some minor shortcomings experienced in Runs 1 and 2, notably:
\begin{itemize}
\item A lack of rigidity in certain aluminium parts that should better be made of steel. This would straighten the movement trajectory and thus allow a more precise transfer of metrology data to detector alignment constants, simplifying the track- and physics-based alignment steps.
\item A better control of the distance between the detector edges and the Roman Pot window facing the beam. Also this improvement would help in the alignment procedure.
\item A revised, more sturdy microswitch and electrical stopper system, less prone to electrical contact problems and hence even more reliable than the present-day system (that had on average one failure per year).
\item The BPMs, mechanically integrated in the Roman Pot beam pipes, have not performed with enough precision to be usable for the detector alignment relative to the beam. Improvements have to be found. An option to be explored is the integration of BPM buttons directly into the XRP jaws as done in the LHC collimators. A problem, however, is the single-jaw nature of the XRPs in contrast to the double-jaw collimators.
\end{itemize}
Also, given the rather small detector dimensions and their small weights, one might explore a lighter, more compact design.

\subsubsection{The ``Cold'' Location 420\,m }
\label{sec:420m-integration}
\begin{figure}[h!]
\begin{center}
\includegraphics[width=\textwidth]{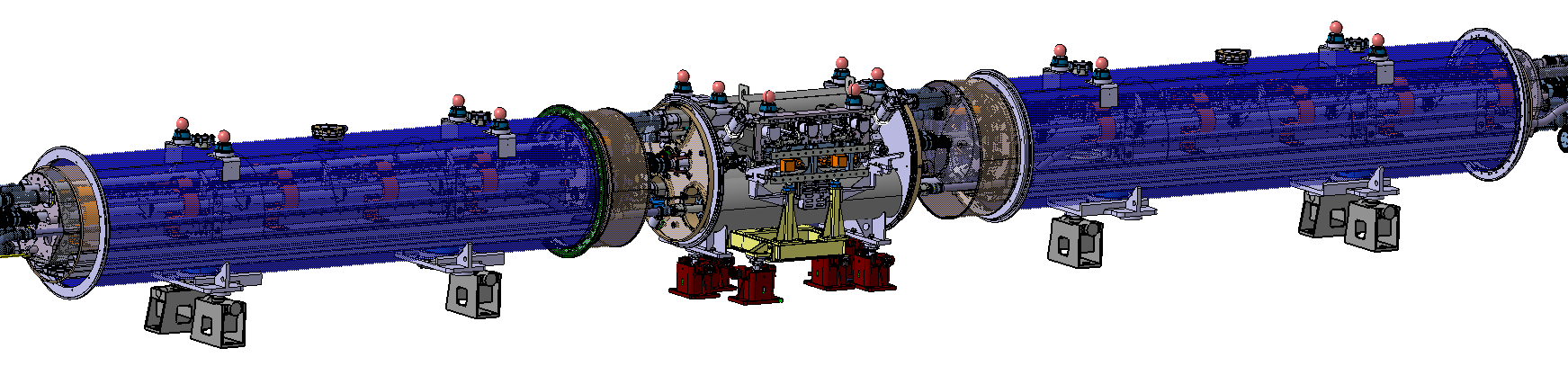}
\end{center}
\caption{Full connection cryostat assembly design, with new TCLD collimator integrated (from Ref.~\cite{connectioncryostat}).}
\label{fig:connectioncryostat}
\end{figure}

The 420\,m location requires the replacement of the present-day empty cryostat LEGR with a connection crystat to create some free space for a detector system. An additional complication is the positive dispersion deflecting protons with $\xi<0$ into the narrow space between the two beam pipes, which precludes the use of Roman Pots with present-day dimensions. The first problem could be solved by reusing or adapting the already existing design of a connection cryostat for the new TCLD collimators to be installed in equivalent locations in IR2 and IR7. Figure~\ref{fig:connectioncryostat} shows this cryostat assembly, already with the TCLD integrated. A zoom on the TCLD itself is shown in Fig.~\ref{fig:tcld_in_cryostat}. For PPS purposes, the TCLD could be replaced with any movable detector vessel fitting in the same space (see Fig.~\ref{fig:tcld_dimensions} for the dimensions). An alternative solution to be explored could be the connection cryostat designed for the old FP420 project~\cite{fp420} (Section~6 therein). The compatibility with the present HL-LHC layout needs to be investigated.

Possible technology options for the movable detector vessel presently under consideration include: 
\begin{enumerate}
\item A modified TCLD Collimator: most of the collimator design would be maintained but the solid jaws replaced with detector housings of the same geometry. Such a device would have the same advantages as the collimator: very stable movement on two axes per jaw, the vertical movement for irradiation dilution, the redundant LVDTs (Linear Variable Differential Transformer~\cite{lvdt}), the button BPMs integrated in the jaws, and optimized impedance properties. A difficulty, on the other hand, would be the significant amount of material in front of the detectors due to the flat-angle taper of the jaw.

\begin{figure}[h!]
\begin{center}
\includegraphics[width=0.59\textwidth]{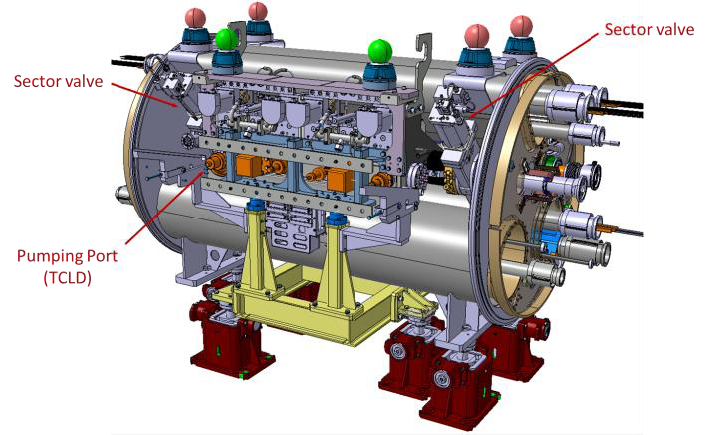}
\includegraphics[width=0.39\textwidth]{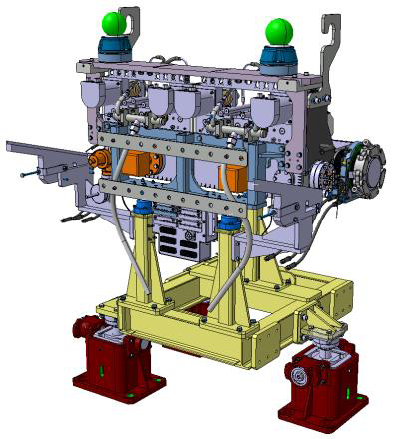}
\end{center}
\caption{Left: TCLD integrated in the bypass cryostat; right: TCLD alone (from Ref.~\cite{tcld}).}
\label{fig:tcld_in_cryostat}
\end{figure}

\begin{figure}[h!]
\begin{center}
\includegraphics[width=\textwidth]{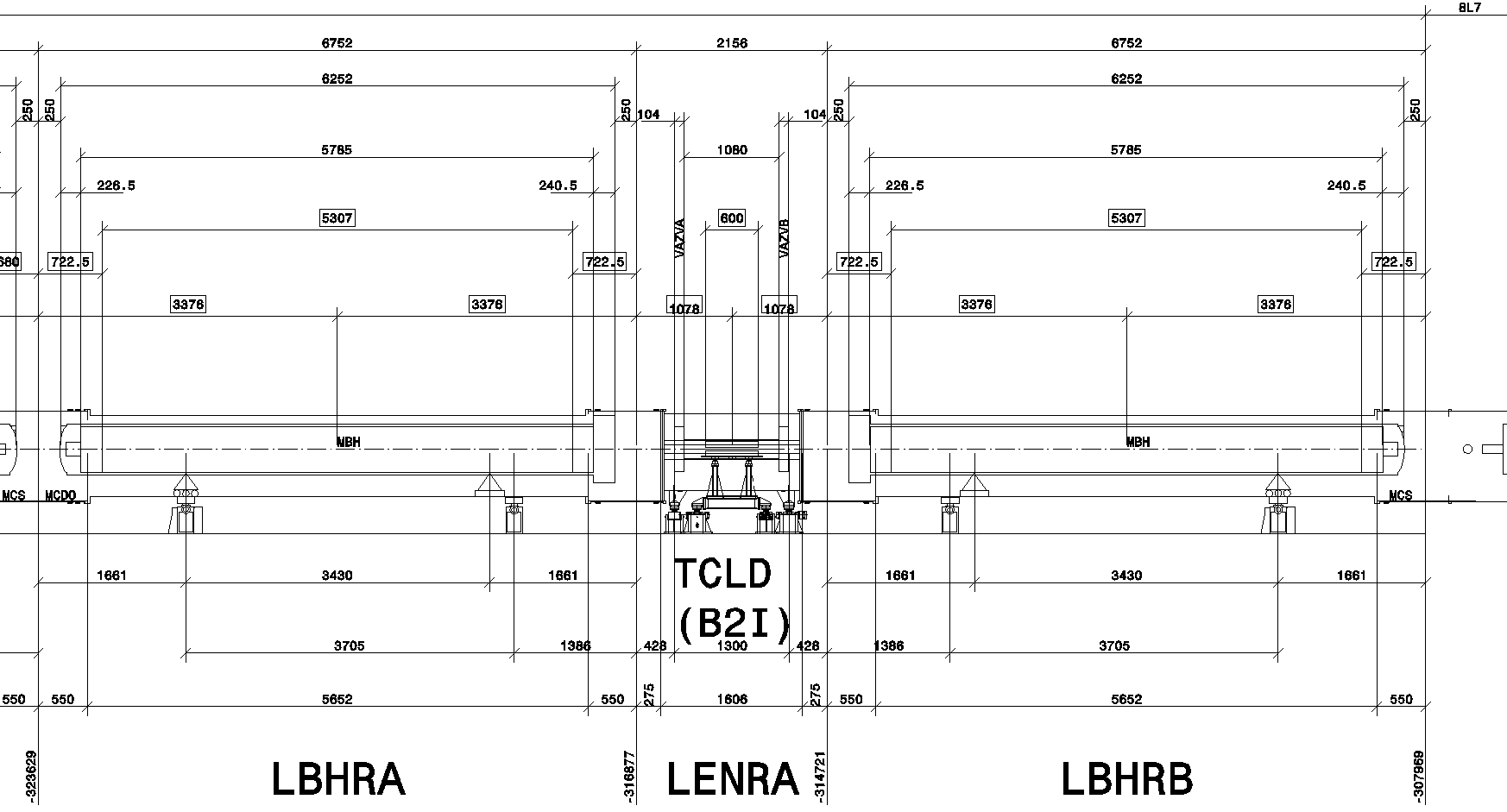}
\end{center}
\caption{Dimensions of the assembly of TCLD and bypass cryostat (from Ref.~\cite{tcld}).}
\label{fig:tcld_dimensions}
\end{figure}

\item New compact Roman Pots could be designed to fit even in the narrow space between the two beam pipes. Such a design could have the advantages of the well-established impedance properties of the present-day Roman Pots, and a very thin window in front of the detectors.
\item Movable Beam Pipe: the design of the ``Hamburg Beam Pipe'' that had once been considered for PPS, could be revisited and improved. The main problem, the impossibility of complete retraction implying a higher impedance before collisions, might be overcome by adding an 11$^{\circ}$ taper and a copper coating; see also the discussion in Ref.~\cite{ctpps} (Section~3.6).
\end{enumerate}

\subsection{Movement and Interlock System}
\label{sec:movementsystem}
The present-day XRP movement system~\cite{movementsystem} is based on the functionality principles and on hardware and software components as used by the LHC collimation system. This implies in particular that all components exposed to radiation are as radiation tolerant as their identical counterparts in the collimation system.

However, in the system design phase, certain simplifications have been made because of the originally intended limited use of the XRPs only for special runs of the TOTEM experiment. After operational experience further modifications were made during LS1 and in year-end technical stops. For the HL-LHC phase the movement system should be fundamentally redesigned taking into account the lessons learnt in Runs 1$-$3, but keeping it as close as possible to the collimator system. The vertical movement capability of the horizontal detector units, allowing the dilution of the most irradiated region, should be integrated in the main movement system. Such an option is already present in the collimator system, called ``5th axis''.

The functionality and implementation of the interlock system~\cite{interlocksystem}, preventing beam operation while any movable device is in a forbidden position, are expected to be very similar to the present ones. A particular upgrade to be envisaged is the introduction of redundant LVDTs (measuring the XRP positions for interlock purposes) to be protected against malfunctions and hence false interlocks.

The design will proceed in continued close cooperation with the Machine Protection Panel.

\subsection{Services}
\begin{sidewaysfigure}[p]
\includegraphics[width=1.0\textwidth]{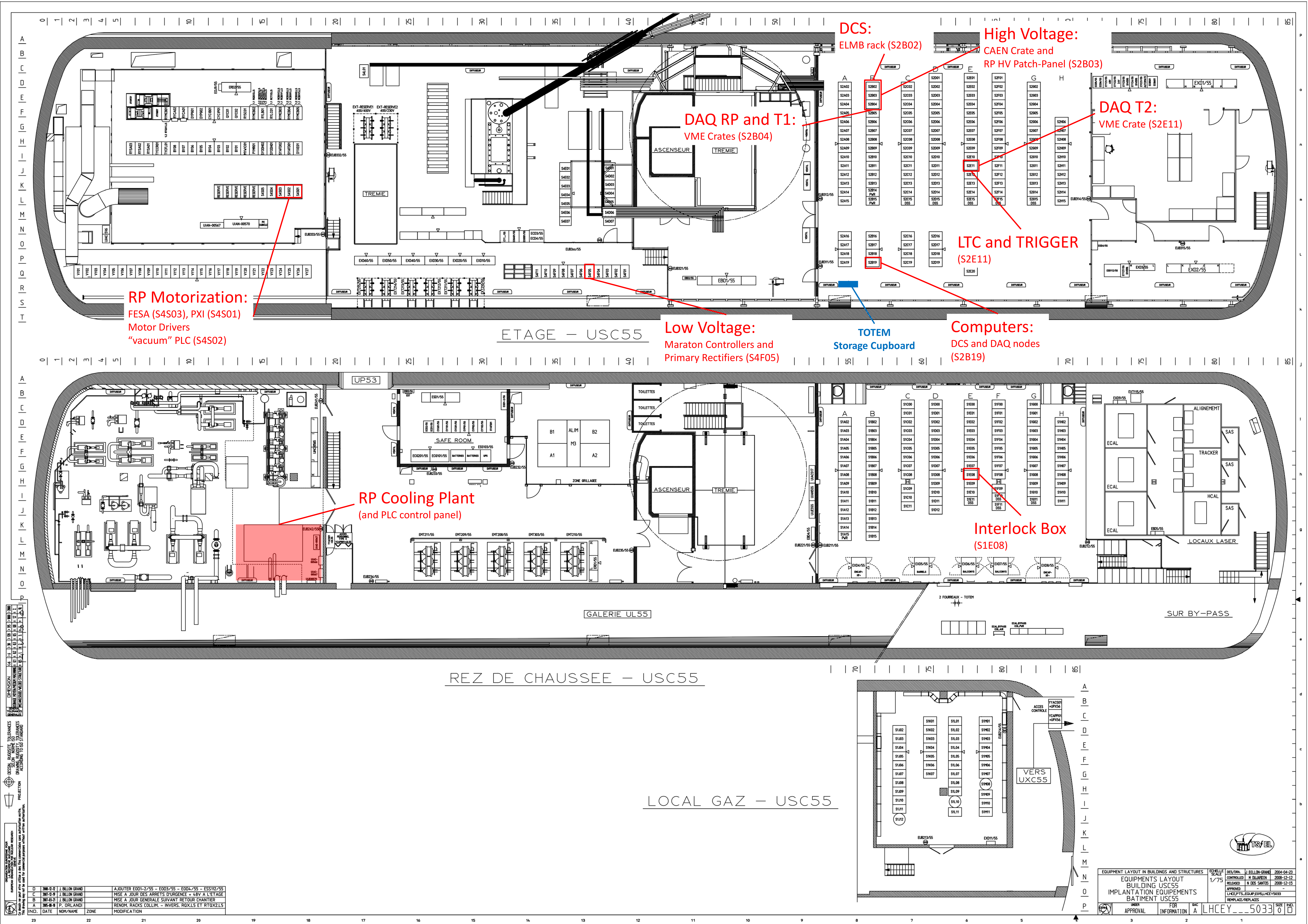}
\caption{Layout of the USC55 service cavern (adapted from Ref.~\cite{usc55}) showing the rack space occupied by TOTEM / PPS equipment in Runs 1$-$3.}
\label{fig:usc55layout}
\end{sidewaysfigure}

\begin{sidewaysfigure}[p]
\includegraphics[width=1.0\textwidth]{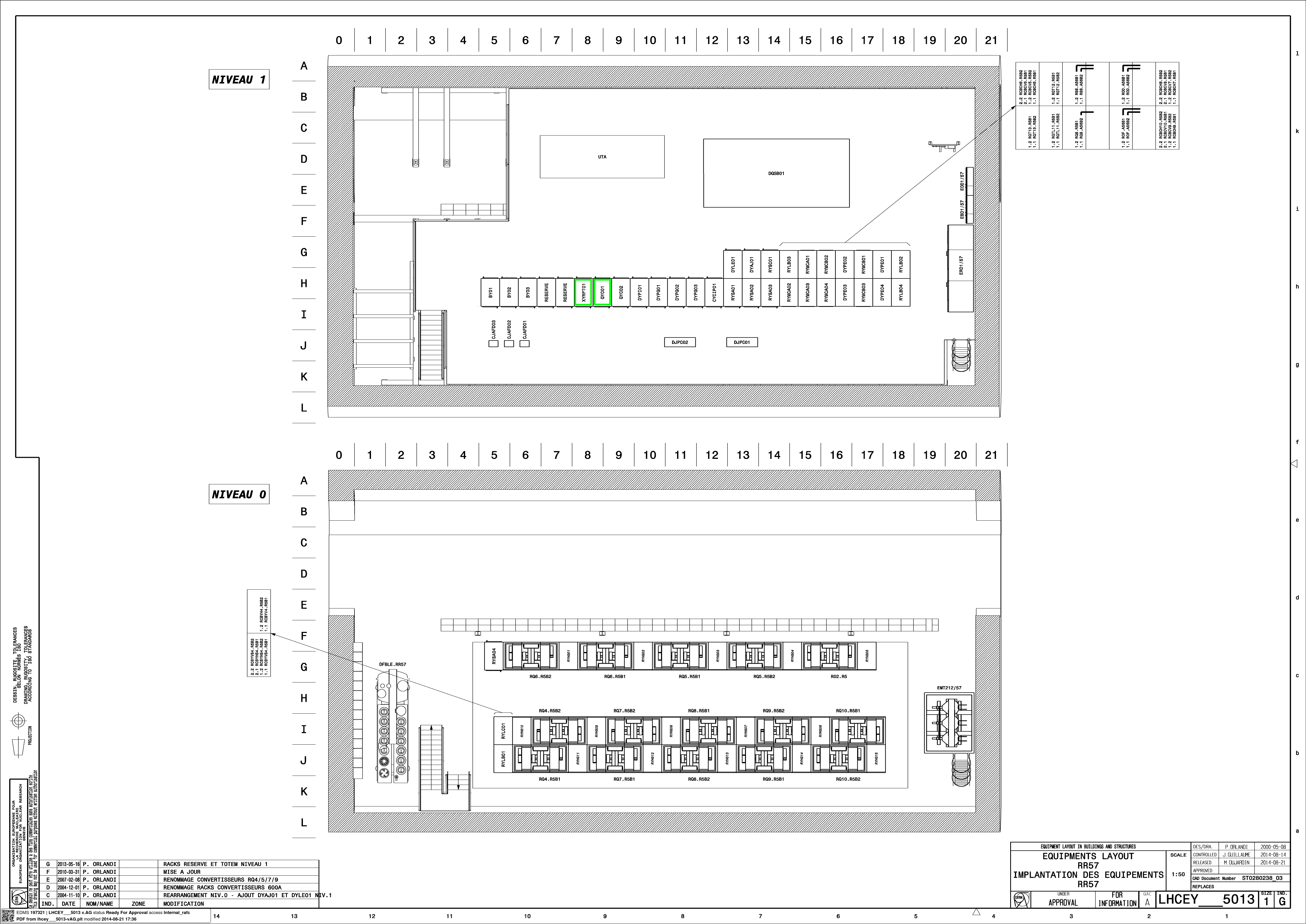}
\caption{Layout of the RR57 tunnel alcove in Sector 5-6~\cite{rr57}. The racks presently used by TOTEM are marked with green frames.}
\label{fig:rr57layout}
\end{sidewaysfigure}

\begin{sidewaysfigure}[p]
\includegraphics[width=1.0\textwidth]{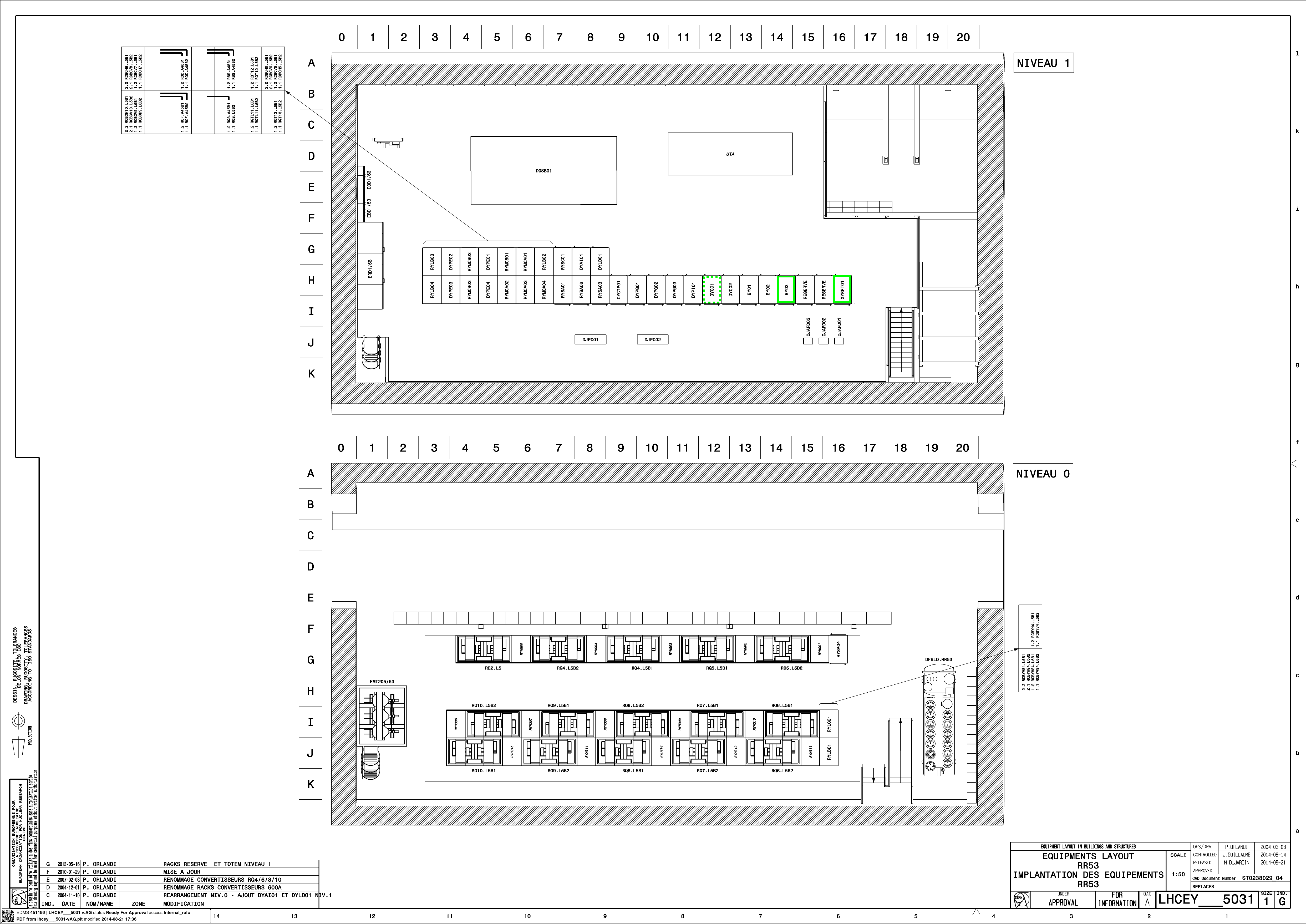}
\caption{Layout of the RR53 tunnel alcove in Sector 4-5~\cite{rr53}. The racks presently used by TOTEM are marked with green frames.}
\label{fig:rr53layout}
\end{sidewaysfigure}

\begin{itemize}
\item Cooling System:\\
The evaporative cooling system based on $\rm C_{3}F_{8}$~\cite{cooling}, successfully used for the TOTEM/PPS XRPs, is a likely candidate technology to be adapted for the new HL-LHC spectrometer. Because of the increased impedance heating power (factor of four, see Section~\ref{sec:impedance}), the cooling power needs to be increased accordingly, which is reachable with a new generation of more efficient heat exchangers at the detector units~\cite{cooling-upgrade,vacek-talk201909}.
Because of the high radiation environment of the LHC tunnel, the main part of this cooling system is installed in the shielded underground service cavern USC55. From there the cooling fluid is transported at ambient temperature to the detector stations. The related services (cooling plant in USC55, cooling fluid lines) have to be foreseen. The capacity and performance of the cooling system presently installed at IP5 are compatible with the requirements and demands of the new RP installation for HL-LHC. It will not need more space, but all components (pumps, valves, DCS elements) need to be exchanged.

Alternatively, the Aircooler system~\cite{vacek-talk201909} might be an option. The ongoing development of innovative heat exchangers will allow the use 
of detector package cooling based on the VORTEX principle even during 
HL-LHC operation. The main advantage is the maintenance free 
operation in the LHC tunnel; in addition, the overhead on the cooling system at IP5 is reduced to the compressor, which has to provide compressed air at around 10\,bar.

\item Secondary Vacuum: \\
Like in the present Roman Pot system, the vessels housing the detectors will have to be under vacuum, separated from the primary machine vacuum. This requires remote-controlled pumps and gauges located in the alcoves RR53 and RR57. 
\item Summary of Rack Space:\\
The following space needs to be reserved (see Fig.~\ref{fig:usc55layout} for USC55, Fig.~\ref{fig:rr57layout} for RR57, and Fig.~\ref{fig:rr53layout} for RR53):
\begin{itemize}
\item The detector electronics and the vacuum equipment together will require 4 racks in each of the tunnel alcoves RR53 and RR57. 
\item The motor control needs 2 racks in USC55.
\item The vacuum control needs 1 rack in USC55.
\item The interlock logic box and the interface to the LHC BIS system (CIBUs) are located in one rack together with other machine interlock components (e.g.\ the main CMS CIBUs) in USC55.
\item Additional racks for detector power supplies, trigger, DAQ, and DCS elements (still to be determined) will be needed in the CMS sections of USC55.
\end{itemize}
\end{itemize}

\newpage
\section{Summary}
We propose to build a new Precision Proton Spectrometer for operation in the full HL-LHC phase. 
\begin{itemize}
\item The physics goal is to extend the central exclusive production studies of the present-day PPS system to processes with lower cross sections and to a wider mass range.
\item Four locations have been identified to be suitable for installing movable proton detectors: 196\,m, 220\,m, 234\,m, and 420\,m, each on both sides of IP5. Preliminary space reservations have been made.
\item The locations at 196\,m, 220\,m, and 234\,m can be instrumented with Roman Pot devices similar to the ones presently used. The 420\,m location requires a bypass cryostat (which has been developed for other locations in the LHC) and a movable detector vessel approaching the beam from the narrow space between the two beam pipes (still to be developed, with some candidate technologies existing).
\item Acceptance studies resulted in a strong preference for a vertical crossing-angle in IP5. The HL-LHC Executive Committee decided to implement it for HL-LHC.
\item With this vertical beam crossing the central mass range covered will be:
\begin{itemize} 
\item ~133\,GeV$-$2.7\,TeV with only the stations at 196\,m, 220\,m, and 234\,m;
\item ~43\,GeV$-$2.7\,TeV with all stations, including 420\,m.
For comparison: the Run-2 configuration covered approximately the mass range 350\,GeV$-$2\,TeV. 
\end{itemize}
\item The intense radiation background implies radiation hardness requirements on all components installed in the tunnel. Service work during short Technical Stops will not be possible.
\item The irradiation dose rate is very strongly peaked near the beam. Hence the most irradiated spot on a detector needs to be periodically shifted with a remote-controlled vertical detector movement system. A shift by 0.5\,mm every 20\,fb$^{-1}$ is sufficient.
\item The mean pileup multiplicity up to 200 makes a longitudinal vertex identification via time-of-flight measurement necessary. The time resolution requirement is not a discrete ``all or nothing'' criterion. A single-plane resolution of 50$-$60\,ps is already adequate.
\item The detector technologies envisaged are those that are already used and continuously improved in the pre-LS3 PPS system:
\begin{itemize} 
\item Diamond detectors for time-of-flight measurements,
\item 3D pixel detectors for tracking, possibly also for timing,
\item UFSD detectors as an alternative for timing if the ongoing developments are successful.
\end{itemize}
\item The stations at 196, 220, and 234\,m are similar to those currently in use at PPS, both in their design and in the technology used. Based on this, a preliminary budget estimation suggests that they could be built at a total cost between 1 and 2\,MCHF.

Conversely, the 420\,m station requires the modification of a cryostat and a novel approach for the detector vessel. More work is needed for an estimate of the cost in this case.
\end{itemize}

\clearpage

\appendix
\section{Glossary}
\begin{description}
\item[ALP] Axion-Like Particles
\item[AQGC] Anomalous Quartic Gauge Coupling
\item[BBLR] Long-Range Beam-Beam compensation unit (machine acronym for ``Beam Beam Long Range'')
\item[BIS] Beam Interlock System
\item[BLM] Beam Loss Monitor
\item[BPM] Beam Position Monitor
\item[BSM] Beyond the standard model
\item[CEP] Central Exclusive Production
\item[CIBU] User interface to the interlock system (``Controls Interlocks Beam User'')
\item[DCS] Detector Control System
\item[DD] Double Diamond, i.e.\ double layer of diamond detectors
\item[HPTDC] High Performance Time-to-Digital Converter
\item[IP] Interaction Point
\item[LSn] n$^{\rm th}$ Long Shutdown of the LHC
\item[LSS5] Long Straight Section of the LHC around IP5
\item[LVDT] Linear Variable Differential Transformer~\cite{lvdt}: used for an XRP position measurement.
\item[MIP] Minimum Ionising Particle
\item[PPS] Precision Proton Spectrometer
\item[Qn] Quadrupole magnet in LHC cell number n counted from the nearest IP, e.g.\ Q6.
\item[RF shield] Radio Frequency shield
\item[RR53 and RR57] Tunnel alcoves for electronics and services at about $\pm$240\,m from IP5.
\item[TAN] Neutral beam absorber (machine acronym for ``Target Absorber Neutrals'')
\item[TCL] Debris collimator (machine acronym for ``Target Collimator Long'')
\item[TCLn or TCL.n] TCL collimator in cell number n counted from the nearest IP, e.g.\ TCL5. In cell 4 it is called TCLX.4.
\item[TCLD] Machine acronym for ``Target Collimator Long Dispersion suppressor''
\item[TCT] Tertiary collimator (machine acronym for ``Target Collimator Tertiary'')
\item[UFSD] Ultra Fast Silicon Detector
\item[USC55] Underground Service Cavern number 55 at IP5.
\item[XRP] Roman Pot (machine acronym for ``eXperiment Roman Pot'')
\item[XRPH] Horizontal Roman Pot
\item[XRPV] Vertical Roman Pot
\end{description}

\clearpage
\section{PPS Station Overview}
\subsection{Pre-LS3 Configuration}
\label{sec:pre-ls3-config}
\begin{sidewaysfigure}
\includegraphics[width=1.0\textwidth]{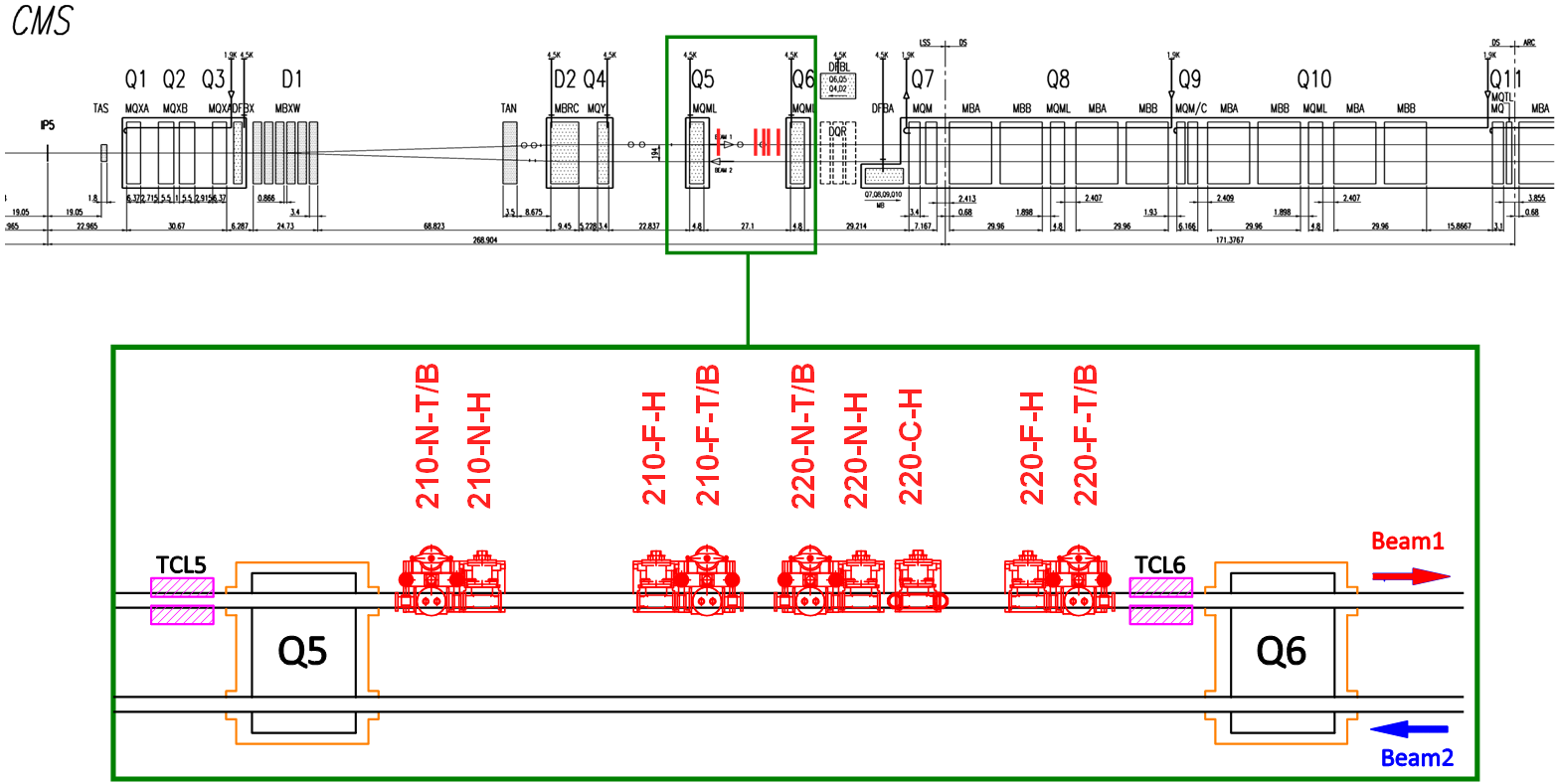}
\caption{Schematic overview of the PPS units in Sector 5-6 (outgoing Beam 1) in the pre-LS3 configuration (top drawing adapted from Ref.~\cite{edms-layout-lss5-prels3}). The instrumentation in Sector 4-5 (outgoing Beam 2) is mirror-symmetric.}
\label{fig:stationoverview-prels3}
\end{sidewaysfigure}

The PPS XRP units in the pre-LS3 configuration of Sector 5-6 are shown in Fig.~\ref{fig:stationoverview-prels3} and listed in Table~\ref{tab:stationoverview-prels3}. The configuration of Sector 4-5 is mirror-symmetric to the one in Sector 5-6. Not all units were used every year, and their detector instrumentation varied.

\begin{table}[h!]
  \begin{center}
    \caption{List of PPS XRP units in the pre-LS3 configuration of Sector 5-6. The units in Sector 4-5 have the same distances from IP5; in their names ``56'' is replaced with ``45''.}
    \label{tab:stationoverview-prels3}
    \begin{tabular}{|l|c|l|}
    \hline  
    Unit & Distance from IP5 [m] & Comments \\
    \hline
    56-210-N-T/B & 203.377 & vertical XRP pair for calibrations \\
    56-210-N-H & 203.826 & horizontal XRP for tracking detectors, removed in LS2 \\
    56-210-F-H & 212.551  & horizontal XRP for tracking detectors\\
    56-210-F-T/B & 213.000 & vertical XRP pair for calibrations and TOTEM special runs\\
    56-220-N-T/B & 214.628 & vertical XRP pair for calibrations and TOTEM special runs\\
    56-220-N-H & 215.077 & horizontal XRP for timing detectors after LS2\\
    56-220-C-H & 215.710 & horizontal XRP for timing detectors\\
    56-220-F-H & 219.551 & horizontal XRP for tracking detectors\\
    56-220-F-T/B & 220.000 & vertical XRP pair for calibrations and TOTEM special runs\\
    \hline
    \end{tabular}
  \end{center}
\end{table}

\subsection{HL-LHC Configuration}
\label{sec:hllhc-config}
\begin{sidewaysfigure}
\includegraphics[width=1.0\textwidth]{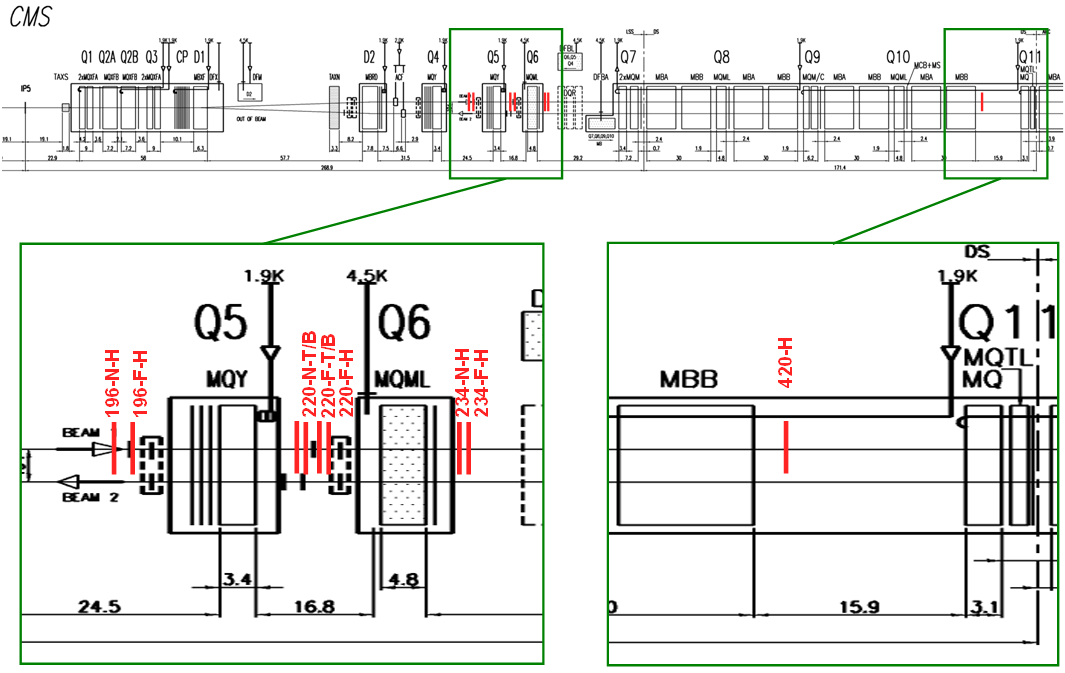}
\caption{Schematic overview of the planned PPS stations in Sector 5-6 (outgoing Beam 1) in the HL-LHC configuration (adapted from Ref.~\cite{edms-layout-lss5}). The instrumentation in Sector 4-5 (outgoing Beam 2) is mirror-symmetric.}
\label{fig:stationoverview-hllhc}
\end{sidewaysfigure}

The PPS units tentatively foreseen for Sector 5-6 at HL-LHC are shown in Fig.~\ref{fig:stationoverview-hllhc} and listed in Table~\ref{tab:stationoverview-hllhc}. The configuration of Sector 4-5 is mirror-symmetric to the one in Sector 5-6. 

\begin{table}[h!]
  \begin{center}
    \caption{Tentative list of PPS detector vessel units in Sector 5-6 foreseen for HL-LHC. The units in Sector 4-5 have the same distances from IP5; in their names ``56'' is replaced with ``45''.}
    \label{tab:stationoverview-hllhc}
    \begin{tabular}{|l|c|l|}
    \hline  
    Unit & Distance from IP5 [m] & Comments \\
    \hline
    56-196-N-H   & 192$-$198 & horizontal unit for tracking and timing detectors\\
    56-196-F-H   & 198$-$199 & horizontal unit for tracking and timing detectors\\
    56-220-N-H   & $\sim 216$ & horizontal unit for tracking and timing detectors\\
    56-220-N-T/B & $\sim 217$ & vertical pair for calibrations\\
    56-220-F-T/B & $\sim 219$ & vertical pair for calibrations\\
    56-220-F-H   & $\sim 220$ & horizontal unit for tracking and timing detectors\\
    56-234-N-H   & $\sim 234$ & horizontal unit for tracking and timing detectors\\
    56-234-F-H   & $\sim 237$ & horizontal unit for tracking and timing detectors\\
    56-420-H   & $\sim 420$ & horizontal unit for tracking and timing detectors\\
    \hline
    \end{tabular}
  \end{center}
\end{table}

\clearpage

\AddNote{This Expression of Interest cites CERN Engineering Data Management Service (EDMS) documents and sources that are relevant to the layout, conditions, and movable devices of the LHC and HL-LHC. In some cases access to these documents is restricted to CERN. These are cited when necessary, to provide the most complete set of technical references.}


\begin{thebibliography}{99}
\addcontentsline{toc}{section}{References}
\bibitem{hllhc-tdr} G. Apollinari et al., ``High-Luminosity Large Hadron Collider (HL-LHC), Technical Design Report V. 0.1'', CERN-2017-007-M.
\bibitem{totem-tdr} TOTEM Collaboration, ``Technical Design Report'', CERN-LHCC-2004-002; addendum CERN-LHCC-2004-020.
\bibitem{totem-jinst} G. Anelli et al., ``The TOTEM Experiment at the CERN Large Hadron Collider'', 2008 JINST 3 S08007.
\bibitem{totem-upgrade} TOTEM Collaboration, ``TOTEM Upgrade Proposal'', CERN-LHCC-2013-009.
\bibitem{ctpps} The CMS and TOTEM Collaborations, ``CMS-TOTEM Precision Proton
Spectrometer Technical Design Report'', CERN-LHCC-2014-021, TOTEM-TDR-003, CMS-TDR-13.
\bibitem{ipac2016} M. Deile et al., ``Roman Pot Insertions in High-Intensity Beams for the CT-PPS Project at LHC'', Proceedings of IPAC2016, Busan, Korea; contribution TUPMW021:\\ 
\verb|http://accelconf.web.cern.ch/AccelConf/ipac2016/papers/tupmw021.pdf |.
\bibitem{hllhc-27coordinationgroup} 27th HL-LHC Coordination Group Meeting, 20.12.2018, non-public meeting.   
\bibitem{hllhc-executivecommittee20200622} HL-LHC Executive Committee, 22.06.2020, non-public meeting.
\bibitem{hllhc-optics1.3} LHC Optics Web: HLLHCV1.3 optics:\\
\verb|http://abpdata.web.cern.ch/abpdata/lhc_optics_web/www/hllhc13/ |.\\
HLLHCV1.3 Optics repository:\\
\verb|http://lhc-optics.web.cern.ch/lhc-optics/HLLHCV1.3/| .
\bibitem{cms-totem-prospects} The CMS and TOTEM diffractive and forward physics working group (M. Albrow et al.), ``Prospects for Diffractive and Forward Physics at the LHC'', CERN-LHCC-2006-039-G-124.
\bibitem{fp420} M.G. Albrow et al., ``The FP420 R\&D Project: Higgs and New Physics with forward protons at the LHC'', JINST 4(2009) T10001, doi:10.1088/1748-0221/4/10/T10001 [arXiv:0806.0302 [hep-ex]].


\bibitem{Khachatryan:2015qba}
V.~Khachatryan \textit{et al.} [CMS],
``Search for diphoton resonances in the mass range from 150 to 850 GeV in pp collisions at $\sqrt{s} =$ 8 TeV'',
Phys. Lett. B \textbf{750}, 494-519 (2015)
doi:10.1016/j.physletb.2015.09.062
[arXiv:1506.02301 [hep-ex]].
\bibitem{Khachatryan:2016hje}
V.~Khachatryan \textit{et al.} [CMS],
``Search for Resonant Production of High-Mass Photon Pairs in Proton-Proton Collisions at $\sqrt s$ =8 and 13 TeV'',
Phys. Rev. Lett. \textbf{117}, 051802 (2016)
doi:10.1103/PhysRevLett.117.051802
[arXiv:1606.04093 [hep-ex]].
\bibitem{Khachatryan:2016yec}
V.~Khachatryan \textit{et al.} [CMS],
``Search for high-mass diphoton resonances in proton\textendash{}proton collisions at 13 TeV and combination with 8 TeV search'',
Phys. Lett. B \textbf{767}, 147-170 (2017)
doi:10.1016/j.physletb.2017.01.027
[arXiv:1609.02507 [hep-ex]].
\bibitem{Aad:2015mna}
G.~Aad \textit{et al.} [ATLAS],
``Search for high-mass diphoton resonances in $pp$ collisions at $\sqrt{s}=8$ TeV with the ATLAS detector'',
Phys. Rev. D \textbf{92}, 032004 (2015)
doi:10.1103/PhysRevD.92.032004
[arXiv:1504.05511 [hep-ex]].
\bibitem{Aaboud:2016tru}
M.~Aaboud \textit{et al.} [ATLAS],
``Search for resonances in diphoton events at $\sqrt{s}$=13 TeV with the ATLAS detector'',
JHEP \textbf{09}, 001 (2016)
doi:10.1007/JHEP09(2016)001
[arXiv:1606.03833 [hep-ex]].
\bibitem{Aaboud:2017yyg}
M.~Aaboud \textit{et al.} [ATLAS],
``Search for new phenomena in high-mass diphoton final states using 37 fb$^{-1}$ of proton--proton collisions collected at $\sqrt{s}=13$ TeV with the ATLAS detector'',
Phys. Lett. B \textbf{775}, 105-125 (2017)
doi:10.1016/j.physletb.2017.10.039
[arXiv:1707.04147 [hep-ex]].

\bibitem{jhep2018} The CMS and TOTEM Collaborations, ``Observation of proton-tagged, central (semi)exclusive production of high-mass lepton pairs in pp collisions at 13 TeV with the CMS-TOTEM precision proton spectrometer'', J. High Energ. Phys. 2018, 153 (2018).\\ 
\verb|https://doi.org/10.1007/JHEP07(2018)153| .

\bibitem{DiphotonPAS}
  CMS and TOTEM Collaborations,
  ``First search for exclusive diphoton production at high mass with intact protons in proton-proton collisions at $\sqrt{s}=13$ TeV at the LHC'', CMS-PAS-EXO-18-014, TOTEM-NOTE-2020-003


%
\bibitem{deFavereaudeJeneret:2009db} 
  J. de Favereau de Jeneret, V. Lemaitre, Y. Liu, S. Ovyn, T. Pierzchala, K. Piotrzkowski, X. Rouby, N. Schul, M. Vander Donckt, 
  ``High energy photon interactions at the LHC'',
  [arXiv:0908.2020 [hep-ph]].

\bibitem{Landau:1948kw}
L.~D.~Landau,
``On the angular momentum of a system of two photons'',
Dokl. Akad. Nauk SSSR \textbf{60}, 207-209 (1948)
doi:10.1016/B978-0-08-010586-4.50070-5
\bibitem{Yang:1950rg}
C.~N.~Yang,
``Selection Rules for the Dematerialization of a Particle Into Two Photons'',
Phys. Rev. \textbf{77}, 242-245 (1950)
doi:10.1103/PhysRev.77.242
\bibitem{Olsson:1983mf}
J.~E.~Olsson,
``Resonance production in $\gamma \gamma$ reactions'',
Lect. Notes Phys. \textbf{191}, 45-73 (1983)
doi:10.1007/3-540-12691-0\_2


\bibitem{denterria:2013} D. d'Enterria, G. da Silveira, ``Observing Light-by-Light Scattering at the Large Hadron Collider'', Phys. Rev. Letters {\bf 111}, 080405 (2013), doi:10.1103/physrevlett.111.080405 , [arXiv:1305.7142 [hep-ph]].
\bibitem{Csaki:2015vek} 
  C.~Cs\'{a}ki, J.~Hubisz and J.~Terning,
  ``Minimal model of a diphoton resonance: Production without gluon couplings'',
  Phys.\ Rev.\ D {\bf 93}, 035002 (2016),
  doi:10.1103/PhysRevD.93.035002,
  [arXiv:1512.05776 [hep-ph]].
\bibitem{Harland-Lang:2016qjy} 
  L.~A.~Harland-Lang, V.~A.~Khoze and M.~G.~Ryskin,
  ``The production of a diphoton resonance via photon-photon fusion'',
  JHEP {\bf 1603}, 182 (2016),
  doi:10.1007/JHEP03(2016)182 ,
  [arXiv:1601.07187 [hep-ph]].
\bibitem{Lebiedowicz:2016lmn} 
  P.~Lebiedowicz, M.~Luszczak, R.~Pasechnik and A.~Szczurek,
  ``Can the diphoton enhancement at 750 GeV be due to a neutral technipion?'',
  Phys.\ Rev.\ D {\bf 94}, 015023 (2016),
  doi:10.1103/PhysRevD.94.015023 ,
  [arXiv:1604.02037 [hep-ph]].
\bibitem{Abel:2016pyc} 
  S.~Abel and V.~V.~Khoze,
  ``Photo-production of a 750 GeV di-photon resonance mediated by Kaluza-Klein leptons in the loop'',
  JHEP {\bf 1605}, 063 (2016),
  doi:10.1007/JHEP05(2016)063 ,
  [arXiv:1601.07167 [hep-ph]].
\bibitem{Fichet:2015vvy} 
  S.~Fichet, G.~von Gersdorff and C.~Royon,
  ``Scattering light by light at 750 GeV at the LHC'',
  Phys.\ Rev.\ D {\bf 93}, 075031 (2016),
  doi:10.1103/PhysRevD.93.075031 ,
  [arXiv:1512.05751 [hep-ph]].
\bibitem{Baldenegro:2018hng} 
  C.~Baldenegro, S.~Fichet, G.~von Gersdorff and C.~Royon,
  ``Searching for axion-like particles with proton tagging at the LHC'',
  JHEP {\bf 1806}, 131 (2018),
  doi:10.1007/JHEP06(2018)131 ,
  [arXiv:1803.10835 [hep-ph]].
\bibitem{Fichet:2016pvq} 
  S.~Fichet, G.~von Gersdorff and C.~Royon,
  ``Measuring the Diphoton Coupling of a 750 GeV Resonance'',
  Phys.\ Rev.\ Lett.\  {\bf 116}, 231801 (2016),
  doi:10.1103/PhysRevLett.116.231801 ,
  [arXiv:1601.01712 [hep-ph]].
\bibitem{Ohnemus:1993qw}
  J.~Ohnemus, T.~F.~Walsh and P.~M.~Zerwas,
  ``gamma gamma production of nonstrongly interacting SUSY particles at hadron colliders'',
  Phys.\ Lett.\ B {\bf 328}, 369 (1994),
  doi:10.1016/0370-2693(94)91492-3 ,
  [hep-ph/9402302].
\bibitem{Drees:1994zx}
  M.~Drees, R.~M.~Godbole, M.~Nowakowski and S.~D.~Rindani,
  ``gamma gamma processes at high-energy p p colliders'',
  Phys.\ Rev.\ D {\bf 50}, 2335 (1994),
  doi:10.1103/PhysRevD.50.2335 ,
  [hep-ph/9403368].
\bibitem{Bhattacharya:1995id}
  G.~Bhattacharya, P.~Kalyniak and K.~A.~Peterson,
  ``Photon and Z induced heavy charged lepton pair production at a hadron supercollider'',
  Phys.\ Rev.\ D {\bf 53}, 2371 (1996),
  doi:10.1103/PhysRevD.53.2371 ,
  [hep-ph/9512255].
\bibitem{Beresford:2018pbt}
  L.~Beresford and J.~Liu,
  ``Search Strategy for Sleptons and Dark Matter Using the LHC as a Photon Collider'',
  Phys.\ Rev.\ Lett.\  {\bf 123}, 141801 (2019),
  doi:10.1103/PhysRevLett.123.141801 ,
  [arXiv:1811.06465 [hep-ph]].
\bibitem{Harland-Lang:2018hmi}
  L.~A.~Harland-Lang, V.~A.~Khoze, M.~G.~Ryskin and M.~Tasevsky,
  ``LHC Searches for Dark Matter in Compressed Mass Scenarios: Challenges in the Forward Proton Mode'',
  JHEP {\bf 1904}, 010 (2019),
  doi:10.1007/JHEP04(2019)010 ,
  [arXiv:1812.04886 [hep-ph]].
\bibitem{Godunov:2019jib}
  S.~I.~Godunov, V.~A.~Novikov, A.~N.~Rozanov, M.~I.~Vysotsky and E.~V.~Zhemchugov,
  ``Quasistable charginos in ultraperipheral proton-proton collisions at the LHC'', 
 JHEP {\bf 2001}, 143 (2020),
  doi:10.1007/JHEP01(2020)143 ,
\bibitem{Schul:2008sr}
  N.~Schul and K.~Piotrzkowski,
  ``Detection of two-photon exclusive production of supersymmetric pairs at the LHC'',
  Nucl.\ Phys.\ Proc.\ Suppl.\  {\bf 179-180}, 289 (2008),
  doi:10.1016/j.nuclphysbps.2008.07.036 ,
  [arXiv:0806.1097 [hep-ph]].
\bibitem{Han:2007bk}
  T.~Han, B.~Mukhopadhyaya, Z.~Si and K.~Wang,
  ``Pair production of doubly-charged scalars: Neutrino mass constraints and signals at the LHC'',
  Phys.\ Rev.\ D {\bf 76}, 075013 (2007),
  doi:10.1103/PhysRevD.76.075013,
  [arXiv:0706.0441 [hep-ph]].
\bibitem{Babu:2016rcr}
  K.~S.~Babu and S.~Jana,
  ``Probing Doubly Charged Higgs Bosons at the LHC through Photon Initiated Processes'',
  Phys.\ Rev.\ D {\bf 95}, 055020 (2017),
  doi:10.1103/PhysRevD.95.055020,
  [arXiv:1612.09224 [hep-ph]].
\bibitem{You:2014npa}
  Y.~You, Y.~Chong-Xing and X.~Yun,
  ``Pair Production of the Doubly Charged Leptons via Electroweak Vector Boson Fusion at the Large Hadron Collider'',
  Chin.\ Phys.\ Lett.\  {\bf 31}, 021201 (2014),
  doi:10.1088/0256-307X/31/2/021201 
\bibitem{Baines:2018ltl}
  S.~Baines, N.~E.~Mavromatos, V.~A.~Mitsou, J.~L.~Pinfold and A.~Santra,
  ``Monopole production via photon fusion and Drell–Yan processes: MadGraph implementation and perturbativity via velocity-dependent coupling and magnetic moment as novel features'',
Eur.\ Phys.\ J.\ C {\bf 78}, 966 (2018),
  Erratum: [Eur.\ Phys.\ J.\ C {\bf 79}, 166 (2019)],
  doi:10.1140/epjc/s10052-018-6440-6, 10.1140/epjc/s10052-019-6678-7 ,
  [arXiv:1808.08942 [hep-ph]].
\bibitem{Dougall:2007tt}
  T.~Dougall and S.~D.~Wick,
  ``Dirac magnetic monopole production from photon fusion in proton collisions'',
  Eur.\ Phys.\ J.\ A {\bf 39}, 213 (2009),
  doi:10.1140/epja/i2008-10701-8,
  [arXiv:0706.1042 [hep-ph]].
\bibitem{Kurochkin:2006jr}
  Y.~Kurochkin, I.~Satsunkevich, D.~Shoukavy, N.~Rusakovich and Y.~Kulchitsky,
  ``On production of magnetic monopoles via gamma gamma fusion at high energy p p collisions'', Mod.\ Phys.\ Lett.\ A {\bf 21}, 2873 (2006),
  doi:10.1142/S0217732306022237.
\bibitem{Khoze:2017igg} 
  V.~A.~Khoze, A.~D.~Martin and M.~G.~Ryskin,
  ``Can invisible objects be `seen' via forward proton detectors at the LHC?'',
  J.\ Phys.\ G {\bf 44}, 055002 (2017),
  doi:10.1088/1361-6471/aa6457,
  [arXiv:1702.05023 [hep-ph]].
\bibitem{Belotsky:2004ex} 
  K.~Belotsky, V.~A.~Khoze, A.~D.~Martin and M.~G.~Ryskin,
  ``Can an invisible Higgs boson be seen via diffraction at the LHC?'',
  Eur.\ Phys.\ J.\ C {\bf 36}, 503 (2004),
  doi:10.1140/epjc/s2004-01967-1,
  [hep-ph/0406037].
  
\bibitem{Pierzchala:2008xc}
  T.~Pierzchala and K.~Piotrzkowski,
  ``Sensitivity to anomalous quartic gauge couplings in photon-photon interactions at the LHC'',  Nucl.\ Phys.\ Proc.\ Suppl.\  {\bf 179-180}, 257 (2008),
  doi:10.1016/j.nuclphysbps.2008.07.032,
  [arXiv:0807.1121 [hep-ph]].
\bibitem{Chapon:2009hh}
  E.~Chapon, C.~Royon and O.~Kepka,
  ``Anomalous quartic W W gamma gamma, Z Z gamma gamma, and trilinear WW gamma couplings in two-photon processes at high luminosity at the LHC'',
  Phys.\ Rev.\ D {\bf 81}, 074003 (2010),
  doi:10.1103/PhysRevD.81.074003,
  [arXiv:0912.5161 [hep-ph]].  
\bibitem{Maniatis:2008zz}
  M.~Maniatis, A.~von Manteuffel and O.~Nachtmann,
  ``Anomalous couplings in gamma gamma $\rightarrow$ W+ W- at LHC and ILC'',  Nucl.\ Phys.\ Proc.\ Suppl.\  {\bf 179-180}, 104 (2008),
  doi:10.1016/j.nuclphysbps.2008.07.012
\bibitem{Gupta:2011be} 
  R.~S.~Gupta,
  ``Probing Quartic Neutral Gauge Boson Couplings using diffractive photon fusion at the LHC'',
  Phys.\ Rev.\ D {\bf 85}, 014006 (2012),
  doi:10.1103/PhysRevD.85.014006,
  [arXiv:1111.3354 [hep-ph]].
\bibitem{Baldenegro:2017aen} 
  C.~Baldenegro, S.~Fichet, G.~von Gersdorff and C.~Royon,
  ``Probing the anomalous $\gamma$$\gamma$$\gamma$Z coupling at the LHC with proton tagging'',
  JHEP {\bf 1706}, 142 (2017),
  doi:10.1007/JHEP06(2017)142,
  [arXiv:1703.10600 [hep-ph]].
\bibitem{Petrov:2004nx}
V.~A.~Petrov and R.~A.~Ryutin,
``Exclusive double diffractive events: Menu for LHC'',
JHEP \textbf{08}, 013 (2004)
doi:10.1088/1126-6708/2004/08/013
[arXiv:hep-ph/0403189 [hep-ph]].
\bibitem{Fichet:2014uka} 
  S.~Fichet, G.~von Gersdorff, B.~Lenzi, C.~Royon and M.~Saimpert,
  ``Light-by-light scattering with intact protons at the LHC: from Standard Model to New Physics'',
  JHEP {\bf 1502}, 165 (2015),
  doi:10.1007/JHEP02(2015)165,
  [arXiv:1411.6629 [hep-ph] (latest update)].
\bibitem{Fichet:2013ola} 
  S.~Fichet and G.~von Gersdorff,
  ``Anomalous gauge couplings from composite Higgs and warped extra dimensions'',  JHEP {\bf 1403}, 102 (2014),
  doi:10.1007/JHEP03(2014)102,
  [arXiv:1311.6815 [hep-ph]].


\bibitem{Espriu:2014jya} 
  D.~Espriu and F.~Mescia,
  ``Unitarity and causality constraints in composite Higgs models'',
  Phys.\ Rev.\ D {\bf 90}, 015035 (2014),
  doi:10.1103/PhysRevD.90.015035,
  [arXiv:1403.7386 [hep-ph]].
\bibitem{Delgado:2014jda} 
  R.~L.~Delgado, A.~Dobado, M.~J.~Herrero and J.~J.~Sanz-Cillero,
  ``One-loop $\rm \gamma\gamma \to$ W$_{L}^{+}$ W$_{L}^{-}$ and $\rm\gamma\gamma \to Z_{L} Z_{L}$ from the Electroweak Chiral Lagrangian with a light Higgs-like scalar'',
  JHEP {\bf 1407}, 149 (2014),
  doi:10.1007/JHEP07(2014)149,
  [arXiv:1404.2866 [hep-ph]].
\bibitem{Chatrchyan:2013akv} 
  S.~Chatrchyan {\it et al.} [CMS Collaboration],
  ``Study of Exclusive Two-Photon Production of $W^+W^-$ in $pp$ Collisions at $\sqrt{s} = 7$ TeV and Constraints on Anomalous Quartic Gauge Couplings'',
  JHEP {\bf 1307}, 116 (2013),
  doi:10.1007/JHEP07(2013)116,
  [arXiv:1305.5596 [hep-ex]].
\bibitem{Khachatryan:2016mud} 
  V.~Khachatryan {\it et al.} [CMS Collaboration],
  ``Evidence for exclusive $\rm\gamma\gamma \to W^+ W^-$ production and constraints on anomalous quartic gauge couplings in pp collisions at $\sqrt{s}=7$ and 8 TeV'',
  JHEP {\bf 1608}, 119 (2016),
  doi:10.1007/JHEP08(2016)119,
  [arXiv:1604.04464 [hep-ex]].
\bibitem{Aaboud:2016dkv} 
  M.~Aaboud {\it et al.} [ATLAS Collaboration],
  ``Measurement of exclusive $\gamma\gamma\rightarrow W^+W^-$ production and search for exclusive Higgs boson production in $pp$ collisions at $\sqrt{s} = 8$ TeV using the ATLAS detector'',
  Phys.\ Rev.\ D {\bf 94}, 032011 (2016),
  doi:10.1103/PhysRevD.94.032011,
  [arXiv:1607.03745 [hep-ex]].

\bibitem{hllhc-sm-yellowreport} HL-LHC and HE-LHC Working Group Collaboration, ``Standard Model Physics at the HL-LHC and HE-LHC'', CERN-LPCC-2018-03, arXiv:1902.04070 .

\bibitem{Abdallah:2003xd}
  DELPHI Collaboration,
  ``Study of tau-pair production in photon-photon collisions at LEP and limits on the anomalous electromagnetic moments of the tau lepton'',
  Eur. Phys. J. C \textbf{35}, 159-170 (2004),
  doi:10.1140/epjc/s2004-01852-y,
  [arXiv:hep-ex/0406010 [hep-ex]].
\bibitem{Achard:2004jj}
P.~Achard \textit{et al.} [L3],
``Muon pair and tau pair production in two photon collisions at LEP'',
Phys. Lett. B \textbf{585}, 53-62 (2004)
doi:10.1016/j.physletb.2004.02.012
[arXiv:hep-ex/0402037 [hep-ex]].

\bibitem{Atag:2010ja}
  S.~Atag and A.~A.~Billur,
  ``Possibility of Determining $\tau$ Lepton Electromagnetic Moments in ${\gamma\gamma \to \tau^{+}\tau^{-}}$ Process at the CERN-LHC'',  JHEP {\bf 1011}, 060 (2010),
  doi:10.1007/JHEP11(2010)060,
  [arXiv:1005.2841 [hep-ph]].

\bibitem{Harland-Lang:2020veo}
L.~A.~Harland-Lang, M.~Tasevsky, V.~A.~Khoze and M.~G.~Ryskin,
``A new approach to modelling elastic and inelastic photon-initiated production at the LHC: SuperChic 4'',
Eur. Phys. J. C \textbf{80}, 925 (2020),
doi:10.1140/epjc/s10052-020-08455-0,
[arXiv:2007.12704 [hep-ph]].
\bibitem{Boonekamp:2011ky} 
  M.~Boonekamp, A.~Dechambre, V.~Juranek, O.~Kepka, M.~Rangel, C.~Royon and R.~Staszewski,
  ``FPMC: A Generator for forward physics'',
  arXiv:1102.2531 [hep-ph].
\bibitem{Khoze:2001xm} 
  V.~A.~Khoze, A.~D.~Martin and M.~G.~Ryskin,
  ``Prospects for new physics observations in diffractive processes at the LHC'',
  Eur.\ Phys.\ J.\ C {\bf 23}, 311 (2002),
  doi:10.1007/s100520100884.
\bibitem{Budnev:1974de} 
  V.~M.~Budnev, I.~F.~Ginzburg, G.~V.~Meledin, V.~G.~Serbo,
  ``The Two photon particle production mechanism. Physical problems. Applications. Equivalent photon approximation'',
  Phys.\ Rept.\  {\bf 15}, 181 (1975),
  doi:10.1016/0370-1573(75)90009-5 .
\bibitem{Petrov:2004hh}
  V.~A.~Petrov, R.~A.~Ryutin, A.~E.~Sobol and J.~P.~Guillaud,
  ``Azimuthal angular distributions in EDDE as spin-parity analyser and glueball filter for LHC'',
  JHEP \textbf{06}, 007 (2005),
  doi:10.1088/1126-6708/2005/06/007,
  [arXiv:hep-ph/0409118 [hep-ph]].
\bibitem{Ryutin:2014eua}
  R.~A.~Ryutin,
  ``Visualizations of exclusive central diffraction'',
  Eur. Phys. J. C \textbf{74}, 3162 (2014),
  doi:10.1140/epjc/s10052-014-3162-2,
  [arXiv:1404.7678 [hep-ph]].
\bibitem{Harland-Lang:2015cta}
L.~A.~Harland-Lang, V.~A.~Khoze and M.~G.~Ryskin,
``Exclusive physics at the LHC with SuperChic 2'',
Eur. Phys. J. C \textbf{76}, 9 (2016)
doi:10.1140/epjc/s10052-015-3832-8
[arXiv:1508.02718 [hep-ph]].

\bibitem{ProtonRecoDPS} ``Proton reconstruction with the Precision Proton Spectrometer (PPS) in Run 2'', CMS DP-2020/047
\bibitem{Harland-Lang:2016apc} 
  L.~A.~Harland-Lang, V.~A.~Khoze and M.~G.~Ryskin,
  ``The photon PDF in events with rapidity gaps'',
  Eur.\ Phys.\ J.\ C {\bf 76}, 255 (2016),
  doi:10.1140/epjc/s10052-016-4100-2 ,
  [arXiv:1601.03772 [hep-ph]].
\bibitem{Khoze:2000cy} 
  V.~A.~Khoze, A.~D.~Martin and M.~G.~Ryskin,
  ``Can the Higgs be seen in rapidity gap events at the Tevatron or the LHC?'',
  Eur.\ Phys.\ J.\ C {\bf 14}, 525 (2000),
  doi:10.1007/s100520000359 ,
  [hep-ph/0002072].
\bibitem{Petrov:2003yt}
  V.~A.~Petrov and R.~A.~Ryutin,
  ``Exclusive double diffractive Higgs boson production at LHC'',
  Eur.\ Phys.\ J.\ C \textbf{36}, 509-513 (2004),
  doi:10.1140/epjc/s2004-01972-4,
  [hep-ph/0311024].
\bibitem{Khoze:2002py} 
  V.~A.~Khoze, A.~D.~Martin and M.~G.~Ryskin,
  ``Diffractive Higgs production: Myths and reality'',
  Eur.\ Phys.\ J.\ C {\bf 26}, 229 (2002),
  doi:10.1140/epjc/s2002-01067-4 ,
  [hep-ph/0207313].
\bibitem{DeRoeck:2002hk} 
  A.~De Roeck, V.~A.~Khoze, A.~D.~Martin, R.~Orava and M.~G.~Ryskin,
  ``Ways to detect a light Higgs boson at the LHC'',
  Eur.\ Phys.\ J.\ C {\bf 25}, 391 (2002),
  doi:10.1007/s10052-002-1032-9 ,
  [hep-ph/0207042].
\bibitem{Cudell:2010cj} 
  J.~R.~Cudell, A.~Dechambre and O.~F.~Hernandez,
  ``Higgs Central Exclusive Production'',
  Phys.\ Lett.\ B {\bf 706}, 333 (2012),
  doi:10.1016/j.physletb.2011.11.022 ,
  [arXiv:1011.3653 [hep-ph]].
\bibitem{Maciula:2010tv} 
  R.~Maciula, R.~Pasechnik and A.~Szczurek,
  ``Central exclusive quark-antiquark dijet and Standard Model Higgs boson production in proton-(anti)proton collisions'',
  Phys.\ Rev.\ D {\bf 83}, 114034 (2011),
  doi:10.1103/PhysRevD.83.114034 ,
  [arXiv:1011.5842 [hep-ph]].
\bibitem{Coughlin:2009tr} 
  T.~D.~Coughlin and J.~R.~Forshaw,
  ``Central Exclusive Production in QCD'',
  JHEP {\bf 1001}, 121 (2010),
  doi:10.1007/JHEP01(2010)121 ,
  [arXiv:0912.3280 [hep-ph]].
\bibitem{Ryutin:2012np}
  R.~A.~Ryutin,
  ``Exclusive Double Diffractive Events: general framework and prospects'',
  Eur. Phys. J. C \textbf{73}, 2443 (2013),
  doi:10.1140/epjc/s10052-013-2443-5,
  [arXiv:1211.2105 [hep-ph]].
\bibitem{Dechambre:2011py} 
  A.~Dechambre, O.~Kepka, C.~Royon and R.~Staszewski,
  ``Uncertainties on exclusive diffractive Higgs and jets production at the LHC'',
  Phys.\ Rev.\ D {\bf 83}, 054013 (2011),
  doi:10.1103/PhysRevD.83.054013 ,
  [arXiv:1101.1439 [hep-ph]].
\bibitem{HarlandLang:2013jf}
L.~A.~Harland-Lang, V.~A.~Khoze, M.~G.~Ryskin and W.~J.~Stirling,
``Latest Results in Central Exclusive Production: A Summary'',
[arXiv:1301.2552 [hep-ph]].
\bibitem{Khoze:2004rc} 
  V.~A.~Khoze, A.~D.~Martin and M.~G.~Ryskin,
  ``Hunting a light CP violating Higgs via diffraction at the LHC'',
  Eur.\ Phys.\ J.\ C {\bf 34}, 327 (2004),
  doi:10.1140/epjc/s2004-01729-1,
  [hep-ph/0401078].
\bibitem{Alwall:2014hca}
  J. Alwall et al.,
  ``The automated computation of tree-level and next-to-leading order differential cross sections, and their matching to parton shower simulations
'',
  CERN-PH-TH-2014-064, CP3-14-18, LPN14-066, MCNET-14-09, ZU-TH-14-14,
  JHEP {\bf 07}, 79 (2014),
  doi:10.1007/JHEP07(2014)079
  [arXiv:1405.0301 [hep-ph]].
\bibitem{Shifman:1979eb}
M.~A.~Shifman, A.~I.~Vainshtein, M.~B.~Voloshin and V.~I.~Zakharov,
``Low-Energy Theorems for Higgs Boson Couplings to Photons'',
Sov. J. Nucl. Phys. \textbf{30}, 711-716 (1979)
ITEP-42-1979.
\bibitem{Dawson:1993qf}
S.~Dawson and R.~Kauffman,
``QCD corrections to Higgs boson production: nonleading terms in the heavy quark limit'',
Phys. Rev. D \textbf{49}, 2298-2309 (1994)
doi:10.1103/PhysRevD.49.2298
[arXiv:hep-ph/9310281 [hep-ph]].
\bibitem{Kniehl:1995tn}
B.~A.~Kniehl and M.~Spira,
``Low-energy theorems in Higgs physics'',
Z. Phys. C \textbf{69}, 77-88 (1995)
doi:10.1007/s002880050007
[arXiv:hep-ph/9505225 [hep-ph]].

\bibitem{dEnterria:2009cwl}
D.~d'Enterria and J.~P.~Lansberg,
``Study of Higgs boson production and its b anti-b decay in gamma-gamma processes in proton-nucleus collisions at the LHC'',
Phys. Rev. D \textbf{81}, 014004 (2010)
doi:10.1103/PhysRevD.81.014004
[arXiv:0909.3047 [hep-ph]].
\bibitem{Sumino:2010bv} 
  Y.~Sumino and H.~Yokaya,
  ``Bound-state effects on kinematical distributions of top quarks at hadron colliders'',
  JHEP {\bf 09}, 34 (2010),
  doi:10.1007/JHEP06(2016)037, 10.1007/JHEP09(2010)034,
  [Erratum: JHEP06,037(2016)],
  [arXiv:1007.0075 [hep-ph]].


%
\bibitem{edms-layout-lss5} J. Oliveira, ``Layouts of HL-LHC Insertions - IR5, IR6, IR7, IR8'', drawing LHCLSXGH0002  v.0, EDMS 1557086.
\bibitem{MADX} L. Deniau et al., ``The MAD-X Program, Version 5.03.07, User's Reference Manual'', CERN, October 20, 2017, \verb| http://www.cern.ch/madx/ |.
\bibitem{hllhc-coordinationgroup202002} M. Deile, ``CMS: Precision proton spectrometer developments for HL-LHC'', 28th HL-LHC Coordination Group meeting, 25.02.2020, non-public meeting. 
\bibitem{slide_cabana} Adaptation of figures in a presentation prepared by Maria Amparo Gonzalez De La Aleja Cabana, with modified XRP locations.
\bibitem{bblr} A. Rossi et al., ``HL-LHC Space Reservation,
Modifications to the IR1 and IR5 of the LHC for Beam-Beam Long-Range Compensator Devices'', EDMS 2037987.
\bibitem{edms-layout-420} H. Prin, ``IR5 Right, Cells C8.R5 to C11.R5'', EDMS 202351  v.AJ.
\bibitem{tcld} M. Gonzalez de la Aleja et al., ``HL-LHC Integration Report for Installation Approval, WP5: TCLD Integration Study'', EDMS 1903950.
\bibitem{connectioncryostat} M. Gonzalez de la Aleja et al., ``HL-LHC Integration Report for Installation Approval, WP11: Point2 Connection Cryostat Full Assembly Integration Study'', EDMS 1904996.
\bibitem{fluka} FLUKA: \verb| http://www.fluka.org | .
\bibitem{ats-optics} S. Fartoukh, ``Achromatic telescopic squeezing scheme and application to the LHC and its luminosity upgrade'', 
Phys. Rev. ST Accel. Beams, vol. 16, 111002, Nov. 2013: \\
\verb|https://journals.aps.org/prab/pdf/10.1103/PhysRevSTAB.16.111002 |.\\
S. Fartoukh, ``The right optics concept for the right dimension of the High Luminosity LHC project'', ICFA Beam Dynamics Newsletter \#71, pp. 116--134 , ed. Jie Gao (IHEP), 2017:\\ 
\verb|http://icfa-bd.kek.jp/Newsletter71.pdf| .\\
S. Fartoukh et al., ``Experimental validation of the Achromatic Telescopic Squeezing (ATS) scheme at the LHC'', 2017 J. Phys.: Conf. Ser. 874 012010: 
\verb|http://iopscience.iop.org/article/10.1088/1742-6596/874/1/012010/pdf| .
\bibitem{dn-19-026} F. Nemes, K. Shchelina, ``Optics and reconstruction formulae studies for PPS'', CMS DN-19-026, 2019.
\bibitem{levellingtalk} N. Karastathis et al., ``Field quality to achieve the required lifetime goals (with beam-beam)'', 7th HL-LHC Collaboration Meeting, 15.11.2017,\\ \verb|https://indico.cern.ch/event/647714/contributions/2646093/ |. 
\bibitem{collimation} D. Mirarchi et al., ``TCL/TCTs setting scenarios for HL-LHC'', Collimation Upgrade Specification Meeting \#83, 24.02.2017, \verb|https://indico.cern.ch/event/614887/ |.
\bibitem{thesis-niewiadomski} H. Niewiadomski, ``Reconstruction of Protons in the TOTEM Roman Pot Detectors at the LHC'', PhD thesis, University of Manchester, 2008, CERN-THESIS-2008-080.
\bibitem{opticspaper} TOTEM Collaboration, ``LHC Optics Measurement with Proton Tracks Detected by the Roman Pots of the TOTEM Experiment'', New J. Phys. 16 (2014) 103041.
\bibitem{ctpps-opticstalk20180821} F. Nemes, ``2017 optics update -- Validation of optical functions'', CT-PPS Physics meeting, 21 August 2018, 
\verb| https://indico.cern.ch/event/750869/ |.
\bibitem{collimator-bba} C. Bracco, ``Commissioning Scenarios and Tests for the LHC Collimation System'', PhD thesis, Ecole Polytechnique Federale de Lausanne, 2009, Section 7.2.3.\\
G. Valentino et al., ``Semi-automatic Beam-based Alignment Algorithm for the LHC Collimation System'', 2nd International Particle Accelerator Conference, San Sebastian, Spain, 4 - 9 Sep 2011, article THPZ034.
\bibitem{performancepaper} G. Antchev et al. (TOTEM Collaboration), ``Performance of the TOTEM Detectors at the LHC'', Int. J. Mod. Phys. A 28 (2013) 1330046.
\bibitem{totemnote2017001} J. Ka\v{s}par, ``Alignment of CT-PPS detectors in 2016, before TS2'', CERN-TOTEM-NOTE-2017-001.
\bibitem{totemnote2015001} F. Nemes, H. Niewiadomski, ``LHC Optics Measurement with Proton Tracks Detected by the Roman Pots of the TOTEM Experiment'', CERN-TOTEM-NOTE-2015-001.
\bibitem{totemnote2017002} F. Nemes, ``LHC optics determination with proton tracks measured in the CT-PPS detectors in 2016, before TS2'', CERN-TOTEM-NOTE-2017-002.
\bibitem{frici-collabo201909} F. Nemes, ``Optics development for PPS and TOTEM'', TOTEM Collaboration Meeting, 11 Sept. 2019: 
\verb| https://indico.cern.ch/event/845195/contributions/3549539/ |.
\bibitem{frici-ppsgeneral201909} F. Nemes, ``Optics developments for PPS, 2016 postTS2'', PPS General Meeting, 24 Sept. 2019:
\verb| https://indico.cern.ch/event/849095/contributions/3568021/ |.
\bibitem{radiationmaps-adorisio} C. Adorisio, ``HL-LHC Residual Dose Rate Estimations in the LSS1 and LSS5 (from TAXS up to Q7)'', Presentation given at the 7th HL-LHC annual collaboration meeting, 13-16 November 2017, EDMS 1868872.
\bibitem{bunchprofile} S. Papadopoulou et al., ``Modelling and Measurements of Bunch Profiles at the LHC'', Proceedings of IPAC2017, Copenhagen, Denmark; contribution TUPVA044; CERN-ACC-2017-329; 
 \verb| http://dx.doi.org/10.1088/1742-6596/874/1/012008  |.
\bibitem{cms-mtd-tdr} CMS Collaboration, ``A MIP Timing Detector for the CMS Phase-2 Upgrade, Technical Design Report'', CERN-LHCC-2019-003, CMS-TDR-020. 
\bibitem{diamond-detectors} G. Antchev et al. (TOTEM Collaboration), ``Diamond Detectors for the TOTEM Timing Upgrade'', JINST 12 (2017) P03007.
\bibitem{upgradeWS-timeresol} N. Turini, E. Bossini, contributions to the
upgrade workshop, 22 - 23 Jan. 2019 : \verb| https://indico.cern.ch/event/791510/ |.
\bibitem{diamond-dpgnote} E. Bossini et al., ``Time resolution of the diamond sensors used in the Precision Proton Spectrometer'', CMS-DP-2019-034 ; CERN-CMS-DP-2019-034.
\bibitem{diamond-dpgnote2} ``Precision Proton Spectrometer timing detector efficiencies and two-arm timing resolution in 2018 data'', CMS-DP-2020-046 ; CERN-CMS-DP-2020-046.
\bibitem{jinst-berretti} M. Berretti et al., ``Timing Performance of a Double Layer Diamond Detector'', JINST 12 (2017) P03026, doi:10.1088/1748-0221/12/03/P03026.
\bibitem{sampic} C. Royon, ``SAMPIC: a readout chip for fast timing detectors in particle physics and medical imaging'', J. Phys.: Conf. Ser. 620 (2015) 012008:\\ \verb|https://doi.org/10.1088/1742-6596/620/1/012008| .\\
E. Delagnes et al., ``The SAMPIC Waveform and Time to Digital Converter'' IEEE Nuclear Science Symposium and Medical Imaging Conference (NSS/MIC), Seattle, WA, 2014, pp. 1-9:\\
\verb|https://doi.org/10.1109/NSSMIC.2014.7431231| .
\bibitem{bossini-Siena} E. Bossini, ``The CMS Precision Proton Spectrometer timing system performance in Run 2, future upgrades and sensor radiation hardness studies''", Proceedings of IPRD2019 15th Topical Seminar on Innovative Particle and Radiation Detectors, 2 Dec. 2019, Siena (I); CMS Note CMS CR-2019/294.
\bibitem{diamond-desytestbeam} E. Bossini, D. Figueiredo, L. Forthomme, F. Garcia Feuntes, ``Test beam results of irradiated single-crystal CVD diamond detectors at DESY-II'', CMS-NOTE-2020-007 ; CERN-CMS-NOTE-2020-007.
\bibitem{upgradeWS-irradiation} T. Naaranoia, ``Diamond irradiation tests'', 
upgrade workshop, 22 - 23 Jan. 2019 : \verb| https://indico.cern.ch/event/791510/ | .
\bibitem{hptdc} J. Christiansen, ``HPTDC High Performance Time to Digital Converter'', CERN, 2004:
\verb|https://cds.cern.ch/record/1067476| . 
\bibitem{nino} F. Anghinolfi et al., ``NINO: an ultra-fast and low-power front-end amplifier/discriminator ASIC designed for the multigap resistive plate chamber'', Nucl. Instr. and Meth. A 533 (2004) 183.
\bibitem{pixel-detectors} F. Ravera, ``The CT-PPS tracking system with 3D pixel detectors, JINST 11 (2016) C11027.\\
F. Ravera, ``3D silicon pixel detectors for the CT-PPS tracking system'', PhD thesis, Universit\`{a} degli Studi di Torino, 2017, CERN-THESIS-2017-473.
\bibitem{pixel-dpgnote} A. Bellora et al., ``Efficiency of the Pixel sensors used in the Precision Proton Spectrometer: radiation damage'', CMS-DP-2019-036; CERN-CMS-DP-2019-036, 2019.
\bibitem{3d-active_edges} C. Da Via et al., ``3D Active Edge Silicon Detector Tests With 120 GeV Muons'', IEEE Trans. Nucl. Sci. 56 (2009) 505-518, and references therein.
\bibitem{3d-timing-sherwood} S. Parker et al., ``Increased Speed: 3D Silicon Sensors; Fast Current Amplifiers'', IEEE Trans. Nucl. Sci. 58 (2011) 404-417.
\bibitem{kramberger_2019} G. Kramberger et al., "Timing performance of small cell 3D silicon detectors", NIM A 934, 26 (2019) doi:10.1016/j.nima.2019.04.088.
\bibitem{3d-timing-vertex19} A. Lai, ``Recent results of the TIMESPOT project on sensors and electronics developments for future vertex detectors'', VERTEX 2019, The 28th International Workshop on Vertex Detectors:
\verb|https://indico.cern.ch/event/806731/contributions/3515305/ |.
\bibitem{3d-timing-IPRD19} R. Mulargia, ``Tracking with timing: the TIMESPOT project'', IPRD19, 15th Topical Seminar on Innovative Particle and Radiation Detectors:
\verb|https://indico.cern.ch/event/843258/contributions/3610762/ |.
\bibitem{ufsd-detectors} Section 5.6 of \cite{ctpps} and references therein.
\bibitem{electricaltrigger} Section 8.2.2 of \cite{totem-jinst}.\\
J. Kopal, Development of the TOTEM Trigger System, PhD thesis, University of West Bohemia, Pilsen, 2014, CERN-THESIS-2014-468.
\bibitem{cmstriggerupgrade} CMS Collaboration, ``The Phase-2 Upgrade of the CMS L1 Trigger -- Interim Technical Design Report'', CERN-LHCC-2017-013; CMS-TDR-017.\\
CMS Collaboration, ``The Phase-2 Upgrade of the CMS Level-1 Trigger -- Technical Design Report'', CERN-LHCC-2020-001, CMS-TDR-20-001.
\bibitem{impedance-run1} M. Deile et al., ``Beam Coupling Impedance Measurement and Mitigation for a TOTEM Roman Pot'', Proceedings of EPAC08, arXiv:0806.4974.
\bibitem{impedance-run2} N. Minafra, ``RF Characterization of the New TOTEM Roman Pot'', CERN-TOTEM-NOTE-2013-1351 003.\\
O. Berrig et al., ``RF Measurements of the New TOTEM Roman Pot'', CERN-TOTEM-NOTE-2015-002.\\
Section 5 of \cite{totem-upgrade}.\\
Section 3.4 of \cite{ctpps}.
\bibitem{impedance-benoit-upgradeWS} B. Salvant, ``Impedance considerations for HL-LHC operation'', upgrade workshop, 22 January 2019, 
\verb| https://indico.cern.ch/event/791510/ |.
\bibitem{lvdt} Macro Sensors: LVDT Basics, Technical Bulletin 0103, 2003; \verb| www.macrosensors.com | .
\bibitem{teofili-talk} L. Teofili et al., ``Thermal Modelling of Totem Roman Pot for Impedance Induced Heating'', 
\verb|https://indico.cern.ch/event/849095/contributions/3577255/ |.
\bibitem{movementsystem} M. Deile et al., ``Movement Control of the TOTEM Roman Pots'', EDMS 873014, 2008.\\
M. Dutour et al., ``TOTEM Roman Pots Control System Use Cases Specification'', EDMS 937276, 2010.\\
S. Ravat, X. Pons, ``Roman Pot Position Calibration Procedure'', EDMS 994972, 2009.
\bibitem{interlocksystem} M. Oriunno et al., ``Operation of the TOTEM Roman Pots'', EDMS 863466, 2007.\\
M. Deile et al., ``Results from the TOTEM Interlock Tests in February - March 2010'', EDMS 1066465, 2010.\\
M. Deile et al., ``The TOTEM Interlock Logic in 2011: Specification and Test Results'', EDMS 1141699, 2011.\\
M. Deile et al., ``Revision of the Movement Control for the TOTEM and ALFA Roman Pots'', EDMS 1141700, 2011.\\
M. Deile et al., ``The Movement Control of the TOTEM and ATLAS-ALFA Roman Pots - Revision 2012'', EDMS 1203969, 2012.\\
M. Deile et al., ``The TOTEM Interlock Logic in 2012: Specification and Test Results'', EDMS 1204523, 2012.\\
M. Deile et al., ``The TOTEM Interlocks in 2015: Test Procedure and Results'', EDMS 1501228, 2015.\\
M. Deile et al., ``The TOTEM Interlocks in 2016: Test Procedure and Results'', EDMS 1607383, 2016.\\
M. Deile et al., ``The TOTEM Interlocks in 2017: Test Procedure and Results'', EDMS 1803133, 2017.\\
M. Deile et al., ``The TOTEM Interlocks in 2018: Test Procedure and Results'', EDMS 1936098, 2018.
\bibitem{cooling} V. Vacek et al., ``Perfluorocarbons and their use in cooling systems for semiconductor particle detectors'', Fluid Phase Equilibr. {\bf 174} (2000) 191.\\
Section 4.5.2 of \cite{totem-jinst}.\\
V. Vacek, ``Cooling System for the Roman Pots of TOTEM experiment'', Engineering Forum: experiences from cooling systems for LHC detectors, CERN, 30 Oct. 2008, \verb|https://indico.cern.ch/event/41288/ |.
\bibitem{cooling-upgrade} V. Vacek, ``Test of the New Type of Heat Exchangers'', TOTEM - PPS Technical Board, 11 March 2019, \verb| https://indico.cern.ch/event/805180/ |, \\
V. Vacek, ``Improvements of Heat Exchangers Suited for the High-Tech Electronics'', Proceedings of HEFAT 2019, Wicklow, Ireland.
\bibitem{vacek-talk201909} V. Vacek, ``Problems in Design of Cooling Systems for Particle Detectors and Relevant Electronics'', PPS General Meeting, 24 Sept. 2019: \\
\verb|https://indico.cern.ch/event/849095/contributions/3568026/ |.
\bibitem{usc55} J. Billon Grand, ``Equipments Layout - Building USC55'', EDMS 463022 v.AL.
\bibitem{rr57} P. Orlandi, ``Equipments Layout - RR57'', EDMS 197321 v.AG.
\bibitem{rr53} P. Orlandi, ``Equipments Layout - RR53'', EDMS 451186 v.AG.
\bibitem{edms-layout-lss5-prels3} J. Rakotoarison, ``Layouts of LHC Insertions IR5, IR6 ,IR7 ,IR8 , Optics Version 6.501'', drawing LHCLSXG\_0003  v.AJ, EDMS 202341  v.AJ.
\end{thebibliography}
\end{document}